\documentclass[12pt,a4paper]{book}
\usepackage{amsmath}
\usepackage{amssymb}
\usepackage{graphicx} 
\def\pmb#1{\setbox0=\hbox{#1}%
  \kern-.025em\copy0\kern-\wd0
  \kern.05em\copy0\kern-\wd0
  \kern-.025em\raise.0433em\box0}
\def\pmbn#1{\setbox0=\hbox{#1}
  \kern-.15em\copy0\kern-\wd0
  \kern-.05em\copy0\kern-\wd0
  \kern.1em\copy0\kern-\wd0\box0}

\def\DD{\hbox{$D$\kern-.9em\raise1.7ex\hbox{$\leftrightarrow$}}}
\def\partiallr{\hbox{$\partial$\kern-.8em\raise1.7ex\hbox{$\leftrightarrow$}}}
\def\Dlr{\hbox{$D$\kern-.9em\raise1.7ex\hbox{$\leftrightarrow$}}}
\def\DeltaDelta{\hbox{$\Delta$\kern-.9em\raise1.7ex\hbox{$\leftrightarrow$}}}
\def\Deltalr{\hbox{$Delta$\kern-.9em\raise1.7ex\hbox{$\leftrightarrow$}}}

\def\bfepsilon{\pmb{$\epsilon$}}

\def\bfnabla{\pmbn{$\nabla$}}

\def\bfsigma{\pmb{$\sigma$}}

\def\DAlemb{\hbox{$\sqcup$\kern-0.66em\lower0.03ex\hbox{$\sqcap$}}}
\def\psl{\hbox{$p$\kern-0.38em\raise0.0ex\hbox{/}}\thinspace}
\def\dsl{\hbox{$\partial$\kern-0.55em\raise0.0ex\hbox{/}}\thinspace}
\newbox\one\newdimen\oned
\def\youngsimm#1#2#3{\def\vrw{\vrule height.4pt width #3pt}%
\def\vrh{\vrule height #3pt}%
{\offinterlineskip\vbox to 0pt{\kern-#2pt\hbox{\kern#1pt\vrw\vrw\vrw}%
\hbox{\kern#1pt\rlap\vrh\vrw\rlap\vrh\vrw\rlap\vrh\vrw\llap\vrh}\vss}}}
\def\younganti#1#2#3{\def\vrw{\vrule height.4pt width #3pt}%
\def\vrh{\vrule height #3pt}%
{\offinterlineskip\vbox to 0pt{\kern-#2pt\hbox{\kern#1pt\vrw}%
\hbox{\kern#1pt\rlap\vrh\vrw\llap\vrh}%
\hbox{\kern#1pt\rlap\vrh\vrw\llap\vrh}%
\hbox{\kern#1pt\rlap\vrh\vrw\llap\vrh}\vss}}}
\def\youngmistoa#1#2#3{\setbox\one=\hbox{1}\oned=\ht\one
\advance\oned by -#3pt\divide\oned by 2%
\def\vrw{\vrule height.4pt width #3pt} \def\vrh{\vrule height #3pt}%
{\offinterlineskip\vbox to0pt{\kern-#2pt%
\hbox{\kern#1pt\vrw\vrw}\hbox{\kern#1pt\rlap\vrh\vrw%
\kern-#3pt\raise-\oned\hbox to #3pt{\hss 1\hss}\rlap\vrh\vrw%
\kern-#3pt\raise-\oned\hbox to #3pt{\hss 2\hss}\llap\vrh}%
\hbox{\kern#1pt\rlap\vrh\vrw%
\kern-#3pt\raise-\oned\hbox to #3pt{\hss 3\hss}\vrh}\vss}}}
\def\youngmistob#1#2#3{\setbox\one=\hbox{1}\oned=\ht\one
\advance\oned by -#3pt\divide\oned by 2%
\def\vrw{\vrule height.4pt width #3pt}\def\vrh{\vrule height #3pt}%
{\offinterlineskip\vbox to 0pt{\kern -#2pt%
\hbox{\kern#1pt\vrw\vrw}\hbox{\kern#1pt\rlap\vrh\vrw%
\kern-#3pt\raise-\oned\hbox to #3pt{\hss 1\hss}\rlap\vrh\vrw%
\kern-#3pt\raise-\oned\hbox to #3pt{\hss 3\hss}\llap\vrh}%
\hbox{\kern#1pt\rlap\vrh\vrw%
\kern-#3pt\raise-\oned\hbox to #3pt{\hss 2\hss}\vrh}\vss}}}
\begin{document}

\begin{titlepage}

\begin{flushright} IFUP--TH/2017\\ 
\end{flushright}
~

\vskip 4 cm

\begin{center} 

\Large{\bf LECTURES ON GRAVITATION}

\vskip 2cm

Pietro Menotti

Dipartimento di Fisica, Universit\`a di Pisa, Italy

\vskip 2cm

March 2017

\end{center}

\vfill

\end{titlepage}

\eject

~

\eject

{\bf INDEX}

\bigskip

\bigskip

\begin{tabbing}

Foreword~~~~~~~~~~~~~~~~\=
~~~~~~~~~~~~~~~~~~~~~~~~~~~~~~~~~~~~~~~~~~~~~~~~~~~~~~~~~~~~~~~~~~~~~~~~~~~~~~~~~~~~~~~~~~~\=7\\

Acknowledgments~~~~~\=
~~~~~~~~~~~~~~~~~~~~~~~~~~~~~~~~~~~~~~~~~~~~~~~~~~~~~~~~~~~~~~~~~~~~~~~~~~~~~~~~~~~~~~~~~~~\=7\\

\>                      \\

Chapter 1: \>The Lorentz group \>9\\

\> 1. The group of inertial transformations: the 
       three solutions \>\\

\> 2. Lorentz transformations \>\\

\> 3. $SL(2,C)$ is connected \>\\

\> 4. $SL(2,C)$ is the double covering of the restricted Lorentz group \>\\

\> 5. The restricted Lorentz group is simple\>\\

\> 6. $SL(2,C)$ is simply connected \>\\

\> 7. Little groups\>\\

\> 8. There are no non-trivial unitary finite dimensional
representation\>\\      
\> ~~~~of the Lorentz group \>\\

\> 9. The conjugate representation \>\\

\> 10. Relativistically invariant field equations \>\\

\> 11. The Klein-Gordon equation\>\\

\> 12.The Weyl equation \>\\

\> 13. The Majorana mass\>\\

\> 14. The Dirac equation \>\\

\> 15. The Rarita- Schwinger equation \>\\

\> 16. Transformation properties of quantum fields \>\\

\>{~}\>\\

Chapter 2: \>Equivalence principle and the path to general
              relativity\> 35\\
\> 1. Introduction \>\\

\> 2. Motion of a particle in a gravitational field \>\\

\> 3. Hamiltonian of a particle in a gravitational field\>\\

\>{~}\>\\

Chapter 3: \>Manifolds\>41\\

\> 1. Introduction \>\\

\> 2. Mappings, pull-back and push-forward \>\\

\> 3. Vector fields and diffeomorphisms\>\\

\> 4. The Lie derivative\>\\

\> 5. Components of the Lie derivative\>\\

\> 6. Commuting vector fields\>\\

\> 7. Stokes's theorem\>\\

\>{~}\>\\

Chapter 4: \> The covariant derivative\>49\\

\> 1. Introduction\>\\

\> 2. Transformation properties of the connection\>\\

\> 3. Parallel transport\>\\

\> 4. Geodesics\>\\

\> 5. Curvature \>\\

\> 6. Torsion\>\\

\> 7. The structure equations\>\\

\> 8. Non-abelian gauge fields on differential manifolds\>\\

\> 9. Non compact electrodynamics\>\\

\> 10. Meaning of torsion\>\\

\> 11. Vanishing of the torsion  \>\\

\> 12. Metric structure\>\\

\> 13. Isometries\>\\

\> 14. Metric compatibility \>\\

\> 15. Contracted Bianchi identities \>\\

\> 16. Orthonormal reference frames\>\\

\> 17. Compact electrodynamics\>\\

\> 18. Symmetries of the Riemann tensor\>\\

\> 19. Sectional curvature and Schur theorem\>\\

\> 20. Spaces of constant curvature\>\\

\> 21. Geometry in diverse dimensions\>\\

\> 22. Flat and conformally flat spaces\>\\

\> 23. The non abelian Stokes theorem\>\\

\>24. Vanishing of torsion and of the Riemann tensor\>\\

\>25. The Weyl-Schouten theorem\>\\

\>26. Hodge * operation\>\\

\>27. The current\>\\

\>28. Action for gauge fields\>\\

\>{~}\>\\

Chapter 5: \> The action for the gravitational field\>79\\

\> 1. The Hilbert action \>\\

\> 2. Einstein $\Gamma-\Gamma$ action \>\\

\> 3. Palatini first order action\>\\

\> 4. Einstein-Cartan formulation\>\\

\> 5. The energy momentum tensor\>\\

\> 6. Coupling of matter to the gravitational field\>\\

\> 7. Dimensions of the Hilbert action \>\\

\> 8. Coupling of the Dirac  field to the gravitational field\>\\

\> 9. The field equations\>\\

\> 10. The generalized energy- momentum tensor\>\\

\> 11. Einstein-Eddington affine theory\>\\

\>{~}\>\\

Chapter 6: \> Submanifolds\>101\\

\> 1. Introduction\>\\

\> 2. Extrinsic curvature\>\\

\> 3. The trace-K action \>\\

\> 4. The boundary terms in the Palatini formulation \>\\

\> 5. The boundary terms in the Einstein-Cartan formulation \>\\

\> 6. The Gauss and Codazzi relations\>\\

\>{~}\>\\

Chapter 7: \> Hamiltonian formulation of gravity\>111\\

\> 1. Introduction\>\\

\> 2. Constraints and dynamical equations of motion\>\\

\> 3. The Cauchy problem for the Maxwell and Einstein \>\\
\> ~~~~equations in vacuum\>\\

\> 4. The action in canonical form\>\\

\> 5. The Poisson algebra of the constraints\>\\

\> 6. Fluids\>\\

\> 7. The energy conditions\>\\

\> 8. The cosmological constant\>\\

\>{~}\>\\

Chapter 8: \> The energy of a gravitational system \>125\\

\> 1. The non abelian charge conservation\>\\

\> 2. Background Bianchi identities \>\\

\> 3. The superpotential\>\\

\> 4. The positive energy theorem \>\\

\>{~}\>\\

Chapter 9: \> The linearization of gravity\>135\\

\> 1. Introduction\> \\

\> 2. The Fierz-Pauli equation\> \\

\> 3. The massless case \> \\

\> 4. The VDVZ discontinuity \> \\

\> 5. Classical general relativity from field theory in Lorentz space \> \\

\>{~}\>\\
 
Chapter 10: \> Quantization in external gravitational fields \>149\\

\> 1. Introduction  \>\\

\> 2. Scalar product of two solutions \>\\

\> 3. The scalar product for general stationary metric \>\\

\> 4. Schwarzschild solution and its maximal extension \>\\

\> 5. Proper acceleration of stationary observers \>\\

\> ~~~~in Schwarzschild metric and surface gravity \>\\

\> 6. The accelerated detector \>\\

\> 7. Elements  of causal structure of space-time \>\\

\> 8. Rindler  metric \>\\

\> 9. Eigenfunctions  of K in the Rindler metric \>\\

\> 10. The Bogoliubov transformation \>\\

\> 11. Quantum field theory in the Rindler \>\\

\> 12. Rindler space with a reflecting wall \>\\

\> 13. Resolution in discrete wave packets \>\\

\> 14. Hawking radiation in 4-dimensional \> \\
\> ~~~~~Schwarzschild space               \>\\

\>{~}\\

Chapter 11: \> $N=1$ Supergravity \> 189\\

\> 1. The Wess-Zumino model\>\\

\> 2. Noether currents\>\\

\> 3. $N=1$ supergravity\>\\

\>{~}\\

Appendix: \> Derivation of the hamiltonian equations of motion and \>\\
\>         of the Poisson algebra of the constraints              \> 201\\

\end{tabbing}

\vfill

\eject

FOREWORD

\bigskip

The suggested textbook for this course are, in the order

\bigskip

R.M. Wald, General Relativity, The Universit of Chicago Press, Chicago
and London, 1984 referred to as [Wald]

\bigskip

S.W.Hawking and G.F.R. Ellis, The large scale structure of space-time,
Cambridge University Press, 1974 referred to as [HawkingEllis]

\bigskip

S. Weinberg, Gravitation and Cosmology, John Wiley and Sons, New York,
1972, referred to as [WeinbergGC]

\bigskip

S. Weinberg, The quantum theory of fields, Cambridge University Press,
1995, referred to as [WeinbergQFT]

\bigskip

Reference to other books and papers are given at the end of each section.

\vskip 3cm

ACKNOWLEDGMENTS

I am grateful to Damiano Anselmi, Giovanni Morchio, Carlo Rovelli and
Domenico Seminara for useful discussions.

\vfill


\eject

\chapter{The Lorentz group}

\section{The group of inertial transformations: the three solutions}

In this section we examine a group theoretical argument which shows
that the Lorentz transformations are a solution of a symmetry problem
in which the independence of the velocity of light from the inertial
frame does not play a direct role (see [1],[2] and reference therein).
Actually there might be no signal with such a property and still
Lorentz transformation would maintain their validity. As we shall see
the solutions of the symmetry problem are three: One is Galilei
transformations, the other is Lorentz transformations while the third
solution has to be discarded of physical grounds.

We work for simplicity in $1+1$ dimensions $(x,t)$. Extension to $1+n$
dimensions is trivial. We accept Galilean invariance as a symmetry group in
the following sense: in addition to space-time translations we assume that the
equivalent frames form a continuous family which depends on a single
continuous parameter $\alpha$.

In setting up the rods and synchronizing the clocks we must respect
homogeneity; e.g. for synchronizing the clocks we can use any signal
(elastic waves, particles etc.) provided we divide by $2$ the time of
come back.

The most general transformation for the coordinates of the events is
$$ 
x' = f(x,t,\alpha) 
$$ 
$$ 
t' = g(x,t,\alpha). 
$$ 
\bigskip

We require:

1) \underline{Homogeneity of space time under space time translations}

 i.e. in 
$$ 
dx' = \frac{\partial
f}{\partial x}dx +\frac{\partial f}{\partial t}dt 
$$ 
$$ dt' =
\frac{\partial g}{\partial x}dx +\frac{\partial g}{\partial t}dt 
$$
the partial derivatives do not depend on $x$ and $t$. Thus we have a
linear transformation. 

Choosing the origin of space and time properly we
have 
$$ 
x' = a(\alpha) x+ b(\alpha)t 
$$ 
$$ 
t'= c(\alpha)x +d(\alpha)t. 
$$
If $v$ is the speed of $O'$ in the frame $L$ we have 
$$
0=
a(\alpha)x+b(\alpha)t = a(\alpha)vt+b(\alpha)t;~~~~ i.e.~~~~b(\alpha)
= -v a(\alpha).  
$$
From now on we shall use as parameter $v$ instead of $\alpha$ i.e. write
$$ 
x' = a(v) x+ b(v)t 
$$
$$ 
t'= c(v)x +d(v)t. 
$$
and the equivalent frames move one with respect to the other with
constant speed.

\bigskip

2) \underline{Isotropy of space}

Inverting $x\rightarrow y=-x$ and $x'\rightarrow y'=-x'$ we admit to have
also a transformation of the group with some speed $u$
$$ 
y' = a(u) y+ b(u)t 
$$ 
$$ 
t'= c(u) y +d(u)t. 
$$ 
Substituting the previous into the above we get
$$
a(v) = a(u);~~~~b(v)=-b(u);~~~~c(v)=-c(u);~~~~d(v)=d(u)
$$
and thus
$$
u=-v;~~~~a=a(|v|);~~~~d=d(|v|);~~~~ c(-v)=-c(v).
$$

3) \underline{Group law}

The equivalence of all inertial frames implies that the transformation
form a group. Thus there must exist the inverse (unique)
$$
\begin{pmatrix}
a(w) & -wa(w) \\ c(w) & d(w) 
\end{pmatrix}
\begin{pmatrix}
a(v) & -va(v) \\ c(v) & d(v) 
\end{pmatrix}
=I
$$
which provides
$$
w= -v\frac{a(v)}{d(v)}= - v {\rho(v)}
$$
where
$$
\rho(v)\equiv\frac{a(v)}{d(v)}.
$$
We obtain also
$$
\frac{a(v)}{c(v)} = -\frac{d(w)}{c(w)}.
$$
But as $A^{-1}A = I = A A^{-1}$ we must also have
$$
\frac{a(w)}{c(w)} = -\frac{d(v)}{c(v)}
$$
and dividing
$$
\frac{a(v)}{d(v)} =\frac{d(w)}{a(w)}; ~~~~i.e.~~~~ \rho(w) =
\frac{1}{\rho(v)}
$$
that is we have reached the functional equation
$$
\rho(- v \rho(v)) = \frac{1}{\rho(v)} =
\rho(v \rho(v))
$$
where in the last passage we took into account that $a$ and $d$ and
thus $\rho$ are even functions.
To solve the equation set
$$
x\rho(x) = \zeta(x);~~~~\rho(x) = \frac{\zeta(x)}{x}
$$
and we have
$$
\rho(\zeta(v)) = \frac{1}{\rho(v)} 
$$
i.e.
$$
\rho((\zeta(v))= \frac{v}{\zeta(v)} 
$$
or
\begin{equation}\label{functional}
\zeta(\zeta(v))= v.
\end{equation}
As $a(0)=d(0)=1$ we have the additional information $\zeta(0)=0$ and
$\zeta'(0)=1$. Expand in power series
$$
\zeta(x) = x +a_2 x^2 + a_3 x^3 + \dots
$$ 
Eq.(\ref{functional}) becomes
$$
x +a_2 x^2 + a_3 x^3 + \dots +a_2 (x +a_2 x^2 + a_3 x^3 + \dots)^2 +
a_3 (x +a_2 x^2 + a_3 x^3 + \dots)^3 + \dots =x
$$
which gives
$$
2 a_2=0.
$$
Then the equation becomes
$$
a_3 x^3 + \dots + a_3 (x + a_3 x^3 + \dots)^3 + a_4 (x + + a_3 x^3 +
\dots)^4+\dots =0
$$
which gives
$$
2 a_3=0
$$
and so on. Finally we have $\zeta(v) = v$, with the consequences
$a(v)=d(v)$ and $w=-v$.

For a proof of $\zeta(v) = v$ without recourse to a power series
expansion see [1].

We now exploit the relation
$$
\begin{pmatrix}
a(v) & va(v) \\ -c(v) & a(v) 
\end{pmatrix}
\begin{pmatrix}
a(v) & -va(v) \\ c(v) & a(v) 
\end{pmatrix}
=I
$$
obtaining
$$
a^2(v)+v a(v)c(v) = 1.
$$
We notice that
$$
\begin{pmatrix}
a(v) & -va(v) \\ \frac{1-a^2(v)}{va(v)} & a(v) 
\end{pmatrix}
\in SL(2,R).
$$
Composing two transformations of the above type, recalling that $d=a$,
we must obtain a transformation of the same type
$$
\begin{pmatrix}
a' & b' \\ c' & a' 
\end{pmatrix}
\begin{pmatrix}
a & b \\ c & a 
\end{pmatrix}
=
\begin{pmatrix}
a'a+b'c & a'b+ b'a \\ c' a+a'c & c'b+a'a 
\end{pmatrix}
$$
i.e. $b'c = c'b$ or
$$
\frac{c'}{b'}=\frac{c}{b} = k~~~{\rm independent~ of~}v.
$$
From the expression of $c(v)$ and $b(v)$ we have
$$
a(v) = \frac{1}{\sqrt{1-kv^2}}
$$
where the $+$ determination of the square root has to be chosen, as we
know that $a(0)=1$.
Substituting we have
$$
\begin{pmatrix}
\frac{1}{\sqrt{1-kv^2}}& -v \frac{1}{\sqrt{1-kv^2}}
\\ 
-\frac{kv}{\sqrt{1-kv^2}}& \frac{1}{\sqrt{1-kv^2}}
\end{pmatrix}
$$

\bigskip
1. $k=0$

In this case we have the Galileo transformations
$$
\begin{pmatrix}
1& -v
\\ 
0& 1
\end{pmatrix}
$$

\bigskip

2. $0<k\equiv 1/c^2$

In this case we have the Lorentz transformations; setting $x^0=ct$
$$
\begin{pmatrix}
\frac{1}{\sqrt{1-\frac{v^2}{c^2}}} &
-\frac{\frac{v}{c}}{\sqrt{1-\frac{v^2}{c^2}}} 
\\ 
-\frac{\frac{v}{c}}{\sqrt{1-\frac{v^2}{c^2}}} &
\frac{1}{\sqrt{1-\frac{v^2}{c^2}}} 
\end{pmatrix} =
\begin{pmatrix}
\cosh \alpha&
-\sinh \alpha
\\ 
-\sinh \alpha&
\cosh\alpha 
\end{pmatrix}
\in SO(1,1) 
$$
with $v/c= \tanh \alpha$. The product of two transformation
characterized by $\alpha$ and $\beta$ gives the transformation
characterized by $\alpha+\beta$.

\bigskip
Composition of velocities 
\begin{equation}\label{compvelocities}
\frac{v}{c}\oplus \frac{w}{c}=\frac{\frac{v}{c}+\frac{w}{c}}{1+\frac{vw}{c^2}}
\end{equation}
shows that $c$ is a limit
velocity. 
Eq.(\ref{compvelocities}) can also be written as 
$$
\frac{v}{c}=\tanh \alpha,~~~~\frac{w}{c}=\tanh \beta,~~~~
\frac{v}{c}\oplus \frac{w}{c}=
\tanh (\alpha+\beta).
$$
With respect to the event $(0,0)$, events can be invariantly
classified as past present and future.

\bigskip
3. $0>k\equiv-1/c^2$

In this case setting $x^0=ct$ we have the transformation
$$
\begin{pmatrix}
\frac{1}{\sqrt{1+\frac{v^2}{c^2}}} &
-\frac{\frac{v}{c}}{\sqrt{1+\frac{v^2}{c^2}}} 
\\ 
\frac{\frac{v}{c}}{\sqrt{1+\frac{v^2}{c^2}}} &
\frac{1}{\sqrt{1+\frac{v^2}{c^2}}} 
\end{pmatrix}
\in SO(2).
$$
Combining more of these transformations we can turn time into space,
and there is no more distinction between present past and
future. There are obvious problems with causality.

\bigskip

References

\smallskip

[1] J-M. Levy-Leblond, ``One more derivation of the Lorentz
transformation'' American Journal of Physics, 44 (1976) 271.

\smallskip

[2] B. Preziosi, ``Inertia principle and transformation laws'' 
Nuovo Cim. 109 B (1994) 1331.

\vfill


\eject

\section{Lorentz transformations}

Combinations of rotations, boosts and inversions gives rise to a
transformation 
\begin{equation}\label{lambda}
x' = \Lambda x
\end{equation} 
with the property $\langle x,\eta x\rangle =\langle x',\eta x'\rangle$,
where $x^0 =ct$ and $\eta={\rm Diag}(-1,1,1,1)$ i.e.
\begin{equation}\label{lorentz}
\eta = \Lambda^T\eta\Lambda.
\end{equation}
The topology is defined by the metric ${\rm tr}((\Lambda'-\Lambda)^T
(\Lambda'-\Lambda))$. From (\ref{lorentz}) we have also
$\Lambda^{-1}\eta^{-1} = \eta^{-1} \Lambda^T$, i.e. 
\begin{equation}\label{lorentz2}
\eta^{-1} = \Lambda\eta^{-1}\Lambda^T. 
\end{equation}
Obviously $\det \Lambda=\pm 1$ and from Eq.(\ref{lorentz}), $-1 =
-(\Lambda^0_{~0})^2 +(\Lambda^j_{~0})^2$ and thus
$|\Lambda^0_{~0}|\geq 1$. Real 
transformations satisfying (\ref{lorentz}) and with $\det \Lambda=1$
and $ \Lambda^0_{~0} \geq 1$ are the restricted Lorentz group. They form a
group because the determinant is a representation and we have that
$\Lambda^0_{~0}\geq 1$ is conserved in the composition. In fact
$$
(\Lambda'\Lambda)^0_{~0}=\Lambda'~^0_{~0} \Lambda^0_{~0} +\Lambda'~^0_{~j}
\Lambda^j_{~0}  
$$
and  from Eq.(\ref{lorentz}) 
$$
(\Lambda'~^0_{~0})^2 > \Lambda'~^j_{~0}\Lambda'~^j_{~0}
$$
while from Eq.(\ref{lorentz2})
$$
(\Lambda~^0_{~0})^2 > \Lambda~^0_{~j}\Lambda~^0_{~j}
$$
and using Schwarz inequality we have
$$
|\Lambda'~^0_{~j} \Lambda^j_{~0}| < \Lambda'~^0_{~0}  \Lambda^0_{~0}  
$$
which provides $(\Lambda'\Lambda)^0_{~0}>0$.
The elements of the Lorentz group can be classified by the sign of the
determinant $(+,-)$ and by the sign of $\Lambda^0_{~0}$,
$(\uparrow,\downarrow)$,
$L^\uparrow_+, L^\uparrow_-,L^\downarrow_+,L^\downarrow_-$. 
 Due to the discontinuity in the determinant
and in $\Lambda^0_{~0}$ they form disjoint sets. $L^\uparrow_+$ is
called the restricted Lorentz group. The other sets are not groups and
can be reached from the elements of $L^\uparrow_+$ by applying $I_s$
(parity) $I_t$ (time inversion), $I_{st} = I_s I_t$ (strong
reflections). The sets $L^\uparrow_+\cup L^\uparrow_-$,
$L^\uparrow_+\cup L^\downarrow_+$, $L^\uparrow_+\cup L^\downarrow_-$, 
are subgroups, respectively the orthochronous, the proper and the
 orthocorous subgroups.

We have 16 elements with 10 conditions thus 6 real degrees of
freedom. Given a $\Lambda$ belonging to the restricted Lorentz group
it can be 
always written as $RZS$ where $Z$ is a boost along $z$
and $R,S$ rotations. This can be done by deconstruction. Such
decomposition is not unique as seen by counting the degrees of freedom.

Lorentz transformations can be defined also as the diffeomorphisms
preserving the distance
\begin{equation}\label{distance}
ds^2 = dx^\mu \eta_{\mu\nu}dx^\nu.
\end{equation}
and such that $x'^\mu(0)=0$.
If Eq.(\ref{distance}) is left invariant under $x'^\mu = x'^\mu(x)$
then the transformation is linear. In fact from
$$
0=\frac{\partial^2 x'^\lambda}{\partial x^\sigma \partial
x^\mu} \eta_{\lambda \rho} \frac{\partial x'^\rho}{\partial
x^\nu}
+ \frac{\partial x'^\lambda}{\partial x^\mu} \eta_{\lambda \rho} 
\frac{\partial^2 x'^\rho}{\partial x^\nu \partial x^\sigma}  \eqno(I)     
$$
we have
$$
0=\frac{\partial^2 x'^\lambda}{\partial x^\mu \partial
x^\nu} \eta_{\lambda \rho} \frac{\partial x'^\rho}{\partial
x^\sigma}
+ \frac{\partial x'^\lambda}{\partial x^\nu} \eta_{\lambda \rho} 
\frac{\partial^2 x'^\rho}{\partial x^\sigma \partial x^\mu}  \eqno(II)     
$$
$$
0=\frac{\partial^2 x'^\lambda}{\partial x^\nu \partial
x^\sigma} \eta_{\lambda \rho} \frac{\partial x'^\rho}{\partial
x^\mu}
+ \frac{\partial x'^\lambda}{\partial x^\sigma} \eta_{\lambda \rho} 
\frac{\partial^2 x'^\rho}{\partial x^\mu \partial x^\nu}  \eqno(III)     
$$
$$
0=I-II+III = 2 \frac{\partial x'^\lambda}{\partial x^\mu
}\eta_{\lambda \rho} \frac{\partial^2 x'^\rho}{\partial x^\nu \partial
x^\sigma} 
$$
and as $x\rightarrow x'$ is a diffeomorphisms and thus invertible 
$$
\frac{\partial^2 x'^\rho}{\partial x^\nu \partial x^\sigma}=0
$$
which implies $x'$ linear functions of the $x$.

\bigskip

References

\smallskip

[1] R.F. Streater and A.S. Wightman,`` PCT, spin and statistics and all
that'' W.A.Benjamin, New York, 1964

\section{$SL(2,C)$ is connected} 

$A\in SL(2,C)$ can be written as $A = a\sigma_0 +{\bf b}\cdot
\bfsigma$ with $a$ complex number and ${\bf
b}$ complex vector related by $\det A \equiv a^2 -{\bf b}^2=1$. We have
$A^{-1} = a\sigma_0 -{\bf 
b}\cdot \bfsigma$. To see the connectedness of $SL(2,C)$ bring
${\bf b}$ to $0$ avoiding ${\bf b}^2=-1$. $a$ 
is 
constrained to move by continuity. At the end we have $a=\pm 1$. If
$a=-1$ bring the element to the identity by
$$
\begin{pmatrix}
e^{i\alpha}& 0\\0&e^{-i\alpha}
\end{pmatrix} \in SL(2,C).
$$

\bigskip

\section{$SL(2,C)$ is the double covering of the restricted Lorentz group}

Let $B=\sigma_\mu x^\mu$ with $\sigma_0$ the identity and $\sigma_k$
the Pauli matrices. $B$ is hermitean iff $x^\mu$ is real. $\det B =
-x^T\eta x$. If $A\in SL(2,C)$
\begin{equation}\label{lambdamunu}
A BA^+ = B' = \sigma_\mu x'^\mu=\sigma_\mu \Lambda^\mu_{~\nu} x^\nu 
\end{equation}
defines a Lorentz transformation; in fact $B'$ is hermitean and $\det
B'=\det B$. If $A_1$ induces $\Lambda_1$ and $A_2$ induces
$\Lambda_2$, $A_2 A_1$ induces $\Lambda_2 \Lambda_1$.

If $A$ is unitary i.e. an element
of $SU(2)$, $\Lambda$ is a rotation. If $A$ is hermitean $\Lambda$ is a
boost e.g.
$$
\cosh\frac{\phi}{2} + \sigma_3 \sinh\frac{\phi}{2} 
$$
induces the Lorentz boost
$$
{x'}^0 = \cosh \phi ~x^0 + \sinh \phi~ x^3,~~{x'}^3 = \sinh \phi~
x^0 + \cosh \phi~ x^3,~~{x'}^1 = x^1,~~{x'}^2 = x^2
$$
Thus $\Lambda$ covers all the restricted Lorentz group. It is a double
covering. In fact if for all hermitean $B$
$$
ABA^+ = CBC^+ 
$$
then
$$
C^{-1}A B = B C^+ (A^+)^{-1}
$$
Putting $B=1$ we have
$$
C^{-1}A =C^+ (A^+)^{-1}
$$
and then $C^{-1}A$ has to commute with all hermitean matrices and thus
with {\it all} ~$2\times 2$ matrices. By
Schur lemma $C^{-1}A =\alpha$. But $SL(2,C)$ tells us $\alpha^2=1$ thus
$\alpha = \pm 1$.

Thus due to the discontinuity in the determinant and in
$\Lambda^0_{~0}$, $A$ cannot describe elements of $L^\uparrow_-$,
$L^\downarrow_+$, 
$L^\downarrow_-$ and thus $SL(2,C)$ is exactly the double covering of
$L^\uparrow_+$. 

\bigskip

\section{The restricted Lorentz group is simple}

Here we follow Wigner [1] writing explicitly the
transformations that in Wigner paper are described by words.

To start: the rotation group is simple.

In fact suppose ${\cal G}$ to be an invariant subgroup of
$SO(3)$ and $g\in {\cal G}$, $g\neq I$ described by $A\in SU(2)$ 
$$
A = \cos \alpha/2 - i {\bf n}\cdot \bfsigma
\sin\alpha/2;~~~~\alpha\neq 0,~~\alpha\neq \pm\pi.
$$
Then with $U\in SU(2)$ representing a rotation by an angle $\beta$
around the axis ${\bf k}$
$$
UAU^+ = UAU^{-1}
$$
induces a new transformation belonging to ${\cal G}$
i.e. the rotations induced by
$$
B = \cos \alpha/2 - i {\bf m}\cdot \bfsigma \sin\alpha/2
$$
with ${\bf m}$ the rotation of the unit vector ${\bf n}$ 
by the angle $\beta$ around the axis ${\bf k}$. Thus to ${\cal G}$
belong all rotations by $\alpha$ around any axis.
$BA$ also induces a rotation belonging to
${\cal G}$
$$
BA = \cos^2\alpha/2-{\bf m}\cdot{\bf n}\sin^2\alpha/2 
$$
$$
-i({\bf m}+{\bf n})\cdot \bfsigma \sin\alpha/2 \cos\alpha/2
-i{\bf m}\wedge{\bf n} \cdot \bfsigma \sin^2\alpha/2
$$
which is a rotation by an angle $\gamma$ with $\cos \gamma/2=
\cos^2\alpha/2-{\bf m}\cdot{\bf n}\sin^2\alpha/2$
and thus we have rotations of any angle $\gamma$ with 
$0<\gamma<2\alpha$ and from the above results rotations by $\gamma$
around any axis. 
Composing such rotations 
we have rotations by any angle around any axis and thus the 
full rotation group. On the other hand $SU(2)$ is not simple due to
the presence of the invariant subgroup $I,-I$.

\bigskip

%

We come now to the Lorentz group. Given an element $\Lambda\neq I$ 
of the 
restricted Lorentz group belonging to the invariant subgroup
${\cal G}$ let $A$ be a representative of $\Lambda$ in $SL(2,C)$ 
(the other is $-A$). By means of a similitude transformation 
$S AS^{-1}$ in is possible to reduce it (Jordan reduction)
to the one of the two forms
$$
\begin{pmatrix}
\rho &0\\
0&\rho^{-1}
\end{pmatrix}
~~~~{\rm with}~~~~\rho^2\neq 1~~~~{\rm or}~~~~
\begin{pmatrix}
\pm 1 &1\\
0&\pm 1
\end{pmatrix}
$$
with $S\in SL(2,C)$ and thus as $A\neq \pm I$, ${\cal G}$ must
contain the transformations induced by one of the two 
above elements and also their inverses. In the first case by taking the
conjugation with
$$
\begin{pmatrix}
1 & b\\
0& 1
\end{pmatrix}
$$
we obtain
$$
\begin{pmatrix}
1 & -b\\
0& 1
\end{pmatrix}
\begin{pmatrix}
\rho & 0\\
0& \rho^{-1}
\end{pmatrix}
\begin{pmatrix}
1 & b\\
0& 1
\end{pmatrix} = 
\begin{pmatrix}
\rho & b(\rho - \rho^{-1})\\
0& \rho^{-1}
\end{pmatrix}
$$
and thus
$$
\begin{pmatrix}
\rho^{-1} & 0\\
0& \rho
\end{pmatrix}
\begin{pmatrix}
\rho & x\\
0& \rho^{-1}
\end{pmatrix}
=
\begin{pmatrix}
1 & y\\
0& 1
\end{pmatrix}
$$
for any $y$; taking the conjugation with $i\sigma_1$ we also obtain
$$
\begin{pmatrix}
1 & 0\\
y& 1
\end{pmatrix}.  
$$
In the second case by conjugation with elements
$$
\begin{pmatrix}
\alpha & 0\\
0& \alpha^{-1}
\end{pmatrix} \in SL(2,C)
$$
we obtain
the same matrices where $1$ 
has to be substituted with $k =\pm 1$. 
Then keeping in mind that ${\cal G}$ is a group
$$
\begin{pmatrix}
k & i\\
0& k 
\end{pmatrix}
\begin{pmatrix}
k & 0\\
i& k
\end{pmatrix}
\begin{pmatrix}
k & i\\
0& k
\end{pmatrix}
=
\begin{pmatrix}
0 & i\\
i& 0
\end{pmatrix}
\Rightarrow
X_\pi \in {\cal G}. 
$$
$X_\pi$ is the rotation by $\pi$ around $x$. Thus due to the previous
theorem all rotations belong
to ${\cal G}$. 
Then denoting by $Z$ a boost along $z$ and keeping in mind that $X_\pi
Z^{-1} X_\pi = Z$  we have
$$
{\cal G}\owns Z X_\pi Z^{-1}\cdot  X_\pi =Z \cdot X_\pi Z^{-1} X_\pi 
= Z^2
$$ 
and as any boost along $z$ can be written as $Z^2$, all boosts along
$z$ belong to ${\cal G}$ and combining with rotations all elements of the
Lorentz group belong to ${\cal G}$. We conclude that the restricted
Lorentz group is simple. On the other hand $SL(2,C)$ 
is not simple due to the presence of the invariant subgroup $I,-I$.


\bigskip

References

\smallskip

[1] E.P. Wigner, `` On unitary representations of the inhomogeneous
Lorentz group'' Ann. Mathematics, {\bf 40} (1939) 149, pag. 167.

\section{$SL(2,C)$ is simply connected}

Consider a closed contour of ${\bf b(\xi)}$ in $C^3$ and a corresponding
contour $a(\xi)$ with the restriction $a$ continuous and $a^2-{\bf
b}^2=1$, $0\leq \xi\leq 1$ and $a(1)=a(0)$ (see Figure 1.1). 
Look at the closed contour
in the complex ${\bf b}^2$ plane. If such a contour  does not enclose
the point $-1$ we can deform it to a point e.g. by ${\bf b}(\xi,\lambda) =
\lambda{\bf b}(\xi)$, $1\geq \lambda \geq
0$. $a(\xi,\lambda)=\sqrt{1+{\bf b}(\xi,\lambda)^2}$ is
determined by continuity. At the end for $\lambda=0$ we have ${\bf
b}\equiv 0$ and $a\equiv \pm 1$ where the $+$ or $-$ depends
on the original determination of the square root for $\lambda=1$
and we succeeded in shrinking the contour to a point (see Figure 1.2).

If the contour in the complex ${\bf b}^2$ plane encloses the point
$-1$ it has to do it an even number of times (see Figure 1.3); 
otherwise $a(\xi)$ does
not follow a closed contour i.e. $a(1)=-a(0)$. 
In this case perform the same shrinking as in
the previous case and stop when the contour touches the point $-1$
i.e. when $\lambda_0^2 {\bf b}^2(\xi_0)=-1$ (see Figure 1.4). 
Then continue the contraction as follows
$$ 
{\bf b}(\xi,\rho) = {\lambda_0}{\bf b}(\xi_0)+ [\lambda_0{\bf
b}(\xi)-\lambda_0{\bf b}(\xi_0)]\rho 
$$
with $\rho$ varying from $1$ to $0$.
We have for all $\rho$, ${\bf b}(\xi_0,\rho) \equiv\lambda_0{\bf b}(\xi_0)$
and thus $a(\xi_0,\rho)\equiv 0$.
For $\rho=0$ we have that the contour has shrunk to the point
$\lambda_0{\bf b}(\xi_0)$ and $a=0$. 

On the other hand $L^\uparrow_+$
is not simply connected. In fact consider a path $A(\xi)$ in $SL(2,C)$
for $0\leq \xi \leq 1$ such that $A(0)=I$ and $A(1)=-I$. To such a
path there corresponds a closed path in $L^\uparrow_+$. However such a
path in $L^\uparrow_+$ cannot be shrunk to a point because the
corresponding path in $SL(2,C)$ is constrained to have $A(0)=-A(1)$. 
$SL(2,C)$ is the universal covering of $L^\uparrow_+$. 

\vfill


\begin{figure}[htb]
\begin{center}
\includegraphics{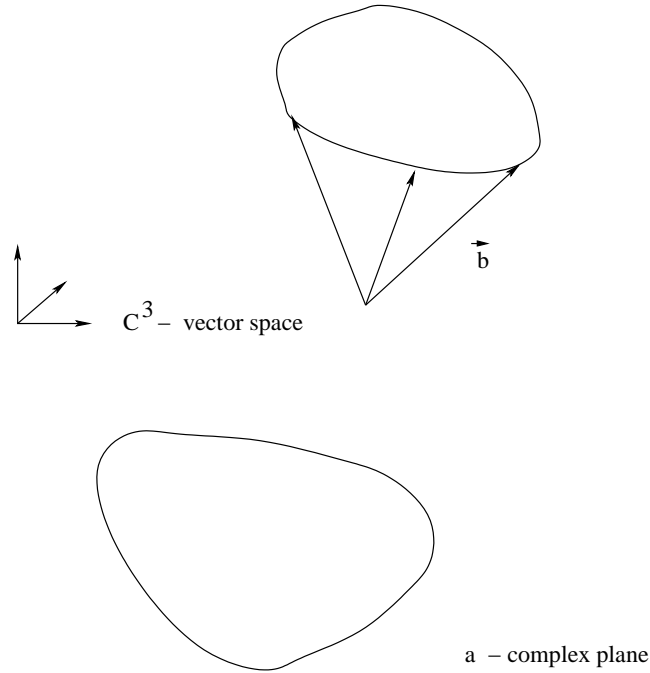}
\end{center}
\caption{Contour in $C^3$ and in the $a$ complex plane}
\end{figure}

\begin{figure}[htb]
\begin{center}
\includegraphics{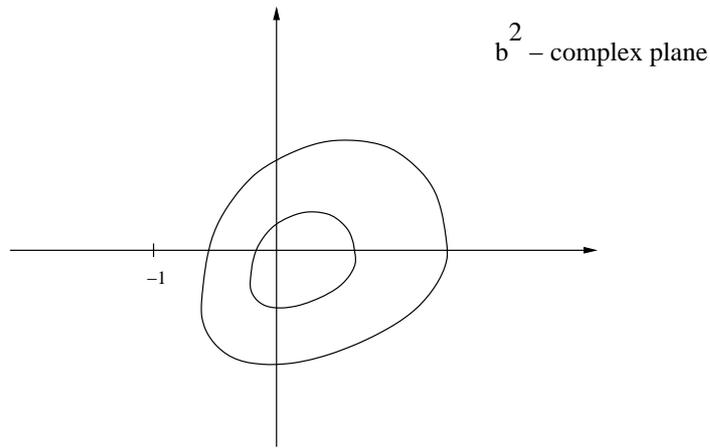}
\end{center}
\caption{Contour of $b^2$}
\end{figure}

\begin{figure}[htb]
\begin{center}
\includegraphics{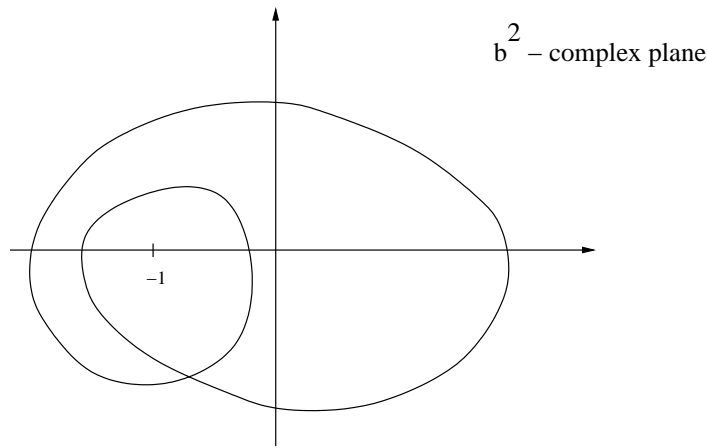}
\end{center}
\caption{First deformation}
\end{figure}\begin{figure}[htb]

\begin{center}
\includegraphics{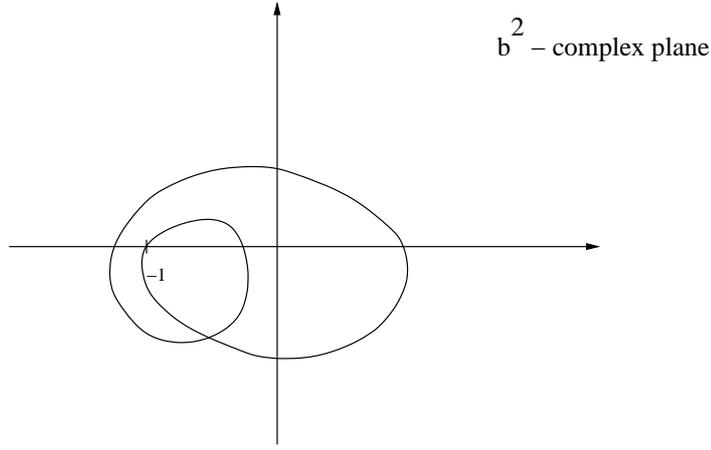}
\end{center}
\caption{Second deformation}
\end{figure}

\section{Little groups}

These are the subgroups of the restricted Lorentz group which leave
unchanged a given four vector.

\bigskip
1. Time like four vector: we can always choose 
$(1,0,0,0)$ gives $SU(2)$ the group of rotations and vice versa. (3
parameter group). In fact if
$$
A\sigma_0 A^+ = \sigma_0 \equiv I
$$
we have that $A$ is unitary. It classifies the particle of non zero mass.

\bigskip

2. Light-like four vector: we can always choose
$(1,0,0,1)$. It gives the little group of the photon.
In fact
$$
A\begin{pmatrix}
1 & 0\\ 0 &0
\end{pmatrix} A^+=
\begin{pmatrix}
a & b\\ c &d
\end{pmatrix}
\begin{pmatrix}
1 & 0\\ 0 &0
\end{pmatrix}
\begin{pmatrix}
a^* & c^*\\ b^* &d^*
\end{pmatrix}
= 
\begin{pmatrix}
1&0\\ 
0 &0
\end{pmatrix}
$$
gives
$$
A=
\begin{pmatrix}
e^{i\alpha/2}&b\\ 
0 &e^{-i\alpha/2}
\end{pmatrix}
\in
E(2)
$$
$E(2)$ being the inhomogeneous two dimensional euclidean group. In
fact defined
$$
T_x = \begin{pmatrix}
1&x\\ 
0 &1
\end{pmatrix}~~~~{\rm and}~~~~R_\alpha =\begin{pmatrix}
e^{i\alpha/2}&0\\ 
0 &e^{-i\alpha/2}
\end{pmatrix}
$$
we have
$$
T_x T_y = T_{x+y},~~~~R_\alpha R_\beta = R_{\alpha+\beta},~~~~T_x
R_\alpha= R_\alpha T_{e^{-i\alpha} x} 
$$
which are the composition relations of $E(2)$. It classifies the
particles of zero mass. Barring the existence of a continuous degree
of freedom [1], the states of the particle are classified by the eigenvalues
of the operator of rotation around the $z$ axis which is the direction
of the space part of the momentum i.e. the helicity. The irreducible
representation of the little group is then given by a single state;
if we introduce parity to this the state of opposite helicity state
has to be added.

\bigskip

3. Space-like four vector: we can always choose $(0,0,1,0)$.
It gives the little group of the tachyon, given by
$SL(2,R)$ as seen by
$$
\sigma_2 \det A = A\sigma_2 A^T= \sigma_2 = A\sigma_2 A^+
$$
from which $A^+=A^T$.
As one can expand the scattering amplitude at 
{\it fixed energy} (little group (1)) in
representations of $SU(2)$ (spherical harmonics), similarly one can
expand the scattering amplitude at {\it fixed momentum transfer} 
(little group (3)) in
representations of $SL(2,R)$. Regge poles are the singularities of the
Fourier transform of the scattering amplitude, in the unitary
irreducible representations of $SL(2,R)$. 

\bigskip

References

\smallskip

[1] [WeinbergQFT] I Chap. 2


\section{There are no non-trivial unitary finite dimensional
representations of the restricted Lorentz group}

Consider
$$
HPH^{-1}
$$
with
$$
H=
\begin{pmatrix}
c+s&0
\\0&c -s
\end{pmatrix}
$$
with $c=\cosh\frac{\phi}{2}$ and $s=\sinh\frac{\phi}{2}$
and
\begin{equation}\label{1P}
P=
\begin{pmatrix}
1 &1\\0&1
\end{pmatrix}
\end{equation}
we have 
$$
HPH^{-1}=
\begin{pmatrix}
1 &e^\phi\\0&1
\end{pmatrix}\in SL(2,C).
$$
For $e^\phi=n$ positive integer we have
$$
\begin{pmatrix}
c+s&0 \\
0&c-s
\end{pmatrix}
\begin{pmatrix}
1 &1\\0&1
\end{pmatrix}
\begin{pmatrix}
c-s&0\\
0& c+s 
\end{pmatrix}
=
\begin{pmatrix}
1 & n\\0&1 
\end{pmatrix}
=
\begin{pmatrix}
1 & 1\\0&1 
\end{pmatrix}^n
$$
thus we have matrices $H_n$ and $P$ such that
$$ 
H_n P H_n^{-1}= P^n.
$$
The induced transformations of $L^\uparrow_+$ satisfy
$$
L_n L_P L_n^{-1} = L_P^n.
$$
For a unitary representation of $L_+^\uparrow$ we have
$$
V_n U  V_n^+ = U^n
$$
with $V_n$ and $U$ unitary operators.
If the representation is finite dimensional ($D$ dimensional) we have
that $U$ and $U^n$ must have the same eigenvalues, for any integer $n$.
Thus is $\lambda_1$ is an eigenvalue of $U$ also $\lambda_1^2,
\lambda_1^3,\dots \lambda_1^{D+1}$ must be eigenvalues of $U$ and as
there are at most $D$ different eigenvalues we have that
$\lambda_1^{k_1} = 1$ for some $k_1\leq D$ and thus $\lambda_1 =
e^{2\pi i n_1/k_1}$ and the same holds for all eigenvalues. As a
result $U^N$ for $N= k_1 k_2 \dots k_D$ has all eigenvalues equal to
$1$ and $U^N$ being unitary we have $U^N=I$. Thus
$$
U = V_N^+ U^N V_N =V_N^+ I V_N =I.
$$
The kernel of the representation, i.e. the elements which are
represented by $I$ form an invariant subgroup of $L_+^\uparrow$. But
the restricted Lorentz group is simple and as to $P$ there corresponds an
element of $L_+^\uparrow$ different from the identity, the kernel of
the representation is the whole $L_+^\uparrow$ and thus the
representation is trivial. 

The result holds also for the finite dimensional unitary
representations of the universal covering group of $L_+^\uparrow$
i.e. $SL(2,C)$. In fact the only invariant subgroup of $SL(2,C)$ is
$(I,-I)$ and $P$ given by (\ref{1P}) does not belong to  this subgroup.

\bigskip

References

\smallskip

[1] E.P. Wigner, Ann. Mathematics,``On unitary representations of the
inhomogeneous Lorentz group'' {\bf 40} (1939) 149, pag. 167.


\section{The conjugate representation}

$A^+$, $A^{-1}$, $A^T$ are not representations. $(A^+)^T\equiv A^*$ is
a representation, called the conjugate representation. $(A^+)^{-1}$ is
also a representation.

In $SU(2)$ the conjugate representation is equivalent to the
fundamental representation $A$ as seen from 
$$ U = a-i\bfsigma\cdot
{\bf b};~~~ a,{\bf b}~~~{\rm real~and}~~~~ a^2+{\bf b}^2 =1.
$$ 
$$ 
U^* = \sigma_2 U \sigma_2. 
$$ 
This is a peculiarity of $SU(2)$. Moreover
for unitary representations $(U^+)^{-1}=U$. For
$SL(2,C)$, $A^*$ is inequivalent to $A$ as seen by taking the trace
e.g. of 
$
\begin{pmatrix}
i &-2\\1&i
\end{pmatrix}
$.  On the other hand $(A^+)^{-1}$ is equivalent to
$A^*$ as seen by the $\det A  =1$ relation written as 

$$ 
\sigma_2 =
A^T\sigma_2A
$$ 
which gives 
$$ 
\sigma_2 = A^+\sigma_2A^*;~~~~(A^+)^{-1}= \sigma_2 A^*\sigma_2. 
$$

Let $p^\mu$ be a four vector (in the following $\displaystyle{p_\mu =
-i \hbar
\frac{\partial}{\partial x^\mu}}$, $p^\mu=\eta^{\mu\nu}p_\nu$).
We have
$$
A\sigma_\mu p^\mu A^+ = \sigma_\mu {p}'^\mu = \sigma_\mu
\Lambda^\mu_{~\nu}p^\nu  
$$
and also
$$
A^*\sigma^*_\mu p^\mu A^T = \sigma^*_\mu {p}'^\mu = \sigma^*_\mu
\Lambda^\mu_{~\nu}p^\nu  
$$
equivalent to
$$
\tilde A\tilde\sigma_\mu p^\mu {\tilde A}^+ = \tilde\sigma_\mu
\Lambda^\mu_{~\nu}p^\nu  
$$
with $\tilde A \equiv  \sigma_2 A^* \sigma_2 = (A^+)^{-1}$ and
$\tilde\sigma_\mu= 
\sigma_2\sigma_\mu^*\sigma_2$, i.e. $\tilde \sigma_\mu=(\sigma_0,
-{\bfsigma})$. Being the representation $\tilde A$ more elegant we
shall use it instead of $A^*$.

It is not too difficult to prove that $A$ and $\tilde A$ are the only
inequivalent two-dimensional continuous representations of $SL(2,C)$. 


\bigskip
\section{Relativistically invariant field equations}

By this we mean linear differential equations invariant under space
time translations i.e. whose coefficients do not depend on $x$ and
invariant under the local linear transformation
$$
\psi'(x') = S(\Lambda_1)\psi(x),~~~~~x'=\Lambda_1 x
$$ 
$S(\Lambda_1)$ being a linear transformation. They can always be
reduced to first order differential equations. 
As
$$
\psi''(x'')= S(\Lambda_2)\psi'(x') =
S(\Lambda_2)S(\Lambda_1)\psi(x)=S(\Lambda_2\Lambda_1)\psi(x) 
$$
$S(\Lambda)$ has to be a representation of the Lorentz group (or
better of its universal covering $SL(2,C)$).

Simplest cases which will be relevant in the future.

\bigskip

References

\smallskip

[1] G.Ya. Lyubarskii, ``The applications of group theory in physics''
Pergamon Press, New York 1960, Chap. XVI.

\section{The Klein-Gordon equation}
$$
\phi'(x')=\phi(x)
$$
In the following we shall use the notation $\displaystyle{p_\mu
= -i\hbar 
\frac{\partial}{\partial x^\mu}}$ 
and $p^\mu = \eta^{\mu\nu}p_\nu$.
The invariant Klein-Gordon equation is
$$
-p_\mu p^\mu\phi(x)= m^2 c^2\phi(x)
$$ 
i.e.
$$
\partial_\mu \partial^\mu\phi(x) = \frac{m^2 c^2}{\hbar^2}\phi(x).
$$ 
The Klein-Gordon equation can be derived from the invariant action
$$
S = \int d^4 x~(\eta^{\mu\nu} \partial_\mu \phi^+
\partial_\nu\phi+\frac{m^2c^2}{\hbar^2} \phi^+\phi)
$$

\bigskip
\section{The Weyl equation}

We look for an equation whose field is represented by a two
component vector which transforms accordingly to the $(0,1/2)$
representation
$$
\psi'(x') = \tilde A\psi(x)
$$ 
where $ A\sigma_\mu x^\mu A^+ = \sigma_\mu {x'}^\mu$ equivalent to $
\tilde A\tilde \sigma_\mu x^\mu\tilde A^+ = \tilde \sigma_\mu {x'}^\mu$.

Such an equation is given by
\begin{equation}\label{1weyl}
\sigma_\mu p^\mu \psi(x) =0.
\end{equation}
In fact if (\ref{1weyl}) holds we have
\begin{equation}
\sigma_\mu {p'}^\mu \tilde A\psi(x) =\sigma_\mu {p'}^\mu \psi'(x') =0
\end{equation}
as seen by rewriting (\ref{1weyl}) as
$$
0=A\sigma_\mu p^\mu A^+ \tilde A \psi = \sigma_\mu {p'}^\mu\tilde A \psi.
$$
Multiplying  on the left by $\tilde\sigma_\mu p^\mu$ we have
$$
- p_\mu p^\mu \psi(x) =0
$$
i.e. Weyl equation describes particles of zero mass.

It is not possible to add to it a mass term
\begin{equation}\label{1weylmass}
\sigma_\mu p^\mu \psi(x) = M\psi(x)
\end{equation}
$M$ being a $2\times 2 $ matrix, without violating Lorentz
invariance. In fact if $\psi(x)\in(0,1/2)$ we have 
$\sigma_\mu {p}^\mu \psi(x)\in(1/2,0)$ because
$$
\sigma_\mu {p'}^\mu \psi'(x') =A\sigma_\mu {p}^\mu A^+\tilde \psi(x)=
A\sigma_\mu {p}^\mu\psi(x)
$$
from which
\begin{equation}
M\psi(x) = A^{-1}M\tilde A\psi(x);~~~~\forall A\in SL(2,C). 
\end{equation}
or
\begin{equation}\label{1mass2}
M\psi(x) = A^+MA\psi(x);~~~~\forall A\in SL(2,C). 
\end{equation}
Choose a point $x$ such that $\psi(x)\neq 0$ and replacing $A$ with
$AU$, $A\in SL(2,C)$, $U\in SU(2)$ we have
$$
M\psi(x) = U^+ A^+ M A U \psi(x)~.
$$
Taking the scalar product
$$
(U\psi(x),A^+ M A U \psi(x)) = (\psi(x),M\psi(x)) \equiv c
(\psi(x),\psi(x))= c (U\psi(x),U\psi(x)) 
$$
for all $U\in SU(2)$ and as $U\psi(x)$ covers the whole two dimensional
space we have $A^+ M A =cI$, which has to hold for  all $A\in SL(2,C)$. For
$A=I$ we have $M=c I$ so that $cI = c A^+A$ for all $A\in SL(2,C)$, which
implies it implies $c=0$.  

\bigskip

Going over to the Fourier transform we have
\begin{equation}\label{1weylfourier}
\sigma_\mu p^\mu\psi_f(p^0,{\bf p})=0,~~~~{\rm with}~~~~(p^0)^2-{\bf p}^2=0.
\end{equation}
Equation (\ref{1weyl}) is not invariant under parity. In fact let us
look  for an invertible matrix $P$ such that 
$$
\psi'(x') = P\psi(x)~~~~{\rm with}~~~~x' = (x^0,-{\bf x})
$$
and such that
$$
\sigma_\mu {p'}^\mu \psi'(x')=0
$$
i.e.
$$
(\sigma_0 p^0 - \bfsigma\cdot {\bf p})P\psi(x)=0.
$$
Going over to Fourier transform we have 
\begin{equation}\label{weylfourier}
(\sigma_0 p^0 + \bfsigma\cdot {\bf p})\psi_f(p^0,{\bf p})=0
~~~~{\rm and}~~~~
(\sigma_0 p^0 - \bfsigma\cdot {\bf p})P\psi_f(p^0,{\bf p})=0.
\end{equation}
i.e. we must have
$$
-\bfsigma \cdot {\bf p}\psi_f(p^0,{\bf p}) = P^{-1}\bfsigma \cdot {\bf
p}P\psi_f(p^0,{\bf p}) ~.
$$
Taking ${\bf p}\rightarrow -{\bf p}$ we have also
$$
-\bfsigma \cdot {\bf p}\psi_f(p^0,-{\bf p}) = P^{-1}\bfsigma \cdot {\bf
p}P\psi_f(p^0,-{\bf p}). 
$$
$\psi_f(p^0,{\bf p})$ and $\psi_f(p^0,-{\bf p})$ span the two
dimensional space otherwise from $\psi_f(p^0,-{\bf p}) =
\alpha\psi_f(p^0,{\bf p})$  it follows
$$
\sigma_0 p^0 \psi_f(p^0,{\bf p})=0.
$$
Thus
$$
-\bfsigma \cdot {\bf p}= P^{-1}\bfsigma \cdot {\bf p}P,~~~\forall {\bf p}.
$$
But such a matrix $P$ does not exists as seen from
$$
\{\bfsigma\cdot {\bf p}, ~a\sigma_0 + \bfsigma\cdot{\bf b}\}= 2 a~
\bfsigma\cdot {\bf p} + 2 {\bf b}\cdot{\bf p}~.
$$
Similar results hold for a $\psi(x)$ which transforms as
$
\psi'(x') = A\psi(x)
$
and obeys
$
\tilde\sigma_\mu p^\mu \psi(x) = 0
$.
Weyl equation can be derived from the invariant action
$$
S = \int d^4x ~\psi^+(x) \sigma_\mu(-i\partial^\mu)\psi(x)
$$
Such an action is hermitean as seen by taking the hermitean conjugate and
integrating by parts.


\section{The Majorana mass}

It is possible to give a mass to a two component field transforming
according to the $(1/2,0)$ representation of the restricted Lorentz group as
follows (similar considerations can be done for the $(0,1/2)$ representation).
Let us consider the equation
\begin{equation}\label{1majorana}
\tilde \sigma_\mu p^\mu \phi(x) = -i mc~  \sigma_2 \phi^*(x)~.
\end{equation}
We saw already that if $\phi'(x') =A\phi(x)$ then
$\tilde \sigma_\mu p'^\mu \phi'(x')=\tilde A \sigma_\mu p^\mu \phi(x)
~\in(0,1/2)$. Thus we have simply to prove that $\sigma_2 \phi^*(x)
~\in(0,1/2)$. In fact
\begin{equation}
\sigma_2(A\phi(x))^* = \sigma_2A^*\sigma_2 \sigma_2\phi^*(x) =
\tilde A ~\sigma_2~\phi^*(x)~.
\end{equation}
Multiplying Eq.(\ref{1majorana}) on the
left by $\sigma_\mu p^\mu$ and using again Eq.(\ref{1majorana}) we obtain,
taking into account that $(p^\mu)^* = -p^\mu$,
$$
-p^2 \phi(x) = m^2c^2 \phi(x)
$$ 
proving that both component of the spinor $\phi(x)$ satisfy the massive
Klein-Gordon equation. 

Equation (\ref{1majorana}) can be derived by varying the action
\begin{equation}\label{1majoranaaction}
S = \int d^4x~[\phi^+(x)i\tilde
\sigma_\mu\partial^\mu\phi(x) +\frac{imc}{2}(\phi^T\sigma_2\phi(x)-
\phi^+\sigma_2\phi^*)] 
\end{equation}

Again integrating by parts it is proved that the action $S$ is hermitean.

The action (\ref{1majoranaaction}) is invariant under the parity
transformation 
\begin{equation}
\phi'(x') = P\phi(x) =\sigma_2 \phi^*(x),~~~~{\rm
  with}~~~~x'=(x^0,-{\bf x})
\end{equation}
and thus also the equations of motion are invariant under such a
transformation.


\section{The Dirac equation}

Consider again a $\psi(x)$ which transforms according to
$$
\psi'(x') = \tilde A \psi(x).
$$
Posing
\begin{equation}\label{dirac1}
\sigma_\mu p^\mu \psi(x) = \phi(x)
\end{equation}
we have that $\phi(x)$ transforms like
$$
\phi'(x') = A \phi(x).
$$
In fact
$$
A\sigma_\mu p^\mu A^+ \tilde A\psi(x) = A \phi(x)
$$
i.e.
$$
\sigma_\mu {p'}^\mu\psi'(x') = \phi'(x')
$$
We have now four components and we must supply other two equations,
invariant under Lorentz transformations.
These are given by
$$
\tilde\sigma_\mu p^\mu \phi(x) = \kappa \psi(x)
$$
which is invariant as
$$
\tilde A \tilde\sigma_\mu p^\mu \tilde A^+ A\phi(x) = \kappa \tilde A \psi(x)
$$
i.e.
$$
\sigma_\mu {p'}\phi'(x') = \kappa \psi'(x').
$$
Multiplying now (\ref{dirac1}) on the left by $\tilde\sigma_\mu p^\mu$
we have 
$$
-p_\mu p^\mu \psi(x) = \kappa \psi(x)
$$
i.e. the two equations
$$
\sigma_\mu p^\mu \psi(x) = \phi(x)
$$
$$
\tilde\sigma_\mu p^\mu \phi(x) = \kappa \psi(x)
$$
describe particles of mass $m$ such that $m^2 c^2=\kappa$. It is better to
distribute in more symmetrical way the $\kappa$
$$
\sigma_\mu p^\mu \psi(x) = mc~ \phi(x)
$$
$$
\tilde\sigma_\mu p^\mu \phi(x) = mc~ \psi(x).
$$
Thus the
$$\Psi(x)=
\begin{pmatrix}
\psi(x) \\ \phi(x)
\end{pmatrix}
$$
transforms according to the $(1/2,0)\oplus(0,1/2)$ representation.
From $\tilde\bfsigma = -\bfsigma$ it is immediately seen that the
parity transformation is
$$
\Psi'(x') = P\Psi(x), ~~~~{\rm with}~~~ x'=(x^0, -{\bf x})
$$
and 
$$P=
\begin{pmatrix}
0&1 \\ 1& 0
\end{pmatrix}.
$$
We can also write
$$
\sigma^\mu p_\mu \psi(x) = mc~ \phi(x)
$$
$$
\tilde\sigma^\mu p_\mu \phi(x) = mc~ \psi(x)
$$
with $\sigma^\mu = \eta^{\mu\nu}\sigma_\nu$
or
$$
\left[-
\begin{pmatrix}
1&0 \\ 0& 1
\end{pmatrix}
p_0+
\begin{pmatrix}
\sigma^k&0 \\ 0& -\sigma^k
\end{pmatrix}
p_k\right]\Psi(x) = mc
\begin{pmatrix}
0&1 \\ 1& 0
\end{pmatrix}
\Psi(x)
$$
which can also be written as
$$
(\gamma^\mu\partial_\mu +\frac{mc}{\hbar})\Psi(x)=0
$$
with
\begin{equation}\label{1gammazero}
\gamma^0 =
\begin{pmatrix}
0&-i \\ -i& 0
\end{pmatrix}
~~~~{\rm antihermitean}
\end{equation}
\begin{equation}\label{1gammaj}
\gamma^j =
\begin{pmatrix}
0&-i \sigma^j\\ i\sigma^j& 0
\end{pmatrix}
~~~~{\rm hermitean}.
\end{equation}
These are the same $\gamma$ matrices as adopted in [1].
Notice that defined
\begin{equation}
\Sigma_\mu =
\begin{pmatrix}
\tilde\sigma_\mu& 0\\ 0 & \sigma_\mu
\end{pmatrix}
\end{equation}
and
\begin{equation}
{\cal A}=
\begin{pmatrix}
\tilde A& 0\\ 0 & A
\end{pmatrix}
\end{equation}
we have obviously
$$
{\cal A}\Sigma_\mu {\cal A}^+ = \Sigma_\nu \Lambda^\nu_{~\mu}.
$$
Taking into account that $\gamma_\mu = i \Sigma_\mu P$ we have
$$
{\cal A}\Sigma_\mu (i P) (-iP) {\cal A}^+ (iP) = \Sigma_\nu (iP)
\Lambda^\nu_{~\mu}
$$
i.e.
\begin{equation}\label{1gammarepresentation}
{\cal A}\gamma_\mu {\cal A}^{-1} = \gamma_\nu 
\Lambda^\nu_{~\mu}.
\end{equation} 
In addition it will be useful to define $\gamma_5 = -i
\gamma^0\gamma^1\gamma^2\gamma^3$ which in the same representation
assumes the form
$$
\gamma_5 =
\begin{pmatrix}
1&0\\0 & -1
\end{pmatrix}
~~~~{\rm hermitean}.
$$
In the following also the symbol $\dsl = \gamma^\mu\partial_\mu$ will
be used.
\bigskip

References

\smallskip

[1] [WeinbergQFT] I pag. 216.

\section{The Rarita- Schwinger equation}

In developing supergravity we shall need the field theoretical
description of spin $3/2$ particles. These are described by a vector-
spinor $\psi^\mu$. We recall that a $4-$vector $V^\mu$ has spin content
$0\oplus 1$ because $\partial_\mu V^\mu$ is a scalar. On the other
hand a spinor describes a particle of spin $1/2$. As a consequence
the spin content
of $\psi^\mu$ is $(0\oplus 1)\otimes 1/2 = 1/2\oplus 1/2 \oplus 3/2$ and
thus we have two unwanted spin $1/2$ particles. These can be
eliminated by imposing the following supplementary conditions
$\gamma^\lambda\psi_\lambda=0$  and $\partial_\lambda\psi^\lambda=0$,
which set to zero two spinors. For massive particles both conditions
are contained in the Rarita- Schwinger equation, which in addition
contains the dynamical equations for the $\psi_\lambda$. Such equation
is
\begin{equation}\label{RS}
\varepsilon^{\mu\nu\rho\sigma}\gamma_\nu\gamma_5(\partial_\rho +\frac{m}{2}
\gamma_\rho)\psi_\sigma =0
\end{equation}
where
$$
\varepsilon_{0123} =1,~~~~\varepsilon^{0123} =-1.
$$
The only properties of the $\gamma^\mu$ we shall use are the Clifford
algebra
$$
\{\gamma^\mu,\gamma^\lambda\}=2 \eta^{\mu\lambda}
$$
and the ensuing relation $\{\gamma^\lambda,\gamma_5\}=0$.

By taking the divergence of Eq.(\ref{RS}) we have, for $m\neq 0$
\begin{equation}\label{1diverg}
\varepsilon^{\mu\nu\rho\sigma}\gamma_\nu\gamma_5 \gamma_\rho
\partial_\mu\psi_\sigma=0. 
\end{equation}
But
$$
\varepsilon^{\mu\nu\rho\sigma}\gamma_\nu\gamma_5 \gamma_\rho=
-i(\gamma^\mu\gamma^\sigma-\gamma^\sigma\gamma^\mu) 
$$
and thus
$$
(\gamma^\mu\gamma^\sigma-\gamma^\sigma\gamma^\mu)
\partial_\mu\psi_\sigma=0. 
$$
Rewriting the above as
$$
(2\gamma^\mu\gamma^\sigma-2 \eta^{\mu\sigma})
\partial_\mu\psi_\sigma=0
$$
we obtain
\begin{equation}\label{div}
\dsl (\gamma^\sigma\psi_\sigma) - \partial_\mu\psi^\mu =0
\end{equation}
with $\dsl = \gamma^\mu\partial_\mu$. We contract now Eq.(\ref{RS})
with $\gamma_\mu$ to obtain
$$
\varepsilon^{\mu\nu\rho\sigma}\gamma_\mu\gamma_\nu\gamma_5
\partial_\rho\psi_\sigma+
\frac{m}{2} \varepsilon^{\mu\nu\rho\sigma}\gamma_\mu \gamma_\nu\gamma_5
\gamma_\rho \psi_\sigma=0.
$$
The first term vanishes due to Eq.(\ref{1diverg}) and thus we have
$$
0=(\gamma^\rho \gamma^\sigma -\gamma^\sigma \gamma^\rho) 
\gamma_\rho\psi_\sigma= -2 (\gamma^\sigma\gamma^\rho
-\eta^{\sigma\rho}) \gamma_\rho\psi_\sigma = -6
\gamma^\sigma\psi_\sigma=0. 
$$
Due to Eq.(\ref{div}) it implies also $\partial_\mu\psi^\mu=0$.
Now we prove that each component of the Rarita- Schwinger field
satisfies the Dirac equation. The term of the Rarita- Schwinger
equation proportional to $m$ can be written as
$$
i\frac{m}{2}
(\gamma^\mu \gamma^\sigma - \gamma^\sigma \gamma^\mu)\psi_\sigma =
im
(\gamma^\mu \gamma^\sigma - \eta^{\mu\sigma})\psi_\sigma = -im \psi^\mu.
$$
With regard to the derivative term, it can be rewritten as
$$
\varepsilon^{\mu\nu\rho\sigma}\gamma_\nu\gamma_5
\partial_\rho\psi_\sigma =
\varepsilon^{\mu\nu\rho\sigma}\gamma_\nu\gamma_5\delta^\lambda_\sigma
\partial_\rho\psi_\lambda =
\frac{1}{2}\varepsilon^{\mu\nu\rho\sigma}\gamma_\nu\gamma_5
(\gamma^\lambda\gamma_\sigma+\gamma_\sigma\gamma^\lambda)
\partial_\rho\psi_\lambda = 
$$
$$
=  \frac{1}{2}\varepsilon^{\mu\nu\rho\sigma}
\gamma_\nu\gamma_5 \gamma^\lambda\gamma_\sigma
\partial_\rho\psi_\lambda. 
$$
But the following equality holds
$$
\frac{1}{2}\varepsilon^{\mu\nu\rho\sigma}
\gamma_\nu\gamma_5 \gamma^\lambda\gamma_\sigma = i(\eta^{\rho\lambda}
\gamma^\mu -\eta^{\mu\lambda} \gamma^\rho) 
$$
which substituted in the previous expression gives
$$
-i\dsl \psi^\mu.
$$
Summing the two terms we have
$$
\dsl \psi^\mu + m \psi^\mu=0.
$$
Obtaining the constraints from the Rarita- Schwinger equation is
similar to what happens in the massive vector theory in which the
equations of motion are
$$
\partial_\mu F^{\mu\nu} -m^2 A^\nu =0,
$$
with $F_{\mu\nu}=\partial_\mu A_\nu-\partial_\nu A_\mu$.
Taking the divergence of the above equation we have $\partial_\nu
A^\nu=0$.

We come now to the massless Rarita- Schwinger equation
\begin{equation}\label{1masslessRS}
\varepsilon^{\mu\nu\rho\sigma}\gamma_\nu\gamma_5\partial_\rho
\psi_\sigma=0.
\end{equation}
Such equation is invariant under the gauge transformation
$$
\psi_\sigma \rightarrow \psi_\sigma + \partial_\sigma\eta
$$
with $\eta$ an arbitrary spinor.
Contracting Eq.(\ref{1masslessRS}) with $\gamma_\mu$ gives
$$
0=(\gamma^\rho\gamma^\sigma-
\gamma^\sigma\gamma^\rho)\partial_\rho\psi_\sigma
=2(\gamma^\rho\gamma^\sigma-\eta^{\sigma\rho})\partial_\rho\psi_\sigma=
2 \dsl (\gamma^\sigma\psi_\sigma) -2 \partial_\sigma\psi^\sigma.
$$
A useful gauge choice is $\gamma^\sigma\psi_\sigma=0$, which induces
$\partial_\sigma\psi^\sigma=0$. Such a gauge is at least locally
attainable because the equation
$$
\gamma^\sigma\psi_\sigma+\dsl \eta=0
$$
is solvable in $\eta$; the inverse of $\dsl$ is given by $\dsl
\partial^{-2}$.

\bigskip

\bigskip
\section{Transformation properties of quantum fields}

We saw that classically the transformation properties of (local)
fields are
\begin{equation}\label{classtransf}
\psi'(x') = S(\Lambda)\psi(x),~~~~{\rm with}~~~~x'=\Lambda x.
\end{equation}
Eq.(\ref{classtransf}) can be rewritten as
$$
\psi'(x) = S(\Lambda)\psi(\Lambda^{-1}x).
$$
This is the notation universally adopted. It would be more proper to
write instead of $S(\Lambda)$, $S(A)$ where $A$ is an element of
$SL(2,C)$.

In the quantum treatment we shall adopt the Heisenberg picture where the
state vector is given {\it sub specie aeternitatis} i.e. constant in
time and unchanged under Lorentz transformations.
The transformation of field is induced by a unitary transformation
which is a representation of the Lorentz group, more properly of
$SL(2,C)$ i.e. the one valued representation of the universal covering of
the Lorentz group.
\begin{equation}\label{quantumtransf}
\psi'(x) = U^+ \psi(x) U = S(\Lambda) \psi(\Lambda^{-1}x).
\end{equation}
In order to be a representation $U$ has to stay on the right. In fact if
$$
U^+_1 \psi(x) U_1 = S(\Lambda_1) \psi(\Lambda_1^{-1}x)
$$
and
$$
U^+_2 \psi(x) U_2 = S(\Lambda_2) \psi(\Lambda_2^{-1}x)
$$
we have
$$
U^+_2 U_1^+ \psi(x) U_1 U_2  = S(\Lambda_1) U^+_2
\psi(\Lambda_1^{-1}x)U_2=S(\Lambda_1) S(\Lambda_2)\psi(\Lambda_2^{-1}
\Lambda_1^{-1}x)=
$$
$$
S(\Lambda_1) S(\Lambda_2)\psi(\Lambda_2^{-1}
\Lambda_1^{-1}x)    
=S(\Lambda_1 \Lambda_2)\psi((\Lambda_1
\Lambda_2)^{-1}x).    
$$
Transformation (\ref{quantumtransf}) can also be rewritten as
$$
U^+\psi(\Lambda x)U = S(\Lambda)\psi(x)
$$
or
$$
U \psi(x)U^{-1} = S(\Lambda)^{-1} \psi(\Lambda x)
$$
which is the form given in [1]
$$
U_2U_1 \psi(x)U^+_1 U^+_2= S(\Lambda_1)^{-1} U_2\psi(\Lambda_1
x)U_2^+=S(\Lambda_1)^{-1} S(\Lambda_2)^{-1}\psi(\Lambda_2\Lambda_1 x)= 
$$
$$
=S((\Lambda_2\Lambda_1)^{-1}\psi(\Lambda_2\Lambda_1 x). 
$$

\bigskip
One could work also in the Schroedinger picture in which the fields in
different reference frames are represented by the same operator
$\psi(x)$ and the state vector changes according to $\Omega\rightarrow
U\Omega$, $U$ being a unitary representation of the Lorentz
group. Then
$$
(U\Omega,\psi(x)U\Omega) =
S(\Lambda)(\Omega,\psi(\Lambda^{-1}x)\Omega)=
(\Omega,U^+\psi(x)U\Omega).
$$

\bigskip

References

\smallskip

[1] [WeinbergQFT] I p.192.


\chapter{Equivalence principle and the path to general relativity}

\section{Introduction}

We saw that special relativity does not differentiate from classical
mechanics with regard to the invariance under inertial
transformations, except for the existence of a limit velocity. If a
field $A_\mu$ exists whose equation of motion are
$$
\partial_\mu F^{\mu\nu}=0~~~~{\rm with}~~~~ F_{\mu\nu}=\partial_\mu
A_\nu-\partial_\nu A_\mu 
$$
we have that such a limit velocity is actually reached by 
the signals described by the field $F^{\mu\nu}$.

One may attempt the inclusion of gravitational phenomena within the
framework of special relativity. The simplest idea is to describe
gravitational phenomena by means of a scalar field e.g. by changing the
metric from $\eta_{\mu\nu}$ to $\eta_{\mu\nu} e^\phi(x)$ (Nordstr\"om theory).
A scalar field
provides an attractive force and if such a field is of zero mass one
has the Newtonian behavior $1/r^2$ of the force at large distances. On
the other hand such a theory does not forecast the correct value for
the advance of the perihelion of Mercury and most important does not
satisfy the equivalence principle as seen most clearly in the absence
of the deflection of light [1]. Equivalence principle on material bodies
is verified with great accuracy (E\"otv\"os $10^{-9}$; Dicke $10^{-11}$;
Shapiro et at. (free fall)  $10^{-13}$ and more accurate experiments are
under way). 

The equivalence principle shows that the avenue of extending the
invariance of the laws of physics to more general coordinate
transformations than Lorentz transformations, is a promising path
toward a theory of gravitation. 

Beware that the principle of equivalence can hold only ``ultralocally'',
i.e. in the abstract case of point bodies; there are phenomena related
to the gravitational field which do not vanish even when the size of
the laboratory is shrunk to zero. For example a quadrupole formed by two
masses $m$ joined by a rod of length $2 l$ posed in the gravitational
field described by the acceleration $g>0$ for 
which $\frac{\partial g}{\partial z}=k^2\neq 0$ when
aligned near the $z$ axis is subject to oscillations of pulsation
$\omega = k$, independent of $l$, i.e. a frequency which does not
vanish for $l\rightarrow 0$.

Given a general metric $g_{\mu\nu}$ describing a gravitational field
the principle of equivalence requires that given one event $x$, it must
be possible to reduce by means of a congruence the metric in $x$ to
the Lorentzian form, i.e. a non singular real matrix $A$ must exist such
that
\begin{equation}\label{congruence}
A^T g A = \eta~~~~{\rm where}~~~~ A^\mu_{~\nu} = \frac{\partial
x^\mu}{\partial y^\nu} 
\end{equation}
Thus the signature of $g$ has to be $(3,1)$ and as a
consequence  $\det g<0$. Obviously given $x$ there are infinite
diffeomorphisms $x^\mu=x^\mu(y)$ which realize (\ref{congruence}).

\bigskip

References 

\smallskip

[1] [HawkingEllis] Chap. 3.

\smallskip

[2] A. Einstein,``Die Grundlage der allgemeinen
Relativit\"atstheorie'' Ann.Phys. 49 (1916). Translated in ``The
principle of relativity" Dover Pub. Inc.

\smallskip

[3] [WeinbergGC] Chap. 3 par. 1, 2.

\section{Motion of a particle in a gravitational field}

In absence of gravitational fields we know that the action of a
particle is given by
$$
S = -mc \int_{x_1}^{x_2} \sqrt{-ds^2}= \int_{q_1 t_1}^{q_2 t_2}
L(q,\dot q)dt 
$$
with
$$
L(q,\dot q) = 
-mc\sqrt{c^2- \dot q^i\dot q^j \eta_{ij}}~.
$$
From the variational principle of analytical mechanics we know that the
correct action in presence of gravitational fields generated by a
change of coordinates is
\begin{equation}\label{particlemotion}
S = -mc \int_{x_1}^{x_2} \sqrt{-g_{\mu\nu}dx^\mu dx^\nu}
\end{equation}
and this gives a coordinate independent formulation of the law of motion.

If we accept the idea that a gravitational field is described by a
generic metric $g_{\mu\nu}$ the laws of motion have to be postulated,
but the form Eq.(\ref{particlemotion}) appears the most natural [1]. 
Thus we shall assume that also in presence of gravitational fields
which cannot be removed over a finite region of space time by means of
a coordinate transformation, 
the law of motion for a material body is given by Eq.(\ref{particlemotion}).

The principle is to render maximum the length of the trajectory when
keeping the initial and final events fixed. The principle of
equivalence is then satisfied as we have reduced the problem of motion
to a geometrical problem, where the mass, the shape, the substance of
the body do not intervene. Thus one has a geometrization of the
gravitational field and of the motion of particles under the influence
of a gravitational field.

\bigskip

References

\smallskip

[1] A. Einstein,``Die Grundlage der allgemeinen Relativit\"atstheorie''
Ann.Phys. 49 (1916) paragraphs 1, 2, 3, 4. Translated in ``The
principle of relativity" Dover Pub. Inc.

\section{Hamiltonian of a particle in a gravitational field}.

The Lagrangian is
$$
L=-mc\sqrt{-g_{\mu\nu}\frac{dx^\mu}{dt}\frac{dx^\nu}{dt}}= 
-mc\sqrt{-c^2 g_{00}-g_{ij}\dot q^i\dot q^j -2c g_{0i}\dot q^i}~.
$$
It is instrumental to write the metric in the ADM form (see section
7.1 for more details)
$$
ds^2 = -N^2 dt^2 +h_{ij}(dx^i+N^i dt)(dx^j+N^j dt).
$$
We shall give a thorough interpretation of $N$ and $N^i$ in the
future. The hypersurfaces $t={\rm const.}$ are space like i.e. the
vector $(0, dx^i)$ has positive norm i.e. $h_{ij}dx^idx^j$ is definite
positive. $h_{ij}$ is usually called the space metric even if it
coincides with the space metric obtained by the coincidence method only
for $N^i=0$. In terms of $N, h_{ij}, N^i$ the $g_{\mu\nu}$ is given by
$$
g_{\mu\nu}=
\begin{pmatrix}
\frac{N^i h_{ij}N^j - N^2}{c^2}& \frac{h_{nl}N^l}{c}\\
\frac{h_{ml}N^l}{c}& h_{mn}
\end{pmatrix}
~~~~{\rm with~inverse}~~~~
g^{\mu\nu}=
\begin{pmatrix}
-\frac{1}{N^2} & \frac{N^n}{N^2}\\
\frac{N^m}{N^2} &h^{mn}-\frac{N^m N^n}{N^2}
\end{pmatrix}.
$$
The Lagrangian becomes
$$
L = -mc\sqrt{Z}
$$
with
$$
Z = N^2 - N^i h_{ij}N^j  - h_{ij}\dot q^i\dot q^j - 2 N^i h_{ij}\dot q^j.  
$$
We have
\begin{equation}\label{2pi}
p_i =\frac{mc (\dot q_i+N_i)}{\sqrt{Z}}
\end{equation}
where we have lowered the indices of $\dot q^i$ and $N^i$ with the
metric $h_{ij}$. Keep in mind that the conjugate momenta are the $p_i$
(lower indices).  

We have the identity
$$
p_i p^i = \frac{m^2c^2 (-Z+N^2 )}{Z}
$$
or
$$
Z = \frac{m^2c^2 N^2}{p_ip^i+m^2c^2}
$$
which allows, using Eq.(\ref{2pi}) to solve in the $\dot q^i$
$$
\dot q_i + N_i = \frac{\sqrt{Z}}{mc}p_i
$$
from which we have
$$
H = p_i \dot q^i -L = 
\frac{\sqrt{Z}}{mc}p_ip^i -p_iN^i+mc\sqrt{Z}  =
$$
$$
=\frac{\sqrt{Z}}{mc}(p_ip^i+m^2c^2)-p_iN^i=
N\sqrt{p_ih^{ij}p_j+m^2c^2} - p_iN^i 
$$
$h^{ij}$ being the inverse of $h_{ij}$. $N,N^i, h^{ij}$ are functions
of $t$ and $q^i$.

The relevant Hamilton equations are
$$
\dot p_i = -\frac{\partial H}{\partial q^i}.
$$

We can now write the Hamilton-Jacobi equation for $S(q,q_0,t,t_0)$
$$
N\sqrt{\partial_iS h^{ij}\partial_jS+m^2c^2}-N^j\partial_jS =
-\partial_t S~. 
$$
Removing the square root by squaring we have
$$
g^{\mu\nu}\partial_\mu S\partial_\nu S+m^2c^2=0 
$$
which has a covariant form being $S$ a scalar. 
If we denote by $q_0^i$ the coordinates of the initial event,
the relevant equation of motion are
$$
\frac{\partial S}{\partial q_0^j} = {\rm const}
$$
which give implicitly $q^j$ as a function of $t$. We shall see an application
of such  equation in computing the geodesic motion  in a time independent
metric. 

For a ``static'' metric i.e. $N^i\equiv 0$ and small velocities we
have
$$
\dot p_i = -mc\frac{\partial N}{\partial q^i}. 
$$
Let us write for the static metric $-g_{00}=N^2/c^2 = 1+2\phi/c^2$
which for $N^2/c^2\approx 
1$ (weak gravitational field) gives $N = c + \phi/c$ and thus
$$
\dot p_i = -m\frac{\partial \phi}{\partial q^i} 
$$
i.e. $\phi$ is the Newtonian gravitational potential.
\vfill

\eject

\chapter{Manifolds}

\section{Introduction}

The most important characteristics of a manifold are its dimension $n$
and its degree of smoothness, given by the degree of smoothness of the
transition functions. The three most important categories are $TOP_n$
where the transition functions are simply continuous and they define
the topological manifolds; $PL_n$ where the transition functions are
piece-wise linear and the $DIFF_n$ in which the transition functions
are $C^\infty$. For the relations between them and the possibility and
number of inequivalent ways of {\it smoothing} a manifold i.e. of
extracting a compatible $C^\infty$ atlas see [1,2,3].

In two dimensions one can consider holomorphic transition function
giving rise to the concept of Riemann surface. Sometime real-analytic
transition function in $n$ dimensions, i.e. transition function
given locally by convergent power expansions, are considered [4].

\bigskip

References

\smallskip

[1] C.P. Rourke and B.J. Sanderson, ``Introduction to Piecewise-linear
topology'', Springer Verlag, Berlin

\smallskip

[2] M. Freedman and F. Quinn, ``Topology of 4-manifolds'', Princeton
University Press, (1990)

\smallskip

[3] C. Nash, ``Differential topology and quantum field theory'',
Academic -Press, London (1991); Chap. I

\smallskip

[4] S. Klainerman,``Recent results in mathematical GR'' The 13th
Marcel Grossmann meeting, vol I pag.93-104, World Scientific

\bigskip

\section{Mappings, pull-back and push-forward}

For any regular mapping $\phi$ between manifolds
$$
\begin{matrix}\phi\\M \longrightarrow M'
\end{matrix}
$$
$M$ and $M'$ not necessarily of the same dimensions, it is possible to
define the pull-back of a function $f$ defined on $M'$, given by
$f(\phi(p))$ and 
denoted by $\phi^*f(p)$. It is also possible to define
the push-forward of a vector ${\bf V}$ denoted by $\phi_*{\bf
V}(p')$ as follows: Given a vector ${\bf V}\in T(p)$, $p\in M$, defined
by the motion 
$\lambda(t)$ with $\lambda(0)=p$, one considers the vector in
$T(p')$, $p'=\phi(p)\in M'$ defined 
by the motion on $M'$, $\phi(\lambda(t))$ again at 
$t=0$. Such a vector which belongs to $T(p')$ i.e. to the tangent
space of $M'$ at $p'$ will be called the push- forward of the vector
${\bf V}(p)$ defined at $T(p)$ and written as $\phi_*{\bf V}(p')$.

Using the push-forward of vectors, forms are naturally pulled back
e.g. for the one-form $w$, $\phi^*w$ is defined by
$$
\langle {\bf V}(p),\phi^*w(p)\rangle\equiv \langle {\phi_*\bf V}(p'), 
w(p')\rangle.
$$
 
In the case of diffeomorphisms between manifolds the
inverse of $\phi$ exists and it is possible to define the push-forward of
forms and the pull-back of vectors. For vectors the pull-back
$\phi^*{\bf V}(p)$ is defined by $(\phi^{-1})_*{\bf V}(p)$, where
$p\in M$, while for forms the push-forward $\phi_*\omega(p')$ is
defined by $(\phi^{-1})^*\omega(p')$, where $p'\in M'$.  One verifies
that $(\phi^{-1})_* = \phi^*$ and $(\phi^{-1})^* = \phi_*$.

Beware that in the pull-back process, the argument of the result is
the starting point of the mapping, while in the push-forward process
it is the end point of the mapping.

This notation, which is the most useful, can be source of confusion in
the case of the push-forward. For this reason it useful to define also
the linear operator $\hat\phi$ from the space $T(p)$ to $T(p')$
\begin{equation}\label{checknotation}
\phi_*{\bf V}(p')=\hat\phi{\bf V}(p)~.
\end{equation}

For a product of two mappings
$$
\begin{matrix}
&\alpha & ~ &\beta\\
M &\longrightarrow & M'
&\longrightarrow & M''
\end{matrix}
$$
one has with $p'=\alpha(p)$, $p''=\beta(p')=\beta\circ\alpha(p)$
$$
(\beta\circ\alpha)_*{\bf V}(p'')= \hat\beta\alpha_*{\bf V}(p')=
\hat\beta\hat\alpha{\bf V}(p).
$$
If $\alpha$ and $\beta$ are diffeomorphisms and thus invertible 
transformations we can write
$$
(\beta\circ\alpha)_*{\bf V}(p'')= \hat\beta\alpha_*{\bf V}(\beta^{-1}p'')=
\hat\beta\hat\alpha{\bf V}(\alpha^{-1}\beta^{-1}p'')~.
$$

\bigskip

References

\smallskip

[1] [HawkingEllis] Chap. 2 [2.1,2.2,2.3,2.4,2.5,2.6,2.7].

\smallskip

[2] [Wald] Chap. 2, Chap. 3, Appendix A, Appendix B.

\smallskip

[3] H. Flanders, ``Differential forms" Academic Press,  New York (1963)
par. Chap.III  3.1, 3.2, 3.3, 3.4, 3.5, 3.6, 3.7, 3.8


\section{Vector fields and diffeomorphisms}

\bigskip

We state now an important theorem [1]:

If ${\bf X}$ is a $C^\infty$ vector field with  compact support (in
particular if the manifold is compact) then there is a unique
family of diffeomorphisms $\phi_t$ which for all $t\in R$ satisfy
the properties

(1) $\phi_t$ is $C^\infty$

(2)
$
\phi_{s+t}(q)=\phi_{s}\circ\phi_{t}(q)
$

(3) ${\bf X}(q)$ is the tangent vector at $t=0$ of the motion
$\phi_t(q)$.

(4) Every collection of diffeomorphisms $\psi_t$ satisfying (3)
must coincide with $\phi_t$.

\bigskip

The outcome is that for compact manifolds there is a one to
one correspondence between one parameter abelian groups of
diffeomorphisms and $C^\infty$ vector fields. Such a result will allow the
definition of the Lie derivative.

\bigskip

If the vector field has a non compact support we can still obtain a
local result by considering, given a point $p$, an open neighborhood of
it $U_p$ and working with the vector field $\rho(x) {\bf X}$ where
$\rho(x)$ is a $C^\infty$ function which is equal to $1$ in a
neighborhood of $p$ and vanishes outside a compact support.

\bigskip

References 

\smallskip

[1] M. Spivak, ``A comprehensive introduction to differential geometry
I'', Publish or Perish, Inc. Boston (1970)

\section{The Lie derivative}

This is a concept related to a vector field i.e. a vector ${\bf X}(p)$
defined at each point of the manifold; it operates on functions,
forms, vector and tensor fields, giving as a result functions, forms,
vector and tensor fields of the same order.

In the following we shall denote by $\phi_t$ the abelian group of
diffeomorphisms generated by the vector field ${\bf X}$.

\bigskip

\underline{Lie derivative of a function}

$$
L_{\bf X}f(p) =\lim_{t\rightarrow 0}\frac{1}{t}
[\phi^*_t f(p)-f(p)]= X^i \partial_i f(p)\equiv X(f)(p) 
$$

\bigskip

\underline{Lie derivative of a form field}

\bigskip

Forms $\omega$ are naturally pulled-back; the definition of the Lie
derivative is 
$$
L_{\bf X}\omega (p) =\lim_{t\rightarrow 0}\frac{1}{t}[\phi^*_t
  \omega(p)-\omega(p)]
$$

\bigskip

\underline{Lie derivative of a vector field}

\bigskip

$\phi_t$ being a diffeomorphisms it has inverse
$\phi^{-1}_t = \phi_{-t}$ and so the pull-back of a vector is well
defined.
$$
L_{\bf X} {\bf Y}(p) =\lim_{t\rightarrow 0}\frac{1}{t}[\phi^*_t {\bf
Y}(p)-{\bf Y}(p)] 
=\lim_{t\rightarrow 
0}\frac{1}{t}[(\phi_t^{-1})_* {\bf Y}(p)-{\bf Y}(p)]=
$$
\begin{equation}\label{vectorLie}
=\lim_{t\rightarrow 0}\frac{1}{t}[\phi_{-t*}
{\bf Y}(p)-{\bf Y}(p)]=\lim_{t\rightarrow 0}\frac{1}{t}[{\bf
Y}(p)-\phi_{t*} {\bf Y}(p)].          
\end{equation}

\section{Components of the Lie derivative}

We write down now the components of the Lie derivatives of a form
$\omega = \omega_k dx^k$.
Denoting on a given chart the diffeomorphism $\phi_t$ by
$$
x'=\phi(x,t),~~~~\phi(x,0) = x
$$
the pull-back of $\omega$ is
$$
\omega(\phi(x,t))_m \frac{\partial \phi^m(x,t)} {\partial x^k} dx^k.
$$
Taking the derivative with respect to $t$ at $t=0$ and recalling that
$$
\left . \frac{\partial \phi^m(x,t)} {\partial t}\right |_{t=0} = X^m(x)
$$
we obtain
\begin{equation}\label{liedercompform}
L_X(\omega)_k = \frac{\partial \omega_k}{\partial x^m}X^m
+\omega_m \frac{\partial X^m}{\partial x^k}
\end{equation}
which easily generalizes to a $\Lambda^n$ form
$$
L_X(\omega)_{k_1\dots k_n} = \frac{\partial
\omega_{k_1\dots k_n}}{\partial x^m} X^m 
+\omega_{m k_2\dots k_n} \frac{\partial X^m}{\partial x^{k_1}}+
\dots +\omega_{k_1 k_2\dots m} \frac{\partial X^m}{\partial x^{k_n}}.
$$
In computing the components of the Lie derivative of a vector one has
to keep in mind that the argument of the 
push-forward is the end point. We shall use a chart with coordinates
$x$ and for clearness sake we shall write for the diffeomorphism
$$
(\phi_t(p))^k = \phi^k(x,t)
$$
and
$$
\partial_m\phi^k(x,t) = \frac{\partial \phi^k(x,t)}{\partial x^m}.
$$
Notice that $\phi^k(x,0) = x^k$ and
$$
\partial_m\phi^k(x,0) = \delta^k_m.
$$

We have
$$
[\phi_{t*}{\bf Y}(x)]^k =  Y^m(\phi(x,-t))~ \partial_m \phi^k (\phi(x,-t),t)
$$
and taking the derivative with respect to $t$, we have  
$$
\frac{d}{dt} [\phi_{t*}{\bf Y}(x)]^k =
\frac{d}{dt}\left
[Y^m(\phi(x,-t))\partial_m\phi^k(\phi(x,-t),t)\right] |_{t=0}=
$$
$$
-\partial_l Y^m(x) X^l(x)\delta^k_m+ Y^m(x)(\frac{\partial}{\partial
t}\delta^k_m +\partial_m X^k(x)) =
$$ 
$$
=Y^m(x) \partial_m X^k(x) - X^m(x) \partial_m
Y^k(x) . 
$$
Taking into account Eq.(\ref{vectorLie}) we have
\begin{equation}\label{liedercompvector}
(L_{\bf X} {\bf Y})^k = X^m \partial_m Y^k 
-Y^m \partial_m X^k
\end{equation}
Thus $L_{\bf X} {\bf Y}=-L_{\bf Y} {\bf X} = [{\bf X}, {\bf Y}]$. The
last is not simply a symbol but a real commutator of the operations.
$$
{\bf X}({\bf Y}(f)) =
X^m\frac{\partial}{\partial x^m}(Y^n\frac{\partial}{\partial x^n}f) 
$$
and antisymmetrizing
$$
{\bf X}({\bf Y}(f)) -{\bf Y}({\bf X}(f)) = (X^m \frac{\partial
Y^k}{\partial x^m}-Y^m \frac{\partial X^k}{\partial
x^m})\frac{\partial f}{\partial x^k} 
$$
which by the way, proves independently that the commutator of two
vector fields is a vector field.

A simpler method to obtain the components of the Lie derivative of a
vector field is to notice that given any form $\omega=\omega_m
dx^m$ we have that $s=\omega_m Y^m$ is a scalar and the Lie
derivative of a scalar is the simple derivative. Then we have
$$
L_{\bf X}s = \partial_m \omega_k X^m Y^k+
\omega_k\partial_m Y^k X^m =
(L_{\bf X}\omega)_k Y^k +\omega_k (L_{\bf X}{\bf Y})^k
$$
and from the expression of the components of the Lie derivative of
1-forms (\ref{liedercompform}) we have again (\ref{liedercompvector}). 
 

\section{Commuting vector fields}

We recall that there is a bijection between $C^\infty$ vector fields
with compact support and one-parameter abelian groups of
diffeomorphisms. Given the vector field ${\bf X}$ we shall call
$\phi_t$ the associated group of diffeomorphisms. Let now be $\alpha$
an other diffeomorphism which commutes with $\phi_t$
$$
\alpha\circ\phi_t\circ \alpha^{-1} = \phi_t.
$$
By taking the derivative with respect to $t$ at $t=0$ we have
$$
\alpha_*{\bf X}(p) = {\bf X}(p)
$$
(always recall that the argument of the push-forward is the end point
of the mapping). 
Vice-versa given a whatever diffeomorphism $\alpha$ also
$\alpha\circ\phi_t\circ \alpha^{-1}$ is a one-parameter group of 
diffeomorphisms and according to the previous results it is generated by
$\alpha_* {\bf X}$. If it turns out that 
$\alpha_*{\bf X} = {\bf X}$ then due to the bijection we have
$$
\alpha\circ\phi_t\circ \alpha^{-1} = \phi_t.
$$
Thus we have at present the following result: necessary and sufficient
condition for $\alpha$ to commute with $\phi_t$ is that $\alpha_*{\bf
X} = {\bf X}$.

Given now the field ${\bf Y}$ and the diffeomorphisms $\psi_s$
generated by it, if $\psi_s$ commutes with $\phi_t$ we have
$$
\phi_t\circ \psi_s\circ \phi_t^{-1} = \psi_s
$$
and thus
$$
\phi_{t*}{\bf Y} ={\bf Y} 
$$
which through the definition of Lie derivative implies
$$
L_{\bf X}{\bf Y} =0.
$$
We prove now the reverse, that is if 
$$
L_{\bf X}{\bf Y}\equiv [{\bf X},{\bf Y}]=0
$$
then the diffeomorphism $\phi_t$ generated by ${\bf X}$ commutes with 
the diffeomorphism $\psi_s$ generated by ${\bf Y}$.

Let us consider, given a point $p$,  the vector ${\bf C}$ defined by
$$
{\bf C}(t) =\phi_{t*}{\bf Y}(p).
$$
We shall see in a moment that ${\bf C}'(t)=0$ and thus 
$$
\phi_{t*}{\bf Y}(p)={\bf Y}(p)
$$
which implies from the above reasonings that $\phi_t$ and $\psi_s$
commute.

To prove that ${\bf C}'(t)=0$ let us write
$$
{\bf C}'(t)=\lim_{h\rightarrow 0} \frac{1}{h}(\phi_{t+h*}{\bf Y}(p) -
\phi_{t*}{\bf Y}(p))=
\lim_{h\rightarrow 0}\frac{1}{h}\left[\hat\phi_{t}\hat\phi_{h}{\bf
Y}(\phi^{-1}_h \circ \phi^{-1}_t(p))- 
\hat\phi_{t}{\bf
Y}(\phi^{-1}_t (p))\right]
$$
where we used the notation of Eq.(\ref{checknotation}). Then
$$
{\bf C}'(t)=
\hat\phi_{t}\lim_{h\rightarrow 0}\frac{1}{h}\left[\hat\phi_{h}{\bf
Y}(\phi^{-1}_h \circ \phi^{-1}_t(p))- 
{\bf Y}(\phi^{-1}_t (p))\right]=
$$
$$
=\hat\phi_{t}\lim_{h\rightarrow 0}\frac{1}{h}\left[\phi_{h*}{\bf
Y}(\phi^{-1}_t(p))- 
{\bf Y}(\phi^{-1}_t (p))\right]=
$$
$$
=-\hat\phi_{t}L_{\bf X}{\bf Y}(\phi_t^{-1}(p))=-\phi_{t*}L_{\bf X}{\bf Y}(p)=0
$$
due to the vanishing of $L_{\bf X}{\bf Y}$. 
Summing up: Necessary and sufficient condition for the two abelian
groups of diffeomorphisms $\phi_t$ and $\psi_s$ to commute is that
their associated vector fields ${\bf X}(p)$, ${\bf Y}(p)$ commute.

\bigskip

Given two commuting one-parameter abelian groups  of diffeomorphisms
$\phi_t$ and 
$\psi_s$, we can construct a two dimensional surface by moving a given
point $p$ as follows
$$
\psi_s\circ\phi_t (p)
$$
and on this surface, set the coordinates system $t,s$. The coordinate
basis of the tangent space then contain
$$
\frac{\partial}{\partial t} ={\bf X}(p);~~~
\frac{\partial}{\partial s} ={\bf Y}(p)
$$

The above results can be extended to an arbitrary number of commuting
vector fields.  

\bigskip

References

\smallskip

[1] M. Spivak, ``A comprehensive introduction to differential geometry
I'', Publish or Perish, Inc. Boston (1970); from pag.5-30 to pag.5-39

\section{Stokes' theorem}\label{3stokes}
Given an $n$-dimensional manifold $M$, a local system of coordinates
$x^1\dots x^n$ defines, if the manifold is orientable, an orientation.
This means that we associate to the symbol
$$
\int_V f~ dx^1\wedge \dots dx^n
$$
the value
$$
\int_V f  ~dx^1\dots dx^n
$$
where $dx^1\dots dx^n$ is the usual measure.

The orientation on $V$ induces an
orientation on the boundary $\partial V$ as follows: Choose around a
point of $\partial V$ a local  system of coordinates $u^1\dots u^n$
which is equioriented with $x^1\dots x^n$ i.e.
$$
\frac{\partial(x^1\dots x^n)}{\partial(u^1\dots u^n)}>0
$$ 
and such that $\partial V$ is described by $u^1=0$ and $u^1$ is negative
inside and positive outside $V$. Then the orientation on $\partial V$
is given by
$$
\int_{\partial V} g ~du^2\wedge \dots du^n=
\int_{\partial V} g~ du^2\dots du^n~.
$$
This convention is equivalent to the one adopted in [1]. Once
defined the orientation on $\partial V$ we have Stokes' theorem for
an $n-1$ form on $M$
$$
\int_V d\omega = \int_{\partial V} \omega~.
$$
Thus if
$$
\omega =\omega_{\nu_2\dots \nu_n} \frac{dx^{\nu_2}\wedge\dots dx^{\nu_n}}{(n-1)!}
$$
we have
$$
\int_V \partial_\lambda \omega_{\nu_2\dots \nu_n} 
\frac{dx^\lambda \wedge dx^{\nu_2}\wedge\dots dx^{\nu_n}}{(n-1)!}
=\int_{\partial V} \omega_{\nu_2\dots \nu_n} \frac{\partial
  x^{\nu^2}}{\partial u^{\sigma_2}} \dots \frac{\partial
  x^{\nu^n}}{\partial u^{\sigma_n}} 
\frac{du^{\sigma_2}\wedge \dots \wedge du^{\sigma_n}}{(n-1)!}
$$

\bigskip

References

\smallskip

[1] S.S. Chern, W.H. Chen, K.S. Lam,``Lectures in differential
geometry'' World Scientific, 1999



\chapter{The covariant derivative}\label{4thecovariantderivative}

\section{Introduction}

We give the defining properties of the covariant derivative.
Given a vector field ${\bf Y}$, ${\bf d}_{\bf
X}{\bf Y}$ is a vector field with the properties
$$
{\bf d}_{f{\bf X}+g{\bf Z}}{\bf Y}=f{\bf d}_{\bf X}{\bf Y}+g{\bf d}_{\bf
Z}{\bf Y}\eqno(I) 
$$
$$
{\bf d}_{\bf X}(\alpha {\bf Y}+\beta {\bf Z})=
\alpha{\bf d}_{\bf X}{\bf Y}+\beta {\bf d}_{\bf X}{\bf
Z}~~~~~~~~~\alpha,~\beta = {\rm const.} \eqno(II)
$$
$$
{\bf d}_{\bf X}(f{\bf Y})= df({\bf X}){\bf Y} +f{\bf d}_{\bf X}{\bf Y}.\eqno(III)
$$
From the property $(I)$
we see that such covariant derivative can be written as the result of a
vector valued 1-form ${\bf d} {\bf Y}$ applied to ${\bf X}$ i.e.
$$
 {\bf d}_{\bf X} {\bf Y}=\langle {\bf d} {\bf Y},{\bf X}\rangle
\equiv {\bf d} {\bf Y}({\bf X}). 
$$
${\bf d} {\bf Y}$ is called the covariant differential.
From the property $(III)$
we have that 
$$
{\bf d} (f{\bf Y}) = df {\bf Y} + f {\bf d} {\bf Y}.
$$
Writing ${\bf Y}= {\bf e}_a Y^a$ we have
$$
{\bf d} {\bf Y} = {\bf d} {\bf e}_a ~~Y^a + {\bf e}_a dY^a.
$$
Thus the covariant derivative is completely defined by the 1-forms
$\Gamma^b_{~a}$
$$
{\bf d} {\bf e}_a = {\bf e}_b \Gamma^b_{~a};~~~~\Gamma^b_{~a}\in \Lambda^1.
$$
By calling $e^b$ the basis in the dual space dual to ${\bf e}_a$ i.e. 
$$
e^b({\bf e}_a)=\delta^b_a
$$
we can write
$$
\Gamma^b_{~a} = \Gamma^b_{~ac}e^c
$$
or equivalently
$$
\Gamma^b_{~ac} =\langle\Gamma^b_{~a},{\bf e}_c\rangle
\equiv  \Gamma^b_{~a}({\bf e}_c).
$$
We have
$$
{\bf d} {\bf Y} = {\bf d} {\bf e}_a ~~Y^a + {\bf e}_a dY^a=
{\bf e}_a(dY^a+\Gamma^a_{~b} Y^b)\equiv{\bf e}_a \nabla Y^a
$$
where $\nabla Y^a$ is defined by
$$
\nabla Y^a \equiv dY^a+\Gamma^a_{~b} Y^b\in\Lambda^1_{(1,0)}.
$$

\bigskip

References

\smallskip

[1] [HawkingEllis] Chap. 2

\section{Transformation properties of the connection}

We have imposed ${\bf d}_{\bf X}{\bf Y} $ to be a vector field, or
equivalently ${\bf d} {\bf Y}$ to be a vector valued 1-form. This imposes the
transformation properties of the connection components
under local changes of the frame ${\bf e}_a$.
Under ${\bf e}_b ={\bf e}'_a\Omega^a_{~b}$ with $\Omega \in GL(n,R)$ we
have 
$$
{\bf d}{\bf e}_b\equiv {\bf e}_a \Gamma^a_{~b}= {\bf e}'_a
\Gamma'^a_{~c}~\Omega^c_{~b}+ {\bf e}'_a d\Omega^a_{~b}= {\bf e}'_a
\Omega^a_{~c} \Gamma^c_{~b} 
$$
i.e.
\begin{equation}\label{affinetr}
\Gamma = \Omega^{-1}d\Omega +\Omega^{-1}\Gamma'\Omega.
\end{equation}
or
\begin{equation}
\Gamma' = \Omega d\Omega^{-1} +\Omega\Gamma\Omega^{-1}.
\end{equation}
Writing
$$
\Gamma^a_{~b}=\Gamma^a_{~bc} e^c
$$
we have
\begin{equation}\label{transfconn}
\Gamma^a_{~bc}=  (\Omega^{-1})^a_{~d} d{\Omega}^d_{~b}({\bf e}_c)
+(\Omega^{-1})^a_{~d}{\Gamma'}^d_{~fk} {\Omega}^f_{~b}  \Omega^k_{~c}. 
\end{equation}
In coordinate base under a change of coordinates we have 
$$
{\bf u}_\mu
= {\bf u}'_\nu \Omega^\nu_{~\mu} = {\bf u}'_\nu\frac{\partial
x'^\nu}{\partial x^\mu}~~~{\rm and}~~~\Omega^\mu_{~\nu}=
\frac{\partial x'^\mu}{\partial x^\nu}.
$$
Substituting in Eq.(\ref{transfconn}) we obtain the familiar
transformation of the connection in the coordinate frame
where the argument of $\Gamma^a_{~bc}$ on the l.h.s. is $x$ 
and the argument of ${\Gamma'}^d_{~fk}$ on the r.h.s. is $x'$ and
they represent the same event.

Notice that for the Lorentz transformation of Chapter 1, 
$\Lambda = \Omega = {\rm const}$.

\section{Parallel transport}

Given a path $\gamma(t)$ and a vector ${\bf Y}(\gamma(t))$ defined along
$\gamma$ we 
say that the vector ${\bf Y}$ is parallel transported along $\gamma(t)$ if,
being ${\bf X}(t)$ the tangent vector to $\gamma(t)$, we have ${\bf d} {\bf
Y}({\bf X}(t)))=0$ for all $t$ i.e. 
$$
X^\nu \partial_\nu Y^\mu +  \Gamma^\mu_{~\beta\nu}Y^\beta X^\nu =0
$$
It appears that the above expression involves values of $Y^\mu$
outside the curve 
$\gamma(t)$. Thus one extends the vector ${\bf Y}$ to a smooth field
in a neighborhood of $\gamma(t)$. We have
$$
X^\nu \partial_\nu Y^\mu = \frac{dY^\mu(\gamma(t))}{dt}
$$
showing that such term does not depend on the particular extension.

Giving the connection at a point $p$ is equivalent to giving the
parallel transport of a frame of $n$ independent vectors ${\bf
Y}_{(k)}$ in $n$ 
independent directions starting from $p$
$$
dY^a_{(k)}({\bf e}_b) + \Gamma^a_{~c}({\bf e_b}) Y^c_{(k)}=0
$$
which can be solved as
$$
\Gamma^a_{~c}({\bf e_b}) = - dY^a_{(k)}({\bf e}_b) R^{(k)}_{~c}
$$
being $R^{(k)}_{~c}$ the inverse matrix of $Y^c_{(k)}$ which exists
being the vectors ${\bf Y}_{(k)}$ independent. These result are
immediately extended to a general gauge theory by 
replacing the tangent space $T_p$ with a general $N$ dimensional
vector space $V$.

\section{Geodesics}

A curve $\gamma(t)$ is called a geodesic curve if its tangent vector
${\bf X}(\gamma(t))$ has along $\gamma(t)$ a covariant derivative
proportional to ${\bf X}(\gamma(t))$ itself i.e.
$$
{\bf d}{\bf X}({\bf X}) = \alpha(\gamma(t)) {\bf X}.
$$
In component form
$$
\nabla X^a({\bf X}) = dX^a({\bf X}) + \Gamma^a_{~b}({\bf X}) X^b=
\alpha(\gamma(t)) X^a 
$$
without any commitment up to now about the reference frame in the
tangent space.
Using the coordinate basis ${\bf u}_\mu$ we have
$$
X^\nu \partial_\nu X^\mu + \Gamma^\mu_{~\beta\nu}X^\beta X^\nu =
\alpha(\gamma(t)) X^\mu 
$$
If we write $\gamma(t)$ in the form $x(t)$ and $\displaystyle{X^\mu =
\frac{dx^\mu}{dt}}$ we have
$$
\frac{d^2
x^\mu}{dt^2}+\Gamma^\mu_{\beta\nu}\frac{dx^\beta}{dt}\frac{dx^\nu}{dt}=
\alpha(t)\frac{dx^\mu}{dt}.
$$
The r.h.s can be put to zero by a change in the parametrization
$\gamma_s(s)= \gamma(t(s))$.
In fact we obtain
$$
\frac{d^2
x^\mu}{ds^2}+\Gamma^\mu_{\beta\nu}\frac{dx^\beta}{ds}\frac{dx^\nu}{ds}=
\left(\alpha(t(s)) \big(\frac{dt}{ds}\big)^2
+ \frac{d^2t}{ds^2}\right)\frac{dx^\mu}{dt}.
$$
Going over to the inverse function, using
$s=s(t(s)),~1=s'(t(s))t'(s),~0=
s''(t(s)) (t'(s))^2 + s'(t(s)) t''(s)$ we have to solve
$$
\alpha(t) = \frac{s''(t)}{s'(t)}
$$
whose general solution is
$$
s(t) = c_0 +c_1\int_0^t dt' \exp\left(\int_0^{t'}\alpha(t'') dt''\right).
$$
Thus $s$ is defined up to an affine transformation and as such called the
affine parameter. A geodesic curve parametrized by an affine parameter
is called simply a geodesic.

\section{Curvature}

Take the exterior covariant differential of ${\bf d} {\bf e}_a$

$$
{\bf d}{\bf d} {\bf e}_a={\bf d}({\bf e}_b \Gamma^b_{~a})= 
{\bf e}_b d \Gamma^b_{~a} + {\bf e}_c\Gamma^c_{~b}\wedge
\Gamma^b_{~a}\equiv
{\bf e}_bR^b_{~a}
$$
$R^b_{~a}\in\Lambda^2$ is the curvature 2-form.
In matrix form
$$
R = d\Gamma + \Gamma\wedge \Gamma
$$
(Notice: being $\Gamma$ a matrix $\Gamma \wedge\Gamma $ is not necessarily
zero; similarly being ${\bf d}$ the covariant differential
${\bf d}{\bf d}$ is not necessarily zero).
Given a vector ${\bf Y} = {\bf e}_a Y^a$ we have
$$
{\bf d}{\bf d} ({\bf e}_a Y^a) ={\bf d}({\bf e}_b \Gamma^b_{~a}Y^a +
{\bf e}_a dY^a)= 
$$
$$
=
{\bf e}_a\Gamma^a_b\wedge\Gamma^b_c Y^c+ {\bf e}_b ~d \Gamma^b_{~a} Y^a
- {\bf e}_b \Gamma^b_{~a}\wedge dY^a + {\bf e}_a \Gamma^a_{~b} 
\wedge dY^b={\bf e}_bR^b_{~a} Y^a
$$
thus the curvature $R^a_{~b}$ is a linear mapping of $T_p$ into itself
which does not depend on the field ${\bf Y}$.

We derive now the transformation properties of the curvature. 
Given 
${\bf e}_a = {\bf e}'_b\Omega^b_{~a}$ 
with $\Omega\in GL(n)$ using the above formula we
have 
\begin{eqnarray}
{\bf d}{\bf d} {\bf e}_a &\equiv&  {\bf e}_b R^{b}_{~a} = {\bf
e}'_c\Omega^c_{~b} R^{b}_{~a} = 
{\bf d}{\bf d}({\bf e}'_b\Omega^b_{~a})= 
{\bf d}({\bf d}{\bf e}'_b~\Omega^b_{~a}+ {\bf e}'_b ~d\Omega^b_{~a})\\
&=& {\bf e}'_c~ R'^c_{~b}~\Omega^b_{~a}- {\bf d}{\bf e}'_b \wedge
d\Omega^b_{~a}+ {\bf d}{\bf e}'_b \wedge d\Omega^b_{~a}+ {\bf
e}'_b~dd\Omega^b_{~a} =    {\bf e}'_c~ R'^c_{~b}~\Omega^b_{~a} 
\end{eqnarray}
i.e.
$$
R = \Omega^{-1}R'~\Omega~~~~~~~~{\rm or}~~~~~~~~ R'=\Omega~ R~\Omega^{-1}
$$
and thus $R^a_{~b}$ are the components of a (1,1) tensor $2$-form.

\section{Torsion}

We shall introduce torsion formally and then go over to its physical
meaning. The torsion $S^a$ is a (1,0) valued 2-form defined by
$$
S^a =\nabla e^a.
$$
We recall that being $e^a({\bf e}_b) = \delta^a_b$, $e^a$ transforms
under local changes of frames according to $\Omega$ i.e. like the
components of a vector: $e'^a = \Omega^a_{~b}e^b$. Thus the
explicit expression of $S^a$ is 
$$
S^a =\nabla e^a = de^a+\Gamma^a_{~b}\wedge e^b \in \Lambda^2_{(1,0)}.
$$
We have
\begin{eqnarray}
S'^a &\equiv& \nabla e'^a = de'^a +\Gamma'^a_{~b}\wedge e'^b=
d[\Omega^a_{~b} e^b] +\Gamma'^a_{~b}\wedge e'^b \\
&=& d\Omega^a_b~\wedge e^b +
\Omega^a_{~b} d e^b + 
(\Omega~d \Omega^{-1}~\Omega)^a_{~b}\wedge e^b 
+\Omega^a_{~b} \Gamma^b_{~c} \wedge e^c\\
&=& \Omega^a_{~b} S^b
\end{eqnarray}
or $S'=\Omega S$ i.e $S^a$ transforms like the components of a vector.
\bigskip

Given an $n$ component field  $W_a$ which transforms like $W_a =
W'_b\Omega^b_{~a}$ we shall call it of type $(0,1)$ while those $V^a$
which transform like $V'^a = \Omega^a_{~b} V^b$ we shall call
of type $(1,0)$. We shall define $\nabla$ acting on $W_a$, by imposing
the validity of Leibniz rule $\nabla(W_a V^b) = \nabla(W_a) V^b
+W_a\nabla(V^b)$, linearity and that on the invariant function $W_a
V^a$ we have  $\nabla(W_a V^a) = d(W_a V^a)$. As a result
$$
\nabla W_a = d W_a - W_b \Gamma^b_{~a}.
$$
We can extend such definition to the case in which $W_a$ and $V^a$ are
differential form by imposing
$$
\nabla(W_a\wedge V^a) = d(W_a\wedge V^a)
$$
We cannot write for the l.h.s.
$$
\nabla(W_a\wedge V^a) = \nabla W_a \wedge V^a + W_a \wedge \nabla V^a
$$
because this is inconsistent with
$$
d(W_a\wedge V^a) = d W_a \wedge V^a + (-)^w W_a \wedge d V^a
$$
being $w$ the order of $W_a$, when e.g. the connection is identically
zero. Thus we shall write
$$
\nabla(W_a\wedge V^a) = \nabla W_a \wedge V^a + (-)^w W_a \wedge \nabla V^a
$$
from which we obtain
$$
\nabla W_a = dW_a - (-)^w W_b \wedge \Gamma^b_{~a}=dW_a -
\Gamma^b_{~a} \wedge W_b
$$
being $\Gamma^b_{~a}$ a one form.

By taking linear combinations of fields $W_b V^a$ one extends the above
rules to tensors $T^a_b$ and actually to any tensorial form of any order 
$T^{a_1,\dots a_m}_{b_1\dots b_n}\in \Lambda^t_{(m,n)}$ e.g.
$$
\nabla T^a_b = dT^a_b + \Gamma^a_{~c}\wedge T^c_b -(-)^t ~ T^a_c\wedge
\Gamma^c_{~b}
$$
which can be written in matrix form as
$$
\nabla T= dT + \Gamma\wedge T -(-)^t ~ T\wedge \Gamma.
$$

\section{The structure equations}

First Bianchi identity: compute the $\Lambda^3$ form

$$
\nabla R^a_{~b} = \nabla (d\Gamma + \Gamma\wedge\Gamma)^a_{~b} 
$$
$$
\nabla R = dd\Gamma+ d\Gamma\wedge\Gamma - \Gamma\wedge d\Gamma+
\Gamma\wedge R-R\wedge\Gamma=
$$
$$
d\Gamma\wedge\Gamma - \Gamma\wedge d\Gamma+
\Gamma\wedge(d\Gamma+
\Gamma\wedge\Gamma)-(d\Gamma+\Gamma\wedge\Gamma)\wedge\Gamma=0.
$$
These are identities i.e. satisfied by any connection $\Gamma^a_{~b}$.

\bigskip
Second Bianchi identity: take the covariant differential of $S^a$
$$
\nabla  S^a = dde^a + d\Gamma^a_b\wedge e^b - \Gamma^a_b\wedge de^b +
\Gamma^a_b\wedge(de^b+\Gamma^b_{~c}\wedge e^c) = R^a_{~b}\wedge e^b\in
\Lambda^3_{(1,0)}. 
$$
These are identities i.e. satisfied by any connection $\Gamma^a_{~b}$
and any forms $e^a$.

Summarizing in matrix form
$$
\nabla R=0;  ~~~~~\nabla S = R\wedge e.
$$

\section{Non abelian gauge fields on differential
manifolds}\label{4gaugedifferential}

One replaces the tangent space $T_p$ with an abstract $N$
dimensional vector space
$V_p$. The covariant differential is defined similarly;  to avoid
confusion we shall use the symbol ${\bf D}$. If ${\bf v}_m$
is a base of the space $V_p$ 
$$
{\bf D}{\bf v}_m = {\bf v}_n A^n_{~m}
$$
where $ A^m_{~n}$ is the gauge field one-form.
The field strength is defined as the curvature induced by  $A^m_{~n}\in
\Lambda^1$ 
$$
{\bf D}{\bf D}{\bf v}_n = {\bf v}_m F^m_{~n}
$$
$$
F =dA+A\wedge A.
$$
The expanded version is the following: putting $A=A_\mu dx^\mu$ and
$F=\frac{1}{2}F_{\mu\nu} dx^\mu \wedge dx^\nu$ we have
$$
F_{\mu\nu} = \partial_\mu A_\nu - \partial_\nu A_\mu +[A_\mu,A_\nu].
$$
The first Bianchi identity is proven in exactly the same way; denoting
with $D$ the covariant differential acting on the 
components, we have
$$
\Lambda^3  \ni DF =0.
$$
The expanded form of the above equation is
$$
D_\lambda F_{\mu\nu}+~{\rm cyclic}~=0
$$
with
$$
D_\lambda F_{\mu\nu} = \partial_\lambda F_{\mu\nu} +[A_\lambda,F_{\mu\nu}].
$$
Under ${\bf v}_n= {\bf v}'_mU^m_{~n}$ one has 
\begin{equation}\label{4gaugetransf}
A' = UdU^{-1} + UAU^{-1}; ~~~~ F'= UFU^{-1}.
\end{equation}

The above described is the gauge theory of the group $GL(N)$.
We can restrict however the connection to the Lie algebra of 
a less general group $G$ and the gauge transformations to the elements
of $G$.

It is very easy to see that if $U\in G$ both $UdU^{-1}$ and
$UAU^{-1}$ are elements of the Lie algebra ${\cal L}$ of $G$. Consider 
in fact a trajectory $\gamma(t)\in G$ with $\gamma(0)=I$
$$
d\gamma = A \in {\cal L}~.
$$
Then also the trajectory $U\gamma(t)U^{-1} \in G$ goes through $I$ for
$t=0$ and thus we have 
$$
U d\gamma U^{-1} = UAU^{-1} \in {\cal L}~.
$$
Similarly given a $U(t)\in G$ with $U(0)$ not necessarily $I$ 
we have $U(0) U^{-1}(t)\in G$ and $U(0) U^{-1}(0)=I$ 
and thus $U(0)dU^{-1}(0)\in {\cal L}$~.

In the following we shall often make use of the non degeneracy of the trace
form of a faithful representation of a given Lie algebra.

If the group $G$ is compact we know that, as all its representations
are equivalent to unitary representations (i.e. antihermitean
generators), any faithful representation of its Lie algebra has non
degenerate trace form i.e. if ${\rm tr}AB=0$, for any $B$ we have $A=0$.
 
Such non degeneracy holds also for faithful representations of non
compact groups which have a semi-simple Lie algebra. E.g. the non
compact group $SL(2,C)$ is not semi-simple due to the presence of the
invariant subgroup $I,-I$
but a simple calculation shows that the Lie algebra of $SL(2,C)$ is
semi-simple and thus the trace form of a faithful representation of the
Lie algebra of $SL(2,C)$ is non degenerate.

\bigskip

References

\smallskip

[1] N. Jacobson, ``Lie algebras'' Dover Publication Chap.II 


\section{Non compact electrodynamics}

Consider the case in which the vector space is one dimensional on the
reals. Then the general vector is ${\bf V} = {\bf v} V$; the
connection is 
$$
{\bf D}{\bf v} = {\bf v} A
$$
with $A$ real 1-form. The field strength is given by
$$
{\bf D}{\bf D}{\bf v} = {\bf v} (dA+A\wedge A) ={\bf v} dA ={\bf v}F.
$$
Explicitly
$$
F= d A = d(A_\nu dx^\nu) = \partial_\mu A_\nu dx^\mu\wedge dx^\nu =
( \partial_\mu A_\nu- \partial_\nu A_\mu) \frac{dx^\mu\wedge dx^\nu}{2}= 
$$
$$
=F_{\mu\nu}\frac{dx^\mu\wedge dx^\nu}{2}.
$$
Consider the general linear transformation on ${\bf v}$,  ${\bf v}=
{\bf v}'e^{-\Lambda(x)}$. 
We have 
$$
A' = e^{-\Lambda(x)} d e^{\Lambda(x)} + e^{-\Lambda(x)} A e^{\Lambda(x)}
=d\Lambda + A
$$
and
$$
F' = e^{-\Lambda(x)} F e^{\Lambda(x)} = F.
$$
$\Lambda$ is the usual gauge transformation.

\section{Meaning of torsion}

As already stated torsion is a space related concept. Consider two
infinitesimal vectors ${\bf v}$ and ${\bf w}$ $\in T_p$. 
Let ${\bf v}_1$ be the
vector ${\bf v}$ parallel transported along ${\bf w}$. In the coordinate
basis ${\bf v}={\bf u}_\mu v^\mu$ and ${\bf w}={\bf u}_\mu w^\mu$ we
have
$$
v_1^\mu = v^\mu -\Gamma^\mu_{\nu\lambda} v^\nu w^\lambda.
$$
Moving now by ${\bf v}_1$ we reach the point of coordinates $x^\mu_0
+w^\mu+v_1^\mu$. Reversing the process the two final points agree iff
$$
w^\mu+v_1^\mu=w_1^\mu+v^\mu.
$$
If this has to happen for all ${\bf v}$  and ${\bf w}$ we have
$$
\Gamma^\mu_{\nu\lambda}=\Gamma^\mu_{\lambda\nu}.
$$ 
Thus a connection symmetric in the lower indices in the 
coordinate basis is equivalent to the
commutativity of infinitesimal displacements.
Writing $e^a$ in terms of the vierbeins $e^a_\mu$, $e^a \equiv e^a_\mu
dx^\mu$ we have
$$
S^a = \partial_\lambda e^a_\mu~ dx^\lambda \wedge dx^\mu+
\Gamma^a_{~b\lambda} ~dx^\lambda \wedge e^b_\mu dx^\mu.
$$
Vanishing of the torsion means
$$
\partial_\lambda e^a_\mu + \Gamma^a_{b\lambda}e^b_\mu
=\partial_\mu e^a_\lambda +\Gamma^a_{b\mu}e^b_\lambda.
$$
Multiplying both members by $e^\nu_a$ and recalling that ${\bf u}_\mu
={\bf e}_a e^a_\mu$ as seen from $e^a({\bf
u}_\mu)= e^a_\mu $ we have, using Eq.(\ref{transfconn})
$$
\Gamma^\nu_{\mu\lambda}= \Gamma^\nu_{\lambda\mu}.
$$
Thus the vanishing of the torsion implies symmetry of the connection in the
lower indices in any {\it coordinate basis}. This can be more simply
achieved by 
$$
S^\mu=0~~~~{\rm implies}~~~~ 0= \nabla dx^\mu = d dx^\mu+
\Gamma^\mu_{~\nu}\wedge dx^\nu = \Gamma^\mu_{\nu\lambda}~ dx^\lambda
\wedge dx^\nu. 
$$

\begin{figure}[htb]
\begin{center}
\includegraphics{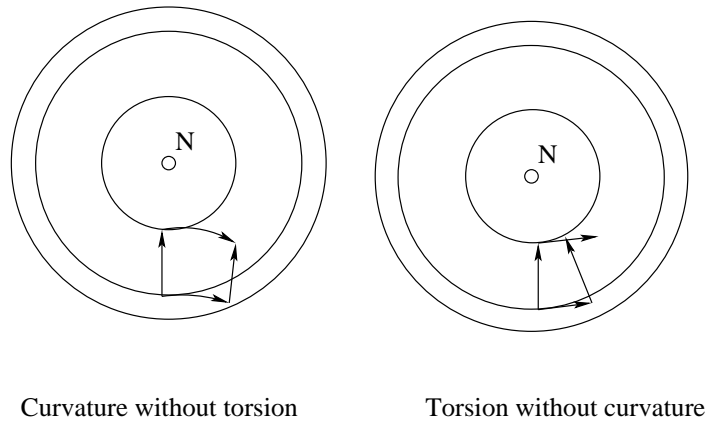}
\end{center}
\caption{Torsion vs. Curvature}
\end{figure}

\section{Vanishing of the torsion}

If the torsion $S^a$ vanishes identically we have from the second
Bianchi identity
$$
0=R^a_{~b}\wedge e^b=\frac{1}{2}R^a_{~bcd}e^c\wedge e^d\wedge e^b
$$
i.e.
$$
R^a_{~[bcd]}=0.
$$

\section{Metric structure}

The metric is a bilinear real symmetric functional on $T_p\times T_p$
$$
g({\bf W},{\bf V})\in R.
$$
Writing ${\bf V} = {\bf e}_a V^a = {\bf u}_\mu V^\mu$ and similarly
for ${\bf W}$ we have 
$$
g({\bf W},{\bf V})= W^a g_{ab}V^b=W^\mu g_{\mu\nu}V^\nu.
$$
Under ${\bf e}_a = {\bf e}'_b \Omega^b_{~a}$ we have
$$
g = \Omega^T g'\Omega ~~~~~{\rm or}~~~~~g' = {\Omega^{-1}}^T g\Omega^{-1}~.
$$
All manifolds (Hausdorf, paracompact) admit a positive definite
metric. Necessary and sufficient condition for a manifold to support a
Lorentzian metric is that it supports a smooth never vanishing line
element. All non compact manifolds support a Lorentzian metric.

\bigskip

References

\smallskip

[1] [HawkingEllis] Chap. 2

\section{Isometries}

Given a mapping $\alpha$ between two manifolds $M$ and $M'$
i.e. $M -{\alpha}\rightarrow M'$ or a
diffeomorphism mapping the manifold in itself, and given a metric on
$M'$ it is possible to pull back the metric from $M'$ to $M$ as we did
with the one-forms. We shall again denote the pulled back metric  by
$\alpha^*g$. If $M=M'$ we shall call $\alpha$ an isometry if
$\alpha^*g(p) = g(p)$. If we have an abelian group of diffeomorphisms
$\phi_t$ generated by the vector field ${\bf X}$ we can compute the
Lie derivative of $g$
$$
L_{\bf X}~g(p) = \lim_{t\rightarrow 0}\frac{1}{t}(\phi_t^*g(p) -g(p)).
$$
If $\phi_t$ is a group of isometries we have $L_{\bf X}~g(p)=0$ and
${\bf X}$ is called a Killing vector field. The explicit form of
$\alpha^*g(p)$ in the coordinate basis is, with $x'=\alpha(x)$
$$
\alpha^*g_{\mu\nu}(x) = \frac{\partial x'^\rho}
{\partial x^\mu} g_{\rho\sigma}(x') \frac{\partial
x'^\sigma}{\partial x^\nu}~.
$$

\section{Metric compatibility}

We impose that parallel transported vectors maintain the value of
their scalar product. The result is
\begin{equation}\label{metriccomp}
0= dg -\Gamma^T g - g \Gamma.
\end{equation}
The above equation can also be read as
$$
\nabla g_{ab} = dg_{ab} - \Gamma^c_{~a} g_{cb} - g_{ac}\Gamma^c_{~b}=0
$$
i.e. the vanishing of the covariant differential.

By taking the differential of Eq.(\ref{metriccomp}) and using again
Eq.(\ref{metriccomp})
we have
$$
0 = - d\Gamma^T g + \Gamma^T \wedge dg - dg\wedge \Gamma  - g d\Gamma=
- d\Gamma^T g + \Gamma^T \wedge (\Gamma^T g + g\Gamma)- (\Gamma^T g +
g\Gamma)\wedge \Gamma  - g d\Gamma= 
$$
\begin{equation}\label{Rcompatibility}
=- R^T g -g R~.
\end{equation}
Written explicitly
$$
g_{ac}R^c_{~b}+R^c_{~a}g_{cb}=R_{ab}+R_{ba}=0~.
$$

\section{Contracted Bianchi identities}

The Bianchi identities for the curvature 2-from $R^a_b\in\Lambda^2_{(1,1}$
$$
\nabla R^a_{~b}=0
$$
can be expanded as
$$
0=\nabla (R^a_{~bcd}e^c\wedge e^d)= \nabla(R^a_{~bcd})e^c\wedge
e^d+R^a_{~bcd}\nabla e^c\wedge e^d-R^a_{bcd}e^c\wedge\nabla e^d  ~.
$$
In absence of torsion i.e $S^a = \nabla e^a=0$ we have 
$$
\nabla(R^a_{~bcd}) e^c\wedge e^d =
\nabla_f(R^a_{~bcd}) e^f\wedge e^c\wedge e^d =0
$$
i.e.
$$
\nabla_f R^a_{~bcd}+\nabla_c R^a_{~bdf}+\nabla_d R^a_{~bfc}=0~.
$$
The Ricci tensor is defined by
$$
R_{bd}=R^a_{~bad}~.
$$
In presence of metric and metric compatibility we can contract indices to have
$$
0= \nabla_f R^a_{~bad}+\nabla_a R^a_{~bdf}+\nabla_d R^a_{~bfa}=
\nabla_f R_{~bd}+\nabla_a R^a_{~bdf}-\nabla_d R_{bf}
$$
and contracting again
$$
0= \nabla_f R -\nabla_a R^a_{~f}-\nabla_b R^b_{~f}=
\nabla_f R -2\nabla_a R^a_{~f}
$$
with $R= R_{bd}~g^{bd}$ the Ricci scalar. These are the contracted Bianchi 
identities.

\section{Orthonormal reference frames}

In presence of a metric we can choose as basis vectors on the tangent
space, an orthonormal frame for each point, i.e. ${\bf
g}({\bf e}_a,{\bf e}_b) = \eta_{ab}$. Metric compatibility now becomes
$$
\Gamma^T\eta + \eta\Gamma = d\eta=0
$$
or
$$
\Gamma = -\eta^{-1}\Gamma^T\eta~~~~{\rm i.e.}~~~~\Gamma\in so(3,1).
$$
Also from Eq.(\ref{Rcompatibility})
$$
R = -\eta^{-1}R^T\eta~~~~{\rm i.e.}~~~~R\in so(3,1).
$$

Thus $\Gamma$ and $R$ belong to the algebra of $SO(3,1)$. Now the changes of
frame are elements of the Lorentz group, ${\bf e}_a={\bf e}'_b
\Omega^b_{~a}$, 
with $\Omega^T\eta\Omega =\eta$ and we have
$$
\Gamma'= \Omega d \Omega^{-1}
+\Omega\Gamma\Omega^{-1},~~~~R'=\Omega R~\Omega^{-1}~. 
$$

\section{Compact electrodynamics}

We consider the case of gauge theory in which space $V_p$ is
$1$-dimensional on the complex.
Given the
basic vector ${\bf v}$, all vectors are given by ${\bf v} V$ with
$V\in C$. Choosing ${\bf v}$ normalized we have
$$
({\bf W},{\bf V}) = W^* V.
$$ 
The connection is given by
$$
{\bf D}{\bf v} = {\bf v} A
$$
with $A$ complex valued 1-form. Imposition of ``metric compatibility''
gives
$$
W'^*V' = (W- \varepsilon A({\bf X}) W)^*(V- \varepsilon A({\bf X}) V) +
O(\varepsilon^2) 
= W^* V
$$
for any ${\bf X}\in T_p$. As a consequence
$$
A^* + A=0
$$
i.e. $A$ is a pure imaginary 1-form. Thus it is useful to change the
notation for the connection to
$$
{\bf D}{\bf v} =  {\bf v}i k A
$$
with $k$ real constant and $A$ real 1-form. Keep in mind that now the
connection is $ikA$.
The gauge transformation now are ${\bf v} = {\bf v}'~ e^{-i k\Lambda}
\equiv {\bf v}'~U$
$$
ikA' = e^{-ik\Lambda} d e^{ik\Lambda} + e^{-ik\Lambda} ikA e^{ik\Lambda}=
i k d\Lambda + ik A~~~
$$
i.e. $A'=A+d\Lambda$. Define now $F$ as
$$
{\bf D}{\bf D}{\bf v}={\bf v}ikF= ik (dA+A\wedge A)~~~~{\rm i.e.}~~~~F=dA~.
$$
The gauge group in now compact i.e. for $\Lambda =2\pi/k$
we have the identity transformation on the vector space $V_p$. We
recall now the transformation properties of the Schr\"odinger
wave function describing a particle of charge $e$
$$
\psi(x) = e^{-\frac{ie}{\hbar c}\Lambda(x)}\psi'(x).
$$ 
For $\Lambda = 2\pi/k$ we must have the identity transformation i.e.
$$
\frac{e}{k\hbar c}= n
$$
with integer $n$ which implies the all charges are integer multiple of a
fundamental charge $k \hbar c$.


\bigskip

References

\smallskip

[1] T.T. Wu, C.N. Yang, 	
``Concept of nonintegrable phase factors and global formulation of gauge
fields'', Phys. Rev. D12 (1975) 3845

\section{Symmetries of the Riemann tensor}\label{4symmetriesofRiemann}
\begin{equation}\label{antisymmetry}
R^a_{bcd}=-R^a_{bdc}
\end{equation}
is simply due to the definition.
We have $n^3(n-1)/2$ independent components.
In the torsionless case we have
\begin{equation}\label{zerotorsion}
R^a_{[bcd]}=0
\end{equation}
and we have to subtract to the above number $n^2(n-1)(n-2)/3!$ which
is the number of independent identities given by
Eq.(\ref{zerotorsion}), to obtain 
$n^2(n^2-1)/3$ independent components.
Metric compatibility gives
\begin{equation}\label{metriccompatibility}
R_{abcd} = -R_{bacd}~.
\end{equation}
If we combine Eq.(\ref{antisymmetry}), Eq.(\ref{zerotorsion}) with
Eq.(\ref{metriccompatibility}) we have the further symmetry
\begin{equation}\label{exchangesymmetry}
R_{abcd} = R_{cdab}.
\end{equation}
In fact let us define
$$
S_{abcd}=R_{abcd} +R_{acdb} +R_{adbc} 
$$
which are identically zero as a consequence of the second Bianchi
identities in absence of torsion. 
We have identically
$$
0= S_{abcd}-S_{bcda}-S_{cdab}+S_{dabc} = 2R_{abcd}-2R_{cdab}~. 
$$
Eq.(\ref{exchangesymmetry}) has been derived without exploiting
the explicit form of the connection.
We can now compute the  number of independent components of the Riemann
tensor in presence of metric compatibility and absence of torsion. 
The components with only $2$ different indices are
$n(n-1)/2$; with three different indices we have $n$ choices for the
double index and $(n-1)(n-2)/2$ choices for the other two thus giving
$n(n-1)(n-2)/2$; with $4$  different indices we have
$n(n-1)(n-2)(n-3)/4!$ choices which should be multiplied by $3$ due
to the three different pairings. However due to Eq.(\ref{zerotorsion})
only $2$ of the three pairings are independent. Notice that the second
Bianchi identity does not play any role when only two or three indices
are different as they are identically satisfied. Summing the
contributions  we have $n^2(n^2-1)/12$.

\bigskip

\section{Sectional curvature and Schur theorem}

We refer to the {\it torsionless and metric compatible} case in which the
Riemann tensor has the symmetries
$$
R_{abcd}=-R_{bacd}=-R_{abdc}=R_{cdab}~.
$$

By simple algebra [1] from the knowledge of 
$$
R_{abcd} v_1^a v_2^b v_1^c v_2^d
$$
one determines completely $R_{abcd}$.
The quantity
\begin{equation}\label{sectional}
\frac{R_{abcd} v_1^a v_2^b v_1^c v_2^d}{{\bf v}_1\cdot {\bf v}_1
~~{\bf v}_2\cdot 
{\bf v}_2 - ({\bf v}_1\cdot {\bf v}_2)^2}
\equiv K(p,{\rm plane}) 
\end{equation}
with ${\bf v}_i\cdot {\bf v}_j = v^a_ig_{ab}v^b_j$
is invariant under linear invertible substitutions of ${\bf v}_1$ and
${\bf v}_2$. In fact given the transformation ${\bf w}_j = {\bf v}_l
a^l_j$, 
due to the antisymmetry in the first two indices we have
$$
R_{abcd} w_1^a w_2^b w_1^c w_2^d = \det(a) R_{abcd} v_1^a v_2^b
w_1^c w_2^d  
$$ 
and similarly for the second pair of indices. The denominator can be
written as
$$
\varepsilon_{ij}({\bf v}_i\cdot {\bf v}_l) ({\bf v}_j\cdot {\bf v}_k) =
\varepsilon_{lk}[({\bf v}_1\cdot {\bf v}_1) ({\bf v}_2\cdot {\bf
v}_2) -({\bf v}_1\cdot {\bf v}_2) 
({\bf v}_2\cdot {\bf v}_1)]    
$$
and thus
$$
\varepsilon_{ij}({\bf w}_i\cdot {\bf w}_l) ({\bf w}_j\cdot {\bf w}_k) =
\det(a) \varepsilon_{ij}({\bf v}_i\cdot {\bf w}_l) ({\bf v}_j\cdot {\bf w}_k) =
$$
$$
=(\det(a))^2
[({\bf v}_1\cdot {\bf v}_1) ({\bf v}_2\cdot {\bf v}_2) -({\bf
v}_1\cdot {\bf v}_2) 
({\bf v}_2\cdot {\bf v}_1)] \varepsilon_{lk}.
$$
$K(p,{\rm plane})$ is called the sectional curvature;  it is a functional of
a plane. In the case of positive definite metric, the denominator in
Eq.(\ref{sectional}) (Gram determinant) is non vanishing if the two
vectors ${\bf v}_1,~{\bf v}_2$  are independent. This is not
necessarily true for Lorentz metric. In this case we define the
sectional curvature 
only for those pairs of vectors which have non vanishing Gram
determinant. However any couple of vectors which have a vanishing Gram 
determinant can be reached continuously by pairs of vector with non 
vanishing Gram determinant.
    
If at a point $p$ 
$$
\frac{R_{abcd} v^a_1 v^b_2 v^c_1 v^d_2}{{\bf v}_1\cdot {\bf v}_1
~~{\bf v}_2\cdot 
{\bf v}_2 - ({\bf v}_1\cdot {\bf v}_2)^2} =c(p)  
$$
i.e. $K$ is independent of the plane in the case of positive definite
metric, and independent of the plane with non vanishing Gram
determinant in the case of Lorentz metric, we have due to the previous 
result
$$
R_{abcd}= c(p)(g_{ac}g_{bd}- g_{ad}g_{bc})
$$
which can be rewritten also as
\begin{equation}\label{constsect}
R_{abcd}= \frac{R(p)}{n(n-1)}(g_{ac}g_{bd}- g_{ad}g_{bc})
\end{equation}
and the Ricci tensor is
$$
R_{bd} =\frac{R(p)}{n}g_{bd}~.
$$
Computing the contracted Bianchi identity of Eq.(\ref{constsect}) we have
$$
0=\nabla_a(\frac{R}{n}\delta^a_b-\frac{1}{2}R \delta^a_b) =
(\frac{1}{n}-\frac{1}{2}) \nabla_b R .
$$
Thus we have Schur theorem: For $n>2$,
if $K$ is independent of the plane $K$ does not depend on the
point.

Such  manifolds are said to have {\it constant curvature}.

Notice that the proof of Schur theorem fails for $n=2$. In two
dimensions for each point there is only one plane and the curvature scalar
can well depend on the point.

\bigskip

References

\smallskip

[1] S. Kobayashi, K. Nomitzu ``Foundations of differential
geometry'', J. Wiley\& Sons, New York, vol. I p. 198

\section{Spaces of constant curvature}

The spaces of constant curvature i.e. those for which
$$
R_{abcd}= k (g_{ac}g_{bd}- g_{ad}g_{bc})
$$
coincide with the maximally symmetric spaces, which in $4$ dimensions
with signature $(-+++)$ are Minkowski, de Sitter and anti-de Sitter.
They admit the maximum number of Killing vectors i.e. $n(n+1)/2$
independent Killing vectors.

\section{Geometry in diverse dimensions}

$n=2$

The Riemann tensor has only one independent component; set in a well
defined frame and at the point $p$
$$
R_{1212} = c(p) (g_{11}g_{22}-g_{12}g_{21})
$$
$c(p)$ is well defined as $\det g\neq 0$.
The tensor
$$
T_{abcd} = c(p) (g_{ac}g_{bd}-g_{ad}g_{bc})
$$
has all the component equal to those of $R_{abcd}$ at the point $p$ in
the given frame and thus in all frames. We have
$$
R_{abcd} = \frac{R(p)}{2} (g_{ac}g_{bd}-g_{ad}g_{bc}).
$$
The Ricci tensor is
$$
R_{bd} = \frac{R(p)}{2} g_{bd}.
$$
\bigskip

$n=3$

Here the Riemann tensor has 6 independent components and the Ricci
tensor also 6 independent components.
We can write
$$
R_{abcd} = \epsilon_{abk}\epsilon_{cdl}F^{kl}
$$
where $F^{kl}$ is a symmetric tensor. We have
$$
g^{ac}\epsilon_{abk}\epsilon_{cdl} = - g_{bd}g_{kl}+ g_{dk}g_{bl}
$$
with $\epsilon_{cdl}=\sqrt{-g}\varepsilon_{cdl}$ and
$\varepsilon_{cdl}$ the usual antisymmetric symbol (see Section 4.26).
Thus we have
$$
R_{bd} = F_{bd} -g_{bd} F; ~~~~ R= - 2 F
$$
from which
$$
R_{abcd} = \epsilon_{abk}\epsilon_{cdl}G^{kl}
$$
being $\displaystyle{G^{kl}= R^{kl} - \frac{g^{kl}}{2} R}$ the
Einstein tensor.  The Weyl tensor (see below) vanishes 
identically in $n=3$.

\bigskip
$n\geq 4$

We can write
$$
R_{ijkl} = W_{ijkl} -\frac{2}{n-2}(g_{i[l} R_{k]j}+g_{j[k} R_{l]i})-
\frac{2}{(n-1)(n-2)}R~g_{i[k}g_{l]j}
$$
being $W_{ijkl}$ the Weyl tensor. $W_{ijkl}$ is traceless and
$W^i_{~jkl}$ is invariant under Weyl transformations $g_{ij}\rightarrow
e^{2f} g_{ij}$ (see Section 4.25).

\section{Flat and conformally flat spaces}

A manifold with metric $g_{ij}$, is said to be flat if there
exists a coordinate system such that in such a system the metric takes
the form $ \hat g_{ij}$ with $\hat g_{ij}$ a constant metric.

In Section 4.23 it is proven that, with a torsionless and metric
compatible connection, necessary and sufficient condition for a space
to be locally flat is the vanishing of the Riemann tensor.

\bigskip

A manifold with metric $g_{j}$, is said to be conformally flat if there
exists a 
coordinate system such that in such a system the metric takes the form
$$
g_{ij}= e^{2f} \hat g_{ij}
$$
with $\hat g_{ij}$ a constant metric. 

In Section 4.24 it is proven the Weyl-Schouten theorem:

For $n \geq 4$ necessary and sufficient condition
for a metric to be locally conformally flat is the vanishing of the Weyl
tensor. 

The necessity of the condition is simple to prove. Suppose that by
means of a diffeomorphism we are able to bring the metric in the form
$$
g_{ij}(x)=e^{2f}\hat g_{ij}
$$
with $\hat g_{ij}$ constant. We have $\hat R_{ijkl}=0$,
$\hat R_{jl}=0$, $\hat R=0$ and thus $\hat W^i_{~jkl}=0$. But
$W^i_{~jkl}$ is invariant under Weyl transformations (see Section
4.25) and thus $W^i_{~jkl}=0$ and being $W$ a tensor it
implies the vanishing of the original Weyl tensor. 
For the sufficient condition see Section 4.25.

\bigskip

For $n=3$ the Weyl tensor vanishes identically and necessary and
sufficient condition for a geometry to be locally conformally flat
is the vanishing of the Cotton tensor as proven in Section 4.25.
 
\bigskip

In dimension $2$ the curvature transform under a Weyl
transformation as (see Section 4.24) 
$$
\tilde R = e^{-2f}(R - 2 \Delta f)~. 
$$
Thus by solving locally the equation 
\begin{equation}\label{4laplace}
\Delta f = \frac{R}{2}
\end{equation}
we obtain the vanishing of the Riemann tensor and thus we have 
local conformal
flatness. Eq.(\ref{4laplace}) can always be locally solved. 
Given a point $p$ we consider a $C^\infty$ function $\rho(x)$
with
compact support which around $p$ is identically $1$ and the solution
is given by
$$
f(x) = \frac{1}{8\pi}\int \rho(x') R(x')  \log[(x_1-x'_1)^2+
(x_2-x'_2)^2]d^2x'~.
$$

At the global level in $n=2$ for
genus $g=0$ i.e. the topology of the sphere we can reduce the metric
to $g_{ij} = \delta_{ij} e^{2f}$; for $g=1$ i.e. the torus topology,
we can reduce the metric to $g_{ij} = \hat g _{ij} e^{2f}$ where $\hat
g$ is a constant metric which depends on two parameters; for $g>1$ we
can reduce the metric to $g_{ij} = \hat g _{ij} e^{2f}$ where $\hat g$
is a metric of constant negative curvature taken equal to $-1$ and
$\hat g$ depends on $6 g-6$ parameters.


\begin{figure}[htb]
\begin{center}
\includegraphics{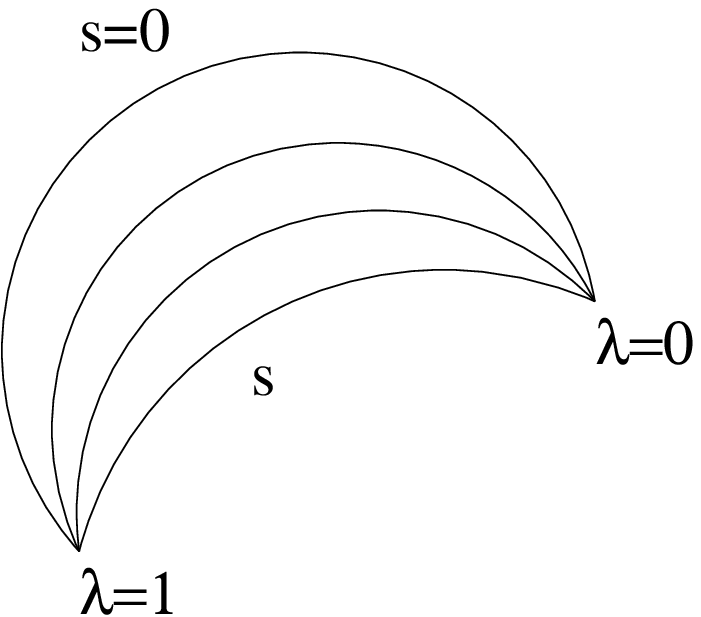}
\end{center}
\caption{The non abelian Stokes theorem-1}
\end{figure}

\begin{figure}[htb]
\begin{center}
\includegraphics{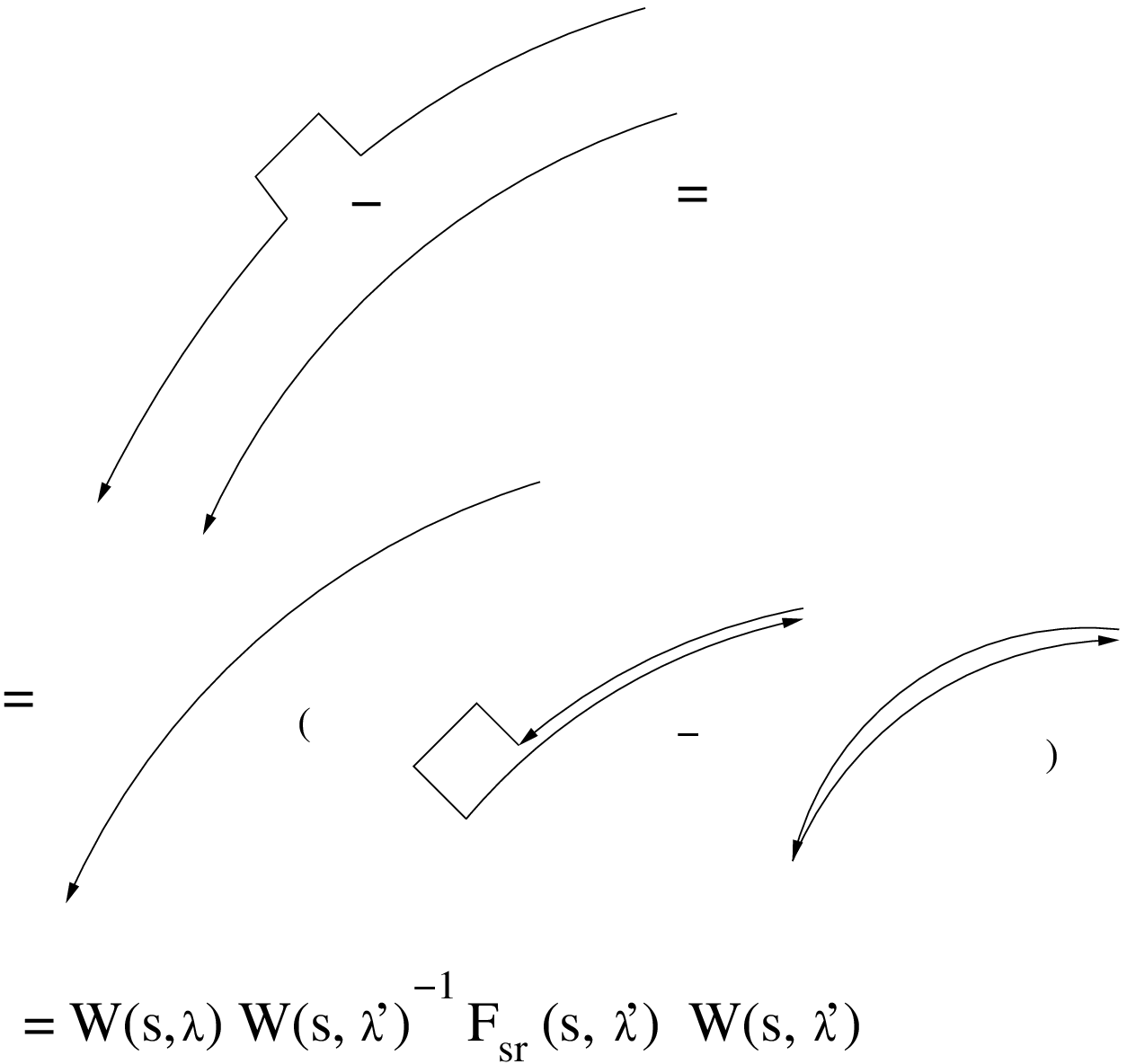}
\end{center}
\caption{The non abelian Stokes theorem-2}
\end{figure}

\section{The non abelian Stokes theorem}

Given a family of paths $x(s,\lambda)$ (see fig. 4.2) 
we want to write the variation
of the line integral $W(s,1)$, called the 
Wilson line,
$$
W(s,\lambda) = {\rm Pexp}\left(-\int_0^\lambda A_\lambda(s,\lambda')
d\lambda'\right)=
{\rm Pexp}\left(-\int_0^\lambda A_\mu(x(s,\lambda')) \frac{dx^\mu}{d\lambda'}
d\lambda'\right)
$$
when the parameter $s$ changes from $0$ to $s$ and the end points are
fixed, i.e. $x(s,0)=x_0$ $x(s,1)=x_1$, as a surface integral containing
the field strength $F_{s\lambda}$.
It is possible to consider also the case in which the end points are
moving in $s$.

We recall that the Wilson line $W(s,\lambda)$ provides the parallel
transport of a vector along the line $s={\rm const.}$
$$
v(s,\lambda) = W(s,\lambda) v(s,0)
$$
as $W(s,\lambda)$ satisfies the equation
\begin{equation}\label{4Weq}
\frac{\partial W(s,\lambda)}{\partial \lambda} =
-A_\lambda(s,\lambda)W(s,\lambda)~.
\end{equation}
The derivative of $W(s,\lambda)$ with respect to $s$ is given by
\begin{equation}\label{4Xeq}
X(s,\lambda)= - W(s,\lambda)\int_0^\lambda 
W(s,\lambda')^{-1} \frac{\partial 
A_\lambda(s,\lambda')}{\partial s} W(s,\lambda')d\lambda'~.
\end{equation}
In fact from Eq.(\ref{4Weq}) we have 
$$
\frac{\partial}{\partial \lambda}\frac{\partial W}{\partial s} = 
- \frac{\partial
A_\lambda(s,\lambda)}{\partial s} W(s,\lambda) -
A_\lambda(s,\lambda) \frac{\partial W(s,\lambda)}{\partial s}
$$
and trivially from Eq.(\ref{4Xeq})
$$
\frac{\partial}{\partial \lambda}X(s,\lambda) = 
- \frac{\partial
A_\lambda(s,\lambda)}{\partial s} W(s,\lambda) -
A_\lambda(s,\lambda) X(s,\lambda)~.
$$
Thus $\frac{\partial W}{\partial s}$ and $X$ satisfy the same first
order equation with the same initial conditions
$$
\frac{\partial W(s,0)}{\partial s}=0;~~~ X(s,0)=0~.
$$
We can now add to Eq.(\ref{4Xeq}) 
$$
0= W(s,1)\int^1_0 \frac{\partial}{\partial\lambda}
\left(W(s,\lambda)^{-1} A_s(s,\lambda) W(s,\lambda)\right) d\lambda
$$
due to
$$
A_s(s,1) = A_\mu(x(s,1)) \frac{dx^\mu(s,1)}{ds}=0
$$
to obtain
\begin{equation}\label{4dW}
\frac{\partial W(s,1)}{\partial s} = -W(s,1) \int_0^1
 W(s,\lambda)^{-1} F_{s\lambda}
(s,\lambda)W(s,\lambda) d\lambda\equiv -W(s,1) G(s)~. 
\end{equation}
An intuitive picture of such equation is given in fig.4.3.~ 
Eq.(\ref{4dW}) is
already an interesting result. It can be further integrated to obtain 
$$
W^{-1}(s,1) W(0,1)~= {\rm Pexp}\big(\int_0^sG(s')ds'\big)
$$

\bigskip

References

\smallskip

[1] P. Menotti, D. Seminara,``Energy theorem for (2+1)-dimensional
gravity'' Ann. Phys. 240 (1995) 203, Appendix A

\smallskip

[2] P. M. Fishbane, S. Gasiorowicz, P. Kraus, ``Stokes's theorems for
non-Abelian fields'' Phys. Rev. D 24 (1981)2324

\section{Vanishing of torsion and of the Riemann tensor}

If $R_{abcd}=0$ due to the non abelian Stokes theorem we have that the
space is teleparallel i.e. the parallel transport of a vector does non
depend on the path. 
If in addition we have vanishing
torsion and metric compatibility the space is locally flat.

In fact teleparallelism means that we can define $n$ independent
vector fields ${\bf v}^{(a)}$, $a=1\dots n$ with the
property, using the covariant components $v^{(a)}_\mu ={\bf
v}^{(a)}\cdot {\bf u}_\mu$ 
$$0=\nabla_\nu v^{(a)}_\mu=\partial_\nu  v^{(a)}_\mu
-v^{(a)}_\rho\Gamma^{\rho}_{\mu \nu}.
$$
If we suppose also $S^a=0$ (symmetric $\Gamma^\nu_{\lambda\mu}$) we have
$$
0=\nabla_\nu v^{(a)}_\mu -\nabla_\mu v^{(a)}_\nu =\partial_\nu
v^{(a)}_\mu -\partial_\mu v^{(a)}_\nu~.  
$$
The above equation tells us
that the vector field $v^{(a)}_\mu$ is locally integrable i.e.  there
exists functions $y^a$ of space such that
$$
v^{(a)}_\mu = \frac{\partial y^a}{\partial x^\mu}.
$$
Let us now go over to the coordinates $y^a$. The components of the
vectors ${\bf v}^{(a)}$ in these new coordinates are
$$
v^{(a)}_b = v^{(a)}_\mu \frac{\partial
x^\mu}{\partial y^b}=\frac{\partial y^a}{\partial x^\mu} \frac{\partial
x^\mu}{\partial y^b}= \delta^a_b 
$$
and then ${\bf d}{\bf v}^{(a)}=0$
becomes
$$
0=\nabla_b v^{(a)}_c=\partial_b\delta^a_c -
\Gamma^f_{cb}\delta^a_f=-\Gamma^a_{cb}.
$$
But $\Gamma^a_{bc}=0$ combined with metric compatibility $0 = dg - g\Gamma
- \Gamma^T g$ implies $\partial_cg_{ab}=0$ i.e. $g_{ab}={\rm const}$. 

\section{The Weyl-Schouten theorem}

We give below a proof of the Weyl-Schouten theorem which gives the
necessary and sufficient conditions for a geometry to be locally
conformally flat. The setting is that of the torsionless metric
compatible i.e. Levi-Civita connection.

From the previous result we see that necessary and sufficient
condition for a space to be conformally flat is the existence of a
Weyl transformation $\tilde g_{ij}=e^{2f}g_{ij}$ such that
$\tilde R^i_{jkl}=0$. 

It is useful to introduce the Weyl tensor via the Schouten tensor
$$
S_{ij} = \frac{1}{n-2}(R_{ij}-\frac{1}{2(n-1)} R g_{ij})
$$
and define the Weyl tensor $W$ through
$$
R_{ijkl} = W_{ijkl} + (S_{ik}g_{jl}+S_{jl}g_{ik}-S_{il}g_{jk}-S_{jk}g_{il})
\equiv W_{ijkl} + (S \odot g)_{ijkl}
$$
where for conciseness we introduced  the Kulkarni notation $\odot$.
One immediately verifies that
$$
R^i_{~jil} = R_{jl}= W^i_{~jil} + R_{ij}
$$
and thus the Weyl tensor $W$ is completely traceless and has the same
symmetry properties in the indices as the Riemann tensor.

Under $\tilde g_{ij}=e^{2f}g_{ij}$ we have
$$
\tilde\Gamma^i_{~jl}=\Gamma^i_{~jl}+g^{im}(\partial_j f ~g_{lm}+
\partial_l f ~g_{jm}-\partial_m f ~g_{jl}) = \Gamma^i_{~jl}+\delta\Gamma^i_{~jl}~.
$$
For the Riemann 2-form we have
\begin{equation}\label{4tildeR}
\tilde R =
R+d~\delta\Gamma+\Gamma\wedge\delta\Gamma+\delta\Gamma\wedge\Gamma+
\delta\Gamma\wedge \delta\Gamma = R+\nabla
\delta\Gamma+\delta\Gamma\wedge \delta\Gamma 
\end{equation}
where $\nabla$ is the covariant derivative with the connection $\Gamma$.
Using Eq.(\ref{4tildeR}) one finds
$$
\tilde R^i_{~jkl}= R^i_{~jkl}-g^{im}(a\odot g)_{mjkl} 
$$
and
$$
\tilde S_{ij}=S_{ij}-a_{ij}
$$
where
$$
a_{ij}=\nabla_i\nabla_jf-\nabla_if \nabla_jf+\frac{1}{2}|\nabla f|^2g_{ij}~.
$$
From this it follows that the Weyl tensor $W^i_{~jkl}$ is invariant
because
\begin{eqnarray}\label{4weylinvariant}
&&\tilde R^i_{~jkl} = \tilde W^i_{~jkl} + \tilde g^{im}(\tilde S\odot
  \tilde g)_{mjkl}=\tilde W^i_{~jkl} + g^{im}(S\odot
  g)_{mjkl}-g^{im}(a\odot g)_{mjkl} \nonumber\\
&=& R^i_{~jkl}-g^{im}(a\odot g)_{mjkl}=W^i_{~jkl}+g^{im}(S\odot g)_{mjkl}
-g^{im}(a\odot g)_{mjkl} ~.
\end{eqnarray}

Using 
$$
\nabla_iR^i_{~jkl}-\nabla_k R_{jl}+\nabla_l R_{jk}=0
$$
and the contracted Bianchi identities one finds
\begin{equation}\label{4cotton}
\nabla_i W^i_{~jkl}=(n-3)(\nabla_j S_{kl}-\nabla_kS_{jl})
\equiv 2 (n-3)C_{jkl}~.
\end{equation}
$C_{jkl}$ is the Cotton tensor. This means that in $n>3$ the
identical vanishing of the Weyl tensor implies the vanishing of the Cotton
tensor.

If $W^i_{~jkl}=0$ the condition for having $\tilde R^i_{~jkl}=0$ is from 
(\ref{4weylinvariant})
$$
\tilde S\odot g =0
$$
i.e.
\begin{equation}\label{4aequation}
a_{ij} = S_{ij}
\end{equation}
because for $n>2$
$$
F\odot g = 0 
$$
implies $F=0$ as it is easily verified by taking traces.
Solubility of (\ref{4aequation}) in terms of $f$ 
implies the existence of a vector field
$v_i = \partial_i f$ which solves
\begin{equation}\label{4vectorequation}
\nabla_i v_j= S_{ij} + v_i v_j- \frac{1}{2} ~v^lv_l~g_{ij}~. 
\end{equation}
We are interested in the inverse problem, i.e. given $S_{ij}$ find an
$f$ such that $v_i=\partial_i f$ solves (\ref{4vectorequation}). On
the other hand in we find a vector field $v_i$ which solves
(\ref{4vectorequation}) due to the symmetry in the indices of the
r.h.s. of Eq.(\ref{4vectorequation}) and the absence of torsion we
have $\partial_i v_j-\partial_j v_i=0$, which locally assures the
existence of a generating function $f$. Thus we have simply to find
the necessary and sufficient condition for the existence of a vector
field which satisfies Eq.(\ref{4vectorequation}).

It is
useful to rewrite the previous equation as
\begin{equation}\label{4vectorequationform}
\nabla v^k= S^k + v^k V- \frac{1}{2} ~v^lv_l~ dx^k \equiv A^k = S^k +r^k
\end{equation}
where we introduced the 1-forms
$$
S^k = S^k_i dx^i,~~~~V = v_i dx^i~.
$$
After rewriting Eq.(\ref{4vectorequationform}) as 
$$
dv^k= A^k - \Gamma^k_{m} v^m
$$
we have the integrability condition 
for Eq.(\ref{4vectorequationform})
$$
\nabla A^k -R^k_{~l} v^l = ddv^k=0
$$
where $R^k_{~l}$ is the Riemann two form.
As the Weyl tensor is assumed zero, the above equation becomes
\begin{equation}\label{4weylintegrability}
0 = \nabla A^k -(S\odot g)^k_{~lij} ~v^l~ \frac{dx^i\wedge dx^j}{2}=
\nabla S^k +\nabla r^k-(S\odot g)^k_{~lij} ~v^l~ \frac{dx^i\wedge dx^j}{2}~.
\end{equation}
But we have
\begin{eqnarray}
\nabla r^k &=& S^k\wedge V+ v^k dV -v_l S^l\wedge dx^k =
(S^k_i v_j + v^k \partial_i v_j -v_l S^l_i\delta^k_j)dx^i\wedge dx^j \\
&=&(S^k_i v_j -v_l S^l_i\delta^k_j)dx^i\wedge dx^j
=(S\odot g)^k_{~lij}~v^l~\frac{dx^i\wedge dx^j}{2}
\end{eqnarray}
which cancels the last term in Eq.(\ref{4weylintegrability}).
Thus we are left with the necessary condition for integrability 
\begin{equation}\label{4nablaS}
\nabla S^k=0 = dS^k +\Gamma^k_l\wedge S^l ~.
\end{equation}
Due to zero torsion and metric compatibility it is also  equivalent to
\begin{equation}\label{}
\nabla_i S_{kj}-\nabla_k S_{ij}=0
\end{equation}
i.e. the vanishing of the Cotton tensor.

Due to Frobenius integrability theorem Eq.(\ref{4nablaS}) 
is not only necessary but also sufficient for the integrability of 
Eq.(\ref{4vectorequationform}) [2].

\bigskip

Notice that the Weyl tensor vanishes identically in dimension
$3$ but this does not mean that in dimension $3$ the space is
conformally flat as Eq.(\ref{4cotton}) does not imply the vanishing
of the Cotton tensor. Thus in dimension $3$ necessary and sufficient
condition for being a geometry conformally flat is the vanishing of the
Cotton tensor.

In dimension equal or higher than $4$ being the Weyl tensor an
invariant if it is different from zero the geometry cannot be
conformally flat. If it is zero then Eq.(\ref{4cotton}) implies that
the Cotton tensor is zero and the space is conformally flat.  We
conclude that in dimension equal or higher than $4$ necessary and
sufficient condition for being geometry conformally flat is the
vanishing of the Weyl tensor.

\bigskip

References

\smallskip

[1] [Wald] Appendix D

[2] J.C. Gerretsen,``Lectures on tensor calculus and differential
geometry'', P. Noodhoff N.V. Groningen (1962)


\section{Hodge * operation}

This is a bijection between $\Lambda^p$ and $\Lambda^{(n-p)}$ which can
be established when we have a metric on the manifold. There
are several level of abstraction with which such a concept can be
introduced. Given an orientable manifold i.e. such that there exists an
atlas with all transition functions with positive Jacobian, the
expression (for definiteness we assume a metric with signature 
$(-,++ \dots)$~.
$$
\epsilon_{\mu_1 \dots \mu_n}=\sqrt{-g}~\varepsilon _{\mu_1\dots \mu_n},
~~~~{\rm with}~~~~\varepsilon_{0,1\dots , n-1}=1
$$  
is an invariant tensor under diffeomorphisms with positive Jacobian. One can
raise indices with the $g^{\mu\nu}$ and we have also 
$$
\epsilon^{\mu_1\dots \mu_n}=\frac{1}{\sqrt{-g}}\varepsilon^{\mu_1\dots
  \mu_n},~~~~{\rm with}~~~~\varepsilon^{0,1\dots , n-1}=-1~.
$$ 
Taking the covariant differential of
$$
\epsilon_{\mu_1,\dots\mu_n} \epsilon^{\mu_1,\dots\mu_n} = -n!
$$ 
we have that $\epsilon_{\mu_1,\dots\mu_n}$ is covariantly constant.
Accordingly one can define an invariant form belonging to $\Lambda^n$
given by
$$
\bfepsilon = \frac{1}{n!}\epsilon_{\mu_1 \dots \mu_n}dx^{\mu_1}\wedge\dots
\wedge dx^{\mu_n}. 
$$
More generally we have
$$
\bfepsilon = \frac{1}{n!}\epsilon_{a_1 \dots a_n}e^{a_1}\wedge\dots
\wedge e^{a_n} 
$$
where
$$
\epsilon_{a_1 \dots a_n}=\sqrt{-\det g_{ab}}~\varepsilon_{a_1 \dots a_n}.
$$
Given a $p-$form
$$
f =\frac{1}{p!}f_{\mu_1\dots\mu_p}dx^{\mu_1}\wedge\dots
\wedge dx^{\mu_p}
$$
we define
$$
*f =\frac{1}{p!(n-p)!}f^{\mu_1\dots\mu_p}\epsilon_{\mu_1\dots\mu_n}
dx^{\mu_{p+1}}\wedge\dots \wedge dx^{\mu_n}\in\Lambda^{(n-p)}.
$$
The following relation holds
$$
**f = -(-)^{p(n-p)}f.
$$
The form $*f$ is called the form dual to $f$.

This allows to define product of two $p-$forms $f$ and $h$ which is a
$n$ form as 
$$
f\wedge *h = (f,h)\bfepsilon.
$$ 
Due to the invariant nature of $\bfepsilon$, $(f,h)$ is an invariant
function which is linear in $f,h$. It is immediately shown that
$$
f\wedge *h = h\wedge* f =(-)^{p(n-p)}*h\wedge f
$$
and thus $(f,h)=(h,f)$.

\section{The current}\label{4thecurrentSec}

Given the gauge connection $A$ and $F=dA+A\wedge A$, we recall that
$DF=0$ (Bianchi identity).
For any form $G\in\Lambda^g_{(1,1)}$ we have
\begin{equation}\label{doubleD}
DDG = F\wedge G - G\wedge F\in\Lambda^{g+2}_{(1,1)}. 
\end{equation}
In fact
$$
DG = dG + A\wedge G-(-)^g G\wedge A\in \Lambda^{g+1}_{(1,1)};
$$
$$
DDG =dd G + dA\wedge G-A\wedge dG -(-)^g dG\wedge A-(-)^{2g}G\wedge dA+
$$
$$
+A\wedge(dG + A\wedge G-(-)^g G\wedge A)-(-)^{g+1}(dG + A\wedge G-(-)^g
G\wedge A)\wedge A=
$$
$$
=(dA+ A\wedge A)\wedge G - G\wedge (dA + A\wedge A)
$$
which is Eq.(\ref{doubleD}).
We apply now Eq.(\ref{doubleD}) to $G=*F\in\Lambda^{n-2}_{(1,1)}$.
Defined $J$ by 
$$
D*F = *J
$$
we have
$$
DD*F = F\wedge *F - *F\wedge F =0~~~~{\rm as}~~~F\in\Lambda^2 
$$
i.e.
$$
D*J=0~.
$$
This is the covariant conservation of the Yang-Mills current valid on
a differential manifold in any dimension $n$ and for arbitrary metric.

In the case of the Maxwell field in $n$-dimensions the above formulas
simplify to
$$
F=d A,~~~~dF=0,~~~~d*F=*J,~~~~ 0=dd*F=d*J~.
$$
The last is equivalent to 
\begin{equation}\label{4divergence}
\partial_\lambda(\sqrt{-g}J^\lambda)=0~.
\end{equation}
Integrating Eq.(\ref{4divergence}) on a cylinder $V$ of base $\Sigma_1$ with
$x^0=c_1={\rm const.}$ and of cover $\Sigma_2$ with $x^0=c_2={\rm const.}$
and a mantle at large distances, provided the field $F$ and thus $J$
decreases sufficiently rapidly at large distances we obtain the
conservation law
$$
Q=\int_{\Sigma_2} \sqrt{-g}~J^0 dx^1 \dots dx^{n-1}=\int_{\Sigma_1}
\sqrt{-g}~J^0 dx^1 \dots dx^{n-1}~. 
$$
Using the ADM metric with $c=1$ i.e. using as time coordinate $x^0 =
ct$ we have
$$
g \equiv \det g_{\mu\nu}= -N^2 \det h_{mn}\equiv -N^2 h
$$
Moreover the normal to the space like surface $x^0= {\rm const.}$ is
given by the time-like vector $n$ whose covariant components are
defined by $0=n_\mu dx^\mu$ with $dx^\mu=(0,dx^1,dx^2,\dots)$
i.e. $n_\mu=(n_0,0,0,\dots)$. The normalization is provided by $-1 =
n_\mu g^{\mu\nu}n_\nu = n_0 g^{00}n_0 = - n_0^2 N^{-2}$ and thus
$n_\mu=(N,0,0,\dots)$. We can now rewrite the expression of the charge
as
$$
Q= \int_\Sigma \sqrt{h}~ n_\lambda J^\lambda dx^1 \dots dx^{n-1} 
$$
which in covariant form can be written
$$
Q= \int_\Sigma n_\lambda J^\lambda d\Sigma
$$
being $d\Sigma$ the invariant area element of the surface $\Sigma
=\partial V$.
In case of torsionless metric compatible connection as
$\displaystyle{\Gamma^\rho_{~\lambda\rho}
=\frac{1}{\sqrt{-g}}\partial_\lambda \sqrt{-g}}$ we have
$$
\frac{1}{\sqrt{-g}}\partial_\lambda
(\sqrt{-g}J^\lambda)=\nabla_\lambda J^\lambda
$$
and thus the current conservation can be written as
$$
\nabla_\lambda J^\lambda=0~.
$$

In the non abelian case $D*F=*J$ is explicitly written as
$$
\partial_\rho(\sqrt{-g} F^{\nu\rho})+\sqrt{-g}(A_\rho
F^{\nu\rho}- F^{\nu\rho} A_\rho)=\sqrt{-g}J^\nu 
$$
and $D*J=0$ as
$$
\partial_\rho(\sqrt{-g}J^\rho)+ \sqrt{-g} (A_\rho J^\rho-
J^\rho A_\rho)=0.
$$
This is the covariant conservation of the covariant current; due to
the additional term containing the $A_\rho$ from the previous equation
we cannot derive a conservation law.

\bigskip

\bigskip
\section{Action for gauge fields}

In four dimensions the expression
$$
\int {\rm Tr}(F\wedge F)
$$
is invariant under gauge transformations and under proper
diffeomorphisms. It is however a pseudoscalar.

Under a variation of $A$ we have
$$
\delta (dA + A\wedge A) = d\delta A + \delta A\wedge A + A\wedge
\delta A =  D\delta A
$$
and thus
$$
\delta\int {\rm Tr}(F\wedge F) = 2\int {\rm Tr}(D\delta A\wedge F)=
2 \int {\rm Tr}(\delta A\wedge D F)
$$
where we have integrated by parts, neglecting the surface term,
provided $\delta A$ vanishes sufficiently rapidly at infinity. Due to
the Bianchi identity $DF=0$ such a variation is zero independently of
any equation of motion. A quantity with such a property is called a
topological invariant and in our case is called the second Chern
class. 

It is useful even if not strictly necessary at the classical level, 
to have an action from
which to derive the equations of motion. If we deal with pure gauge
theory we have that
\begin{equation}\label{ymaction}
S_{YM}=\int {\rm Tr}(F\wedge *F)
\end{equation}
is invariant under diffs and under local gauge
transformations. Notice that Eq.(\ref{ymaction}) is meaningful in $n$
dimensions as $F\in\Lambda^2_{(1,1)}$ and
$*F\in\Lambda^{n-2}_{(1,1)}$. Moreover due to the
appearance of two antisymmetric 
structures it is a scalar. It
provides also the equation of motion as under a variation of the gauge
field we have
$$
\delta S_{YM} = 2 \int {\rm Tr}(D\delta A \wedge *F)~.
$$
Then using
$$
d{\rm Tr}(\delta A\wedge *F ) = D{\rm Tr}(\delta A\wedge *F) = 
{\rm Tr}(D\delta A\wedge *F ) - {\rm Tr}(\delta A \wedge D *F)~, 
$$
for a variation of $A$ which vanishes on the boundary we have
$$
\delta S_{YM} = 2 \int {\rm Tr}(\delta A \wedge D*F)~.
$$
For a faithful representation of the Lie algebra of a compact group
or for a faithful representation of a semi-simple Lie algebra the vanishing
of  ${\rm Tr}(\lambda \rho)$
for any $\lambda$, $\lambda$ and $\rho$ belonging to the Lie algebra,
gives $\rho=0$ and thus we have $D*F=0$.
In presence of an external current we add
$$
-2\int {\rm Tr}(A\wedge *J)~.
$$
and we reach the equation of motion $D*F=*J$. $DD*F=0$ imposes
$D*J=0$. The current $J$ as a rule arises from the coupling of the
Yang-Mills field $A$ to the matter fields $\psi$ and the total action is
given by
$$
S = S_{YM} +S_{\rm matter}~.
$$
$S$ is constructed invariant under gauge transformations. The current
is defined as
$$
\delta S_{\rm matter} = - 2\int{\rm Tr}(\delta A\wedge *J)~.
$$

Let us
consider an infinitesimal gauge transformation
$$
\delta A = D\lambda~.
$$
$S_{YM}$ is invariant while
$$
\delta S_{\rm matter} = - 2\int{\rm Tr}(\delta A\wedge *J)+\delta_\psi
S_{\rm matter} 
$$
where $\delta_\psi S_{\rm matter}$ is the variation of $S_{\rm
matter}$ due to the change of the matter fields $\psi$ under the gauge
transformation. But on the equation of motion  $\delta_\psi S_{\rm matter}=0$
as $\delta S_{\rm matter}$ on the equation of motion is zero for any
variation of $\psi$. Thus on the equations of motion we have reached
$$
0= -\int {\rm Tr}(D\lambda\wedge *J) = \int {\rm Tr}(\lambda\wedge
D*J)   
$$
which implies $D*J=0$. This result is in agreement with the
kinematic derivation of the same relation obtained from the identity
$DD*F=0$.

Finally notice that in four dimensions, if we have a configuration
such that $F=\pm *F$ (self-dual (antiself)- dual field), $F$ 
satisfies the equation of motion
$$
D*F = DF\equiv 0.
$$

 
\chapter{The action for the gravitational field}

\section{The Hilbert action}

Hilbert action is given by
\begin{equation}\label{5Hactio}
S_H = \int_V \sqrt{-g}~ R~ d^nx=
\int_V \sqrt{-g}~ g^{\mu\nu}R_{\mu\nu} d^nx.
\end{equation}
From the variational viewpoint this action is not correct as it
contains second derivatives of $g_{\mu\nu}$. It is like
writing the action of a particle in one dimension as
\begin{equation}\label{5dotdotoscillator}
S=\int_{t_0q_0}^{t_1q_1} \left[-\frac{1}{2}q \ddot q
-V(q)\right]dt~. 
\end{equation}
We have
$$
\delta S =\int_{t_0 q_0}^{t_1 q_1} \delta q (-\ddot q -
\frac{\partial V}{\partial q}) dt-\left[\frac{1}{2}q\delta \dot
q\right]^{t_1}_{t_0} 
$$
i.e. in order to obtain the correct equations of motion it is not
sufficient to impose $\delta q=0$ at $t_0$ and $t_1$ but it is
necessary to impose also $\delta \dot q=0$ at $t_0$ and $t_1$. That
would give $4$ conditions at $t_0$ and $t_1$ which is as a rule
inconsistent with the equations of motion. I.e. action
(\ref{5dotdotoscillator}) cannot be used in a variational treatment. 

Such action can be mended by adding a proper boundary term
$$
S_1 = S +\frac{1}{2}(q\dot q)(t_1)-\frac{1}{2}(q\dot q)(t_0).
$$  
A similar cure has to be applied to the Hilbert action by adding a
proper boundary term giving rise to the so called ${\rm Tr} K$-action. 
This will be done in Section \ref{6TrKaction}.

\section{Einstein's $\Gamma\Gamma$ action}\label{5GammaGammaSec}

In performing the variation with respect to $g_{\mu\nu}$ we shall use
the convention to treat $\delta g_{\mu\nu}$ as a tensor i.e.  $$
\delta g^{\mu\nu} = g^{\mu\alpha}\delta g_{\alpha\beta}g^{\beta\nu} $$
which means that $\delta(g^{\mu\nu}) = -\delta g^{\mu\nu}$.  From 
$$
S_H =  \int_{V}d^n x\sqrt{-g}~g^{\mu\nu}R_{\mu\nu}
=\int_{V}d^n x\sqrt{-g}~g^{\mu\nu}[\partial_\lambda\Gamma_\nu-
\partial_\nu\Gamma_\lambda+\Gamma_\lambda\Gamma_\nu-
\Gamma_\nu\Gamma_\lambda]^\lambda_{~\mu} 
$$
we have 
\begin{eqnarray} 
\delta S_H &=&
\int_{V}d^n x\delta(\sqrt{-g}~g^{\mu\nu})[\partial_\lambda\Gamma_\nu-
\partial_\nu\Gamma_\lambda
+\Gamma_\lambda\Gamma_\nu-\Gamma_\nu\Gamma_\lambda]^\lambda_{~\mu}
+\int_{V}d^n x \sqrt{-g}~g^{\mu\nu}[D_\lambda\delta\Gamma_\nu
-D_\nu\delta\Gamma_\lambda]^\lambda_{~\mu} \nonumber\\
&=&- \int_{V}d^n
x\sqrt{-g}~\delta g^{\mu\nu}G_{\mu\nu}
+\int_{V}d^n x \sqrt{-g}~g^{\mu\nu}[D_\lambda\delta\Gamma_\nu
-D_\nu\delta\Gamma_\lambda]^\lambda_{~\mu} 
\end{eqnarray}
being
$G_{\mu\nu}$ the Einstein tensor 
$$ 
G_{\mu\nu} = R_{\mu\nu}
-\frac{g_{\mu\nu}}{2} R~.
$$ 
$D$ is the space analogue of the gauge gauge covariant differential of 
Section \ref{4gaugedifferential} where we had $\delta F = D\delta A$.
Here we have
$$
D_\lambda\delta\Gamma^\rho_{\mu\nu}=\partial_\lambda\delta\Gamma^\rho_{\mu\nu}+
\Gamma^\rho_{\sigma\lambda}\delta\Gamma^\sigma_{\mu\nu}-
\delta\Gamma^\rho_{\sigma\nu}\Gamma^\sigma_{\mu\lambda}~.
$$
Using the $0$-torsion property of the Levi-Civita connection
i.e. $\Gamma^\alpha_{\beta\gamma}=\Gamma^\alpha_{\gamma\beta}$, the
above equation can be rewritten as
$$ \delta S_H = - \int_{V}d^n
x\sqrt{-g}~\delta g^{\mu\nu}G_{\mu\nu} + \int_V d^n
x\sqrt{-g}~g^{\mu\nu}[~\bfnabla_\lambda\delta\Gamma_\nu-
\bfnabla_\nu\delta\Gamma_\lambda]^\lambda_{~\mu}.  
$$ 
where 
$$ 
\bfnabla_\lambda \delta \Gamma^\alpha_{\beta\gamma} 
$$ 
is the
full covariant derivative of the $(1,2)$ tensor $\delta
\Gamma^\alpha_{\beta\gamma}$ taking into account also the covariant
vector index $\gamma$
$$ 
\bfnabla_\lambda \delta
\Gamma^\alpha_{\beta\gamma}=D_\lambda \delta
\Gamma^\alpha_{\beta\gamma}-\delta
\Gamma^\alpha_{\beta\sigma} \Gamma^{\sigma}_{\gamma\lambda}~.
$$ 
We recall that $\delta \Gamma$ is a tensor and thus
\begin{equation}\label{vectorv} 
v^\lambda\equiv \delta
\Gamma^\lambda_{~\mu\nu} g^{\mu\nu}- \delta \Gamma^\nu_{~\mu\nu}
g^{\mu\lambda} \equiv 2 \delta \Gamma^{[\lambda\nu]}_{~~~~\nu}
\end{equation} is a vector.  Thus we found the identity 
$$ 
g^{\mu\nu}\delta
R_{\mu\nu}= \nabla_\lambda v^\lambda
$$ 
which is known as Palatini identity and we have reached the result
\begin{equation}\label{5deltaSH} 
\delta S_H = -\int_{V}\sqrt{-g}~\delta
g^{\mu\nu}G_{\mu\nu}~d^nx + \int_V \partial_\lambda(\sqrt{-g}~
v^\lambda)~ d^nx
= -\int_{V}\sqrt{-g}~\delta
g^{\mu\nu}G_{\mu\nu}~d^nx + \int_{\partial V} v^\lambda \Sigma_\lambda.  
\end{equation} 
where 
$$
\Sigma_\lambda =
\epsilon_{\lambda\sigma_2 \dots \sigma_n}
\frac{\partial x^{\sigma_2}}{du^2} \dots
\frac{\partial x^{\sigma_n}}{du^n} du^2 \dots du^n
$$
and $\partial V$ is given by $x^\mu(0,u^2\dots u^{n-1})$ being
the $u$'s the local coordinates with $u_1=0$ on the boundary,  
$u^1<0$ inside $V$ and $u^1>0$ outside $V$ and 
$\frac{\partial x^\mu}{\partial u^\nu}>0$ (see Section \ref{3stokes}).

We have {\it algebraically}, treating
$\Gamma$ as a tensor 
$$ 
\partial_{\lambda}\Gamma^{\alpha}_{\mu\nu}-
\partial_{\nu}\Gamma^{\alpha}_{\mu\lambda}+
[\Gamma_{\lambda},\Gamma_{\nu}]^\alpha_{~\mu}=
D_{\lambda}\Gamma^{\alpha}_{\mu\nu}-
D_{\nu}\Gamma^{\alpha}_{\mu\lambda}-
[\Gamma_{\lambda},\Gamma_{\nu}]^\alpha_{~\mu}= 
$$ 
$$
=\bfnabla_{\lambda}\Gamma^{\alpha}_{\mu\nu}-
\bfnabla_{\nu}\Gamma^{\alpha}_{\mu\lambda}-
[\Gamma_{\lambda},\Gamma_{\nu}]^\alpha_{~\mu} 
$$ 
and thus 
$$
g^{\mu\nu} R_{\mu\nu}=g^{\mu\nu} \left(\partial_{\lambda}\Gamma^{\lambda}_{\mu\nu}-
\partial_{\nu}\Gamma^{\lambda}_{\mu\lambda}+
[\Gamma_{\lambda},\Gamma_{\nu}]^\lambda_{~\mu}\right)= -g^{\mu\nu}
[\Gamma_{\lambda},\Gamma_{\nu}]^\lambda_{~\mu}+
2~\nabla_\lambda\Gamma^{[\lambda\nu]}_{~~~\nu}= 
$$
\begin{equation}\label{noncovidentity} 
=-g^{\mu\nu}
[\Gamma_{\lambda},\Gamma_{\nu}]^\lambda_{~\mu}+
\frac{2}{\sqrt{-g}}\partial_\lambda(\sqrt{-g}~\Gamma^{[\lambda\nu]}_{~~~\nu})
\equiv -g^{\mu\nu}
[\Gamma_{\lambda},\Gamma_{\nu}]^\lambda_{~\mu}+
\frac{1}{\sqrt{-g}}\partial_\lambda(\sqrt{-g}~W^\lambda)
\end{equation}
with the obvious definition of $W^\lambda$. 
We stress that Eq.(\ref{noncovidentity}) is an
algebraic identity, where the transformation properties of the symbol
$\Gamma^\lambda_{\mu\nu}$ play no role.
Then we define Einstein's $\Gamma\Gamma$-action
$$
S_{\Gamma\Gamma} =-\int_V\sqrt{-g}~g^{\mu\nu}
[\Gamma_{\lambda},\Gamma_{\nu}]^\lambda_{~\mu}d^nx = S_H 
-\int_V \partial_\lambda (\sqrt{-g} ~W^\lambda)d^nx=
S_H-\oint_{\partial V} W^\lambda \Sigma_\lambda~.
$$
$S_{\Gamma\Gamma}$ has only first order derivatives and from
Eq.(\ref{5deltaSH})
its variation keeping $g_{\mu\nu}$ fixed on $\partial V$ is given
simply by 
$$
\delta S_{\Gamma\Gamma}= -\int_{V}\sqrt{-g}~\delta
g^{\mu\nu}G_{\mu\nu}d^nx. 
$$
In fact on $\partial V$ we have
$$
\delta W^\lambda = 2 ~\delta(\sqrt{-g}\Gamma^{[\lambda
\nu]}_{~~~\nu})=\sqrt{-g}~\delta(\Gamma^{\lambda}_{~\nu'\nu}
g^{\nu'\nu}  -\Gamma^{\nu}_{~\nu'\nu} g^{\nu'\lambda})=  
$$
$$
=\sqrt{-g}~(\delta\Gamma^{\lambda}_{~\nu'\nu}
g^{\nu'\nu}  -\delta\Gamma^{\nu}_{~\nu'\nu} g^{\nu'\lambda})= 2
\sqrt{-g}~\delta \Gamma^{[\lambda\nu]}_{~~~\nu} =\sqrt{-g}~v^\lambda.
$$

The algebraic manipulations are not covariant and the resulting action
$S_{\Gamma\Gamma}$ is not invariant. Despite that, it is a good
action. It is an example of a non invariant action which gives rise to
covariant equations. 


\section{Palatini first order action}

The previous approach is called the second order approach as the
fundamental variables $g_{\mu\nu}$ appear with second order
derivatives.

In the Palatini approach we are given a metric $g_{\mu\nu}$ and a
torsionless connection $\Gamma$, independent of the metric. 
In the following all covariant
derivative will refer the connection $\Gamma$. No metric compatibility
of $\Gamma$ is assumed.

$R^\alpha_{~\mu\beta\nu}$ is well defined independently of $g_{\mu\nu}$,
and also the Ricci tensor $R_{\mu\nu}$.
An invariant action is
$$
\int_V \sqrt{-g}~g^{\mu\nu}R_{\mu\nu}~d^nx +~{\rm boundary ~terms}.
$$   
where the covariant boundary terms are given in Section \ref{6palatinibtsec}.

We remark that
$$
\sqrt{-g}~\nabla_\mu v^\mu=\sqrt{-g}~(\partial_\mu v^\mu
+\Gamma^\rho_{\mu\rho}v^\mu) 
=\partial_\mu(\sqrt{-g}~v^\mu)+\sqrt{-g}(\Gamma^\rho_{\mu\rho}
-\frac{1}{\sqrt{-g}}\partial_\mu \sqrt{-g})v^\mu.  
$$

The variation with respect to $g_{\mu\nu}$ gives
$$
0=-\int\sqrt{-g}\delta g^{\mu\nu} G_{\mu\nu} d^nx =0,~~~~{\rm
i.e}~~~~G_{(\mu\nu)}=0,
$$
while the variation with respect to $\Gamma^\lambda_{\mu\nu}$ gives
$$
0=\int \sqrt{-g}~g^{\mu\nu}[D_\lambda\delta \Gamma_\nu - D_\nu\delta
\Gamma_\lambda ]^\lambda_{~\mu} d^nx= 
$$
$$
=\int \sqrt{-g}~g^{\mu\nu}[~\bfnabla_\lambda\delta \Gamma_\nu - 
\bfnabla_\nu\delta
\Gamma_\lambda ]^\lambda_{~\mu} d^nx 
$$
due to the absence of torsion.

Consider the first term
\begin{eqnarray}\label{5divergence}
&&\sqrt{-g}~g^{\mu\nu}~\bfnabla_\lambda\delta
\Gamma^\lambda_{\mu\nu}=\sqrt{-g}~\bfnabla_\lambda~(g^{\mu\nu}\delta
\Gamma^\lambda_{\mu\nu})- \sqrt{-g}~\delta
\Gamma^\lambda_{\mu\nu}\nabla_\lambda g^{\mu\nu}\nonumber\\ 
&=&\partial_\lambda( \sqrt{-g}g^{\mu\nu}~\delta \Gamma^\lambda_{\mu\nu})+
\sqrt{-g}~(\Gamma^\rho_{\lambda\rho} -\partial_\lambda
\log{\sqrt{-g}})g^{\mu\nu}\delta \Gamma^\lambda_{\mu\nu}
-\sqrt{-g} \delta\Gamma^\lambda_{\mu\nu}\nabla_\lambda g^{\mu\nu}\nonumber\\
&\equiv& \partial_\lambda( \sqrt{-g}g^{\mu\nu}\delta \Gamma^\lambda_{\mu\nu})
+\sqrt{-g} ~C^{\mu\nu}_\lambda \delta\Gamma^\lambda_{\mu\nu}
\end{eqnarray} 
with 
\begin{equation}\label{Ceq}
C^{\mu\nu}_\lambda= g^{\mu\nu}(\Gamma^\rho_{\lambda\rho} -\partial_\lambda
\log{\sqrt{-g}})-\nabla_\lambda g^{\mu\nu}.
\end{equation}
The divergence term in (\ref{5divergence}) is canceled by the variation 
of the boundary term.
Adding the second term gives
$$
(C^{\mu\nu}_\lambda
-C^{\mu\rho}_\rho\delta^\nu_\lambda)\delta\Gamma^\lambda_{\mu\nu}  
$$
and due to the symmetry of
$\Gamma^\lambda_{\mu\nu}$ in the lower indices
\begin{equation}\label{5Cidentity}
C^{\mu\nu}_\lambda
-\frac{1}{2}(C^{\mu\rho}_\rho\delta^\nu_\lambda
+C^{\nu\rho}_\rho\delta^\mu_\lambda)=0~. 
\end{equation}
Contracting $\nu$ with $\lambda$ we have
$$
\frac{1-n}{2}C^{\mu\nu}_\nu=0
$$
and substituting im Eq.(\ref{5Cidentity})
\begin{equation}\label{Czero}
C^{\mu\nu}_\lambda=0~.
\end{equation}
In Eqs.(\ref{Ceq},\ref{Czero}) write $\Gamma$ as $\hat\Gamma + \Delta$, where
$\hat\Gamma$ is the metric compatible connection; then we have 
$$
\Delta^\mu_{\mu'\lambda} g^{\mu'\nu}+\Delta^\nu_{\nu'\lambda} g^{\mu\nu'}
-g^{\mu\nu} \Delta^\rho_{\lambda\rho}=0.
$$ 
Taking the trace $$
(2-n)\Delta^\rho_{\lambda\rho}=0
$$ 
which combined with the previous equation gives 
$$
\Delta_{\mu\nu\lambda}+\Delta_{\nu\mu\lambda}=0.
$$ 
Rotating twice and
subtracting 
one from the other two, recalling that
$\Delta_{\mu\nu\lambda}=\Delta_{\mu\lambda\nu}$ we have
$\Delta_{\mu\nu\lambda}=0$ and thus
$\Gamma^\lambda_{\mu\nu}$ is the metric compatible torsionless
connection i.e. the Levi-Civita connection.

The Palatini result renders possible the
first order formulation of gravity. If the matter action depends only
on $g_{\mu\nu}$ the variation with respect to
$\Gamma^\lambda_{\mu\nu}$ is still zero and the same procedure holds.

\section{Einstein-Cartan formulation}

Here again we shall use a first order formalism, with the
difference that now metric compatibility is assumed from the start and
the vanishing of torsion is obtained as the consequence of the
equations of motion in absence of matter; the formalism is apt to
describe the coupling with fermions and here a non vanishing torsion
will appear.

We shall consider a connection on the tangent space in which we have
chosen an orthonormal system of vectors ${\bf e}_a$, ${\bf g}({\bf
e}_a,{\bf e}_b)=\eta_{ab}$ and the connection 1-form due to metric
compatibility must belong to the algebra of $SO(n-1,1)$, $\Gamma^a_{~b}\in
so(n-1,1)$. 
From these we compute the curvature 2-form $R^a_{~b}$. The
action is given by 
$$
S_{EC} = -\int {\rm Tr}(R\eta^{-1}\wedge H) +{\rm boundary~terms}
$$
where the covariant boundary terms are given in Section 
\ref{6EinsteinCartanbtSec}.
$H$ is the $n-2$ form  
$$
H_{ba}= \frac{1}{(n-2)!}\varepsilon_{b a a_3,\dots
a_n}e^{a_3}\wedge\dots \wedge e^{a_n} 
\in \Lambda^{n-2}_{(0,2)}
$$
being $\varepsilon_{a_1\dots a_n}$ the standard 
antisymmetric symbol, with $\varepsilon_{0 1 \dots n-1} =1$.
Such action is obviously invariant under diffeomorphisms; it is
invariant also under local $SO(n-1,1)$ transformations. In fact under
$$
{\bf e}_a = {\bf e}'_b \Omega^b_{~a},~~{\rm from ~which}~~e'^{c} =
\Omega^c_{~d}e^d 
$$
we already know that
$$
R' = \Omega R\Omega^{-1}.
$$
From 
$$
\varepsilon_{a_1\dots a_n} \Omega^{a_1}_{~b_1}\dots
\Omega^{a_n}_{~b_n} = \det(\Omega)\varepsilon_{b_1\dots
b_n}
$$
one obtains
$$
\varepsilon_{b a a_3 \dots a_n} \Omega^{a_3}_{~b_3}\dots
\Omega^{a_n}_{~b_n} = \det(\Omega)~(\Omega^{-1})^{b_1}_{~b}
~(\Omega^{-1})^{b_2}_{~a}~\varepsilon_{b_1\dots b_n}
$$
from which one gets
$$
H' = \det(\Omega)(\Omega^{-1})^T H \Omega^{-1}.
$$
Then
$$
{\rm Tr}(R'\eta^{-1}\wedge H') = \det(\Omega){\rm
Tr}(\Omega R\Omega^{-1} \eta^{-1} 
(\Omega^{-1})^T \wedge H\Omega^{-1}) 
= \det(\Omega){\rm Tr}(R\eta^{-1}\wedge H)
$$
as $\Omega^{-1} \eta^{-1}(\Omega^{-1})^T = \eta^{-1}$. 
Thus strictly speaking the
action is invariant only under proper local Lorentz transformations. 

In the following indices are raised and lowered with the tensors
$\eta^{-1}= \eta = {\rm diag}(-1,1 \dots 1)$.
$S_{EC}$ can also be written as
\begin{equation}\label{ECaction}
S_{EC} = \frac{1}{(n-2)!}\int R^{ab}\wedge e^{a_3}\wedge\dots
e^{a_n}\varepsilon_{a b a_3\dots 
a_n}. 
\end{equation}
Before extracting the equations of motion we want to relate $S_{EC}$
with the Hilbert action. Writing 
$$
R^{ab}=\frac{1}{2!}R^{ab}_{~~fg}e^f\wedge e^g
$$
we have 
$$
S_{EC} = \frac{1}{2(n-2)!}\int
R^{ab}_{~~fg}\varepsilon_{abcd\dots}(-\varepsilon^{fgcd\dots}) \det(e^a_\mu) 
dx^0\wedge dx^1\wedge \dots \wedge dx^{n-1}=
$$
$$ 
= \frac{1}{2}\int R^{ab}_{~~fg}\delta^{fg}_{ab} \bfepsilon 
= \int R \bfepsilon.
$$
$\bfepsilon= \sqrt{-{\rm det}g_{ab}}~e^0\wedge\dots \wedge e^{n-1}$ 
is the volume form as
$$
g_{\mu\nu}=e^a_\mu\eta_{ab}e^b_\nu~,~~~~e^2=\det(e^a_\mu)^2 = -g~.
$$
It is however much better to work directly with action
(\ref{ECaction}).
For simplicity we work from now on in four
dimensions, but the extension to $n$ dimensions is trivial.
Variation of $\Gamma$ gives
$$
\delta \Gamma^{ab} \wedge (D e^c \wedge e^d-e^c \wedge D e^d
)\varepsilon_{abcd}=0 
$$
where to keep in touch with gauge theories we denoted the $\nabla$
appearing in Section \ref{4gaugedifferential} as $D$, 
but they are the same operation.
Thus
\begin{equation}\label{contorsion}
D e^c \wedge e^d \varepsilon_{abcd}=0;~~~~{\rm i.e.}~~~~S^c \wedge
e^d \varepsilon_{b a cd}=0 
\end{equation}
where $S^c$ is the torsion.
But this implies $S^c=0$ i.e. ~vanishing~torsion~in~absence~of~matter.
In fact write
$$
S^c = \frac{1}{2}S^c_{uv}e^u\wedge e^v
$$
and multiply Eq.(\ref{contorsion}) by $\wedge e^f$ to obtain
$$
S^c_{uv}\delta^{uvf}_{abc}=0
$$
i.e.
$$
S^f_{ab}+S^c_{ca}\delta^f_b -S^c_{cb}\delta^f_a=0.
$$
Taking the trace
$$
(n-2)S^c_{cb}=0~~~~{\rm from~which}~~~~S^c_{ab}=0.
$$
Thus is absence of matter $\Gamma$ in addition of being metric
compatible is also torsionless, i.e. it is the Levi-Civita connection.
 
Variation of $e^d$ implies
\begin{equation}\label{ECeq}
R^{ab}\wedge e^c \varepsilon_{abcd}=0
\end{equation}
i.e. Einstein's equations.  
In fact multiplying by $\wedge e^h$ we have
$$
\frac{1}{2}R^{ab}_{~~fg} \varepsilon^{fgch}
\varepsilon_{abcd}~\bfepsilon=0 
$$
or
$$
0= R^{ab}_{~~fg} \delta^{fgh}_{abd}=
2( R \delta^{h}_{d}- 2 R^h_d) =- 4 G^h_d
$$
being $G^h_d$ the Einstein tensor.

\bigskip

References

\smallskip

[1] A. Trautman, ``On the Einstein-Cartan equation''
Bull. Acad. Pol. Sci. Ser. Math. Astron. Phys. 20 (1972) 185, 503, 895

\smallskip

[2] Y. Ne'eman, T. Regge, ``Gauge theory of gravity and supergravity
on a group manifold''Riv. Nuovo Cim. I (1978) 1

\bigskip
\section{The energy momentum tensor}

If the Lagrangian $L$ does not contain explicitly the coordinates we
have
$$
\partial_\mu L  = 
\frac{\partial L}{\partial\phi}\partial_\mu\phi
+\frac{\partial L}{\partial\partial_\lambda\phi}
\partial_\mu\partial_\lambda\phi= 
({\rm via~ eq.~of~motion})=
\partial_\lambda\big(\frac{\partial L}{\partial \partial_\lambda \phi}
\partial_\mu\phi\big) 
$$ 
and thus
$$
\partial_\lambda\Big(-\frac{\partial L}{\partial \partial_\lambda \phi}
\partial_\mu\phi+ \delta^\lambda_\mu L\Big)=0 
$$ 
i.e.
$$
-\frac{\partial L}{\partial \partial_\lambda \phi}
\partial_\mu\phi+\delta^\lambda_\mu L= T^{c\lambda}_\mu 
$$
is conserved.

Example:
$$
L=\frac{1}{2}[-\eta^{\mu\nu}\partial_\mu \phi\partial_\nu
\phi-m^2\phi^2-V(\phi)]
$$
$$
T^c_{\lambda\mu}=\partial_\lambda\phi\partial_\mu\phi+
\eta_{\lambda\mu}\frac{1}{2}[-\eta^{\rho\sigma}\partial_\rho  
\phi\partial_\sigma \phi-m^2\phi^2-V(\phi)].
$$
The conserved energy-momentum vector (covariant components) is given by
$$
P^\mu = \int n_\lambda T^{c\lambda\mu} d^{(n-1)}x
$$
with $n_\mu=(1,0,0,0)$.
$$
T^c_{00}=
\frac{1}{2}[\partial_0\phi\partial_0\phi+
\partial_j\phi\partial_j\phi+m^2\phi^2+V(\phi)]
$$
is the energy density.

\bigskip
\section{Coupling of matter to the gravitational field}

We give an alternative definition of the energy momentum tensor by
means of the coupling to a gravitational field.

It is not always possible to render the matter Lagrangian invariant under
diffeomorphisms by introducing the metric tensor $g_{\mu\nu}$. If it is
possible the
 lagrangian density is given by $\sqrt{-g}~L_s$ where $L_s$ is a {\it
scalar} and the action is 
$$
S_{m}=\int_V\sqrt{-g}~L_s d^n x =\int_V L ~d^nx.
$$
Under diffeomorphisms 
$$
x^\mu=x'^\mu+\varepsilon \xi^\mu(x')
$$
(which obviously include the translations) we have (pull-back from $x$
to $x'$)
$$
g'_{\alpha\beta}(x') = g_{\alpha\beta}(x') +\varepsilon\xi^\lambda(x')
\partial_\lambda g_{\alpha\beta}(x')+ \varepsilon\partial_\alpha\xi^\lambda(x')
g_{\lambda\beta}(x')+ \varepsilon\partial_\beta \xi^\lambda(x')
g_{\alpha\lambda}(x')=
$$
$$
= g_{\alpha\beta}(x')+\varepsilon L_\xi g_{\alpha\beta}(x')
$$
being $L_\xi$ the Lie derivative. In absence of torsion we can compute
the Lie derivative replacing usual derivatives with covariant
derivatives and assuming metric compatibility we obtain 
$$
g'_{\alpha\beta}=g_{\alpha\beta}+\varepsilon\nabla_\beta \xi_\alpha+
\varepsilon\nabla_\alpha\xi_\beta~. 
$$
Putting to zero the change in the action we have
$$
0=\int(\nabla_\mu\xi_\nu+\nabla_\nu\xi_\mu)  \frac{\partial
\sqrt{-g}L_s}{\partial g_{\mu\nu}}d^n x~+
\int \frac{\partial
\sqrt{-g}L_s}{\partial
\partial_\lambda
g_{\mu\nu}}\partial_\lambda(\nabla_\mu\xi_\nu+\nabla_\nu\xi_\mu) ~d^n x+
\delta_\phi S_m
$$
where the last term is the contribution due to the variation
of the matter fields under the diffeomorphisms. 
Such contribution vanishes on the Lagrange equations of motion for the
matter fields. Integrating by parts we have
$$
0=2 \int \nabla_\mu \xi_\nu \left(\frac{\partial
\sqrt{-g}L_s}{\partial g_{\mu\nu}}- \partial_\lambda\frac{\partial
\sqrt{-g}L_s}{\partial
\partial_\lambda
g_{\mu\nu}}\right)d^nx
$$
and again integrating by parts
$$
0= -\int \sqrt{-g}\xi_\nu \nabla_\mu T^{\mu\nu} d^nx
$$
from which
$$
\nabla_\mu T^{\mu\nu}=0
$$
having defined
$$
T^{\mu\nu}=\frac{2}{\sqrt{-g}}\left[ \frac{\partial
\sqrt{-g}L_s}{\partial g_{\mu\nu}}- \partial_\lambda\frac{\partial
\sqrt{-g}L_s}{\partial
\partial_\lambda
g_{\mu\nu}}\right]~.
$$
We can summarize the result as follows
\begin{equation}
\int\sqrt{-g} T^{\mu\nu}\delta g_{\mu\nu} d^nx=
-\int \sqrt{-g} T_{\mu\nu}\delta (g^{\mu\nu}) d^nx=
\int \sqrt{-g} T_{\mu\nu}\delta g^{\mu\nu} d^nx=2 \delta S_m
\end{equation}
where we used the notation of Section \ref{5GammaGammaSec}
$\delta g^{\mu\nu}=g^{\mu\rho}\delta g_{\rho\sigma}g^{\sigma\nu}$.

\bigskip

Example:
The invariant Lagrangian of the previous case is
$$
L_s= \frac{1}{2}[-g^{\mu\nu}\partial_\mu\phi\partial_\nu\phi -m^2\phi^2
-V(\phi)]
$$
and taking into account that $L_s$ does not depend on $\partial_\lambda
g_{\mu\nu}$ we have
$$
T^{\mu\nu} = 2
\frac{1}{\sqrt{-g}}\frac{\partial\sqrt{-g}}{\partial g_{\mu\nu}}L_s+2
\frac{\partial L_s}{\partial g_{\mu\nu}}=\nabla^\mu\phi\nabla^\nu\phi
+g^{\mu\nu}L_s
$$
with $\nabla^\mu\phi = g^{\mu\nu}\partial_\nu\phi$, which agrees with
the previous one for $g_{\mu\nu}=\eta_{\mu\nu}$. 

\section{Dimensions of the Hilbert action}

Giving $g_{\mu\nu}$ the dimensions $l^2$ and $x^\mu$ dimensions $l^0$
we have in $n$ dimensions $\Gamma^\alpha_{\beta\gamma}\sim l^0$,

$R^\alpha_{\beta\gamma\delta}\sim l^0$, $R_{\beta\delta}\sim l^0$,
$R\sim l^{-2}$, $\sqrt{-g}\sim l^n$ and thus $S_H\sim l^{n-2}$.

If we give the dimensions $g_{\mu\nu}\sim l^0$, and $x^\mu\sim l^1$ we
have in $n$ dimensions $\Gamma^\alpha_{\beta\gamma}\sim l^{-1}$,
$R^\alpha_{\beta\gamma\delta}\sim l^{-2}$, $R_{\beta\delta}\sim
l^{-2}$, $R\sim l^{-2}$, $\sqrt{-g}\sim l^0$ and thus again
$S_H\sim l^{n-2}$.

From  (Poisson equation in $n-1$ dimensions)
$$
G\frac{m^2}{l^{n-3}} \sim mc^2
$$
we have
$$
mc^2 \sim \frac{c^4 l^{n-3}}{G}
$$   
and
$$
{\rm action}~ \sim mc^2 t \sim mc^2 \frac{l}{c} \sim \frac{c^3 l^{n-2}}{G}
$$   
and thus we have the action, in all dimensions
$$
\frac{c^3}{16\pi G}S_H +S_{\rm matter}
$$
where the factor $16\pi$ is obtained by fitting Newton's law.
From
$$
\hbar =\frac{c^3 l_P^{n-2}}{G}
$$
we derive the expression of the Planck length
$$
l_P = \left(\frac{G\hbar}{c^3}\right)^{1/(n-2)}
$$ 
which is $n=4$ becomes
$$
l_P = \left(\frac{G\hbar}{c^3}\right)^{1/2}.
$$
Thus the adimensional exponent which appears in Feynman functional
integral in four dimensions is
$$
\frac{1}{16\pi l_P^2}(S_H+{\rm boundary~terms})+\frac{1}{\hbar} S_{\rm
matter}~.
$$ 
The Planck length can be obtained also equating one half of the
Schwarzschild radius (in four dimensions) 
$$
\frac{Gm^2}{\frac{r_s}{2}}=mc^2~~~~{\rm i.e.}~~~~~\frac{r_s}{2} = \frac{Gm}{c^2}
$$
to the Compton wave length
$$
r_c = \frac{\hbar}{mc}.
$$
Solving in $m$ the equation $r_s/2 = r_c$ gives the Planck mass
$$
m_P =\left(\frac{\hbar c}{G}\right)^{1/2} = 1.21~10^{19}GeV/c^2 =2.17~
10^{-8} kg 
$$
whose Compton wave length is the Planck length
$$
l_P = \frac{\hbar}{m_p c} = 1.62 ~10^{-35} m. 
$$

One can also obtain a complete set of units exploiting the electric
charge. From
$$
\frac{e^2}{r} = G \frac{m^2}{r}
$$
we have
$$
m=\sqrt{\frac{e^2}{G}}
$$
where $\hbar$ does not appear. Such $m$ is related to the Planck mass
as follows
$$
m=\sqrt{\frac{e^2}{G}}=m=\sqrt{\frac{e^2}{\hbar c}\frac{\hbar c}{G}}=
\frac{1}{\sqrt{137}}~m_P~.
$$

\section{Coupling of the Dirac field to the gravitational field}
\label{5couplingmatterSec}

First we consider the Dirac Lagrangian in Minkowski space.

We saw how under a Lorentz transformation 
\begin{equation}\label{lorentzdirac}
x'^\mu = \Lambda^\mu_{~\nu}x^\nu
\end{equation}
we have with 
$$
\Psi(x) = 
\begin{pmatrix}
\psi\\
\phi
\end{pmatrix}
$$
$$
\Psi'(x') = 
\begin{pmatrix}
\tilde A&0\\
0&A
\end{pmatrix}
\Psi(x) \equiv
{\cal A}\Psi(x)
$$
and we shall write ${\cal A}={\cal A}[\Lambda]$ where ${\cal A}$ is
defined up to a sign. 

We recall that
$$
A\sigma_\mu x^\mu A^+= 
\sigma_\mu x'^\mu = \sigma_\mu \Lambda^\mu_{~\nu}x^\nu
$$
or
\begin{equation}\label{trA}
A\sigma_\mu A^+ = \sigma_\nu \Lambda^\nu_{~\mu}
\end{equation}
and similarly
\begin{equation}\label{trtildeA}
\tilde A\tilde \sigma_\mu \tilde A^+ = \tilde\sigma_\nu \Lambda^\nu_{~\mu}
\end{equation}
with $\tilde\sigma_\mu=(\sigma_0,-\sigma_m)$ and $\tilde A= (A^+)^{-1}$.

$\Psi^+\Psi$ is not an invariant; instead $\Psi^+\gamma^0\Psi = - i(
\psi^+\phi + \phi^+\psi) \rightarrow -i( \psi^+A^{-1}A\phi +
\phi^+A^+A^{-1+}\psi)$ is invariant. It is usual to define
$$
\bar\Psi = i\Psi^+\gamma^0\equiv \Psi^+\gamma^4
$$ 
and we have
$$
\bar\Psi'(x') = \bar \Psi(x){\cal A}^{-1}.
$$
We recall that
$$
\gamma_\mu =i
\begin{pmatrix}
0&\tilde\sigma_\mu \\
\sigma_\mu &0
\end{pmatrix}
$$
and from Eq.(\ref{trA},\ref{trtildeA}) we have
\begin{equation}\label{5calArepresentation}
{\cal A}\gamma_\mu {\cal A}^{-1} = \gamma_\nu \Lambda^\nu_{~\mu}.
\end{equation}
In general for a vector $v$ we have
\begin{equation}\label{gammavector}
{\cal A}\gamma^\mu {\cal A}^{-1}v_\mu={\cal A}\gamma_\mu {\cal
A}^{-1} v^\mu=
\gamma_\nu\Lambda^\nu_{~\mu}v^\mu=\gamma_\nu v'^\nu = \gamma^\nu v'_\nu
\end{equation}
from which follows the invariance of the Lagrangian
$$
L = \bar\Psi(\gamma^\mu \partial_\mu +m)\Psi.
$$ 
The invariance of the mass term is trivial while for the kinetic term
we have under global Lorentz transformations
$$
\bar\Psi \gamma^\mu \partial_\mu\Psi = \bar\Psi{\cal A}^{-1}{\cal
A}\gamma^\mu {\cal A}^{-1}\partial_\mu {\cal A}\Psi=
\bar\Psi' \gamma^\mu \partial'_\mu \Psi'
$$
having used Eq.(\ref{gammavector}).

We want now to write a Lagrangian invariant under diffeomorphisms and
under local Lorentz
transformations $SO(3,1)$. Let us consider the Lagrangian
\begin{equation}\label{diracgrav}
\bar\Psi e^\mu_{~a}\gamma^a(\partial_\mu+\omega_\mu)\Psi+m\bar\Psi\Psi.
\end{equation}
being $\omega_\mu$ the components of a $4\times 4$ matrix valued $1$-form.
We recall that the vierbeins $e^a_{~\mu}$ are defined by $e^a =
e^a_{~\mu} dx^\mu$, being $e^a$ the forms dual to ${\bf e}_a$; the
$e^\mu_{~a}$ appearing in Eq.(\ref{diracgrav}) are the inverse of
$e^a_{~\mu}$ i.e. $e^a_{~\mu} e^\mu_{~b}= \delta^a_b$. 
The procedure is to treat the $\Psi$ as scalars under diffeomorphisms;
then we have that 
the above written Lagrangian is invariant under diffeomorphisms, due to
the contraction $e^\mu_{~a} \partial_\mu$. 
Let us now consider local $SO(3,1)$ transformations (no diffeomorphism)
${\bf e}_a ={\bf e}'_b
\Lambda^b_{~a}$, which give $v'^a = \Lambda^a_{~b}{v}^b$ 
(cfr. Eq.(\ref{lorentzdirac})), where $\Lambda$ now depends on $x$. 
We recall that 
\begin{equation}\label{gaugetrgamma}
\Gamma'=\Lambda d\Lambda^{-1} +\Lambda\Gamma \Lambda^{-1} 
\end{equation}
and
$$
\Psi'(x') = {\cal A}[\Lambda]\Psi(x)~.
$$
We have 
$$
\bar\Psi
e^\mu_{~a}\gamma^a(\partial_\mu+\omega_\mu)\Psi+m\bar\Psi\Psi=\bar\Psi'
e^\mu_{~a}{\cal A}\gamma^a {\cal A}^{-1}(\partial_\mu+{\cal A}\partial_\mu
{\cal A}^{-1}
+{\cal A}\omega_\mu {\cal A}^{-1})\Psi'+m\bar\Psi'\Psi'=  
$$
$$
=\bar\Psi'
e'^\mu_{~a} \gamma^a (\partial_\mu+{\cal A}\partial_\mu {\cal A}^{-1}
+{\cal A}\omega_\mu {\cal A}^{-1})\Psi'+m\bar\Psi'\Psi'  
$$
due to Eq.(\ref{gammavector}). Thus we have that the
Lagrangian (\ref{diracgrav}) is invariant iff the 1-form $\omega$
transforms as follows
\begin{equation}\label{gaugetromega}
\omega' = {\cal A}d {\cal A}^{-1} +{\cal A}\omega {\cal A}^{-1}.
\end{equation}
Any 1- form which transforms according to Eq.(\ref{gaugetromega}) does the
job. It is important however that one can find a $\omega$ with the
above properties, without introducing a new dynamical field. We recall
that $\Gamma^a_{~b}$ belongs to the algebra of $SO(3,1)$ due to the
metric compatibility
$$
0=d\eta-\eta\Gamma-\Gamma^T\eta =-\eta\Gamma-\Gamma^T\eta~. 
$$
Moreover we saw that there is (up to the sign) a one to one
correspondence between $\Lambda$ and ${\cal A}$. Thus with $\Lambda=
I+\varepsilon \rho$ and ${\cal A}[\Lambda]={\cal A}[1+\varepsilon\rho] 
= I+\varepsilon a$
$$
(I+\varepsilon \rho) \leftrightarrow (I+\varepsilon a)
$$
induced by Eqs.(\ref{trA},\ref{trtildeA}) which as we are in the
neighborhood of the origin, sets a one-to-one 
correspondence between $\rho$ and $a$. 
$a$ is a linear function of $\rho$ whose space is $n(n-1)/2 = 6$
dimensional (as we are referring to the frames ${\bf e}_a$ we
shall use latin indices)
$$
a = \Sigma^{~b}_{a} \rho^a_{~b} = \Sigma_{ab} \rho^{ab}. 
$$
We construct $\omega$ as follows 
\begin{equation}\label{spinconnection}
\omega = \Sigma^{~b}_{a} \Gamma^a_{~b} =\Sigma_{ab} \Gamma^{ab} 
\end{equation}
i.e. $\omega$ are defined as the representative in the representation
$(1/2,0)\oplus(0,1/2)$ of the Lie algebra elements $\Gamma$.
We must now prove that under the transformation (\ref{gaugetrgamma})
of $\Gamma$ the so defined $\omega$ transforms according to
(\ref{gaugetromega}) i.e.
$$
\omega'=\Sigma_a^{~b}(\Lambda
d\Lambda^{-1}+\Lambda\Gamma\Lambda^{-1})^a_{~b}=
{\cal A}d {\cal A}^{-1} +{\cal A}\omega {\cal A}^{-1} ~.
$$
This simply follows from the fact that ${\cal
A}\in (1/2,0)\oplus(0,1/2)$ is a (double valued) representation of
$\Lambda\in SO(3,1)$.

In fact if $\Lambda$ and $\Lambda + d\Lambda$ are two nearby transformations of
$SO(3,1)$ then from
$$
\Lambda(\Lambda^{-1} + d\Lambda^{-1}) = I + \Lambda d\Lambda^{-1}
$$
we have that $\Lambda d\Lambda^{-1} \in so(3,1)$ , i.e. it belongs to
the algebra of $SO(3,1)$. Then we have 
$$
{\cal A}[\Lambda(\Lambda^{-1} + d\Lambda^{-1})] = {\cal A}[I+
\Lambda d\Lambda^{-1}] = I+ \Sigma^{~b}_{a}
(\Lambda d\Lambda^{-1})^a_{~b} = 
$$
$$
={\cal A}[\Lambda] ({\cal A}[\Lambda^{-1}] + d
{\cal A}^{-1})= I + {\cal A}[\Lambda] d{\cal A}^{-1} 
= I+ {\cal A}[\Lambda] d{\cal A}^{-1}    
$$
from which
$$
\Sigma^{~b}_{a}(\Lambda d\Lambda^{-1})^a_{~b}=
{\cal A}[\Lambda] d{\cal A}^{-1}
$$
Similarly
$$
\Lambda(I+\varepsilon \Gamma) \Lambda^{-1}= I + \varepsilon
\Lambda\Gamma \Lambda^{-1}\in SO(3,1)
$$
and thus
$\Lambda \Gamma\Lambda^{-1} \in so(3,1)$. Thus
$$
{\cal A}[\Lambda(I+\varepsilon \Gamma)\Lambda^{-1}] = {\cal A}[I+
\varepsilon \Lambda\Gamma\Lambda^{-1}] = I+ \varepsilon \Sigma^{~b}_{a}
(\Lambda\Gamma\Lambda^{-1})^a_{~b} = 
$$
$$
={\cal A}[\Lambda] {\cal A}[I + \varepsilon \Gamma]{\cal
A}[\Lambda^{-1}]= {\cal A}[\Lambda](I+\varepsilon
\Sigma^{~b}_a\Gamma^a_{~b}){\cal A}[\Lambda^{-1}]= {\cal A}[\Lambda]
(I+\varepsilon\omega){\cal A}[\Lambda^{-1}]= I+ \varepsilon {\cal A}\omega
{\cal A}^{-1}
$$
from which
$$
\Sigma^{~b}_a(\Lambda\Gamma \Lambda^{-1})^a_{~b}={\cal A}\omega {\cal A}^{-1}.
$$
Summing the two contributions we have Eq.(\ref{gaugetromega}).

We compute now explicitly the matrices $\Sigma_{ab}$.
Writing for the infinitesimal Lorentz transformation 
$I+ \varepsilon \rho$, ${\cal A}[I+ \varepsilon \rho]=I+\varepsilon
\Sigma_{a b}\rho^{a b}$, with
$\Sigma_{a b}=-\Sigma_{ba}$, keeping in mind that
$\rho^{ab}=-\rho^{ba}$  we have from Eq.(\ref{5calArepresentation})
$$
{\cal A}[I+ \varepsilon \rho]\gamma_c~{\cal A}[I+ \varepsilon \rho]^{-1} = 
\gamma_d~(I+ \varepsilon \rho)^d_{~c}
$$
i.e.
$$
[\Sigma_{ab}\rho^{ab},\gamma_c] = \gamma_d \rho^{df}\eta_{fc}
$$
or
$$
[\Sigma_{ab},\gamma_c] = \frac{1}{2}(\gamma_a \eta_{bc}-\gamma_b \eta_{ac}).
$$
Such equation is simply solved by
$$
\Sigma_{ab}=\frac{1}{8}[\gamma_a,\gamma_b].
$$
as it is checked by using the Clifford algebra of the $\gamma$'s.
The explicit representation of the $\Sigma_{ab}$ is
$$
\Sigma_{ij}=\frac{1}{4}
\begin{pmatrix}
\sigma_i\sigma_j&0\\
0&\sigma_i\sigma_j
\end{pmatrix},
~~~~~
\Sigma_{i0}=\frac{1}{4}
\begin{pmatrix}
\sigma_{i}&0\\
0&-\sigma_{i}
\end{pmatrix}.
$$
Notice that the two dimensional traces of these $\Sigma_{ab}$ are zero
so that $I+\varepsilon \Sigma_{ab}\rho^{ab}\in
(1/2,0)\oplus(0,1/2)$. As near the identity the correspondence
between the Lorentz transformations and the matrices ${\cal A}$ is a 
bijection, our solution
for the $\Sigma_{ab}$ is the unique solution.

Multiplying by $2i$ due to the different normalization of the
generators, we
have the same result as [WeinbergQFT] I, p. 217 (where the sum in
the analog of Eq.(\ref{spinconnection}) is done for $a>b$ while here
the sums are on all $a$ and $b$).
$\omega$ is called the spin connection and has to be considered as the
fundamental connection.

Summarizing the gauging of the Dirac Lagrangian is given by
$$
\bar\Psi\left(\gamma^a\partial_a\Psi+m\Psi\right)\rightarrow
\bar\Psi\left(\gamma^a e^\mu_a(\partial_\mu+\omega_\mu)\Psi+m\Psi\right) 
$$
$$
\equiv \bar\Psi\left(\gamma^a e^\mu_a D_\mu\Psi+m\Psi\right)\equiv
\bar\Psi\left(\gamma^ a
D_a\Psi+m\Psi\right)=L_s 
$$
with obvious definitions of $D_\mu$ and $D_a$ and 
$$
S_M=\int L_s \bfepsilon=\int e L_s d^4x=\int L d^4x
$$
is the invariant Dirac action. Performing an integration by parts and
using the fact that our connection $\Gamma$ is metric compatible one
proves that the action $S_M$ is hermitean.


\section{The field equations}

We shall put $\kappa= 8 G\pi/c^2$. The total action becomes
$$
\frac{1}{2\kappa}S_{EC}+S_M + {\rm  b.t.}
$$
where, for $n=4$
$$
S_{EC} = \frac{1}{2}\int R^{ab}\wedge e^{c}\wedge e^d
~\varepsilon_{a b c d }~. 
$$
Variation with respect to $e^d$ gives
$$
R^{ab}\wedge e^c \varepsilon_{abcd} = \kappa t_d
$$
where we set
$$
\delta S_M = - \int t_d\wedge \delta e^d
$$
which is always possible.
Variation with respect to $\Gamma^{ab}$ gives
\begin{equation}\label{torsioneq}
D e^c\wedge e^d ~\varepsilon_{abcd}=
 S^c\wedge e^d \varepsilon_{abcd} = s_{ab}\in \Lambda^3
\end{equation}
where
$$
\kappa~\delta S_M = -\int \delta \Gamma^{ab} \wedge s_{ab}.
$$
Due to $\Gamma^{ba}= -\Gamma^{ab}$, $s_{ba}=- s_{ab}$.
We see that in presence of fermions $s_{ab}\neq 0$. In fact in the
case of the Dirac field we have
$$
s_{ab}= -\kappa~e~\bar\Psi\gamma^c e^\mu_c
\Sigma_{ab}\Psi\varepsilon_{\mu\nu\lambda\rho}~
\frac{1}{3!}dx^\nu\wedge dx^\lambda\wedge dx^\rho
$$ 
i.e. the source of the torsion is given by the spin content.
From this
we see that in the Einstein-Cartan formulation in presence of fermions
the torsion is not zero. We shall now solve Eq.(\ref{torsioneq}). The
procedure 
has an immediate extension to $n$ dimensions even if we work with
$n=4$. Writing
$$
S^c = \frac{1}{2}S^c_{ij} e^i\wedge e^j
$$
and 
$$
s_{ab} = \frac{1}{3!}s^k_{ab} \varepsilon_{kmnp} e^m\wedge e^n\wedge e^p
$$
and multiplying by $e^h$ we reach
$$
\frac{1}{2}S^c_{ij}\delta^{ijh}_{abc}= s^h_{ab}
$$
i.e
$$
S^h_{ab}+S^c_{ca}\delta^h_b - S^c_{cb}\delta^h_a = s^h_{ab}.
$$
Taking the trace we have
$$
S^c_{cb} + S^c_{cb}-4 S^c_{cb} = s^c_{cb}
$$
from which
$$
S^c_{cb} =-\frac{1}{2}s^c_{cb}
$$
and
$$
S^h_{ab}= s^h_{ab} + \frac{1}{2}\left( \delta^h_b s^c_{ca} -
\delta^h_a s^c_{cb}\right).  
$$
Thus the torsion is confined in the region where the matter field is
different from zero; we have a non propagating torsion. On the other
hand theories in
which the gravitational action contains quadratic or higher terms in
the Riemann tensor can possess propagating torsion.

In order to find the connection $\Gamma^a_{~b}$, which by assumption is
metric compatible, we shall define
$$
\Gamma^a_{~b} = \Gamma[e]^a_{~b}+ K^a_{~b}
$$
being $\Gamma[e]^a_{~b}$ the Levi-Civita connection and
$K^a_{~b}$ the 
contorsion tensor 1-form. $K^a_b$ is a true tensor being the
difference of two connections. 
To compute $\Gamma[e]$ we could start from $g_{\mu\nu}= e^a_\mu
\eta_{ab}e^b_\nu$, substitute in 
$$
\Gamma[e]^\alpha_{~\beta\gamma}=\frac{1}{2}g^{\alpha\alpha'}(\partial_{\beta}
g_{\alpha'\gamma}+\partial_{\gamma} g_{\beta\alpha'} 
-\partial_{\alpha'} g_{\beta\gamma})
$$
and perform the
transformation (\ref{transfconn}) with $\Omega^\mu_{~a} = e^\mu_{~a}$
to reach the value in our orthonormal 
frame. It is instructive to follow a more direct procedure. Vanishing
torsion for $\Gamma[e]$ means
$$
0= de^a + \Gamma[e]^a_{~b}\wedge e^b. 
$$
Setting 
$$
de^a =\frac{1}{2}E^a_{~cb}e^c\wedge e^b
$$
we have
$$
0=E_{acb}+\Gamma[e]_{abc}-\Gamma[e]_{acb}. 
$$
Rotating the indices twice and subtracting from the first two the
last one, keeping in mind that due to metric compatibility
$\Gamma[e]_{bac} = -\Gamma[e]_{abc}$, we reach
$$
\Gamma[e]_{abc}=\frac{1}{2}\left(E_{cba} + E_{abc}-E_{bac}\right). 
$$ 
We must determine now the contorsion from $S^a$ that we already know.
$$
S^a = de^a+\Gamma[e]^a_{~b}\wedge e^b + K^a_{~b}\wedge e^b =K^a_{~b}\wedge e^b
$$
Setting as usual
$$
S^a = \frac{1}{2}S^a_{~ij}e^i\wedge e^j;~~~~K^a_{~b} =
K^a_{~bi} e^i
$$
we have
$$
S^a_{~ij} =K^a_{~ji}- K^a_{~ij}
$$
and lowering one index
$$
S_{aij} =K_{aji}- K_{aij}.
$$
Keeping in mind that metric compatibility imposes
$K_{aij}=-K_{iaj}$ one can solve in $K$ with the usual method of
rotation of indices and finally we obtain
$$
K_{ija} = \frac{1}{2}(S_{aij}+S_{iaj}-S_{jai})
$$
\begin{equation}\label{fullconnection}
\Gamma_{abc}=\Gamma[e]_{abc}+\frac{1}{2}\left( S_{cab} + S_{acb}-
S_{bca} \right) 
\end{equation}
which correctly is antisymmetric in $a,b$.

Having solved completely the connection $\Gamma$ in terms of
$\Gamma[e]$ and $K$ which is given in terms of the torsion source
$s_{ab}$ we can go back to the first equation
$$
R^{ab}\wedge e^c\varepsilon_{abcd}=\kappa~t_d~.
$$
We have
$$
R(\Gamma[e]+K) = R(\Gamma[e]) + D[e] K + K\wedge K 
$$
where $R(\Gamma[e])$ is the usual metric curvature. 
Thus the equation has now become
\begin{equation}\label{5vierbeinemt}
\left(R(\Gamma[e]) + D[e] K + K\wedge K\right)^{ab}\wedge e^c 
\varepsilon_{abcd}=\kappa~ t_d. 
\end{equation}
Such equation has to be supplemented by the one obtained by varying
the matter field $\Psi$ i.e.  the Dirac equation
\begin{equation}\label{5dirac2}
\gamma^a D_a \Psi + m\Psi=0.
\end{equation}
The independent variables in the system
(\ref{5vierbeinemt},\ref{5dirac2}) are the vierbeins $e^a_{~\mu}$ and
the Dirac field $\Psi$.

Given a solution to equations (\ref{5vierbeinemt},\ref{5dirac2}) we
can apply to it an arbitrary diffeomorphism transformation and an
arbitrary local Lorentz rotation to obtain again a solution. 
Obviously is solving the system 
Eqs.(\ref{5vierbeinemt},\ref{5dirac2}) one can apply a gauge fixing 
procedure.

By multiplying Eq.(\ref{5vierbeinemt}) on the right by $e^f$ we have
$$ 
\left(R(\Gamma[e]) + D[e] K + K\wedge K\right)^{ab}\wedge e^c\wedge
e^f\varepsilon_{abcd}=\kappa~ t_d\wedge e^f. 
$$
Then isolating the $R(\Gamma[e])$ term (which is the metric Riemann
tensor) and proceeding as in the discussion of the pure  Einstein-
Cartan action, Eq.(\ref{5vierbeinemt}) can be rewritten as
$$
G^f_d[e] = \kappa~ T(eff)^f_d
$$
where $T(eff)^f_d$ is an effective symmetric energy momentum tensor,
with $\nabla[e]_f T(eff)^f_g =0$, which contains also the four fermion term
originating from $K\wedge K$. However $T^{ab}$ cannot be obtained from the
variation of the action with respect to $g_{ab}$ because  $g_{ab}$
does not appear in the action, being $e^a_\mu$ and $\Gamma^{ab}_\mu$
the fundamental gravitational variables. Still is better to work
directly with the system (\ref{5vierbeinemt},\ref{5dirac2}).

\bigskip

We saw that in the first order Einstein-Cartan approach when the
gravitational field is coupled with the Dirac field the connection
acquires torsion.

However this is not the only way to couple fermions to gravity. One
could as well consider the torsionless Levi-Civita connection
$\Gamma[e]$ and couple invariantly the fermion to gravity via [2]
$$
\bar\psi\gamma^\rho\partial_\rho \psi\rightarrow
\bar\psi\gamma^\rho(\partial_\rho+\Sigma_{ab}\Gamma[e]^{ab}_\rho) \psi
$$
and as such this way of proceeding can be called a second order approach. 
The action is still invariant under diffeomorphisms and under local
Lorentz transformations. The two Lagrangians differ by a four-fermion 
term. We shall see in Chapter \ref{supergravityChap} that the formulation of
supergravity is simplest in the first order approach in which torsion
is present. 
 
\bigskip

References

\smallskip

[1] A. Trautman, Bull. Acad. Pol. Sci. Ser. Math. Astron. Phys.
````On the Einstein-Cartan equations''20 (1972) 185, 503, 895

\smallskip

[2] P. van Nieuweinhuizen, ``Supergravity'', Physics Reports 68 (1981) 189

\section{The generalized energy- momentum tensor}

We saw that if the coupling of matter to gravity can be achieved by means of
the metric tensor $g_{\mu\nu}$ the energy momentum tensor can be defined by
the relation
$$
\delta S_M = \frac{1}{2}\int d^nx~\sqrt{-g}~{\bf T}^{\mu\nu}\delta g _{\mu\nu}
$$
or
$$
{\bf T}^{\mu\nu}=\frac{2}{\sqrt{-g}}\frac{\delta S_M}{\delta g_{\mu\nu}}={\bf
T}^{\nu\mu} 
$$
where we used the boldface for reason of clearness. By definition ${\bf
T}^{\mu\nu}$ is symmetric and we found that on-shell
$$
\nabla_\mu {\bf T}^{\mu\nu}=0~.
$$
Einstein equations take the form
$$
G^{\mu\nu}=\kappa~ {\bf T}^{\mu\nu}
$$
where we could also add the cosmological constant.

If gravity is described by the vierbeins $e^a_\mu$ as it is unavoidable in
presence of fermions the energy momentum tensor is \underline{defined} by
$$
\frac{\delta S_M}{\delta e^a_{~\mu}(x)}= e T^\mu_{~a}(x)~.
$$
The mixed tensor $T^\mu_{~a}(x)$ will be the fundamental object in 
the following.
In boson theory if we introduce artificially $e^a_\mu$ through $g_{\mu\nu}=
e^a_\mu\eta_{ab}e^b_\nu$ we have
$$
T^\lambda_{~a} = \frac{1}{2}({\bf T}^{\lambda\nu}\eta_{ab}e^b_{~\nu}+
{\bf T}^{\mu\lambda}\eta_{ab}e^b_{~\mu})
$$
and
$$
T^{\lambda a}\equiv T^\lambda_{~b}\eta^{ba}=\frac{1}{2}({\bf
T}^{\lambda\nu}e^a_{~\nu}+{\bf T}^{\mu\lambda}e^a_{~\mu})~.
$$
If we multiply by $e^\rho_{~a}$ we have
$$
T^{\lambda a}e^\rho_a=\frac{1}{2}({\bf T}^{\lambda\rho}+{\bf
T}^{\rho\lambda})= {\bf T}^{\lambda\rho}~.
$$
The action in the fermionic case (and trivially in the bosonic case) is
invariant under local Lorentz transformations
$$
\delta e^a_{\mu}=\alpha^a_{~b}e^b_{~\mu}, ~~~~{\rm
with}~~\alpha^{ab}=-\alpha^{ba}~. 
$$
Then \underline{on the equations of motion} we have
$$
0=\delta S_M =\int e T^\mu_{~a}\alpha^a_{~b}e^b_{~\mu}d^nx=
\int e T^{ba}\alpha_{ab} d^nx
$$
i.e. on the equations of motion the antisymmetric part of $T^{ba}\equiv 
e^b_{~\mu} T^{\mu}_{~c}\eta^{ca}$ has to vanish. 
Thus we have the symmetry of $T^{ba}$ as
a consequence of local Lorentz invariance.

We examine now the consequences of invariance under diffeomorphisms. Under a
diffeomorphisms we have (Lie derivative of a covariant vector)
\begin{equation}\label{Liedervierbein}
\delta_\xi e^a_{~\mu}=\xi^\nu\partial_\nu e^a_{~\mu}+\partial_\mu \xi^\nu
e^a_{~\nu}~.
\end{equation}
But we can rewrite Eq.(\ref{Liedervierbein}) as
$$
\delta_\xi e^a_{~\mu}=\xi^\nu\nabla[e]_\nu e^a_{~\mu}+\nabla[e]_\mu \xi^\nu
e^a_{~\nu}
$$
where $\nabla[e]_\nu$ is the covariant derivative calculated by means of the
metric compatible connection $\Gamma[e]^\mu_{\lambda\nu}$ 
because in the Lie derivative
we can replace the derivatives with covariant derivatives provided the
connection is torsionless. Then, ignoring on the equations of
motion the variation of the matter fields, we have
$$
\delta_\xi S_M = \int d^nx e T^{\mu}_{~a} (\xi^\nu \nabla[e]_\nu e^a_{~\mu}+
e^a_{~\nu} \nabla[e]_\mu\xi^\nu)~. 
$$
Integrating by parts the second term, using the metric compatibility of
$\Gamma[e]$ we have
\begin{equation}\label{interm}
\delta_\xi S_M = \int d^nx ~e ~(T^{\mu}_{~a} \xi^\nu \nabla[e]_\nu e^a_{~\mu}
-\xi^\nu \nabla[e]_\mu T^{\mu}_{~\nu}) 
\end{equation}
where we defined from the fundamental $T^\mu_{~a}$, $T^\mu_{~\nu}=T^\mu_{~a}
e^a_{~\nu}$.

We examine now the first term. We recall that
$$
{\bf e}_a = {\bf u}_\mu e^\mu_{~a}
$$
so that
$$
\Gamma[e]^a_{~b}= e^a_{\mu}de^\mu_{~b}+ e^a_{~\mu}\Gamma[e]^\mu_{~\lambda
}e^\lambda_{~b}= -de^a_{\mu} e^\mu_{~b}+ e^a_{~\mu}\Gamma[e]^\mu_{~\lambda
}e^\lambda_{~b}
$$
i.e.
$$
\Gamma[e]^a_{~b}e^b_{\mu}= -\nabla[e] e^a_{~\mu} 
$$
and thus the first term becomes
\begin{equation}\label{antisymm}
-\int d^nx e T^{ba}\xi^\nu\Gamma[e]_{ab\nu}~.
\end{equation}
As we have that $\Gamma[e]_{ab\nu}$ is antisymmetric in $a,b$, and
 $T^{ab}$ symmetric in $a,b$, which is a consequence of local Lorentz
invariance,  Eq.(\ref{antisymm}) vanishes. Then from
Eq.(\ref{interm}) it follows that
$$
\nabla[e]_\mu T^{\mu\nu}=0
$$
which is a consequence of the combined invariances under local Lorentz
transformations and diffeomorphisms. Such symmetry and covariant conservation
property of the energy momentum tensor are consequences of the
invariance at the classical level. Such properties can be violated at the
quantum level giving rise to the so called gravitational anomalies.

\bigskip

References

\smallskip

[1] L. Alvarez-Gaum\'e, E. Witten,``Gravitational Anomalies''
Nucl. Phys. B234 (1984) 269

\smallskip

[2] R. Bertlmann,``Anomalies in Quantum Field Theory'' Oxford University
Press 1996.

\section{Einstein-Eddington affine theory}

It is remarkable that one can provide an affine formulation of gravity,
i.e. a formulation in which the basic variable is an affine connection
and no metric is present at the beginning.

We start with a symmetric connection (absence of torsion)
$\Gamma^\lambda_{\mu\nu}$ from which the Ricci tensor can be readily
computed. No metric is present. The action is
$$
S_{EE} =  \int L_{EE}~d^4x=k\int \sqrt{|\det R_{(\mu\nu)}|}~ d^4x.
$$ 
where $R_{(\mu\nu)}$ denotes the symmetrized Ricci tensor and $k$ is a
constant.
Such an action is diffeomorphism invariant because
$$
\det R'_{(\rho\sigma)} = \det R_{(\mu\nu)} ~\det\frac{\partial
x^\mu}{\partial x'^\rho} ~\det\frac{\partial
x^\nu}{\partial x'^\sigma} 
$$
while
$$
d^4x'=d^4x ~\det\frac{\partial x'^\mu}{\partial x^\nu}.
$$
The equations of motion are
$$
0= \int\frac{\partial L_{EE}}{\partial R_{(\mu\nu)}}\delta R_{\mu\nu}=
\int\frac{\partial L_{EE}}{\partial R_{(\mu\nu)}}
({\bfnabla}_\lambda \delta\Gamma_{\nu}-{\bfnabla}_\nu
\delta\Gamma_{\lambda})^\lambda_\mu d^4x
$$ 
Define now the symmetric $(2,0)$ tensor $g^{\mu\nu}$
\begin{equation}\label{5defg}
\sqrt{|g|} g^{\mu\nu}\equiv k
\frac{\sqrt{|\det R_{(\cdot\cdot)}|}}{2} [R_{(\cdot\cdot)}^{-1}]^{\mu\nu}=
\frac{\partial L_{EE}}{\partial R_{(\mu\nu)}} 
\end{equation}
where $g$ is the determinant of the inverse of $g^{\mu\nu}$ which will
be denoted by $g_{\mu\nu}$ i.e. by definition
$g^{\mu\lambda}g_{\lambda\nu}=\delta^\mu_\nu$.
$g^{\mu\nu}$ is a $(2,0)$ tensor as seen by taking the determinant of both
sides of Eq.(\ref{5defg}). 
The equations of motion now take the form
$$
0=  \int g^{\mu\nu}
\sqrt{|g|}(\bfnabla_\lambda \delta\Gamma_{\nu}-\bfnabla_\nu
\delta\Gamma_{\lambda})^\lambda_\mu d^4x
$$
and we are faced exactly with the Palatini problem of Section 5.3; 
we deduce that
$\nabla_\lambda g_{\mu\nu}=0$ from which $\Gamma$ is the unique metric
compatible torsionless connection. But now the associated Ricci tensor
is symmetric and going back to Eq.(\ref{5defg}) we have
$$
\frac{2}{k}g_{\mu\nu}=R_{\mu\nu}
$$
which is equivalent to
$$
R_{\mu\nu} -\frac{1}{2}g_{\mu\nu} R +\Lambda g_{\mu\nu}=0
$$
i.e. pure Einstein theory with a cosmological constant with $\Lambda
=2/k$.
Coupling to matter, if attainable through the $g^{\mu\nu}$, presents no
difficulties.

The symmetry constraint on the connections $\Gamma^\mu_{\nu\lambda}$
has also been relaxed in the attempt to interpret the antisymmetric
part of the $\Gamma^\mu_{\nu\lambda}$ as the field-strength tensor of
the electromagnetic field [2].

\bigskip

References

\smallskip

[1] G. Magnano, ``Are there metric theories of gravity other that
General Relativity?'' XI  Italian Congress of General Relativity and
Gravitation, (1994) 213;  e-Print Archive: gr-qc/9511027 and
references within.

\smallskip

[2] A.T. Filippov ``An old Einstein-Eddington generalized gravity and
modern ideas of branes and cosmology'' arXiv:1011.2445v1 [gr-qc] and
references within.


\chapter{Submanifolds}\label{6submanifolds}
 
\section{Introduction}
Given a 4-dimensional manifold with lorentzian metric let us consider
a 3-dimensional sub-manifold $\Sigma$. The tangent space of $\Sigma$ at
any point is 3-dimensional. Let us consider the normal to $\Sigma$ at
a point, i.e. the vector which is orthogonal to all tangent vectors at
$p$.  Such a vector always exists and is unique up to a multiplying
factor.  In fact if ${\bf v}_1,{\bf v}_2,{\bf v}_3$ is a base of the
tangent space at $p$ let us consider a further independent vector
${\bf v}_0$ and construct $g^{0\mu}{\bf v}_\mu$, where $g_{\mu\nu}
=({\bf v}_\mu,{\bf v}_\nu)$. The scalar product
with ${\bf v}_j$ is $\delta^0_j=0$. Any vector orthogonal to the
tangent space at $p$ is $a^\mu{\bf v}_\mu$ with $a^\mu g_{\mu j}\equiv a_j=0$.
Thus the only freedom is the value of $a_0$.

If the normal at any point of $\Sigma$ is time-like that surface is
said space-like. If it is space-like the surface is said time-like. If
it is light-like the surface is said to be null.

Now we prove that the intrinsic metric of a space like surface is
positive definite. That of a time-like surface is of signature
$-++$. The metric of a null surface is degenerate.

In fact for a space-like surface the representation of $g$ in the base
${\bf n},{\bf v}_j$ is $g_{00}<0$, $g_{0j}=0$ 
from which it follows that the $3\times3$ -matrix
$({\bf v}_j,{\bf v}_k)$ has signature $+++$. Similarly for a time-like
surface the signature of the matrix $({\bf v}_j,{\bf v}_k)$ is
$-++$. For a null-surface we have $0=({\bf n},{\bf n})=
(g^{0\mu}{\bf v}_\mu,g^{0\nu}{\bf v}_\nu) = g^{00}$ i.e.
${\bf n}= g^{0j} {\bf v}_j$, ${\bf n}\in T(p)$ and at the same
time orthogonal to $T(p)$, giving $g^{0j}g_{jl}=0$.

\section{Extrinsic curvature}

Given a surface $\Sigma$ embedded in an $n$ dimensional manifold $M$,
i.e. a
$n-1$ dimensional manifold embedded in $M$, let us consider the unit
vector $n$ normal to the surface $\Sigma$. 

We construct the tensor $h_{ab} = g_{ab}\mp n_a
n_b$ for $n_a n^a = \pm 1$. Such a tensor is the metric tensor on the
surface and $h^a_b$ is a projection on the tangent space of the
surface. 

We define extrinsic curvature of the surface
the tensor, belonging to the tangent space on the $n-1$ dimensional
manifold $$ K_{ab} = h^{a'}_a h^{b'}_b \nabla_{a'} n_{b'} = h^{a'}_a
\nabla_{a'} n_{b}.  $$ The last equality is due to metric
compatibility of $\nabla$ and to $n^a\nabla n_a
=0$. For definiteness we work with $n^a n_a=-1$~.

In order to compute $\nabla_a n_b$ we need an extension of $n_b$
outside to surface, but due to the projection $h^{a'}_a$ the defined
$K_{ab}$ does not depend on such extension.

We want to prove that $K_{ab}=K_{ba}$. Let $\rho=0$ be the equation
defining the surface. $n_a$ is given by
$$
n_a =\lambda \nabla_a\rho
$$
being $\lambda$ a proper normalizing function to have $n^2=-1$.

We have
$$
\nabla_c \lambda = \lambda^2 n^a\nabla_a\nabla_c \rho. 
$$
Then $K_{ab}$
$$
K_{ab}= \lambda h^c_a(n^dn_b\nabla_c\nabla_d\rho +
\nabla_c\nabla_b\rho) =
$$
$$
=\lambda (n^dn_b\nabla_a\nabla_d\rho
+n_an_bn^cn^d\nabla_c\nabla_d \rho +\nabla_a\nabla_b\rho
+n^cn_a\nabla_c\nabla_b\rho)   
$$
which is symmetric under the exchange of $a$ and $b$ , due to the
absence of torsion i.e. $\nabla_a\nabla_b\rho=\nabla_b\nabla_a\rho$.

There is an important relation between the extrinsic curvature and a
Lie derivative of the metric $h_{ab}$
$$
L_n h_{ab}=n^c \partial_c h_{ab}+ h_{cb} \partial_a n^c+h_{ac} \partial_b n^c 
$$
$$
=n^c \nabla_c h_{ab}+ h_{cb} \nabla_a n^c+h_{ac} \nabla_b n^c= 
$$
$$
=n^c \nabla_c (n_an_b)+ \nabla_a n_b+ \nabla_b n_a= 
$$
$$
=n^c n_a \nabla_c n_b+n^c n_b \nabla_c n_a+ \nabla_a n_b+ \nabla_b
n_a= K_{ab}+K_{ba} = 2 K_{ab}.
$$

\bigskip

References

\smallskip

[1] [HawkingEllis] Chap. 2


\section{The trace-$K$ action}\label{6TrKaction}

Using the concept of extrinsic curvature we can give a covariant form
of the gravitational action. The problem is that to add to the Hilbert
action a covariant boundary term whose variation cancel the
contribution
\begin{equation}\label{6boundaryv}
\int_{\partial V} v^\lambda \Sigma_\lambda 
\end{equation}
which appears in Eq.(\ref{5deltaSH}).
We start from the contribution to Eq.(\ref{6boundaryv}) 
of the top surface $\Sigma_2$ (see Fig 7.1).
$$
\int_{{\Sigma}_2} \sqrt{-g} ~v^\lambda
\varepsilon_{\lambda\mu_2\dots\mu_n} \frac{dx^{\mu_2}
\wedge \dots dx^{\mu_n}}{(n-1)!}~.
$$
 One should not confuse
the surfaces $\Sigma_1,~\Sigma_2,~\Sigma_t$ with the differential form
$\Sigma_\lambda$ appearing e.g. in
eqs.(\ref{6boundaryv},~\ref{5deltaSH}) and others.

Using $\sqrt{-g}=\sqrt{h}~N$, $N>0$ we can rewrite the above as 
(see eq(\ref{5deltaSH}))
\begin{equation}\label{6vn}
\int_{\Sigma_2} N\sqrt{h} ~v^0 d^{n-1}x
=-\int_{\Sigma_2} \sqrt{h} ~v^\lambda n_\lambda  ~d^{n-1}x
\end{equation}
where $n_\lambda=(-N,0,0,0)$ is the normal to the
space-like surface $\Sigma_2$. $n$ is time-like and being 
$n^0 = g^{0\mu}n_\mu = \frac{1}{N}>0$ this is called the 
future-pointing normal. We recall that
\begin{equation}
v^\lambda n_\lambda =
\delta\Gamma^\lambda_{~\mu\nu} g^{\mu\nu} n_\lambda- \delta \Gamma^\nu_{~\mu\nu}
g^{\mu\lambda}n_\lambda~.
\end{equation}

Let us compute the variation of 
$$
\int_{\Sigma_2}\sqrt{h} ~K ~d^{n-1}x
=\int_{\Sigma_2}\sqrt{h} ~h^\mu_\nu\nabla_\mu n^\nu ~d^{n-1}x~.
$$
In the variational procedure we have on $\partial V$, $\delta
g_{\mu\nu}=0$ and thus also $\delta n_\nu=0$, $\delta n^\nu=0$ on
$\partial V$ but not necessarily outside $\partial V$.
We have
\begin{eqnarray}\label{6first}
\delta\int_{\Sigma_2}\sqrt{h}~ h^\mu_\nu~\nabla_\mu n^\nu ~d^{n-1}x&=&
\int_{\Sigma_2}\sqrt{h}~h^\mu_\nu
(\partial_\mu\delta n^\nu+\delta\Gamma^\nu_{\rho\mu}n^\rho)~d^{n-1}x\nonumber\\
&=&
\int_{\Sigma_2}\sqrt{h} ~(\delta\Gamma^\mu_{\rho\mu}n^\rho+
n_\rho\delta\Gamma^\rho_{\nu\mu} n^\nu n^\mu )~ d^{n-1}x
\end{eqnarray}
because on $\partial V$, $\delta n^\nu=0$ and
as a consequence $\partial_\mu\delta n^\nu = c~ n_\mu$.
We can also write, using metric compatibility 
\begin{eqnarray}\label{6second}
& &\delta\int_{\Sigma_2}\sqrt{h}~K~d^{n-1}x =
\delta\int_{\Sigma_2}\sqrt{h}~ h^{\mu\nu}~\nabla_\mu n_\nu
~d^{n-1}x\\ 
&=&\int_{\Sigma_2}\sqrt{h}~h^{\mu\nu}
(\partial_\mu\delta n_\nu-\delta\Gamma^\rho_{\nu\mu}n_\rho)~d^{n-1}x=
\int_{\Sigma_2}\sqrt{h} ~(-g^{\mu\nu}\delta\Gamma^\rho_{\nu\mu}n_\rho-
n_\rho\delta\Gamma^\rho_{\nu\mu} n^\nu n^\mu )~ d^{n-1}x\nonumber~.
\end{eqnarray}
Addition of the term (\ref{6first}) and (\ref{6second}) gives minus
(\ref{6vn}). Thus
$$
2\int_{\Sigma_2} \sqrt{h} ~K  d^{n-1}x
$$
will cancel in the variation, 
the contribution to Eq.(\ref{6boundaryv}) of $\Sigma_2$ and
similarly
$$
-2\int_{\Sigma_1} \sqrt{h} ~K~d^{n-1}x
$$
will cancel the contribution to Eq.(\ref{6boundaryv}) of $\Sigma_1$.

We come now to the contribution of the mantle $B$ in
Eq.(\ref{6boundaryv}). Denoting by $u_\lambda$ the outward-pointing
unit normal to $B$ we have
$$
g_{\mu\nu}=\gamma_{\mu\nu}+u_\mu u_\nu~.
$$
The extrinsic curvature of $B$ is
$$
\Theta_{\mu\nu}=\gamma^{\alpha}_{~\mu}\nabla_\alpha u_\nu,
~~~~\Theta =\gamma^\alpha_\mu\nabla_\alpha u^\mu ~.
$$
Working as before we have
$$
\int_B\sqrt{-\gamma} ~v^\lambda u_\lambda ~d^{n-1}x = - 2\delta\int_B
\sqrt{-\gamma}~
\Theta ~d^{n-1}x~.
$$
We can now write the trace-$K$ action [1]
$$
S_K = S_H + 2 \bigg(\int_{\Sigma_2}-\int_{\Sigma_1}\bigg) \sqrt{h}~ K ~d^3x +
2\int_B\sqrt{-\gamma}~ \Theta~ d^3x
$$ 
which has a form invariant under diffeomorphisms.

Remark: The above given trace-$K$ action is complete 
for orthogonal boundaries i.e. when on the two dimensional surface
where $B$ meets $\Sigma_t$ we have $n^\lambda
u_\lambda=0$. Otherwise as discovered in [2] and treated in detail in
[1,3] one has to add two contributions belonging to two 2-dimensional
sub-manifolds.

\bigskip

References

\smallskip

[1] J.D. Brown, S.R. Lau, J.W. York ``Action and energy of the
gravitational field'' gr-qc/0010024

\smallskip

[2] G. Hayward, ``Gravitational action for spacetimes with non
smooth boundaries'' Phys. Rev. D 47 (1993) 3275

\smallskip

[3] S.W. Hawking and C.J. Hunter, ``The gravitational Hamiltonian in
the presence of non-orthogonal boundaries'' Class. Quant. Grav. 13
(1996) 2735

\smallskip


\section{The boundary term in the Palatini formulation}\label{6palatinibtsec}

With the developed tools we can give a covariant form for the boundary
terms in the Palatini action [1].

The boundary term in the variation of $\Gamma^{\lambda}_{\mu\nu}$ was
$$
\int d^nx \partial_\lambda(\sqrt{-g}g^{\mu\nu}\delta
\Gamma^\lambda_{\mu\nu}-
\sqrt{-g}g^{\lambda\mu}\delta \Gamma^\nu_{\mu\nu})~.
$$
Let us compute it first on $\Sigma_2$ where it gives
\begin{equation}\label{6palatinibt}
-\int_{\Sigma_2}\sqrt{h}(n_\lambda
g^{\mu\nu}\delta\Gamma^\lambda_{\mu\nu} -
n_\lambda g^{\lambda\mu}\delta\Gamma^\nu_{\mu\nu})d^{n-1}x~. 
\end{equation}
where $n_\lambda$ is the future-pointing time-like unit normal
defined in the previous section.
Working exactly as for the trace-$K$ action we find that the variation
of 
\begin{equation}\label{6palatiniboundary}
-\int_{\Sigma_2}\sqrt{h} ~(h^\mu_\nu\nabla_\mu n^\nu
+h^{\mu\nu}\nabla_\mu n_\nu) ~d^{n-1}x
\end{equation}
using the absence of torsion, i.e. the symmetry of the
connection in the lower indices $\mu\nu$,
is exactly minus (\ref{6palatinibt}). The only difference is that, as
we do not assume at the start metric compatibility, we cannot identify
Eq.(\ref{6palatiniboundary}) with twice the integral of the trace of
the exterior curvature. Moreover we can write
$$
h^\mu_\nu\nabla_\mu n^\nu+h^{\mu\nu}\nabla_\mu n_\nu=
\nabla_\mu n^\mu+g^{\mu\nu}\nabla_\mu n_\nu+n^\mu\nabla_\mu
(n^\nu n_\nu)=\nabla_\mu n^\mu+g^{\mu\nu}\nabla_\mu n_\nu~.
$$
Similarly one
deals with $\Sigma_1$ and $B$. Summarizing we have the complete
Palatini action
\begin{eqnarray}
S_P = \int_V~\sqrt{-g}~g^{\mu\nu}~R_{\mu\nu} d^nx &-&
\bigg(\int_{\Sigma_2}-\int_{\Sigma_1}\bigg)
\sqrt{h}~(\nabla_\nu n^\nu
+g^{\mu\nu}\nabla_\mu n_\nu)~d^{n-1}x\\
&+&\int_B\sqrt{-\gamma}~(\nabla_\nu u^\nu
+g^{\mu\nu}\nabla_\mu u_\nu)~d^{n-1}x
\end{eqnarray}
where $u_\mu$ is the outward-pointing unit normal to $B$ and in the second
line $\gamma_{\mu\nu}=g_{\mu\nu}-u_\mu u_\nu$, $u^\mu u_\mu=1$.

\bigskip

References

\smallskip

[1] Yu.N. Obukhov,``The Palatini principle for manifolds with boundary''
Class. Quantum Grav. 4 (1987) 1085

\section{The boundary term in the Einstein-Cartan
  formulation}\label{6EinsteinCartanbtSec} 

The developed formalism allows to compute also the covariant boundary
terms in the Einstein-Cartan formulation [1]

From Eq.(\ref{ECaction}) we recall that the left-over term in deriving the
equation of motion by varying $\Gamma$ is
$$
\frac{1}{(n-2)!}\int d(\delta \Gamma^{ab}\wedge e^{a_3}\wedge \dots e^{a_n}
~\varepsilon_{ab a_3\dots a_n})~.
$$
Not to overburden the notation we shall deal with the case $n=4$, the
extension to all $n$ being trivial.
We have
$$
\frac{1}{2}\int_V d(\delta \Gamma^{ab}\wedge e^c\wedge e^d~
\varepsilon_{abcd})=\frac{1}{2}
\int_{\partial V}\delta \Gamma^{ab}\wedge e^c\wedge e^d~
\varepsilon_{abcd}~.
$$
Such expression can be rewritten in the form (\ref{6boundarySigma2})
below.

For completeness we give here the derivation. 
In general for a three form $\omega$ we have
$$
\int_{\partial V} \omega=\frac{1}{3!}\int_{\partial V}
\omega_{\mu_2\mu_3\mu_4} ~dx^{\mu_2}\wedge dx^{\mu_3}\wedge
dx^{\mu_4}=
\int_{\partial V} e~\Omega^\beta~
\varepsilon_{\beta\mu_2\mu_3\mu_4}dx^{\mu_2}\wedge
dx^{\mu_3}\wedge dx^{\mu_4}
$$
where
$$
e\Omega^\beta = -\frac{1}{3!}\varepsilon^{\beta\mu_2\mu_3\mu_4}\omega_{\mu_2\mu_3\mu_4}~.
$$
Thus referring to the surface $\Sigma_2$ we have
\begin{eqnarray}\label{6Omegaform}
\frac{1}{3!}\int_{\Sigma_2} e~\Omega^\beta~
\varepsilon_{\beta\mu_2\mu_3\mu_4}dx^{\mu_2}\wedge
dx^{\mu_3}\wedge dx^{\mu_4}&=&-\int_{\Sigma_2} ~\sqrt{h} ~n_\beta~
\Omega^\beta~ d^3x=\nonumber\\
=\frac{1}{3!}\int_{\Sigma_2}\frac{\sqrt{h}}{e}n_\beta~
\varepsilon^{\beta\mu_2\mu_3\mu_4} \omega_{\mu_2\mu_3\mu_4}~d^3x~.
\end{eqnarray}
where $n_\mu=(-N,0,0,0)$, $N>0$. In the present case we have
$$
\omega =\frac{1}{3!}~\omega_{\mu_2\mu_3\mu_4} ~dx^{\mu_2}\wedge dx^{\mu_3}\wedge
dx^{\mu_4}=
\frac{1}{2}~\delta \Gamma^{ab}_{~f} e^f\wedge e^c\wedge e^d
~\varepsilon_{abcd}~.
$$
Substituting into (\ref{6Omegaform}) and using
$$
\varepsilon^{\beta\mu_2\mu_3\mu_4}~e^f_{\mu_2}~e^c_{\mu_3}~e^d_{\mu_4}=
e~e^\beta_k\varepsilon^{kfcd}
$$
we obtain for such a contribution
\begin{equation}\label{6boundarySigma2}
\int_{\Sigma_2}~\sqrt{h}~ n_k~\delta\Gamma^{ab}_{~f}~\delta^{kf}_{ab}~d^3x =
2 \int_{\Sigma_2}~\sqrt{h}~ n_a~\delta\Gamma^{ab}_{~b}~d^3x
\end{equation}
where we took into account the antisymmetry in $a,b$ of $\delta
\Gamma^{ab}_{~c}$.

Consider now the variation of
\begin{equation}\label{6ECK}
2\int_{\Sigma_2}\sqrt{h} ~h^\mu_a\nabla_\mu n^a~d^3x ~.
\end{equation}
where $h^\mu_a= h^\mu_\nu ~e^\nu_a= e^\mu_a+n^\mu n_a$

Following the same procedure as above we have for the variation
$$
2 ~\delta \int_{\Sigma_2}\sqrt{h} ~h^\mu_a\nabla_\mu n^a~d^3x
= 2\int_{\Sigma_2}\sqrt{h}~
(\delta \Gamma^a_{~ba}n^b+n^\mu\delta\Gamma^{ab}_{~\mu} ~n_an_b)d^3x=
-2\int_{\Sigma_2} \sqrt{h}~n_a \Gamma^{ab}_{~b} d^3x
$$
which is minus the boundary contribution (\ref{6boundarySigma2}). 
Also one notices that in Eq.(\ref{6ECK}) due to metric compatibility we
have $h^\mu_a\nabla_\mu n^a= e^\mu_a\nabla_\mu n^a$.
Similarly one
deals with the contribution of $\Sigma_1$ and of $B$. Finally the
complete action for Einstein-Cartan formulation is
$$
S_{EC}= \frac{1}{2}\int_V R^{ab}\wedge e^c\wedge e^d
\varepsilon_{abcd}-
2 \bigg(\int_{\Sigma_2}-\int_{\Sigma_1}\bigg)\sqrt{h}
~e^\mu_a\nabla_\mu n^a~d^3x+
2 \int_B \sqrt{-\gamma} ~e^\mu_a\nabla_\mu u^a ~d^3x
$$
where $n_\mu=(-N,0,0,0)$ and $u_\mu$ is the unit outward-pointing
normal to $B$ and again in the last term $\gamma_{\mu\nu}
=g_{\mu\nu}-u_\mu u_\nu$.

\bigskip

References

\smallskip

[1] Yu.N. Obukhov,``The Palatini principle for manifolds with boundary''
Class. Quantum Grav. 4 (1987) 1085


\bigskip

\section{The Gauss and Codazzi equations}

Given a surface $\Sigma$ embedded in an $n$ dimensional manifold $M$, i.e. an
$n-1$ dimensional manifold embedded in $M$, the are important
relations between the curvature $R$ in $n$ 
dimensions, the intrinsic curvature ${\cal R}$ of the $n-1$
dimensional surface $\Sigma$  and the extrinsic curvature of the $n-1$
dimensional surface $\Sigma$ embedded in the $n$ dimensional
manifold $M$. By $D$ 
here we shall understand the covariant derivative in the sub-manifold,
with zero torsion and compatible with the metric $h_{ab}$ induced by
$g_{ab}$ on the sub-manifold $\Sigma$ i.e. the pull back of $g_{ab}$
to $\Sigma$.

The main equation to be used is
$$
(D_a D_b-D_b D_a)v^c = {\cal R}^c_{~c'ab}v^{c'}
$$
from which it follows
\begin{equation}\label{curvcov}
(D_a D_b-D_b D_a)\omega_c = -{\cal R}^{c'}_{~c ab}\omega_{c'}. 
\end{equation}
We recall that the covariant derivative $D$ can be computed from the
covariant derivative $\nabla$ on the $n$-dimensional manifold through
a projection procedure. For example 
$$
D_a T^b_{cd} = h^{a'}_a h^b_{b'} h_c^{c'} h_d^{d'}\nabla_{a'} T^{b'}_{c'd'}.
$$ 

A result which is useful in proving the following relations is
\begin{equation}\label{auxeq}
h^{a'}_ah^{b'}_b\nabla_{a'} h^{c}_{b'} = h^{a'}_ah^{b'}_b\nabla_{a'}
(n^{c}n_{b'}) =n^cK_{ab}. 
\end{equation}

Then from Eq.(\ref{curvcov})
$$
-{\cal R}^d_{cab}= -h^{a'}_ah^{c'}_ch^{b'}_b h^d_{d'} R^{d'}_{c'a'b'}-
K_{ac} K^d_b +K_{bc} K^d_a  
$$
or
$$
{\cal R}_{dcab}= \Pi_\Sigma (R_{dcab}) + K_{ac} K_{db}
- K_{bc} K_{da} 
$$
where $\Pi_\Sigma$ denotes the projection on $\Sigma$.
This is the Gauss equation.

Taking two traces and keeping in mind the (anti)-symmetry of the Riemann 
tensor (see Section ~\ref{4symmetriesofRiemann}) in the indices one gets
\begin{equation}\label{6gauss}
{\cal R} + K^2 - {\rm Tr}(KK) = 2 n^b n^c G_{cb} 
\end{equation}
where $G_{cb}$ is the Einstein tensor. 

To prove the Codazzi equations take
$$
D_a K^c_{b}=h^c_{c'}h^{b'}_{b}h^{a'}_{a}\nabla_{a'}K^{c'}_{b'}\equiv
h^c_{c'}h^{b'}_{b}h^{a'}_{a}\nabla_{a'}(h^{b''}_{b'}\nabla_{b''}n^{c'}). 
$$
Using Eq.(\ref{auxeq})we find
$$
D_a K^c_{b} = K_{ab} h^c_{c'}n^{b'}\nabla_{b'}n^{c'}+
h^c_{c'}h^{b'}_{b}h^{a'}_{a}\nabla_{a'}\nabla_{b'}n^{c'}.  
$$
Contracting $c$ with $b$
$$
D_a K=  K_{ab} n^{b'}\nabla_{b'}n^{b}+
h^{b'}_{c}h^{a'}_{a}\nabla_{a'}\nabla_{b'}n^{c}.  
$$
Contracting $a$ with $c$
$$
D_a K^a_b= K_{ab} n^{b'}\nabla_{b'}n^{a}+
h^{a'}_{c}h^{b'}_{b}\nabla_{a'}\nabla_{b'}n^{c}  
$$
we have
\begin{eqnarray}\label{6codazzi}
& & D_a K^a_b-D_b K =h^{a'}_{c}h^{b'}_{b}(\nabla_{a'}\nabla_{b'}n^{c}-
\nabla_{b'}\nabla_{a'}n^{c}) =
h^{a'}_{c}h^{b'}_{b}R^c_{~c'a'b'}n^{c'}\nonumber\\
&=&h^{b'}_{b}R_{c b'}n^{c}
= h^{d}_{b}G_{c d}n^{c}.
\end{eqnarray}
These are the Codazzi equations.

The Gauss and Codazzi equations relate intrinsic and extrinsic properties
of an $n-1$-dimensional  surface to geometric properties of the
$n$-dimensional space in which such a surface is embedded. The
vanishing of Eq.(\ref{6gauss}) for any surface i.e. for any $n$, tells us the
$G^{ab}=0$. For a two dimensional surface embedded in three dimensions
$G^{ab}=0$ tells us that the three dimensional space is flat. For an
$n-1$-dimensional surface embedded in an $n$-dimensional space with
$n\geq 4$ $G^{ab}=0$ tells us that the space is Ricci flat.

If we have at our disposal only one normal
$$
n_aG^{ab}n_b=0,~~~~~~~~h^a_b G^{bc}n_c=0
$$
are equivalent to
$$
n_aG^{ab}n_b=0,~~~~~~~~G^{ac}n_c=0
$$
i.e. in the ADM reference system $G^{00}=0,~~G^{j0}=0$.


\bigskip

References

\smallskip

[1] [HawkingEllis] Chap. 2

\smallskip

[2] [Wald] Chap. 10


\bigskip


\chapter{The hamiltonian formulation of gravity}

\section{Introduction}\label{7intoductionSec}

The motivations for a hamiltonian formulation of gravity are:

\bigskip

1. The need to define a Cauchy problem.

2. Separation of physical from gauge degrees of freedom.

3. A hamiltonian formulation is a road to quantization as done in
non relativistic quantum mechanics and also in 
relativistic field theory.

4. Give a framework to perform numerical calculations like black hole
scattering, black hole coalescence, emission of gravitational waves.

\bigskip

\begin{figure}[htb]
\begin{center}
\includegraphics{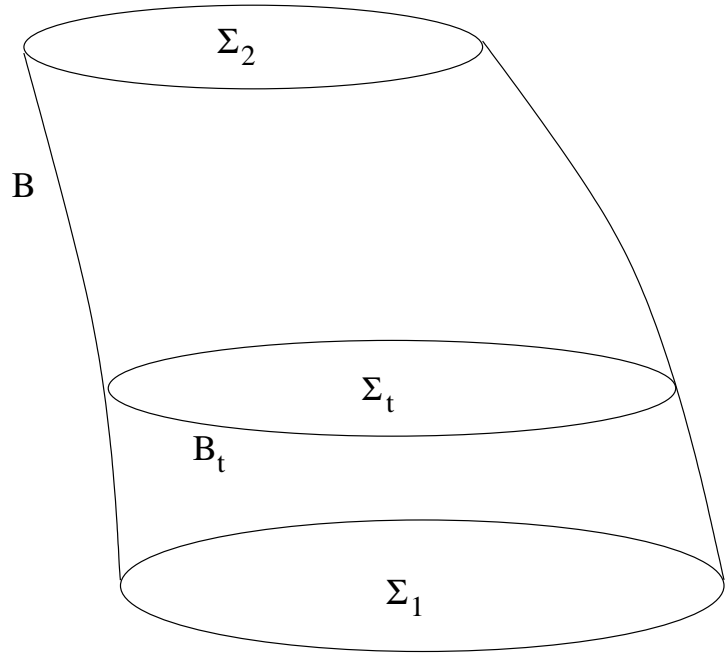}
\end{center}
\caption{Space-time foliation}
\end{figure}

One assumes that space-time can be foliated in space like surfaces
$\Sigma_t$ which will labeled by a real parameter $t$. Space-like
surface means that all vectors belonging to the tangent space of $\Sigma_t$
are space-like vectors; this is equivalent to saying that the vector
$n_a$ normal to $\Sigma_t$ is a time-like vector-field.

One introduces a vector field called ``time flow'' which generates the
diffeomorphisms  which map  $\Sigma_{t_0}$ into $\Sigma_{t_0+t}$ and thus
such that $t^a\nabla_a t=1$. From its definition we see that the $t^a$
is not unique. The metric on $\Sigma_t$ is the one
induced by the metric $g_{ab}$ on $M$ i.e. the pull-back of $g_{ab}$
to $\Sigma_t$.

One can decompose $t^a$ in normal and tangential part
$$
t^a = N^a -n^a t^b n_b \equiv N^a + N n^a.
$$

The space metric is given by the tensor
$$
h_{ab}=g_{ab}+n_a n_b;~~~~n^a n_a=-1.
$$

In a general frame, vectors in $\Sigma_t$ are described by four (non
independent) components. 

\bigskip
The time flow vector field can be used to set up a special
coordinate system $(t,[\phi_{t}^{-1}(p)]^{j})$ i.e. the ADM coordinate
system. In such a coordinate system,
under $\phi_t$ the  $x^j$ do not vary and thus $t^\mu=(1,0,0,0)$. 
Being $N^\mu$ tangent to $\Sigma_t$ we have
$N^\mu=(0,N^j)$ and being $n^\mu$ orthogonal to all tangent vectors of
$\Sigma_t$ we have
$n_\mu=(-N,0,0,0)$ and thus
$$
0 = N^j + N n^j;~~~~ n^j = -\frac{N^j}{N}
$$
$$
-1 =- N n^0+0;~~~~ n^0 = \frac{1}{N}~.
$$
As $n_j=0$ we have $h_{ij}=g_{ij}$. 
Being $n^\mu$ orthogonal to all tangent vectors of $\Sigma_t$
$(0,.,.,.)$ we have
$$
0=n^\mu g_{\mu i} = \frac{1}{N}g_{0i}-\frac{N^j}{N}h_{ij}
$$
from which $g_{0j}=N^i h_{ij}= N_j$~.
$$
-1 = n^\mu n_\mu = \frac{1}{N^2}g_{00}+ h_{ij} \frac{N^iN^j}{N^2} - 2
\frac{N^i}{N}g_{i0} \frac{1}{N}~.
$$
Summarizing
$$g_{\mu\nu}=
\begin{pmatrix}
N^j N_j -N^2& N_n\\
N_m&h_{mn}
\end{pmatrix}
$$
which is the ADM metric and in such coordinate system 
$$
t^\mu=(1,0,0,0);~~~~N^\mu=(0,N^j);~~~~n_\mu=(-N,0,0,0);~~~~
n^\mu=(\frac{1}{N},-\frac{N^j}{N}).
$$  
Sometimes the inverse of $g_{\mu\nu}$ is needed; it is given by
$$
g^{\mu\nu}=
\begin{pmatrix}
-\frac{1}{N^2} & \frac{N^n}{N^2}\\
\frac{N^m}{N^2} &h^{mn}-\frac{N^m N^n}{N^2}
\end{pmatrix}.
$$

\section{Constraints and dynamical equations of motion}

If Einstein equations are satisfied (in absence of matter it means that the
space is Ricci-flat  $G_{ab}=0$) we have

$$
{\cal R}+ K^2 - {\rm Tr}(KK)= 2 n_an_bG^{ab}=0\eqno(I)
$$
and
$$
D_a K^a_b-D_b K =h_{bc}G^{c d}n_{d} =0~.\eqno(II)
$$
Due to $2K_{ab} = L_nh_{ab}$ the four relations contain only first
order time derivative while the system of Einstein equations are (linear)
in second order time derivative. Thus one naively expects that as initial
conditions we must give both $h_{ab}, N^i,N$ and $\dot h_{ab}, \dot
N^i,\dot N$ but such initial conditions are not arbitrary but subject
to (I) and (II). On the other hand if Einstein equations should
predict uniquely the time development of the metric, we would have a
violation of invariance under diffeomorphisms, as given a solution any
other metric given by a diffeomorphism of the first, must also be a
solution. 

Is is remarkable that the validity of the single scalar equation,
which relates the intrinsic and extrinsic curvature
$$
{\cal R} + K^2-{\rm Tr}(KK)=0
$$
on all space like surfaces is equivalent to Einstein
equations i.e. to the whole general relativity, in absence of
matter. In fact the validity of $(I)$ for all time-like $n$ implies
the $G_{ab}=$ i.e. that space time is Ricci flat.  

However given a foliation of the space time $n$ is not
arbitrary. Recalling that $h^a_b =\delta^a_b + n^a n_b$ we have that
the validity of 
$$
{\cal R}+ K^2 - {\rm Tr}(KK)=0
$$
and
$$
D_a K^a_b-D_b K =0.
$$
due to the invertibility of $g_{ab}$ imply the four relations
$G^{ab}n_b=0$, which in the ADM frame assume the form $G^{\mu 0}=0$.  
These are the 4 Einstein equations which do not
contain second order time derivatives and as such are called
instantaneous equations. The other $6$ are
$G^{ij}=0$ and they do contain second order time derivative as will be
shown in developing the hamiltonian approach.

Thus in the initial conditions of the Cauchy problem we cannot choose
the metric $h_{ij}, N^j, N$ and the time derivative of the metric
$\dot h_{ij}, \dot N^j, \dot N$ arbitrarily but such initial
conditions are subject to ($I$) and
($II$).

\section{The Cauchy problem} \hskip 1cm {\bf \Large for the Maxwell and 
Einstein equations in vacuum}

\bigskip

There is a strict analogy between the Cauchy problem for the Maxwell
equations in the $A$ description and the Cauchy problem for the
Einstein equations

1. $A_\mu$ are 4 fields

2. $A\rightarrow A+d\Lambda$ is a transformation which leaves unchanged
the field strength $F_{\mu\nu}$.
 
3. $\partial_\mu F^{\mu\nu}=0$ is a system of 4 equations, 
which is second order in the time derivatives.

4. Not all equations are of second order in the time derivative:
$$
\partial_j F^j_{~~0}= \partial_j( \partial^j A_0 -\partial_0 A^j)=0
$$ 
is first order and contains no time derivative of $A_0$.

5. $\partial_\mu F^\mu_{~ j}=0$ are 3 equations containing second order
time derivatives
\begin{equation}\label{secondeq}
\partial_\mu F^\mu_{~~j}= \partial_0( \partial^0 A_j -\partial_j
A^0)+\partial_i( \partial^i A_j -\partial_j A^i)=0 
\end{equation}
No $\ddot A_0$ appears. 

To proceed in time we need $\dot A^j$ and to extract $\ddot A^j$ from
Eq.(\ref{secondeq}) we need 
to know on $\Sigma$, $A^l$ and also $\partial _0 A^0$.

A consequence of Eq.(\ref{secondeq}) is
$$
\partial_0 ( \partial_\mu F^{\mu 0})= 0.
$$

Thus the position of the Cauchy problem in $A_\mu$ is

\bigskip

I. Choose $A_0$ as an arbitrary function of space time.

II. Choose $A_j({\bf x},0)$ and $\dot A_j({\bf x},0)$ subject to
the condition
\begin{equation}\label{emconstr}
(\partial_j \partial^j A_0 -\partial_j \dot A^j)({\bf x},0)=0
\end{equation}
but otherwise arbitrary. 

III. Solve the 3 second order equations for $A_j$

$\partial_\mu F^\mu_{~~j}=0$ i.e.
\begin{equation}\label{emdyn}
\ddot A^j = \partial^j \dot A_0+\partial_i( \partial^i A^j
-\partial^j A^i).
\end{equation}
As a consequence of Eq.(\ref{emdyn}), Eq(\ref{emconstr}) is satisfied
at all times.

\bigskip

The separation of the gauge degree of freedom is most clearly seen in
the hamiltonian formulation.

Starting from the lagrangian
$$
L = \frac{1}{4}F_{\mu\nu}F^{\mu\nu}=
\frac{1}{4}(\partial_i A_j-\partial_j A_i)(\partial^i A^j-\partial^j
A^i)+\frac{1}{2}(\partial_0 A_j-\partial_j A_0)(\partial^0 A^j-\partial^j A^0)
$$
let us compute the momenta conjugate to $A_j$
$$
E^j = \frac{\partial L}{\partial \dot A_j}=\partial^0 A^j-\partial^jA^0
$$
and rewrite the action in hamiltonian form
\begin{eqnarray}
S&=&\int (E^j \dot A_j -H) d^4x\\
&=&
\int \bigg(E^j \dot A_j +\frac{1}{2}E^jE_j+\frac{1}{4}
(\partial_i A_j-\partial_j A_i)(\partial^i A^j-\partial^jA^i)
+A_0 \partial_j E^j\bigg) d^4x
\end{eqnarray}
where a space integration by parts has been performed. We see that
$A_0$ appears without derivatives, i.e. as a Lagrange multiplier.
Its variation give the constraint
\begin{equation}\label{7divE}
\partial_j E^j=0~.
\end{equation}
The variation of $E^j$ gives
\begin{equation}\label{7dotA}
\dot A_j = - E_j+\partial_j A_0
\end{equation}
while variation of $A_j$ give
\begin{equation}\label{7dotE}
\dot E^j = -\partial_i(\partial^iA^j-\partial^jA^i)~.
\end{equation}
Antisymmetry in the indices in the last equation gives
\begin{equation}
\partial_j \dot E^j=0~.
\end{equation}
To solve the equations of motion choose $A_0$ as an arbitrary function
of space and time. At $t=0$ choose $A_j$ and $E^j$ arbitrarily but
with $E^j$ subject to Eq.(\ref{7divE}). Using
eqs.(\ref{7dotA},\ref{7dotE}) propagate the fields $A_j$ and $E^j$ in
time. As a consequence of Eq.(\ref{7dotE}) the constraint equation 
(\ref{7divE}) will be satisfied at all times.

\bigskip

Similarly for the Einstein equations we have

1. $h_{ij}$ $N$, $N^j$ are 10 fields.

2. The physical content of the solution is unchanged under the
transformation $x'^\mu(x)$ (diffeomorphisms) given by 4 arbitrary
functions.
 
3. $G^{\mu\nu}=0$ is a system of 10 equations, which is second order in
time derivatives. 

4. Not all equations contain second order time derivatives:

$G^{00}=0$ and $G^{0i}=0$ contain only first order time derivatives.

5. The other 6 equations $G^{ij}=0$ contain second order time derivatives

No $\ddot N$, $\ddot N^a$ appear in the equations. These two
statements will be proven in the following section.

\bigskip
Thus the position of the Cauchy problem in $h_{ij}$ $N$, $N^a$ is

I. Choose  $N$, $N^a$ as arbitrary functions of space time.

II.  Choose $h_{ij}({\bf x},0)$ and $\dot h_{ij}({\bf x},0)$
subject to 

\begin{equation}\label{grconstr}
G^{00}({\bf x},0)=0;~~~~G^{0j}({\bf x},0)=0
\end{equation}
but otherwise arbitrary. 

III. Solve the 6 second order equations for $h_{ij}$

\begin{equation}\label{grdyn}
G^{ij}=0.
\end{equation}

As a consequence of Eqs.(\ref{grdyn}), Eqs.(\ref{grconstr}) are satisfied
at all times; also this statement will be proven in the following
section.

Again the separation of the diffeomorphism from the physical degrees
of freedom is most clearly seen in the hamiltonian formalism to which
we turn in the next section.

\section{The action in canonical form}\label{7canonicalactionSec}

In what follows we shall refer to the somewhat simpler case of the
boundary $B$ orthogonal to $\Sigma_t$, i.e. $n^\mu u_\mu =0$ on $B$,
being $u_\mu$ the normal to $B$, and $n^\mu$ as usual the normal to
$\Sigma_t$. 

Using the Gauss equation (\ref{6gauss}) we can rewrite the action as
($D=n-1$)
\begin{eqnarray}\label{action1}
S &=& \int_V\sqrt{-g} ~dt ~d^{D}x~({\cal R}+K^2- {\rm Tr}(KK)
-2n^\mu R_{\mu\nu}n^\nu) + 2\int_B\sqrt{-\gamma}~d^D x~\Theta\nonumber\\
&+&2\bigg(\int_{\Sigma_2}-\int_{\Sigma_1}\bigg)\sqrt{h} ~d^D x ~K
\end{eqnarray}
being $\Theta$ the extrinsic curvature of $B$ and $\sqrt{-\gamma}~ d^Dx$
the area element of $B$.
Using
$$
n^\mu R_{\mu\nu}n^\nu = n^\nu\nabla_\mu\nabla_\nu n^\mu -n^\nu\nabla_\nu\nabla_\mu
n^\mu =
$$
\begin{equation}\label{divergence}
= -\nabla_\mu n^\nu\nabla_\nu n^\mu + \nabla_\mu(n^\nu\nabla_\nu n^\mu)
+\nabla_\nu n^\nu\nabla_\mu n^\mu - \nabla_\nu(n^\nu\nabla_\mu n^\mu) 
\end{equation}
and transforming the two divergence terms in surface integrals,
the action becomes
$$
S = \int dt \int_{\Sigma_t} \sqrt{-g}~d^{D}x({\cal R}-K^2+ {\rm Tr}(KK))
+2\int_B\sqrt{-\gamma} ~d^Dx(\Theta +u_\mu(n^\mu K-a^\mu)) 
$$
where $a^\mu$ is the acceleration vector $a^\mu =
n^\lambda\nabla_\lambda n^\mu$; moreover due to $u_\mu n^\mu=0$ the
part proportional to $K$ vanishes.
Notice that the two surface terms due
to $\Sigma_2$ and $\Sigma_1$ in action (\ref{action1}) have been canceled by
the contribution of the two divergence terms in Eq.(\ref{divergence}) 
computed on $\Sigma_2$ and $\Sigma_1$. The surface contribution on
$B$, i.e. the last integral in
the previous equation, can be expressed in terms
of the extrinsic curvature $k$ of $B_t=B\cap \Sigma_t$ as a
sub-manifold of $\Sigma_t$. 

In fact under our working hypothesis $u_\mu n^\mu=0$, $u_\mu$ coincides
with the unit vector $r_\mu$ belonging $\Sigma_t$ and orthogonal to $B_t$. The
metric on $B_t$ is given by
$$
\sigma_{\mu\nu}=h_{\mu\nu}-u_\mu u_\nu
$$
and thus the extrinsic curvature of $B_t$ is
$$
k_{\mu\nu}=\sigma^\alpha_\mu D_\alpha r_\nu=\sigma^\alpha_\mu D_\alpha
u_\nu=
h^\beta_\mu\nabla_\beta u_\nu-u^\alpha u_\mu h^\beta_\alpha
\nabla_\beta u_\nu
$$
whose trace is given by
$$
k=k^\mu_\mu=\nabla_\mu u^\mu +n^\beta n_\mu\nabla_\beta u^\mu =
\Theta -u_\mu a^\mu.
$$
Thus we obtain the action in Lagrangian
form $S=\int L ~dt$ i.e. 
\begin{equation}\label{7lagragianform}
S = \int dt\left( \int_{\Sigma_t} \sqrt{h}~ d^D x N({\cal R}-K^2+ {\rm
Tr}(KK)) + 2 \int_{B_t}dx^{D-1} \sqrt{-\gamma}~ k\right) ~.
\end{equation}
$\sqrt{-\gamma}$ can also be written as $N\sqrt{\sigma}$, being 
$\sqrt{\sigma}$ the area element of $B_t$. We notice
that the extrinsic curvature of $\Sigma_t$, 
$K_{\mu\nu}=h^\alpha_\mu\nabla_an_\nu$
appears quadratically in the above formula.

The remarkable feature of Eq.(\ref{7lagragianform}) is that the top and bottom
boundary terms have disappeared and thus it is of the form
$\int L dt$ where $L$ is the Lagrangian of the system. Thus we can
develop from it the hamiltonian formalism.

We shall need now a more explicit form of the extrinsic curvature
tensor.
$$
K_{ab} = \frac{1}{2}L_n h_{ab}=\frac{1}{2}(n^c\nabla_c h_{ab}
+h_{cb}\nabla_a n^c +h_{ac}\nabla_b n^c) =
$$
$$
=\frac{1}{2N}(N n^c\nabla_c h_{ab}+h_{cb}\nabla_a (N n^c) +h_{ac}\nabla_b
(N n^c))   
$$
because $h_{ac}n^c=0$. Substitute now $N n^a = t^a - N^a$ to obtain
$$
K_{ab}=\frac{1}{2N}(L_t h_{ab}-L_{\bf N} h_{ab})
$$
which lying in $T_\Sigma$ can be rewritten as
$$
K_{ab} = \frac{1}{2N}(h^{a'}_a h^{a'}_a L_t h_{a'b'}-h^{a'}_a
h^{a'}_a L_{\bf N} h_{a'b'}) =
\frac{1}{2N}(h^{a'}_a h^{a'}_a L_t h_{a'b'}- D_a
N_b- D_b N_a ).
$$
Taking into account that $N^a\in T_\Sigma$ i.e. $N^a n_a=0$ and
going over to ADM coordinate system where $t^a=(1,0,0,0)$, 
$n_i=0$ we have
\begin{equation}\label{hdot}
K_{ij} = \frac{1}{2N}(\partial_0 h_{ij}- D_i
N_j- D_j N_i ).
\end{equation}
These are all the significant components of the tensor $K_{ab}\in
T_\Sigma$ as all the other (redundant) components can be obtained from $n^a
K_{ab}=0$. Moreover in forming scalars on $\Sigma$ i.e. contracting
with vectors or tensors belonging to $T_\Sigma$ only the components
$K_{ij}$ play a role as for such vectors or tensors we have
$t^{0\mu}=0$. 

Notice that of the variables $h_{ij}, N^i, N$ no time derivatives of
$N^i$ and $N$ appear in the action and as such they appear as Lagrange
multipliers. We compute with the usual rules the conjugate momenta to
$h_{ij}$
\begin{equation}\label{7conjugatemom}
\pi^{ij}(t,x)=\frac{\delta L}{\delta \dot h_{ij}(t,x)}=
\sqrt{h}(K^{ij} - K h^{ij})
\end{equation}
which can be inverted in $\dot h_{ij}$ using Eq.(\ref{hdot}). Beware
that $\pi^{ij}$ is not a tensor on $\Sigma_t$ but a tensorial density.
Write now $L$ in hamiltonian form and perform one integration by
parts in $d^Dx$ to obtain
\begin{equation}\label{SKhamiltonian}
S_H = \int dt \int_{\Sigma_t} d^D x (\pi^{ij}\dot h_{ij} - NH - N^i
H_i) + 2\int 
dt\int_{B_t}\sqrt{\sigma}~d^{D-1}x~(N k - r_i
\frac{\pi^{ij}}{\sqrt{h}} N_j) 
\end{equation}
with
$$
H=\sqrt{h}~\left(-{\cal R}+\frac{{\rm
Tr}(\pi\pi)}{h}-\frac{\pi^2}{(D-1)h}\right),    
~~~~H_i= -2\sqrt{h}~D_j\left(\frac{\pi^j_{~i}}{\sqrt{h}}\right)
$$
$r_i$ being the outward pointing unit normal to $B_t$ as a 
sub-manifold of $\Sigma_t$
and $k$ the extrinsic curvature of intersection of $\Sigma_t$
with $B$ considered as a sub-manifold of $\Sigma_t$.

Variation with respect to $N^i$ and with respect to $N$ give
$$
H_i=0,~~~~ H=0.
$$
These are the constraints and they are equivalent to $G^{i0}=0$ and
$G^{00}=0$ as it easily checked using Eq.(\ref{7conjugatemom}).
 
Variation with respect to $\pi^{ij}$ give 
$$
\dot h_{ij}=\frac{2N}{\sqrt{h}}(\pi_{ij} -\frac{\pi}{D-1}h_{ij})+ D_i N_j+D_j
N_i ~.
$$
The hard part is 
$
\displaystyle{\frac{\delta S}{\delta h_{ij}}}
$
which are the real dynamical equations of motion and correspond to
combinations of $G^{ij}=0$.
\begin{eqnarray}\label{7pidotequation}
\dot \pi^{ij}&=& 
-N \sqrt{h}({\cal R}^{ij}-\frac{1}{2}{\cal R}h^{ij}) 
+\frac{N}{2\sqrt{h}}h^{ij}({\rm Tr}(\pi\pi)
-\frac{1}{D-1}\pi^2)\nonumber\\
&-&\frac{2N}{\sqrt{h}}(\pi^{ik}\pi_k^{~j}-\frac{1}{D-1}\pi\pi^{ij})+
\sqrt{h}(D^iD^j N - h^{ij}D^kD_k N)\nonumber\\
&+&\sqrt{h} D_k (\frac{1}{\sqrt{h}} N^k \pi^{ij})
-\pi^{ki}D_k N^j-\pi^{kj}D_k N^i.
\end{eqnarray}
In performing such a calculation one has to keep in mind that the true 
canonical variables are $h_{ij}$
and $\pi^{ij}$ and thus e.g. $\pi^j_i$ has to be understood as
$\pi^{jn}h_{ni}$. The variation of $\sqrt{h}~{\cal R}$ is well known
from the analogous variation in the Hilbert action through the Palatini
identity 
$$
h^{ij}\delta{\cal R}_{ij}= D_b(D_c \delta
h^{c b} -\frac{1}{2} D^b\delta 
h_{c}^c)- 
\frac{1}{2} D_b D^b \delta  h_{c}^c=
D_b D_c \delta  h^{c b}- 
D_b D^b  \delta  h_{c}^c
$$
but when one integrates by parts one has to keep in mind that contrary to
what happens for the Hilbert action, the $N$ which stays in front of $H$
gives rise to the additional contribution in (\ref{SKhamiltonian}).
The details of the calculation are given in the Appendix.

\section{The Poisson algebra of the constraints}

It is of interest to compute the Poisson algebra of the
constraints. The result is 
$$
\left\{H_i({\bf x}),H_j({\bf x}')\right\} = H_i({\bf x}') \partial_j
\delta({\bf x},{\bf x}')- 
H_j({\bf x})\partial'_i\delta({\bf x},{\bf x}')
$$

$$
\left\{H_i({\bf x}),H({\bf x}')\right\} = H({\bf
x})\partial_i\delta({\bf x},{\bf x}') 
$$

$$
\left\{H({\bf x}),H({\bf x}')\right\} =
H^l({\bf x})\partial_l\delta({\bf x},{\bf x}')-H^l({\bf
x}')\partial'_l\delta({\bf x},{\bf x}')  
$$
In smeared form with
$$
{\cal H}[{\bf N},N] =\int (H({\bf x}) N({\bf x}) +H_i({\bf x})
N^i({\bf x}))d^Dx  
$$
we have
\begin{equation}\label{7H1H2}
\{{\cal H}[{\bf N}_1,0], {\cal H}[{\bf N}_2,0]\}= {\cal H}[[{\bf
N}_1,{\bf N}_2],0] 
\end{equation}
\begin{equation}\label{7H1H}
\{{\cal H}[{\bf N}_1,0], {\cal H}[0,N]\}= {\cal H}[0,{\cal L}_{{\bf N}_1}N] 
\end{equation}
\begin{equation}\label{7HH}
\{{\cal H}[0,N], {\cal H}[0,N']\}= {\cal H}[h^{ij}(N D_j N'-N'D_j N),0]~.
\end{equation}

It is simpler to work with the smeared form of the P.B. .
The derivation of the above reported algebra is given in the Appendix.

One notices that:

\smallskip

1) Eqs.(\ref{7H1H2},\ref{7H1H}) give a canonical representation of the space
diffeomorphisms.

2) The only P.B. which contains explicitly the metric is (\ref{7HH}).

3) The P.B. algebra of the constraints closes. This implies that
as
$$
\dot F =\{F,{\cal H}\}
$$
we have that if the constraints are initially zero 
they remain conserved equal to zero in virtue of the equations of motion.

\bigskip

References

\smallskip

[1] C.J. Isham,``Canonical quantum gravity and the problem of
time'' Lectures at NATO Advanced Study Institute, Salamanca 1992,
p. 157.
 
\smallskip

[2] J.Pullin,``Canonical quantization of general relativity: the last 18
years in a nutshell'' gr-qc/0209008

\smallskip

[3] C. Rovelli, ``Quantum gravity'' Cambridge University Press.

\smallskip

[4] T. Thiemann, ``Modern canonical quantum general relativity''
Cambridge University Press, (2007)

\smallskip

\section{Fluids}

We consider only very simple fluids. They will be useful to introduce
the energy conditions, which play a very important role in various
instances, like the positive energy problem and the problem of
closed time like curves.

\bigskip
1. Dust

\begin{equation}\label{dust}
T^{\lambda\nu}= \mu U^\lambda U^\nu;~~~~U^\lambda = {\rm
four~velocity~field},~~~~U_\lambda U^\lambda=-1. 
\end{equation}
From Bianchi identities
\begin{equation}\label{fluideq}
\nabla_\lambda (\mu U^\lambda) U^\nu +\mu U^\lambda \nabla_\lambda U^\nu =0.
\end{equation}
Multiply by $U_\nu$
\begin{equation}\label{matterconserv}
\nabla_\lambda (\mu U^\lambda)= \nabla_\lambda J^\lambda=0 
\end{equation}
we obtain the conservation of ``charge''
$$
Q =\int_\Sigma \sqrt{h}~d^3 x ~n_\lambda (\mu U^\lambda)
=\int_\Sigma \mu U^\lambda \Sigma_\lambda~. 
$$
Eq.(\ref{fluideq}) now becomes
$$
\mu U^\lambda\nabla_\lambda U^\nu=0
$$
i.e. the integral lines of the field $U^\lambda$ are geodesic curves
and the dust follows a geodesic motion.  In this case for the
structure (\ref{dust}) Einstein equations determine the motion of the
particles of the fluid.

Eq.(\ref{matterconserv}) gives 
$$
U^0\nabla_0 \mu +\mu \nabla_0 U^0 = -\nabla_j (\mu U^j)
$$
which combined with the three equations
$$
U^0\nabla_0 U^j = - U^i\nabla_i U^j
$$  
and recalling that $U^0$ is a function of $U^j$
form a system of four differential  equation of the first order in the
time derivative which determine the time evolution of $\mu$ and $U^j$.

\bigskip

2. Perfect fluid: It is assumed that the energy momentum $T^{\lambda\nu}$
is constructed out of a four velocity field $U^\lambda$ and a set of
scalar functions.

One imposes isotropy i.e. that in the rest frame and in an
orthonormal system in the tangent space, $T^{ab}$ is invariant
under $SO(3)$ rotations 
$$
\Lambda=
\begin{pmatrix}
1& 0\\
0&R
\end{pmatrix}
$$
with $R\in SO(3)$, i.e.
$$
\Lambda T\Lambda^T = T.
$$
We have
$$
\Lambda^{-1}=
\begin{pmatrix}
1& 0\\
0&R^T
\end{pmatrix}.
$$
Through Schur lemma this imposes $T^{ab}$ to be of the form
$$
T^{ab}=
\begin{pmatrix}
\mu&0&0&0\\
0&p&0& 0 \\
0&0&p&0\\
0&0&0&p 
\end{pmatrix}
$$
where $\mu$ is the energy density in the rest frame and $p$ is the
pressure.
Written in covariant form using the four velocity we have
$$
T^{\lambda\nu}= c_1 ~U^\lambda U^\nu + c_2 ~g^{\lambda\nu}
$$
i.e.
\begin{equation}\label{perfectEMT}
T^{\lambda\nu}= (\mu+p)U^\lambda U^\nu +g^{\lambda\nu} p.
\end{equation}
The dust is given by the equation of state $p=0$.
Developing $\nabla_\lambda T^{\lambda\nu}$ and multiplying by $U_\nu$ we have
\begin{equation}\label{cons2}
0=\nabla_\lambda((\mu+p) U^\lambda) - U^\lambda  \nabla_\lambda p=
(\mu+p)\nabla_\lambda U^\lambda +U^\lambda \nabla_\lambda \mu.
\end{equation}
We look for a scalar $\rho$, density function, such that the
current $\rho U^\lambda$ is conserved as a consequence of
Eq.(\ref{cons2}) i.e. 
$$
\nabla_\lambda
(\rho U^\lambda)=U^\lambda\nabla_\lambda \rho +\rho\nabla_\lambda
U^\lambda = \alpha\left (U^\lambda \nabla_\lambda \mu +(\mu+p)
\nabla_\lambda U^\lambda\right)
$$
It is satisfied by 
$$
\nabla_\lambda \rho =\alpha \nabla_\lambda \mu;~~~~\rho =\alpha(\mu+p).
$$
Setting $\rho =\rho(\mu)$ (isoentropic fluid) we
have
$$
\frac{\rho'}{\rho}=\frac{1}{\mu+p}.
$$
where the prime denotes the derivative with respect to $\mu$. From
the above we have also $p=p(\mu)$.  Thus
$$
\rho(\mu) = k \exp\int^\mu\frac{dx}{x+p(x)}
$$
and for such $\rho$ we have
$$
\nabla_\nu(\rho U^\nu)=0
$$
and
$$
\int_\Sigma \rho U^\lambda \Sigma_\lambda = Q ~~~~~~~~{\rm conserved}~.
$$
From Eq.(\ref{cons2}) we obtain
\begin{equation}\label{una}
U^0 \nabla_0 \mu +(\mu+p) \nabla_0 U^0 =
-U^i\nabla_i \mu -(\mu+p)\nabla_iU^i
\end{equation}
to which we add
\begin{equation}\label{tre}
(\mu+p) U^0 \nabla_0 U^j + U^j U^0\nabla_0 p = 
-(\mu+p) U^k \nabla_k U^j - U^j U^k\nabla_k p - \nabla^j p
\end{equation}
where $U^0$ is determined by $U^\mu g_{\mu\nu}U^\nu=-1$. The set of
eqs.(\ref{una},\ref{tre}) form a system of four differential equations
of first order in the time derivative for the four unknown $\mu$ and
$U^j$ which determine their time development.

\bigskip

References

\smallskip
[1] [HawkingEllis] Chap. 3

\section{The energy conditions}\label{7energyconditionsSec}

1. The weak energy condition (WEC)

It states that for any time like vector $v^\mu$ and thus also for any
light-like vector the following inequality holds for the energy
momentum tensor
$$
v^\mu T_{\mu\nu}v^\nu\geq 0.
$$
The meaning is that any observer which isolates as small region of
space should associate to the included matter a positive energy.

2. The dominant energy condition (DEC)

It states that the WEC holds and in addition for any time-like and
thus also for any light-like vector $v^\mu$ the vector
$$
T^{\mu\nu}v_\nu
$$
should be time-like or null. The meaning of such a condition is that
any observer which isolates as small region of space should associate
to the included matted a time like or light like energy momentum
vector.

3. The strong energy condition (SEC)

It states that for any time-like and thus also for any light-like vector
$v^\mu$ the following inequality holds
$$
(T_{\mu\nu}-\frac{1}{2}g_{\mu\nu} T)v^\mu v^\nu\geq 0~.
$$
In four dimensions, in absence of cosmological constant it becomes
$$
v^\mu R_{\mu\nu}v^\nu\geq 0~.
$$
Such condition is satisfied for most of the physical fields but does
not hold generally.

The energy conditions are distinguishing features of gravity as compared
to gauge theories where there is no a priory restriction on the
currents.

The WEC and the DEC are the basic inputs for proving the positivity of
the energy of a gravitational system and for establishing theorems with
regard to the absence of closed time like curves (CTC) or the
chronology protection.


\bigskip

References

\smallskip

[1] [HawkingEllis] Chap. 4

\section{The cosmological constant}
 
The cosmological constant $\Lambda$ provides in Einstein equations an
``effective energy momentum tensor'' with $\mu+p=0$ and $p=-\Lambda$, 
whose canonical form is
$$
T^{ab}= \frac{1}{k}
\begin{pmatrix}
\Lambda&0&0&0\\
0&-\Lambda&0& 0 \\
0&0&-\Lambda&0\\
0&0&0&-\Lambda
\end{pmatrix}
$$
Such ``effective energy momentum tensor'' satisfies (barely) the weak
and dominant energy condition only for 
$\Lambda>0$. On the other hand as $\mu+p=0$ the velocity terms 
disappear in Eq.(\ref{perfectEMT}) 
and thus it can hardly be interpreted as a fluid.
Presently the value of $\Lambda$ is found
positive; of the energy content of the universe
$70\%$ is assigned to the cosmological constant; of the remaining
$30\%$, $4\%$ to ordinary matter out of which only one fourth directly
visible and the remaining $26\%$ to ``dark matter'' i.e. a form of matter we
have not yet accessed. The present value of the cosmological constant
is
$$
(c\hbar)^3 \frac{c^4}{8\pi G} \Lambda = (3~ 10^{-3}~eV)^4,~~~~ \Lambda
~l^2_P = 3.9~ 10^{-123}
$$
\vfill

\eject


\chapter{The energy of a gravitational system}

\section{The non abelian charge conservation}

There is an analogy between the problem of conservation of the
non-abelian charge in Yang-Mills theory and the problem of definition
of energy in gravitation.

We already saw in Section \ref{4thecurrentSec} that from the YM equations
$ D*F=*J $ it follows $0=DD*F=D*J=0$, explicitly from
$$
D_\mu F^{\mu\nu} = \partial_\mu F^{\mu\nu}+ [A_\mu, F^{\mu\nu}] = J^\nu
$$
it follows
$$
0= D_\nu J^\nu = \partial_\nu J^\nu +[A_\nu, J^\nu].
$$
This is the covariant conservation of the covariant current
$J^\nu$, which does not imply a conserved quantity. 

It is possible to derive a conservation law as follows: write
$$
\partial_\mu F^{\mu\nu} = J^\nu - [A_\mu, F^{\mu\nu}] \equiv j^\nu.
$$
Then we have $\partial_\nu j^\nu=0$ from which the conserved charge can
be obtained
$$
Q = \int_\Sigma j^0 d^3x.
$$ 
However $j^\mu$ is not a covariant quantity in contrast to $J^\nu$.
We can also write
$$
Q = \int_\Sigma \partial_\mu F^{\mu 0} d^3x= \int_\Sigma \partial_j
F^{j 0} d^3x =\int_{\partial \Sigma} u_j 
F^{j 0} d^2x 
$$
being $\partial \Sigma$ the boundary of $\Sigma$ which will be taken
to infinity. $u_i$ is the outward-pointing unit normal. 
As a result if we consider a gauge transformation
$U$ arbitrary but constant at space infinity we have that $Q$
under such gauge transformation, transform covariantly $Q\rightarrow
U(\infty) Q U(\infty)^{-1}$ due to the covariance property of $F^{\mu\nu}$.

We turn now to the energy of a system which interacts
gravitationally.

\section{The background Bianchi identities}

In this chapter we shall work with the second order formalism; this
means that the connection will be the metric compatible torsionless
Levi-Civita connection. Later in Section \ref{8positivesection} 
vierbeins will occur and we
recall that with the Levi-Civita connection we have $\nabla e^a=0$ as
we discussed in Chapter \ref{4thecovariantderivative}.
Starting from Einstein equations
\begin{equation}\label{8ELambdaeq}
{\cal G}^{\mu\nu}\equiv R^{\mu\nu} -\frac{1}{2}g^{\mu\nu}R+\Lambda g^{\mu\nu}=
\kappa~ T^{\mu\nu}
\end{equation}
where $\kappa=8\pi G_N/c^3$ with $G_N$ Newton's constant, we 
consider a background metric $\bar g_{\mu\nu}$ satisfying Einstein's
equations in 
absence of matter
\begin{equation}\label{8backgroundeq}
\bar{\cal G}^{\mu\nu}\equiv\bar R^{\mu\nu} 
-\frac{1}{2}\bar g^{\mu\nu}\bar R+\Lambda \bar g^{\mu\nu}=0
\end{equation}
From the contracted Bianchi identities we have for any $g_{\mu\nu}$ 
\begin{equation}\label{8identity}
\nabla_\mu (R^{\mu\nu} -\frac{1}{2}g^{\mu\nu}R+\Lambda g^{\mu\nu})= 0
\end{equation}
Put $g_{\mu\nu}=\bar g_{\mu\nu}+\varepsilon h_{\mu\nu}$. $\varepsilon$
is a formal parameter which at the end will be put equal to $1$; it
will be useful to characterize the powers of $h_{\mu\nu}$ which
intervene in an expression.
We consider both $\bar g_{\mu\nu}$ and $h_{\mu\nu}$ as tensors under
diffeomorphisms.

We have e.g.
$$
\Gamma_{\mu\nu\rho} = \frac{1}{2}(\bar g_{\mu\nu,\rho}+\bar
g_{\mu\rho,\nu}-\bar g_{\nu\rho,\mu}) +\frac{\varepsilon}{2}(h_{\mu\nu,\rho}+
h_{\mu\rho,\nu}-h_{\nu\rho,\mu}) 
$$
and raising the index with $g^{\lambda\mu}$ we have
\begin{equation}\label{8newgamma}
\Gamma^\lambda_{\nu\rho} = \frac{1}{2}\bar g^{\lambda\mu}
(\bar g_{\mu\nu,\rho}+\bar
g_{\mu\rho,\nu}-\bar g_{\nu\rho,\mu}) +O(\varepsilon)
=\bar\Gamma^\lambda_{\nu\rho}+O(\varepsilon)~.
\end{equation}
We have
$$
R^{\mu\nu} -\frac{1}{2}g^{\mu\nu}R+\Lambda g^{\mu\nu}=
(\bar R^{\mu\nu} -\frac{1}{2}\bar g^{\mu\nu}\bar R+\Lambda \bar
g^{\mu\nu}) + \varepsilon K^{\mu\nu} +O(\varepsilon^2)=
\varepsilon K^{\mu\nu} +O(\varepsilon^2)
$$
$K^{\mu\nu}$ depends both on $\bar g_{\mu\nu}$ and $h_{\mu\nu}$ and it
is a tensor under diffeomorphisms.

Moreover due to (\ref{8newgamma}) 
we have $\nabla_\mu = \bar\nabla_\mu +O(\varepsilon)$. Then
identity (\ref{8identity}) tells us that the following identity holds
\begin{equation}\label{8backgroundbianchi}
\bar\nabla_\mu K^{\mu\nu}=0
\end{equation}
which is called the background Bianchi identity [1]. This is an 
{\it identity} which holds for any $h_{\mu\nu}$.  

We notice that $\bar\Gamma^\lambda_{\nu\rho}$ and $\bar R_{\mu\nu}$
are defined in terms of the metric $\bar g_{\mu\nu}$ in particular we
have $\bar R^\mu_\nu= \bar g^{\mu\mu'}\bar R_{\mu'\nu}$ and $\bar
R^{\mu\nu}= \bar g^{\mu\mu'}\bar R_{\mu'\nu'}\bar g^{\nu'\nu}$.

The we can write Einstein' equation (\ref{8ELambdaeq}) as
($\varepsilon=1$)
$$
K^{\mu\nu} =\kappa ~T^{\mu\nu}- {\cal G}^{\mu\nu}+K^{\mu\nu}\equiv
8\pi\kappa (T^{\mu\nu}+t^{\mu\nu})=8\pi\kappa {\cal T}^{\mu\nu}
$$
where $t^{\mu\nu}$ is minus the non linear part in $h_{\mu\nu}$ of the
l.h.s. of (\ref{8ELambdaeq}) divided by $8\pi\kappa$ and  
is called the energy momentum tensor of the gravitational field, while
${\cal T}^{\mu\nu}$ is called the total energy momentum tensor. As a
consequence of (\ref{8backgroundbianchi}) we have
\begin{equation}\label{8covariantbackroundcons}
\bar\nabla_\mu {\cal T}^{\mu\nu}=0
\end{equation}
which is the background covariant conservation of the total energy
momentum tensor.

It is of interest the case in which the background metric $\bar g_{\mu\nu}$
possesses Killing vectors, in different words, when we have isometries
i.e. diffeomorphisms $\phi$ for which 
$\phi^*\bar g_{\mu\nu}(x)= \bar g_{\mu\nu}(x)$.
It means that
$$
\phi^*(g_{\mu\nu}-\bar g_{\mu\nu})(x)= \phi^*(g_{\mu\nu})(x) 
-\phi^*(\bar g_{\mu\nu})(x) = \phi^*(g_{\mu\nu})(x) -\bar g_{\mu\nu}(x)
$$
and thus the difference between $g_{\mu\nu}$ and an invariant (under
$\phi$) background, 
transforms like a tensor under the group of the
isometry transformations of $\bar g_{\mu\nu}$.
E.g. for $\bar g_{\mu\nu}=\eta_{\mu\nu}={\rm diag}~(-1,1,1,1)$ we have that
$$
g_{\mu\nu}-\eta_{\mu\nu}
$$
is a tensor under Lorentz transformations.

If $\bar\xi_\nu$ is a Killing vector field of the background we have
$\bar\nabla_\mu \bar\xi_\nu+\bar\nabla_\nu \bar\xi_\mu=0$ and as a consequence
due to (\ref{8covariantbackroundcons}) we have
$$
\bar\nabla_\mu ({\cal T}^{\mu\nu}\bar\xi_\nu) =0
$$
i.e. a conserved quantity
$$
Q = 
\int_\Sigma {\cal T}^{\mu\nu}\bar\xi_\nu~ \sqrt{-\bar g}
~\varepsilon_{\mu\alpha\beta\gamma}\frac{dx^\alpha\wedge dx^\beta\wedge
dx^\gamma}{3!}=\int_\Sigma {\cal T}^{\mu\nu}\bar\xi_\nu~\Sigma_\mu
$$
provided ${\cal T}^{\mu\nu}\bar\xi_\nu$
vanishes sufficiently quickly at space infinity as to make the surface
integral at space infinity to vanish and the integral over $\Sigma$
convergent.  

In general there is no problem for $T^{\mu\nu}$ which can be assumed
to vanish exactly outside a compact space support; for
$t^{\mu\nu}$ to vanish sufficiently quickly at space
infinity $\bar g_{\mu\nu}$ must be a sufficiently accurate background
for $g_{\mu\nu}$ at space infinity [2]. 
Given an isometry transformation $\phi$, connected to the identity we
can compute the push-forward $\phi_*\xi(\lambda)$ of a Killing 
vector $\xi(\lambda)$.
This will be a linear combination of the Killing vectors of $\bar
g_{\mu\nu}$ which form a finite dimensional linear space [3]. Then we
shall have for the associated conserved quantity
$$
Q[\phi_*\xi(\lambda)]=\int_\Sigma {\cal T}^{\mu\nu}\bar
\phi_*\xi_\nu(\lambda)~\Sigma_\mu=\xi(\rho) A^\rho_{~\lambda}(\phi)~.
$$

Now we specialize to the case $\Lambda=0$ and the background
metric chosen to be the Minkowski 
$\bar g_{\mu\nu}=\eta_{\mu\nu}$. Let $\xi(\lambda)$ be one of the four 
Killing vector of space-time translations 
$$
\xi(\lambda) = \frac{\partial}{\partial x^\lambda}~.
$$
Then from 
$$
\partial_\mu ({\cal T}^{\mu\nu}\xi(\lambda)_\nu)=0
$$
provided the matter (not really a problem) and the gravitational
energy momentum tensor vanish sufficiently rapidly at space infinity,
we have the four conserved quantities 
$$
P_\lambda\equiv\int_\Sigma{\cal T}^{\mu\nu}\xi(\lambda)_\nu ~\Sigma_\mu 
$$
where $\Sigma$ is any space-like surface (e.g. a space-like plane).
Let us perform a coordinate transformation which is an isometry of the
background, in our case a Poincar\'e transformation $x'^\mu =
\Lambda^\mu_{~\nu} x^\nu+a^\mu$. The new Killing vectors are given by
$$
\xi'(\lambda) = \frac{\partial}{\partial x'^\lambda}=
\frac{\partial x^\rho}{\partial x'^\lambda}
\frac{\partial}{\partial x^\rho}=\xi(\rho)(\Lambda^{-1})^\rho_{~\lambda}~.
$$
Then we have
$$
P'_\lambda = P_\rho(\Lambda^{-1})^\rho_\lambda
$$
i.e. the $P_\lambda$ transform like the covariant components of a
four vector. This is the total energy momentum four vector of the
gravitational system.

If $\Sigma$ is provided by the plane $x^0={\rm const}$, 
the components of the Killing vector $\xi(\lambda)$ are
$\xi^\nu(\lambda)=\delta^\nu_\lambda$ we can also write
$$
P_\lambda\equiv\int_\Sigma {\cal T}^{\mu}_{~~\nu}~ \xi^\nu(\lambda)
~\Sigma_\mu 
=\int_\Sigma {\cal T}^0_{~~\lambda} dx^1 ~dx^2 ~dx^3 
$$

\bigskip

References

\smallskip

[1] L.F. Abbott and S. Deser, ``Stability of gravity with a cosmological
constant'' Nucl.Phys. B195 (1982) 76

\smallskip

[2] [WeinbergGC] Chap.7

\smallskip

[3] [WeinbergGC] Chap.13

\section{The superpotential}

We now work out the superpotential in the case of Minkowski background.
We recall that Einstein's equations can be written as
$$
-\frac{1}{4}R^{\alpha\beta}_{~~\mu\nu}
\delta^{\mu\nu\rho}_{\alpha\beta\lambda}= \kappa~ T^\rho_\lambda ~. 
$$
Then after defining
$$
\hat\Gamma^\alpha_{\beta\nu}=
\eta^{\alpha\alpha'}\Gamma_{\alpha'\beta\nu} 
$$
which is linear in $h_{\mu\nu}$ we have for the linear part $\hat
G^\rho_{\lambda}$ of the Einstein tensor 
$$
\hat G^\rho_\lambda = -\frac{1}{4}\hat R^{\alpha\beta}_{~~\mu\nu}
\delta^{\mu\nu\rho}_{\alpha\beta\lambda}=
\partial_\mu \hat Q^{\mu\rho}_{~~\lambda} = \kappa~{\cal T}^\rho_{~\lambda}
$$
being the superpotential $\hat Q^{\mu\rho}_{~~\lambda}$ defined by
$$
\hat
Q^{\mu\rho}_{~~\lambda}=
\frac{1}{2}~\hat\Gamma^{\alpha \beta}_{~~~\nu}   
\delta^{\nu\mu\rho}_{\alpha\beta\lambda}~.
$$
Due to the antisymmetry of $\hat Q^{\mu\rho}_{~~\lambda}$ we have
$
\partial_\rho {\cal T}^\rho_{~\lambda}=0~.
$
After defining the two-form
$$
\hat Q_\lambda = \hat Q^{\sigma\alpha}_\lambda\varepsilon_{\sigma\alpha\mu\nu}
\frac{dx^\mu\wedge dx^\nu}{4}=\frac{1}{2}\hat
Q^{\sigma\alpha}_\lambda~\hat S_{\sigma\alpha}
$$
the conserved $P_\lambda$ can be written as
$$
\kappa~ P_\lambda = \kappa~\int_\Sigma {\cal T}^\rho_{~\lambda} \Sigma_\rho= 
\int_\Sigma \partial_\mu \hat Q^{\mu\rho}_{~~\lambda} \Sigma_\rho=
-\int_\Sigma d\hat Q_{\lambda}~.
$$
In fact
$$
-\int_\Sigma d\hat Q_\lambda = -\int_\Sigma
\partial_\rho \hat
Q^{\sigma\alpha}_\lambda \varepsilon_{\sigma\alpha\mu\nu}
\frac{dx^\rho\wedge dx^\mu\wedge dx^\nu}{4}=
\int_\Sigma \partial_\alpha Q^{\alpha\sigma}_{\lambda}~\Sigma_\sigma
$$
where we used
\begin{equation}\label{8swaparea}
\frac{1}{2!}\varepsilon_{\sigma\alpha\mu\nu}~dx^\rho\wedge dx^\mu\wedge dx^\nu=
\frac{1}{3!}(\delta^\rho_\alpha\varepsilon_{\sigma\mu\nu\kappa}-
\delta^\rho_\sigma\varepsilon_{\alpha\mu\nu\kappa})
dx^{\mu}\wedge dx^{\nu}\wedge dx^{\kappa}~.
\end{equation}
But now $P_\lambda$ can be computed as a surface
integral
\begin{equation}\label{8Qhatexpression}
\kappa~ P_\lambda = -\int_\Sigma d\hat Q_\lambda = -\int_{\partial\Sigma}
\hat Q_\lambda =- \int_{\partial\Sigma} \hat Q^{\sigma\alpha}_\lambda 
\varepsilon_{\sigma\alpha\mu\nu}\frac{dx^\mu\wedge dx^\nu}{4}=
-\frac{1}{2}\int_{\partial\Sigma} \hat Q^{\sigma\alpha}_\lambda~\hat S_{\sigma\alpha}
\end{equation}
i.e. the energy momentum four vector can be computed in terms of a two
dimensional integral on the 2-sphere at infinity.

Saying that the energy is positive is the same as saying that given any
future directed time like four vector $u^\lambda$ we have
\begin{equation}\label{8negative}
\kappa~ u^\lambda P_\lambda = -\frac{1}{2}\int_{\partial\Sigma}
u^\lambda \hat Q^{\sigma\alpha}_\lambda ~S_{\sigma\alpha}
=-\int_{\partial\Sigma} u^\lambda \hat Q_\lambda\leq 0.
\end{equation}

It is useful to give Eq.(\ref{8Qhatexpression}) a covariant expression.
Neither $\Gamma^\alpha_{\beta\nu}$ nor $\Gamma^\alpha_{\beta'\nu} g^{\beta'
\beta}$ are linear in $h_{\mu\nu}$ but they are equivalent to
$\hat\Gamma^\alpha_{\beta\nu}$ and $\hat\Gamma^{\alpha\beta}_{~~\nu}$ 
in the integral at infinity as the correction terms vanish more quickly and
thus give a zero contribution [2].  Moreover we can subtract from
$\Gamma^\alpha_{\beta\nu}$ zero written as
$\Gamma^\alpha_{\beta\nu}({\rm vacuum})$ in the Minkowskian metric
$g_{\mu\nu}=\eta_{\mu\nu}$ with the advantage that
$$
\Delta \Gamma^{\alpha}_{\beta\nu}
$$
is a true tensor under all diffeomorphisms.
As asymptotically we have $e =\sqrt{-g}=1$, as far as the computation of $
u^\lambda P_\lambda$ is concerned, 
we can replace in (\ref{8negative}) the form $u^\lambda \hat
Q_\lambda$ with the covariant two-form
\begin{equation}
Q \equiv\frac{1}{2}u^\lambda ~Q^{\sigma\alpha}_{~~\lambda}\epsilon_{\sigma\alpha\mu\nu} 
\frac{1}{2}dx^\mu\wedge dx^\nu =
\frac{1}{2}u^\lambda ~Q^{\sigma\alpha}_{~~\lambda}S_{\sigma\alpha}
\end{equation}
where
$
\epsilon_{\alpha\beta\gamma\delta} = \sqrt{-g} ~\varepsilon_{\alpha\beta\gamma\delta}
$~.
$$
Q^{\sigma\alpha}_{~~\lambda}=\frac{1}{2}\Delta\Gamma^\zeta_{\beta'\nu}g^{\beta'\beta} 
\delta^{\nu\sigma\alpha}_{\zeta\beta\lambda}
$$ 
and $u^\lambda$ is a time-like future directed vector field, constant at
space infinity. Summarizing the conservation law takes the form
$$
0=\int_V ddQ = \int_{\Sigma_2} dQ-\int_{\Sigma_1} dQ
$$
with
\begin{equation}\label{8covariantform}
-\int_{\Sigma} dQ = -\int_{\partial \Sigma} Q =-\int_{\partial
  \Sigma}u^\lambda Q^{\sigma\alpha}_\lambda dS_{\sigma\alpha}=
-\frac{1}{2}\int_{\partial\Sigma} u^\lambda~\Delta\Gamma^{\zeta\beta}_{~~\nu}
~\delta^{\nu\sigma\alpha}_{\zeta\beta\lambda} ~S_{\sigma\alpha}=
\kappa ~u^\lambda P_\lambda
\end{equation}
which is a fully covariant expression. 

\bigskip

[1] L.F. Abbott and S. Deser ``Stability of gravity with a cosmological
constant'' Nucl.Phys. B195 (1982) 76

\smallskip

[2] [WeinbergGC] Chap.7

\smallskip

[3] J. Nester,`` A new gravitational energy expression with a simple
positivity proof'' Phys. Lett. 83 A (1981) 241

\section{The positive energy theorem}\label{8positivesection}

In the proof of the theorem we follow [3] giving more calculational
details.
We start noticing that given any spinor $\psi$,~ 
$u^\lambda =i\bar\psi\gamma^\lambda\psi$ is a time future directed four vector.
In fact $u^0 =i\bar\psi\gamma^0\psi = \psi^+\psi>0$ in all reference
frames.

We now show that we can replace in (\ref{8covariantform}) the integrand
$$
u^\lambda Q^{\sigma\alpha}_{~~\lambda}=u^\lambda
\frac{1}{2}\Delta\Gamma^{\zeta\beta}_{~~\nu}
\delta^{\nu\sigma\alpha}_{\zeta\beta\lambda}
$$
with [2][3]
\begin{equation}\label{8E}
E^{\sigma\alpha}= \epsilon^{\sigma\alpha\delta\beta}
(\bar\psi\gamma_5\gamma_\delta \nabla_\beta\psi-
\nabla_\beta\bar\psi\gamma_5\gamma_\delta\psi) 
\end{equation}
where $\psi$ is a properly chosen spinor.

We recall that $\gamma_5=\gamma^1\gamma^2\gamma^3\gamma^4=
i\gamma^1\gamma^2\gamma^3\gamma^0$ and
$\epsilon^{\sigma\alpha\delta\beta}= e^{-1}\varepsilon^{\sigma\alpha\delta\beta}$
and we notice that (\ref{8E}) is a fully covariant expression.

$\psi$ will be chosen as 
$$
\psi = \psi_c +\phi
$$
with $\psi_c$ a constant spinor such that
$$
i\bar\psi_c\gamma^\lambda\psi_c = u^\lambda
$$
and
\begin{equation}\label{8phibehav}
\phi = O(\frac{1}{r}).
\end{equation}
$E^{\sigma\alpha}$ differs from $u^\lambda Q^{\sigma\alpha}_{~~\lambda}$ by a
divergence and by a term 
$$
O(\frac{1}{r^3}).
$$
In fact treating first the $\partial_\beta$ part of $\nabla_\beta$
we have
$$
\bar\psi M\partial_\beta \psi =\bar\psi M\partial_\beta \phi =
\partial_\beta(\bar\psi M\phi) -(\partial_\beta\bar\phi)M\phi
$$
while
$$
-\partial_\beta\bar\psi M\psi =-\bar\partial_\beta\phi M\psi =
-\partial_\beta(\bar\phi M\psi) +\bar\phi M\partial_\beta\phi
$$
with $(\partial_\beta\bar\phi)M\phi$ and $\bar\phi
M\partial_\beta\phi$ both $O(1/r^3)$. The divergence term being
the boundary of the boundary zero i.e. $\partial \partial
\Sigma=\emptyset$ gives zero contribution; in fact
$$
\int_{\partial \Sigma}
\epsilon^{\sigma\alpha\delta\beta}\partial_\beta Y  
\epsilon_{\sigma\alpha\mu\nu}dx^\mu\wedge dx^\nu =
- 2 \int_{\partial\Sigma}\delta^{\delta\beta}_{\mu\nu}~\partial_\beta
Y~dx^\mu\wedge dx^\nu =4 \int_{\partial\Sigma} d(Y~dx^\delta)= 0 
$$
and the term $O(1/r^3)$
integrated on the two dimensional surface at infinity 
gives also a vanishing contribution.

Thus we are left with the spin-connection terms. Using
$[\gamma_a,\gamma_b]/4=\sigma_{ab}$ we have
\begin{eqnarray}
& &\frac{1}{2}
\epsilon^{\sigma\alpha\delta\beta}(\bar\psi\gamma_5\gamma_\delta
\sigma_{tz}\psi+\bar\psi\sigma_{tz}\gamma_5\gamma_\delta\psi)
\Gamma^{tz}_{~~\beta}=\frac{1}{2}
\epsilon^{\sigma\alpha\delta\beta}\bar\psi\gamma_5\{\gamma_\delta,
\sigma_{tz}\}\psi\Gamma^{tz}_{~~\beta}\nonumber\\
&=&\frac{1}{2}\epsilon^{\sigma\alpha\delta\beta}
\bar\psi\gamma_5\{\gamma_d,   
\sigma_{tz}\}\psi e^d_\delta e^z_\zeta e^t_\tau
\Gamma^{\tau\zeta}_{~~\beta}=-\frac{1}{2}
\epsilon^{\sigma\alpha\delta\beta}i\bar\psi\gamma^\lambda\psi
\epsilon_{\delta\tau\zeta\lambda} \Gamma^{\tau\zeta}_{~~\beta}\nonumber\\
&=&\frac{1}{2}
\delta^{\sigma\alpha\beta}_{\tau\zeta\lambda}  
i\bar\psi\gamma^\lambda\psi
\Gamma^{\tau\zeta}_{~~\beta}=\frac{u^\lambda}{2}
\Gamma^{\tau\zeta}_{\beta}~\delta^{\beta\sigma\alpha}_{\tau\zeta\lambda}=u^\lambda
Q^{\sigma\alpha}_{~~\lambda}~.     
\end{eqnarray}
The great advantage of the obtained expression is that we can compute
the two dimensional integral as a three dimensional integral of a
divergence. In fact, to prove the positivity of $-u^\lambda P_\lambda$ 
it is
necessary to go back to a three dimensional integral; only in this way
we can have a relation with the energy momentum tensor of matter which
can be the only cause of the positivity of the energy through the
energy conditions. Summarizing the present situation is
\begin{eqnarray}\label{8Eexpression}
&&8\pi\kappa ~u^\lambda P_\lambda = -\int_{\partial\Sigma} E^{\sigma\alpha}
S_{\sigma\alpha}=
-\int_{\partial\Sigma} E^{\sigma\alpha}
\epsilon_{\sigma\alpha\mu\nu}\frac{dx^\mu\wedge dx^\nu}{2}\nonumber
\\
&=&-\frac{1}{2}\int_{\Sigma} \partial_\rho (E^{\sigma\alpha}
\epsilon_{\sigma\alpha\mu\nu}) ~dx^\rho\wedge dx^\mu\wedge dx^\nu
=-\frac{1}{2}\int_{\Sigma} 
\nabla_\rho E^{\sigma\alpha}\epsilon_{\sigma\alpha\mu\nu}
 ~dx^\rho\wedge dx^\mu\wedge dx^\nu\nonumber\\
&=&
-2\int_{\Sigma} \nabla_\alpha E^{\sigma\alpha}
\epsilon_{\sigma\mu\nu\kappa}\frac{dx^\mu\wedge dx^\mu\wedge
  dx^\mu}{3!}
= -2\int_{\Sigma} \nabla_\alpha E^{\sigma\alpha} \Sigma_\sigma
\end{eqnarray}
where in the second line we used the zero torsion of the connection and
in the last line we used (\ref{8swaparea}).

We compute now the divergence of $E^{\sigma \alpha}$.

Being $\nabla_\alpha \epsilon^{\sigma\alpha\delta\beta}=0$ and
noticing that due to the absence of torsion
$$
\epsilon^{\sigma\alpha\delta\beta}\nabla_\alpha e^d_\delta=0
$$
we have 
$$
\nabla_\alpha E^{\sigma\alpha} = \frac{1}{2}\epsilon^{\sigma\alpha\delta\beta}
\bar\psi\gamma_5\gamma_\delta[\nabla_\alpha,\nabla_\beta]\psi -\frac{1}{2}
\epsilon^{\sigma\alpha\delta\beta}
[\nabla_\alpha,\nabla_\beta]\bar\psi\gamma_5\gamma_\delta\psi +
$$
$$
+2\epsilon^{\sigma\alpha\delta\beta}
\nabla_\alpha\bar\psi\gamma_5\gamma_\delta\nabla_\beta \psi
$$
First we examine the curvature terms $[\nabla_\alpha,\nabla_\beta]$.
Recalling that $\gamma_5 = i\gamma^1\gamma^2\gamma^3\gamma^0$ we have
$$
\frac{1}{4}\epsilon^{\sigma\alpha\delta\beta}
\bar\psi\{\gamma_5\gamma_\delta,\sigma_{\tau\zeta}\}\psi
R^{\tau\zeta}_{\alpha\beta}= \frac{1}{4}
\epsilon^{\sigma\alpha\delta\beta} i\bar\psi\gamma^\lambda\psi
\epsilon_{\lambda\delta\tau\zeta} R^{\tau\zeta}_{~~\alpha\beta}
=\frac{1}{4} u^\lambda
R^{\tau\zeta}_{~~\alpha\beta}\delta^{\alpha\beta\sigma}_{\tau\zeta\lambda}=
-8\pi\kappa u^\lambda T^\sigma_\lambda = -8\pi\kappa V^\sigma 
$$
with $V^\sigma$, due to the dominant energy is a non-space like vector
with $V^0<0$ (cfr. Section \ref{7energyconditionsSec}). 

This term substituted in Eq.(\ref{8Eexpression}) 
$$
\int_\Sigma\nabla_\alpha E^{\sigma\alpha}d\Sigma_\sigma
$$
gives
\begin{equation}\label{8firstnegative}
-8\pi\kappa\int_\Sigma
V^\sigma\epsilon_{\sigma\mu\nu\kappa}\frac{dx^\mu\wedge dx^\nu\wedge dx^\kappa}
{3!}\geq 0~.
\end{equation}

The remainder is
$$
2\epsilon^{\sigma\alpha\delta\beta}
\nabla_\alpha\bar\psi\gamma_5\gamma_\delta\nabla_\beta \psi
$$
which recalling that
$$
\gamma_5\gamma_\delta\epsilon^{\sigma\alpha\delta\beta} 
=-i\{\gamma^\sigma,\sigma^{\alpha\beta}\} 
$$
becomes
$$
-2i\nabla_\alpha\bar\psi(\gamma^\sigma\sigma^{\alpha\beta}+
\sigma^{\alpha\beta}\gamma^\sigma)\nabla_\beta\psi.
$$
We have, using $x^0={\rm const}$ on three dimensional spatial
integration surface
\begin{eqnarray}\label{8decomposition}
& &-2i \nabla_j\bar\psi
\{\gamma^0,\sigma^{j k}\}\nabla_k \psi = -4 \nabla_j\psi^+
\sigma^{j k}\nabla_k \psi \nonumber\\
&=&
-\nabla_j\psi^+
(\gamma^j\gamma^k -\gamma^k\gamma^j)\nabla_k \psi 
=2\nabla_j\psi^+\nabla_k\psi \eta^{jk}
-2 \nabla_j\psi^+\gamma^j\gamma^k\nabla_k\psi.  
\end{eqnarray}
There is some freedom in choosing a $\psi$ satisfying the property of
becoming a constant spinor at space infinity with the correction 
$\phi$ behaving as (\ref{8phibehav}). One can exploit such a freedom to
go over to the Witten gauge [2],[4]
$$
\gamma^k\nabla_k\psi =0.
$$
Then the last term in Eq.(\ref{8decomposition}) vanishes and one is
left with a positive contribution which summed to
(\ref{8firstnegative}) gives a total positive result and thus from
Eq.(\ref{8Eexpression})
$8\pi\kappa u^\lambda P_\lambda <0$ which is the statement of the
positive energy theorem.


\bigskip

References

\smallskip

[1] R. Schoen , S. T. Yau, ``On the proof of the positive mass conjecture
in general relativity'' Comm. Math. Phys. 65 (1976) 45;
Phys. Rev. Lett. 43 (1979) 1459

\smallskip

[2] E. Witten, ``A new proof of the positive energy theorem''
Comm. Math. Phys. 80 (1981) 381

\smallskip

[3] J. Nester,`` A new gravitational energy expression with a simple
positivity proof'' Phys. Lett. 83 A (1981) 241

\smallskip

[4] T. Parker, C. Taubes, `` On Witten's proof of the positive energy
theorem'' Comm. Math. Phys. (1982)


\chapter{The linearization of gravity}

\section{Introduction}

First we want to understand the field theoretical description of
particle of spin $2$ in Minkowski space. Given a symmetric tensor of
order $2$ in four dimensions, it has $10$ independent components. The
trace $h^\mu_\mu$ is a scalar field while $\partial_\mu h^\mu_\nu$ is
a vector field. We want to write down an equation whose solutions
satisfy the requirements $h^\mu_\mu=0$ and $\partial_\mu h^\mu_\nu=0$
; we are then left with $10-1-4= 5$
independent components which describe the $5$ possible values
$-2,-1,0,1,2$ of the helicity of a massive spin-$2$ particle.

We get inspiration from the linearized Einstein equations
$$
R_{\mu\nu} -\frac{g_{\mu\nu}}{2}R = \kappa~T_{\mu\nu}
$$
with $g_{\mu\nu}=\eta_{\mu\nu}+\kappa h_{\mu\nu}$ and expanding in $\kappa$
$$
R = d\Gamma\in \Lambda^2_{(1,1)}
$$
or explicitly
$$
R^\alpha_{~\beta\mu\nu} = (\partial_\mu \Gamma_\nu-\partial_\nu
\Gamma_\mu)^\alpha_{~\beta}= 
\partial_\mu \Gamma^\alpha_{~\beta\nu}-\partial_\nu \Gamma^\alpha_{~\beta\mu}.
$$
We immediately have the linearized Bianchi identities
$$
dR = dd\Gamma=0
$$
which written explicitly take the form
$$
R^\alpha_{~\beta[\mu\nu,\rho]}=0.
$$
By double contraction we obtain the linearized contracted Bianchi
identities
$$
0= 2 \partial_\rho R^\rho_{~\nu}-\partial_\nu R.
$$
The explicit form of $R_{\alpha\beta\mu\nu}$ is
\begin{equation}\label{9linearRiemann}
R_{\alpha\beta\mu\nu}=\frac{1}{2}(h_{\nu\alpha,\beta\mu}-
h_{\beta\nu,\alpha\mu}-h_{\alpha\mu,\beta\nu}+h_{\beta\mu,\alpha\nu})   
\end{equation}
while
\begin{equation}\label{9linearG}
R_{\beta\nu}-\frac{1}{2}\eta_{\beta\nu} R =
\frac{1}{2}(h^\alpha_{~\nu,\alpha\beta}-h_{\beta\nu,~\alpha}^{~~~\alpha} -
h^\alpha_{~\alpha,\beta\nu}+h^\alpha_{~\beta,\alpha\nu})-
\frac{1}{2}\eta_{\beta\nu}(h^{\alpha\gamma}_{~~,\alpha\gamma}-
h^{\gamma~~\alpha}_{~\gamma,~~\alpha})\equiv G_{\beta\nu}.  
\end{equation}
Under the linearized form of the diffeomorphism
\begin{equation}\label{9gaugegr}
h_{\mu\nu}\rightarrow h_{\mu\nu}+\partial_\mu\xi_\nu+\partial_\nu\xi_\mu
\end{equation}
we have
$$
\delta\Gamma_{\alpha\beta\nu}=\xi_{\alpha,\beta\nu}
$$
or better
$$
\delta\Gamma_{\alpha\beta} = d(\xi_{\alpha,\beta})\in \Lambda^1
$$
\bigskip\bigskip
and thus we have
$$
\Lambda^2_{0,2}\ni \delta R_{\alpha\beta} = dd(\xi_{\alpha,\beta})=0
$$
i.e. the Riemann tensor is invariant and a fortiori $G_{\beta\nu}$ is
invariant.

\bigskip

\section{The Fierz-Pauli equation}

Before discussing the particle content of the equation
$G_{\beta\nu}=0$ we want to
modify it to include a mass term. We consider the equation
\begin{equation}\label{9fierzpauli1}
G_{\beta\nu} +\frac{m^2}{2}(h_{\beta\nu} -c ~\eta_{\beta\nu}
h^\alpha_{~\alpha})=0. 
\end{equation}
By taking the divergence and recalling the linearized contracted
Bianchi identities we have for $m^2\neq 0$
\begin{equation}\label{9divergence}
h^\beta_{~\nu,\beta} - c ~h^\alpha_{~\alpha,\nu}=0.
\end{equation}
But also the trace of Eq.(\ref{9fierzpauli1}) has to be zero i.e.
$$
-2 h^{\beta\alpha}_{~~ ,\beta \alpha}+ 2 h^{\beta~~~\alpha}_{~\beta
,\alpha}+ m^2 (h^\beta_{~\beta} -4 c ~h^\alpha_{~\alpha})=0
$$
By taking the divergence of Eq.(\ref{9divergence}) we have
$$
h^\beta_{~\nu},_\beta^{~~\nu} - c ~h^\alpha_{~\alpha},^\nu_{~\nu}=0
$$
Choose then the free parameter $c$ equal to $1$. In this way from
Eq.(\ref{9fierzpauli1}) we obtain
$$
-3 m^2 h^\alpha_{~\alpha}=0
$$
i.e. we accomplish the condition $h^\alpha_{~\alpha}=0$. Substituting
then in Eq.(\ref{divergence}) we obtain also $h^\beta_{~\nu,\beta}=0$. 

The above equation can be obtained through variation of the following
Lagrangian
\begin{equation}\label{9fplagrangian}
L=L_G +\frac{m^2}{4}(h^\alpha_{~\beta}h^\beta_{~\alpha}
-h^\alpha_{~\alpha} h^\beta_{~\beta})
\end{equation}
where
$$
L_G = \frac{1}{2}(\partial_\lambda h^\lambda_{~\nu}
\partial^\nu h^\mu_{~\mu}- 
\partial_\lambda h^\lambda_{~\nu}    
\partial_\mu h^{\mu\nu})
+\frac{1}{4}(\partial_\lambda h_{\mu\nu}\partial^\lambda
h^{\mu\nu}- \partial_\lambda h^\mu_{~\mu} \partial^\lambda
h^\nu_{~\nu})~.
$$
In order to understand the mass, substitute in Eq.(\ref{9fierzpauli1})
$h^\alpha_{~\alpha}=0$ and $h^\alpha_{~\beta,\alpha}=0$ and thus
$$
-\frac{1}{2}h^{~~~\alpha}_{\beta\nu,~\alpha}+\frac{m^2}{2}h_{\beta\nu}=0
$$
i.e. the mass-shell is given by
$$
k^2+m^2=0
$$
after setting $h_{\mu\nu}=\varepsilon_{\mu\nu}e^{ik\cdot x}$ with
$\varepsilon_{\mu\nu}=\varepsilon_{\nu\mu}$, 
$\varepsilon^\mu_{~\mu}=0$ and $k^\mu
\varepsilon_{\mu\nu}=0$, for $k^2+m^2=0$.
The propagator can be extracted from the Lagrangian (\ref{9fplagrangian})
and is given by
\begin{eqnarray}\label{9FPpropagator}
G^m_{\mu\nu\alpha\beta}&=&\frac{1}{k^2+m^2}[\eta_{\mu\alpha}\eta_{\nu\beta}+
\eta_{\mu\beta}\eta_{\nu\alpha}-\eta_{\mu\nu}\eta_{\alpha\beta} \nonumber\\
&+&\eta_{\mu\alpha}\frac{k_\nu k_\beta}{m^2}+\eta_{\nu\beta}\frac{k_\mu
k_\alpha}{m^2}+\eta_{\mu\beta}\frac{k_\nu
k_\alpha}{m^2}+\eta_{\nu\alpha}\frac{k_\mu k_\beta}{m^2}\nonumber\\
&+&\frac{4}{3}(\frac{1}{2}\eta_{\mu\nu} -\frac{k_\mu k_\nu}{m^2})
(\frac{1}{2}\eta_{\alpha\beta} -\frac{k_\alpha k_\beta}{m^2})]~.
\end{eqnarray}
Notice that on the mass-shell we have the following relation for the
residues ${\rm Res}~G^\mu_{~\mu\alpha\beta}=0$, ${\rm
Res}~k^\mu G_{\mu\nu\alpha\beta}=0$.  

As $k$ is time-like we can choose a frame where $k^0=m,~k^j=0$. Then due
to $k^\mu \varepsilon_{\mu\nu}=0$ we have $
\varepsilon_{0\nu}=\varepsilon_{\mu 0} =0$ and $\varepsilon_{ij}$ is a
three dimensional symmetric traceless tensor and as such describes a
massive particle of spin $2$.

\section{The massless case}

Due to the invariance of the linearized Riemann tensor under the gauge
transformation $h_{\mu\nu}\rightarrow h'_{\mu\nu}=h_{\mu\nu}+
\xi_{\mu,\nu}+\xi_{\nu,\mu}$ we have that
$$
G_{\mu\nu}(\xi_{\alpha,\beta}+\xi_{\beta,\alpha}) \equiv 0
$$ 
which means that the equation
\begin{equation}\label{9sourceeq}
G_{\mu\nu}(h_{\alpha\beta})= \kappa t_{\mu\nu}
\end{equation}
is not invertible.

On the other hand due to the linearized Bianchi identities we have
that the source $t_{\mu\nu}$ is subject to the restriction
$$
t^\mu_{~\nu,\mu}=0~.
$$
Let us consider a solution of
$$
G_{\mu\nu}(h_{\alpha\beta})=0
$$
and perform on a gauge transformation such that
$h'^\alpha_{~\alpha}=0$. This is always possible because set
$n=h^\alpha_{~\alpha}$, the equation
$$
2 \xi^\alpha_{~,\alpha} +n = 0 
$$  
is solved by
$$
\xi_\mu = -\frac{1}{2}\partial_\mu \frac{1}{\partial^2} n.
$$
Substituting now into Eq.(\ref{9linearG}) we have
$$
h'^\alpha_{~\nu,\beta\alpha}-h'^{~~\alpha}_{\beta\nu,~\alpha}+
h'^{\alpha}_{~\beta,\alpha\nu}-\eta_{\beta\nu}
h'^{\alpha\gamma}_{~~~,\alpha\gamma} =0~.  
$$
Taking the trace we have
\begin{equation}\label{9divN}
h'^{\beta\alpha},_{\beta\alpha}=0
\end{equation}
We look now for a further gauge transformation such that
$h''^\alpha_{~~\beta,\alpha}=0$ i.e. defined $N_\beta =
h'^\alpha_{~~\beta,\alpha}$ we need to solve
$$
0= N_\beta +\partial^2 \xi_\beta +\xi^\alpha_{~,\alpha\beta}~.
$$
Notice that as a consequence of Eq.(\ref{9divN}) $N^\alpha_{~,\alpha}=0$.
Then a solution of the above equation is given by
$$
\xi_\nu = -\frac{1}{\partial^2} N_\nu
$$
and still we have $h''^\alpha_{~~\alpha}=0$.
The equation $G_{\beta\nu}=0$ now becomes
$$
{h''}_{\beta\nu,\alpha}^{~~~~~\alpha}=0
$$
and thus after setting $h_{\mu\nu}=\varepsilon_{\mu\nu}(k) e^{ik\cdot x}$
the mass shell is given by $k^2=0$. We can go over to the 
Lorentz frame such that $k^\mu = (\kappa,0,0,\kappa)$. Imposition of
$k^\mu \varepsilon_{\mu\nu}=0$ give now
$\varepsilon_{00}=-\varepsilon_{03}=-\varepsilon_{33}=c_0$ and we have
$\varepsilon_{0j}=-\varepsilon_{3j}=c_j$. We are still free to perform a gauge
transformation with $\xi^\mu_{~,\mu}=0$ i.e. $k^\mu \xi_\mu(k)=0$,
because this does not alter $\varepsilon^\mu_{~\mu}=0$ and $k^\mu
\varepsilon_{\mu\nu}=0$. In fact
$$
k^\mu( k^\mu \xi_\nu(k)+k^\mu \xi_\nu(k))=k^2\xi_\nu(k)+k_\nu k^\mu
\xi_\mu(k)) =0. 
$$
Thus we perform the gauge transformation with $\xi_0(k)
=\xi_3(k)=c_0/2 \kappa$ and $\xi_j(k) = c_j/\kappa$ 
and find that all elements of
$\varepsilon_{\mu\nu}$ are zero except for $\varepsilon_{mn}$ with $m$
and $n$ equal to $1$ or $2$. Keeping in mind that $ \varepsilon_{11}=
-\varepsilon_{22}$ we see that the quantum states are in one to one
correspondence with the complex field. Two independent base vectors
are given by
$$
\varepsilon_+=
\begin{pmatrix}
1& i \\
i&-1
\end{pmatrix}
~~~{\rm and~}~~~
\varepsilon_-=
\begin{pmatrix}
1& -i \\
-i&-1
\end{pmatrix}
$$   
which under rotation around the $3$ axis
$$
R=
\begin{pmatrix}
\cos\theta& -\sin\theta \\
\sin\theta&\cos\theta
\end{pmatrix}
$$
transform like
$$
R~\varepsilon_+ R^T = e^{-2i\theta}\varepsilon_+ ~~~~~
R~\varepsilon_- R^T = e^{2i\theta}\varepsilon_- 
$$
and thus represent the two states of helicity $2$ and $-2$.

We come now to the propagator in the massless case. We already saw
that $G_{\mu\nu}(h_{\alpha\beta})$ is invariant under the
gauge transformation (\ref{9gaugegr})
i.e. $G_{\mu\nu}(\xi_{\alpha,\beta}+\xi_{\beta,\alpha})=0$ for any
$\xi_\alpha$. This means that the equation
\begin{equation}\label{9Gt}
G_{\mu\nu}(h_{\alpha\beta}) =\kappa~t_{\mu\nu}
\end{equation}
which can have solutions only for $t^\mu_{~\nu,\mu}=0$, has no unique
solution i.e. there is no inverse of the operator $G_{\mu\nu}$. We
want to examine if by restricting the $h_{\alpha\beta}$ such inverse
exists. Given $h_{\alpha\beta}$ we look for a gauge transformation
$\xi_\alpha$ such that $h'^\alpha_{~\beta,\alpha}=0$. Define
$h^\alpha_{~\beta,\alpha}= N_\beta$ and solve
$$
\partial^2 \xi_\beta + \partial_\beta \xi^\alpha_{~,\alpha} = -N_\beta
$$
This is solved by 
$$
\xi_\nu = \frac{1}{\partial^2}(-N_\nu +
\frac{1}{2~\partial^2}\partial_\nu
N^\alpha_{~,\alpha})
$$
which shows that the considered gauge fixing is attainable and
complete. Eq.(\ref{9sourceeq}) now becomes in Fourier transform
$$
\frac{1}{2} k^2 \varepsilon_{\beta\nu}(k)  +\frac{1}{2} k_\beta k_\nu
\varepsilon^\alpha_{~\alpha}(k)  -\frac{1}{2} \eta_{\beta\nu} k^2
\varepsilon^\alpha_{~\alpha}(k)  = t_{\beta\nu}(k) 
$$
which has the unique solution
$$
\varepsilon_{\beta\nu}(k) =\frac{1}{k^2} 
(2\eta_{\beta\alpha} \eta_{\nu\gamma}+ \frac{k_\beta
k_\nu}{k^2} \eta_{\alpha\gamma}-
\eta_{\nu\beta}\eta_{\alpha\gamma}) t^{\alpha\gamma}. 
$$
It is simpler, but equivalent, to use the harmonic gauge fixing
$$
h^\alpha_{~\beta,\alpha}- \frac{1}{2} h^\alpha_{~\alpha,\beta}=0
$$
which is also attainable and complete and given by
$$
\xi_\alpha = -\frac{1}{\partial^2} N_\alpha
$$
with
$$
N_\alpha = h^\beta_{~\alpha},_{\beta}-\frac{1}{2}h^\beta_{~\beta},_{\alpha}.
$$
Then Eq.(\ref{9sourceeq}) becomes
$$
\frac{1}{2} k^2 \varepsilon_{\beta\nu}(k) - \frac{k^2}{4}
\varepsilon^\alpha_{~\alpha}(k)\eta_{\beta\nu}= \kappa ~t_{\beta\nu}(k)
$$
which has the unique solution
$$
\varepsilon_{\beta\nu}(k) = \frac{1}{k^2} 
(\eta_{\beta\alpha} \eta_{\nu\gamma}+\eta_{\nu\alpha}
\eta_{\beta\gamma}- \eta_{\beta\nu} \eta_{\alpha\gamma})~\kappa ~t^{\alpha\gamma}(k). 
$$
i.e. the propagator in the harmonic gauge is
\begin{equation}\label{9harmonicpropagator}
G_{\mu\nu\alpha\beta}= \frac{1}{k^2} 
(\eta_{\mu\alpha} \eta_{\nu\beta}+\eta_{\nu\alpha}
\eta_{\mu\beta}- \eta_{\mu\nu} \eta_{\alpha\beta})~.
\end{equation}

A general gauge fixing can be obtained as follows [1]. Construct $4$
(linear) functions $C_\mu$ of $h_{\mu\nu}$ and its derivatives which are not
invariant under the gauge transformations $\delta h_{\mu\nu} =
\partial_\mu\xi_\nu+\partial_\nu\xi_\mu$ i.e.
$$
\delta C_\mu = M_{\mu\nu}\xi^\nu \neq 0~~~~{\rm for}~~~\xi^\nu\neq 0~.
$$
This implies the invertibility of $M_{\mu\nu}$. Add now 
$\frac{1}{2}C^\mu C_\mu$
to the Lagrangian $L$ appearing in the original invariant action. Consider a
solution of the Euler-Lagrange equation derived from
\begin{equation}\label{9gfLagrangian}
\int (L+\frac{1}{2} C^\mu C_\mu) d^4x
\end{equation}
i.e. from
$$
\delta S+\int C^\mu \delta C_\mu d^4x=0~.
$$
On such solution the action (\ref{9gfLagrangian}) is stationary, in particular
stationary under a gauge transformation of $h_{\mu\nu}$ i.e. we must
have
$$
\int C^\mu M_{\mu\nu}\xi^\nu d^4x =0 ~~~~{\rm for~any}~~~\xi^\nu~.
$$
Thus due to the invertibility of $M_{\mu\nu}$ on such solutions we
have $C^\mu=0$. Thus we have achieved two goals: 1) The solution
of the equations of motion derived from (\ref{9gfLagrangian}) are
stationary point of the original action $S$. 2) Such solution satisfy
$C^\mu=0$.

\bigskip

References

\smallskip

[1] G. 't Hooft, ``Quantum gravity: A fundamental problem and some
radical ideas'' Cargese Summer School 1979, 323


\section{The VDVZ discontinuity}

We want now to compute the one graviton exchange between two massive
sources in the static limit, i.e. when the space momenta of the two
bodies go to zero. The two sources will be described by the energy
momentum tensor $T_j^{\mu\nu} = m_j U_j^\mu U_j^\nu$ $j=1,2$ in the
limit $U_j\rightarrow (1,0,0,0)$. If we denote by $g_m$ the coupling
constant in the massive case we have, from Eq.(\ref{9FPpropagator}) 
for the exchange of a massive graviton for small momentum transfer
$$
g_m^2~m_1 m_2 G^m_{0000}(k) = \frac{g_m^2~m_1 m_2}{k^2+m^2}
 (1+\frac{1}{3})~.  
$$
To fit Newton's law $1/m$ must be much larger than the radius of the
solar system and $G_N= 4 g_m^2/3$, being $G_N$ Newton's constant.
 
From the exchange of a massless graviton we have, from
Eq.(\ref{9harmonicpropagator}) 
denoting by $g$ the
new coupling constant
$$
g^2~m_1 m_2 G_{0000}(k) = \frac{g^2~m_1 m_2}{k^2}  
$$
and to have agreement we need
$$
G_N = \frac{4 g_m^2}{3} = g^2
$$
We examine next the deflection of light by $m_1$. The light ray is
described by the energy momentum tensor $T^{\mu\nu}=E U^\mu
U^\nu$ with $U^\mu= (1,0,0,1)$, and as expected from electromagnetism
we have $T^\mu_\mu=0$. In the massive case we have
$$
2 \frac{g_m^2 m_1 E} {k^2+m^2}
$$ 
while in the massless case
$$
2 \frac{g^2 m_1 E } {k^2}
$$
Thus also in the limit $m\rightarrow 0$ the effect of the exchange of
a massive graviton is $3/4$ of the one due to the exchange of a
massless graviton. This is the van Dam-Veltman-Zacharov discontinuity.
Experiments on light deflection agree with the exchange of a massless
graviton, i.e. with general relativity. A less simple computation
gives for the exchange of a massive graviton an advance of the
perihelion of Mercury which is $2/3$ of the value for the exchange of
a zero mass graviton i.e. $2/3$ of the value predicted by general
relativity.

\bigskip

References

\smallskip

[1] H. van Dam, M. Veltman, ``Massive and massless Yang-Mills and
gravitational fields'', Nucl. Phys. B22 (1970) 397

\smallskip

[2] V.I Zakharov,``Linearized gravitation theory and the graviton
mass'', JETP Lett. 12 (1970) 312 

\smallskip

051208

\section{Classical general relativity from field theory} 
\hskip 1.5cm {\bf \Large in Lorentz space}

\bigskip

We outline here an argument which has developed over a number of years
which asserts that the quantum field theory in Minkowski space, of a
massless spin $2$ particle contains, as low energy limit, classical
general relativity.

Initially such argument was developed in the second order
formalism. A notable simplification of the argument has been achieved
by using the first order (Palatini) formalism.

The idea is that all classical forces are mediated at the quantum
level by the exchange of a particle of proper mass and proper
spin. First we review why one chooses a massless spin $2$
particle. 

The simplest choice is the exchange of a spin $0$ particle. The
simplest coupling of such a scalar field $\phi$ with ``mass'', 
i.e. with the energy
momentum tensor is through the trace of such a tensor, i.e. $\kappa \phi
T^\mu_\mu$. However such coupling, being the
trace of the energy momentum tensor of the Maxwell field zero,
would decouple electromagnetism from
gravity and thus we would not observe deflection of light. 
One can investigate other relativistically
invariant couplings in which light is coupled to a spin zero 
gravity, with the result that the deflection of light by the sun 
depends both on the photon momentum and photon polarization [5]; 
both effects are not observed.

The exchange of spin $1$ particle gives rise to repulsion of equal
``charge'' particle, i.e. if two particles are attracted by the same
source they repel each other. 

Fermionic fields lack a classical limit. Thus the next simplest choice
is a spin $2$ field and as gravity is classically compatible 
with infinite range the
choice falls on a massless particle. Moreover we saw in the previous
section that if the particle is of spin $2$  in order not to conflict
with experiment the mass has to be exactly zero.
We already know how to describe it via the massless Fierz-Pauli equation.

We recall that the Lagrangian is given by
$$
L_G = -\frac{1}{4}[2~ \partial_\mu h^\mu_{~\nu} \partial_\nu
  h^\alpha_{~\alpha}
-2~\partial_\alpha h^\alpha_{~\nu} \partial_\beta h^{\beta\nu}
+ \partial_\mu h_{\alpha\beta} \partial^\mu h^{\alpha\beta} 
- \partial_\mu h^\alpha_{~\alpha} \partial^\mu h^\beta_{~\beta}]~.
$$
We need a Lagrangian because to develop a quantum theory an action
is needed either directly in the path integral approach, or indirectly
to build up the hamiltonian formalism.

The purpose now is to build a Lorentz invariant, interacting,
consistent field theory of a massless spin 2 particle.

As gravity is coupled to mass the simplest model is
described by the Lagrangian
\begin{equation}\label{9simplemod}
L_G +\kappa h^{\mu\nu} T^M_{\mu\nu} + L_M
\end{equation}
where $L_M$ is the free Lagrangian the matter field and $T^M$ is
the matter energy momentum tensor derived from $L_M$. We have at this
stage a
Lorentz invariant model whose equation of motion for $h_{\mu\nu}$ are
obtained by varying $h_{\mu\nu}$
$$
\hat G_{\mu\nu} = \kappa~ T^M_{\mu\nu}
$$ 
being $\hat G_{\mu\nu}$ the linearized Einstein tensor. Such equation
however is untenable because due to the linearized Bianchi identities
we have identically $\partial_\mu \hat G^\mu_{~\nu} =0$ and thus
$\partial_\mu T^{M\mu}_{~~~~\nu} =0$. Thus the matter energy momentum
tensor is locally conserved, but in presence of gravitational
interaction this cannot be true: the energy momentum four vector of
a planet is not constant as the interaction provides an
exchange of energy and momentum between the matter field 
and the gravitational
field. Adding to $T^M_{\mu\nu}$ the energy momentum tensor of the
free gravitational field is still not sufficient because also the
interaction term contributes to the generation of the total conserved
energy momentum tensor of the theory. Thus the modification of
(\ref{9simplemod}) must be of the form
$$
L=L_G +\kappa h_{\mu\nu}T^M_{\mu\nu} + L_M +\kappa L_1(\kappa,
h_{\mu\nu},\phi) 
$$
where $L_1$ must perform the following miracle: the variation of the
action w.r.t. $h_{\mu\nu}$ must give the equation
$$
\hat G_{\mu\nu} = \kappa ~~^tT_{\mu\nu}
$$
where $^tT_{\mu\nu}$ now is the total, conserved, energy momentum tensor derived
from $L$. Such problem is not of easy solution.

Once we have realized that a non linear interaction term of the
gravitational field with itself is necessary, we see that the problem
subsists even in absence of matter. Thus from now on we shall forget
about the matter field and the problem now is
$$
L= L_G(h_{\mu\nu})+\kappa L_I(\kappa,h_{\mu\nu})
$$
and we need
$$
\frac{\delta L_I(h_{\mu\nu})}{\delta h_{\mu\nu}} =T^{\mu\nu}[
L_G+\kappa L_I]
$$ 
One can attack the problem by expanding in powers of $\kappa$. 
$$
L= L_G(h_{\mu\nu})+\kappa L_I^{(1)}(h_{\mu\nu})+\kappa^2
L_I^{(2)}(h_{\mu\nu})+\dots 
$$
and imposing
$$
\sum_n\kappa^n \frac{\delta L_I^{(n)}(h)}{\delta h_{\mu\nu}}=\kappa
T^{\mu\nu}[L_G(h)+ \sum_n \kappa^n L_I^{(n)}(h)]
$$
$$
\frac{\delta L_I^{(1)}}{\delta h_{\mu\nu}}=T^{\mu\nu}[L_G]
$$
$$
\frac{\delta L_I^{(n)}}{\delta h_{\mu\nu}}= T^{\mu\nu}[L_I^{(n-1)}]~.
$$
We are working in the second order formulation and such a program has
been carried through by Gupta [1] and by Thirring [3].

In general given the Lagrangian
$$
L_G+\kappa L_I
$$
where $L_G$ is quadratic and $L_I$ at least cubic if
$$
\hat G^{\mu\nu}=\frac{\delta L_G}{\delta h_{\mu\nu}}
$$
satisfies the identity (linearized Bianchi identity)
$$
\partial_\mu \hat G^{\mu\nu} =0
$$
and we have a problem of consistency [6].
In fact given the exact equations
$$
\hat G^{\mu\nu}(h) + \kappa G_I^{\mu\nu}(h)=0
$$
let us expand the solution in power series of $\kappa$
$$
h = h_0 +\kappa h_1 +\kappa^2 h_2+\dots
$$ 
We have
\begin{equation}\label{9zeroeq}
\hat G^{\mu\nu}(h_0) =0
\end{equation}
$$
\hat G^{\mu\nu}(h_1) + G_I^{\mu\nu}(h_0)=0
$$
$$
................
$$
\begin{equation}\label{9nminusoneeq}
\hat G^{\mu\nu}(h_{n-1}) + H_{n-1}^{\mu\nu}(h_0,h_1,\dots h_{n-2})=0
\end{equation}
$$
\hat G^{\mu\nu}(h_n) + H_n^{\mu\nu}(h_0,h_1,\dots h_{n-1})=0
$$
$$
...............
$$
From the linearized Bianchi identity we have that $h_0$, in addition
to (\ref{9zeroeq})  must satisfy the restriction
$$
\partial_\mu G_I^{\mu\nu}(h_0) =0
$$
and similarly $h_0,h_1,\dots h_{n-1}$ in  addition 
(\ref{9nminusoneeq}) have to satisfy the further restrictions
$$
\partial_\mu H_n^{\mu\nu}(h_0,h_1,\dots h_{n-1}) =0
$$
which may well be inconsistent unless $L_I$ is very special. This is
what happens in gravity and in Yang-Mills theory.

Instead of following this path, we shall employ here the first order
approach following Deser [4] and Deser and Boulware [5]; the great
advantage is that due to the implicit definition of the connections a
single iteration will be sufficient to solve the problem.

We start with some technical remarks about the variational derivation
of Einstein equations. Usually one obtains them by varying  the action
$$
\frac{1}{2\kappa}\int \sqrt{-g}~g^{\mu\nu} R_{\mu\nu} d^n x + \int L_M d^nx
$$
w.r.t. $g^{\mu\nu}$. One could as well vary w.r.t. $g_{\mu\nu}$ but we
can vary also w.r.t. $\bar g^{\mu\nu}\equiv \sqrt{-g}~g^{\mu\nu}$. The
variation of the first term gives simply $R_{\mu\nu}$ while we notice
that
$$
\frac{\delta L_M}{\delta g^{\mu\nu}}= \frac{\delta L_M}{\delta \bar
g^{\mu\nu}} \sqrt{-g} -\frac{1}{2} \sqrt{-g} \frac{\delta
L_M}{\delta \bar g^{\rho\sigma}} g^{\rho\sigma}g_{\mu\nu} 
$$
from which defining
$$
\sqrt{-g}\tau_{\mu\nu}\equiv 2\frac{\delta L_M}{\delta \bar g^{\mu\nu}}
$$
we have
$$
T_{\mu\nu} = \tau_{\mu\nu} - \frac{1}{2}g_{\mu\nu} \tau
$$ 
or
\begin{equation}\label{9tauemt}
\tau_{\mu\nu} = T_{\mu\nu} - \frac{1}{n-2} g_{\mu\nu} T
\end{equation}
Thus we obtain Einstein equations in the form
$$
R_{\mu\nu} = \kappa~\tau_{\mu\nu} =\kappa~( T_{\mu\nu}- \frac{1}{n-2}
g_{\mu\nu} T)~. 
$$ 

We give now the first order formulation of the free
gravitational field.
\begin{equation}\label{9linearfirstorder}
L_G = \bar h^{\mu\nu}(\Gamma^\alpha_{\mu\nu,\alpha}-
\Gamma^\alpha_{\mu\alpha,\nu})+\eta^{\mu\nu}(\Gamma^\alpha_{\beta\alpha}
\Gamma^\beta_{\mu\nu}-\Gamma^\alpha_{\beta\nu}\Gamma^\beta_{\mu\alpha})~.
\end{equation}
Here the fundamental independent fields are the {\it Lorentz tensors}
$\bar h^{\mu\nu}$ and the connection
$\Gamma^\alpha_{\mu\nu}$ symmetric in the lower indices 
and indices are raised by $\eta^{\mu\nu}$.
 Notice that changing in Eq.(\ref{9linearfirstorder}) $\bar h^{\mu\nu}$
into $\eta^{\mu\nu}+\bar h^{\mu\nu}$ changes that Lagrangian by a irrelevant
divergence.
Variation w.r.t.  $\bar h^{\mu\nu}$ gives
\begin{equation}\label{9Gammaequation}
\Gamma^{\alpha}_{\mu\nu,\alpha}-\frac{1}{2}\Gamma^{\alpha}_{\alpha\mu,\nu}
-\frac{1}{2}\Gamma^{\alpha}_{\alpha\nu,\mu} =0.
\end{equation}
Notice the necessary symmetrization as starting from the first order
formalism we do not know a priori that the Ricci tensor is
symmetric. Variation w.r.t. $\Gamma^\alpha_{\mu\nu}$ gives
\begin{equation}\label{9hbarequation}
-\eta^{\rho\nu}\Gamma^\mu_{~\rho\alpha}-\eta^{\rho\mu}\Gamma^\nu_{~\rho\alpha}=
\bar h^{\mu\nu},_{\alpha} -\eta^{\mu\nu}\frac{\bar h^\beta_\beta,_{\alpha}}{n-2}
\end{equation}
from which the $\Gamma$'s can be computed
and using (\ref{9hbarequation}) we obtain from (\ref{9Gammaequation}),
in four dimensions
\begin{equation}\label{9Riccibar}
2 R^L_{\mu\nu}\equiv \DAlemb \bar h_{\mu\nu}-\bar
h^\alpha_{\mu},{}_{\nu\alpha}
-\bar h^\alpha_{\mu},{}_{\nu\alpha} -\frac{1}{2}\eta_{\mu\nu}\DAlemb
\bar h^\alpha_\alpha =0~.
\end{equation}
These are not the equations which can be obtained contracting the
linearized Riemann tensor (\ref{9linearRiemann}) which would give
\begin{equation}\label{9Riccilinear}
2 R^L_{\mu\nu}=-\DAlemb h_{\mu\nu} -
h^\alpha_\alpha,_{\mu\nu}+h^\alpha_\mu,{}_{\nu\alpha}+
h^\alpha_\nu,{}_{\mu\alpha}~. 
\end{equation}
However setting
$$
\bar h_{\mu\nu} = -(h_{\mu\nu}-\frac{1}{2}\eta_{\mu\nu}h^\alpha_\alpha)
$$
Eq.(\ref{9Riccibar}) becomes Eq.(\ref{9Riccilinear}). 
This is not surprising as the
Lagrangian (\ref{9linearfirstorder}) is just the quadratic part of the 
Palatini Lagrangian. 

We want now to perform the first step i.e. adding to the Lagrangian
(\ref{9linearfirstorder}) the term 
$\bar h^{\mu\nu}\tau_{\mu\nu}$ where $\tau_{\mu\nu}$ is the
combination (\ref{9tauemt}) of energy momentum tensor derived from the free
Lagrangian (\ref{9linearfirstorder}).  

We know the general rule how to compute the energy momentum tensor,
even if historically such a procedure came after general relativity,
but it is logically independent of it. The procedure is to render the
Lagrangian invariant under diffeomorphisms by properly introducing a
metric $f_{\mu\nu}=\eta_{\mu\nu}+\psi_{\mu\nu}$ whose covariant
metric density $\bar f^{\mu\nu}=\sqrt{-f} f^{\mu\nu}$ will be written as 
$\bar f^{\mu\nu}=\eta^{\mu\nu}+\bar\psi^{\mu\nu}$,  
and changing derivatives into
covariant derivatives. Then one has to take the functional derivative
w.r.t. $\bar\psi^{\mu\nu}$ and then go back to flat space.

The first replacement in (\ref{9linearfirstorder}) is 
$\eta^{\mu\nu}\rightarrow
\eta^{\mu\nu}+\bar\psi^{\mu\nu}$. The functional derivative
w.r.t. $\bar\psi^{\mu\nu}$ gives as contribution to $\tau_{\mu\nu}$
$$
\Gamma^\alpha_{\mu\nu}\Gamma^\beta_{\alpha\beta}-
\Gamma^\alpha_{\beta\mu}\Gamma^\beta_{\alpha\nu} ~. 
$$
The derivative of the {\it tensors} $\Gamma^\alpha_{\mu\nu}$ have to
be replaced by the covariant derivatives; calling $C^\alpha_{\beta\mu}$
the metric torsionless connections generated by $f_{\mu\nu}$ we have e.g.
$$
\Gamma^\alpha_{\mu\nu},{}_{\beta}\rightarrow \Gamma^\alpha_{\mu\nu},{}_{\beta}+
C^\alpha_{\gamma\beta}\Gamma^\gamma_{\mu\nu}-
C^\gamma_{\mu\beta}\Gamma^\alpha_{\gamma\nu}-
C^\gamma_{\nu\beta}\Gamma^\alpha_{\mu\gamma} ~.
$$
However as at the end we must go over to the flat limit the result is
the same as using for $C$ the linearized expression i.e.
$$
C^\alpha_{\beta\mu}=\frac{1}{2}\eta^{\alpha\gamma}[
\psi_{\beta\gamma},{}_\mu+\psi_{\gamma\mu},{}_\beta-
\psi_{\beta\mu},{}_\gamma]~.
$$
As a result the related contributions to $\tau_{\mu\nu}$ are of
divergence type i.e. $\partial_\alpha H^\alpha_{\mu\nu}$~.
On flat space the energy momentum tensor is defined up to divergences
and we shall drop the last contribution. For a detailed discussion see
[4].
Then the action becomes
$$
L_G = (\eta^{\mu\nu}+\bar h^{\mu\nu})(\Gamma^\alpha_{\mu\nu,\alpha}-
\Gamma^\alpha_{\mu\alpha,\nu})+(\eta^{\mu\nu}+\bar h^{\mu\nu}) 
(\Gamma^\alpha_{\alpha\beta}
\Gamma^\beta_{\mu\nu}-\Gamma^\beta_{\mu\alpha}\Gamma^\alpha_{\nu\beta}) 
$$
which is exactly the Palatini Lagrangian; the theory has acquired full
invariance under diffeomorphisms. The reason why in the first order
formalism one step is sufficient while in the second order formalism
and infinite sequence of iterations are necessary, is that in the
first order formalism the connection is implicitly defined by the
action and thus a single change in the action corresponds to an
infinite number of changes in the expression of $\Gamma$ in terms of
the metric.

\bigskip

References

\smallskip

[1] S.N. Gupta,` Quantization of Einstein's gravitational field:
general treatment'', Proc. Phys. Soc. (London) A65 161 (1952) 608;
``Quantum Field Theory in Terms of Ordered Products'',
Rev. Mod. Phys. 29 (1957) 334

\smallskip

[2] E. Corinaldesi, ``Quantum field theory and the two-body problem'',
Nuovo Cim. 1 (1955) 1289; `The Two-body Problem in the Theory of the
Quantized Gravitational Field'', Proc. Phys. Soc. (London) A69 (1956)
189

\smallskip

[3] W. Thirring, Fortschritte der Physik,``Lorentz-invariante
Gravitationstheorien'', 7 (1959) 79; Ann. Phys. (N.Y.),``An
alternative approach to the theory of gravitation'', 16 (1961) 96

\smallskip

[4] S. Deser,``Selfinteraction and gauge invariance'',
J. Gen. Rel. Grav. 1(1970) 9

\smallskip

[5] S. Deser, D. G. Boulware,`` 	
Classical General Relativity Derived from Quantum Gravity'' 
Ann. Phys. (N.Y.) 89 (1975) 193

\smallskip

[6] R.M. Wald,``Spin-2 Fields and General Covariance'', Phys. Rev. D
33 (1986) 3613

\vfill



\chapter{Quantization in external gravitational fields}

\section{Introduction}

The quantization of fields in external gravitational field is a well
defined procedure when the gravitational field possesses a time like
Killing vector field. 

More generally the presence of a Killing vector field allows to derive from the
equation $\nabla_\mu T^{\mu\nu}=0$ a true conservation law
$$
\nabla_\mu(T^{\mu\nu}\xi_\nu)=0
$$
i.e. the quantity
$$
E= \int_\Sigma d^D x~\sqrt{\gamma} ~n_\mu T^{\mu\nu}\xi_\nu
$$
provided the fields decrease sufficiently fast at space infinity, is
conserved. $n_\mu$ is the (future directed) normal to the space-like
surface $\Sigma$ and  $\gamma$ the determinant of the metric induced on
$\Sigma$. 
In an ADM coordinate system
$$
E = -\int_\Sigma d^Dx ~\sqrt{h}~ N~ T^{0\nu}\xi_\nu.
$$
We shall consider first the special situation in which there exists a
time like Killing vector field which is surface orthogonal. We recall
the Frobenius criterion for surface orthogonality of a vector field:
Defined $\xi = \xi_\mu dx^\mu$ necessary and sufficient condition for
being $\xi_\mu$ (locally) surface orthogonal is 
$$
d\xi = \xi\wedge \theta
$$
being $\theta$ a one form.

Starting from a surface $\Sigma_0$ orthogonal to the time like Killing
vector field $\xi$ and which intersects all the integral lines of the
Killing vector field, one can
construct a foliation of space time in terms of space like surfaces,
orthogonal to the Killing vector field. One produces the foliation by 
applying the group of diffeomorphisms $\phi_t$, generated by
$\xi_\mu$, 
to $\Sigma_0$, symbolically $\Sigma_t = \phi_t \Sigma_0$. By $\phi_t$
all vectors tangent to $\Sigma_0$ are pushed forward to 
vectors tangent
to $\Sigma_t$ and we have for $v$ belonging the tangent space of
$\Sigma_0$
$$
\xi^\mu(0,x) g_{\mu\nu}(0,x) v^\nu(0,x)=0~.
$$
By definition of $\phi_t$, $\xi(0,x)$ is pushed forward to $\xi(t,x)$, and
being $\phi_t$ an isometry we have
$$
\xi^\mu(t,x) g_{\mu\nu}(t,x) v^\nu(t,x)=\xi^\mu(0,x) g_{\mu\nu}(0,x)
v^\nu(0,x)=0~. 
$$
Moreover we can choose as time-flow vector just $\xi^\mu$. In fact by
definition of tangent vector we have
$$
\xi^\mu \nabla_\mu t=1~.
$$
The conserved quantity is 
$$
E =\int_\Sigma d^D x~\sqrt{\gamma} ~n_\mu T^{\mu\nu}\xi_\nu
$$
which in the ADM coordinate system becomes
$$
E = -\int_\Sigma d^Dx ~\sqrt{h}~ N ~T^{0}_{~0}
$$
being there the time-flow vector $t^\mu = \xi^\mu=(1,0,0,0)$ (cfr.
Section \ref{7intoductionSec}) and the last one can also be rewritten as
$$
E = \int_\Sigma d^Dx \frac{\sqrt{h}}{N} T_{00}~.
$$
In the ADM coordinate system the orthogonality of $\xi$ to $\Sigma_t$
gives $\xi^\mu g_{\mu j}=g_{0j}=N_j =0$ and thus also
$g^{0j}=0$. Moreover being $\xi^\mu=t^\mu$ a Killing vector the
$g_{\mu\nu}$ is independent of time.
 
We shall refer to an hermitean scalar field. The K.G. equation becomes 
$$
g^{00}\partial_0\partial_0\phi
+\frac{1}{\sqrt{-g}}\partial_l\sqrt{-g}g^{ln}\partial_n\phi -m^2\phi=0
$$
or
$$
-\partial_0\partial_0\phi = K\phi
$$
with
$$
K=
\frac{1}{(-g^{00})}\left[-\frac{1}{\sqrt{-g}}
\partial_l\sqrt{-g}g^{ln}\partial_n + m^2\right]. 
$$
$K$ is hermitean and positive in the measure
\begin{equation}\label{10fullingmeasure}
(-g^{00})\sqrt{-g}~d^Dx. 
\end{equation}
In fact
$$\int \phi^*_2K\phi_1 (-g^{00})\sqrt{-g}~d^Dx=
\int\sqrt{-g}~(\partial_l\phi^*_2~ 
g^{lk}~\partial_k\phi_1+m^2\phi_2^*\phi_1)~ d^Dx.
$$
According to Friedrich's theorem there is always
a self adjoint extension with the same lower bound.
For a self-adjoint extension of $K$ the set $\phi_n({\bf x},\omega)$ 
$$
\omega^2\phi_n(\omega,{\bf x})= K\phi_n(\omega,{\bf x})
$$
where the index $n$ takes into account degeneracy, is complete. Being
$K$ real we can choose a real orthonormal basis of $\phi_n(\omega,{\bf
x})$. 
From the action
$$
S=\int dt\int d^Dx
(-g^{00})\sqrt{-g}\frac{1}{2}\left(\dot\phi^2-\frac{1}{(-g^{00})}(
g^{ij}\partial_i\phi\partial_j\phi +m^2\phi^2)\right) = 
$$
$$
=\int dt\int dx^D L =\int dt~ {\cal L}
$$
we have
$$
\pi = \frac{\delta {\cal L}}{\delta \dot\phi} = \frac{\partial
L}{\partial\dot\phi}=(-g^{00})\sqrt{-g}~\dot\phi 
$$
and we can compute the hamiltonian
$$
H=\int d^D x(\pi\dot\phi-L)=
\int d^Dx \frac{1}{2}\left(
\frac{\pi^2}{(-g^{00})\sqrt{-g}}+ \sqrt{-g}(g^{kl}\partial_k\phi
\partial_l\phi+m^2\phi^2)\right)  
$$
which agrees with
$$
E= \int_\Sigma d^Dx \frac{\sqrt{h}}{N} T_{00}
$$
derived in Section \ref{5couplingmatterSec} 
from a different procedure, i.e. $\sqrt{-g}~T_{\mu\nu}$ as twice the
variational derivative of the action with respect to $g^{\mu\nu}$.
We could proceed to quantization by imposing the equal time
commutation relations
$$
[\phi({\bf x},t),\pi({\bf y},t)]=i\delta({\bf x}-{\bf y}).
$$
It is simpler to compute the Lagrangian (action) using the mode
expansion
$$
\phi({\bf x},t) =\sum_n\int d\omega ~q_n(\omega,t) ~\phi_n({\bf x},\omega)
$$
with the orthonormality relation
$$
\int (-g^{00})\sqrt{-g}~d^Dx~\phi_m({\bf x},\omega)\phi_n({\bf
x},\omega')d{\bf x} = \delta_{mn}~\delta(\omega-\omega')
$$
(real field) reaching for the Lagrangian
$$
L = \sum_n\int d\omega \frac{1}{2}[\dot q^2_n(\omega,t)-\omega^2
q_n^2(\omega,t)]. 
$$
We can compute the conjugate momenta $p_n(\omega,t) = \dot
q_n(\omega,t)$ and the hamiltonian
$$
\sum_n\int d\omega \frac{1}{2}[ p^2_n(\omega,t)+\omega^2
q_n^2(\omega,t)] 
$$
and thus we have an infinity of harmonic oscillators.
Canonical quantization is obtained by imposing
$$
[q_m(\omega,t),p_n(\omega',t)]=i\delta_{mn}~\delta(\omega-\omega').
$$ 
Setting
$$
a_n(\omega,t) =\frac{p_n-i\omega q_n}{\sqrt{2 \omega}}
$$
$$
a^+_n(\omega,t) =\frac{p_n+i\omega q_n}{\sqrt{2 \omega}}
$$
one has the commutation relations
$$
[a_m(\omega,t),a^+_n(\omega',t)] = \delta_{mn}~\delta(\omega-\omega')
$$
and
$$
H= \sum_n\int d\omega~ \omega~ \frac{1}{2}[a^+_n(\omega,t) a_n(\omega,t)
+a_n(\omega,t) a^+_n(\omega,t)] 
$$
or
$$
H= \sum_n\int d\omega~ \omega~ [a^+_n(\omega,t) a_n(\omega,t)
+\frac{1}{2}\delta(0)] 
$$
which gives to $a^+a$ the particle number interpretation  and to
$\omega$ the meaning of energy of the quantum. The infinity is
discarded through normal ordering.
The field now can be written
$$
\phi = \sum_n \int (f_n(\omega,{\bf x}) a_n(\omega,t) +
\bar f_n(\omega,{\bf x}) a^+_n(\omega,t)) d\omega
$$
with
$$
f_n(\omega,{\bf x}) = \frac{i}{\sqrt{2\omega}}\phi_n(\omega,{\bf x}).
$$
The equation of motion $\displaystyle{\dot F = \frac{1}{i}[F,H]}$
gives 
$$
\dot a^+_n(\omega,t) = i\omega a^+_n(\omega,t)
$$
and thus
$$
a^+(\omega,t) = e^{i\omega t} a^+(\omega,0).
$$

\section{The scalar product of two solutions}\label{10staticproductsec}

A different path to introduce the measure (\ref{10fullingmeasure}) is
the following. Consider two solutions $\phi$ and $f$ of the K.G. equation 
$$
\frac{1}{\sqrt{-g}}\partial_\mu(\sqrt{-g}g^{\mu\nu}\partial_{\nu}\phi)
-m^2\phi=0. 
$$
We have
\begin{eqnarray}\label{10difference}
0&=& \bar f\frac{1}{\sqrt{-g}}\partial_\mu(\sqrt{-g}g^{\mu\nu}\partial_{\nu}\phi)
-m^2\bar f\phi
-\phi\frac{1}{\sqrt{-g}}\partial_\mu
(\sqrt{-g}g^{\mu\nu}\partial_{\nu}\bar f)+m^2\bar f\phi= \nonumber\\
&=&\frac{1}{\sqrt{-g}}\partial_\mu(\sqrt{-g}(\bar f g^{\mu\nu}\partial_{\nu}\phi
-\phi g^{\mu\nu}\partial_{\nu}\bar f)) 
\end{eqnarray}
By integrating the r.h.s. of Eq.(\ref{10difference}) on 
the volume $V$ bounded by two space like surfaces $\Sigma_2$ and
$\Sigma_1$ and a mantle at space infinity we have the result
\begin{eqnarray}\label{10constmixed}
0&=&\int_V\partial_\mu(\sqrt{-g}(\bar f g^{\mu\nu}\partial_{\nu}\phi
-\phi g^{\mu\nu}\partial_{\nu}\bar f) d^nx = \\
&=&\int_{\Sigma_2}
(\bar fg^{\mu\nu}\partial_{\nu}\phi 
-\phi g^{\mu\nu}\partial_{\nu}\bar f) \Sigma_\mu-\int_{\Sigma_1}
(\bar f g^{\mu\nu}\partial_{\nu}\phi 
-\phi g^{\mu\nu}\partial_{\nu}\bar f)\Sigma_\mu\nonumber
\end{eqnarray}
with 
$$
\Sigma_\mu = \frac{1}{(n-1)!}\epsilon_{\mu\sigma_1\dots\sigma_{n-1}}
dx^{\sigma_1}\wedge\dots dx^{\sigma_{n-1}}
$$
thus defining an invariant scalar product (not positive
definite). This holds also for a time dependent $g_{\mu\nu}$ . In the
case in which the metric is stationary $g_{\mu\nu}({\bf x})$ and
static  $g_{0 i}({\bf x})=0$ choosing the space section $x^0\equiv t={\rm
const}$, for solutions of the form
$\phi(x) = e^{i\omega t}\varphi({\bf x})$ and $f(x) = e^{i\omega'
t}f_s({\bf x})$ we have defining
\begin{equation}
(f,\phi) = i \int_{\Sigma}
(\bar f g^{\mu\nu}\partial_{\nu}\phi 
-\phi g^{\mu\nu}\partial_{\nu}\bar f)\Sigma_\mu
\end{equation}
\begin{equation}\label{10staticproduct}
(f,\phi)= e^{i(\omega-\omega')t} (\omega+\omega')
\int (-g^{00})\sqrt{-g} \bar f_s({\bf x})\varphi({\bf x}) d^{n-1} x
\end{equation}
which due to the constancy in time makes $(f,\phi)$ to vanish for
$\omega\neq \omega'$. For $\omega=\omega'$ we have
\begin{equation}\label{fphiscalarproduct}
(f,\phi)= 2\omega
\int (-g^{00})\sqrt{-g} \bar f_s({\bf x})\varphi({\bf x}) d^{n-1} x
\end{equation}
obtaining the positive definite invariant conserved scalar
product, on the positive frequency solutions, 
which differs from the one induced by the measure
(\ref{10fullingmeasure}) by the multiplicative factor $2\omega$.
Notice that for $\phi(x) = e^{i\omega t}\varphi({\bf x})$ and $f(x) 
= e^{-i\omega't}f_s({\bf x})$ $\omega>0,~~\omega'>0$ $(f,\phi)$ is
always zero.

\bigskip

References

\smallskip

[1] [Wald] Chap. 14

\smallskip

[2] N.D. Birrel and P.C.W. Davies ``Quantum fields in curved space'' Cambridge
University press, Chap. 3 

\smallskip

\section{The scalar product for the general stationary metric}

The application of Eq.(\ref{10constmixed}) to periodic solutions with
positive frequencies in the
stationary but non static case leads to the scalar product 
$$
i\int_{\Sigma_2}\sqrt{-g}(2 \omega i\phi^*g^{00}\phi+\phi^*g^{0k}\nabla_k \phi
-\phi g^{0k}\nabla_k \phi^*
)) dx^{n-1}
$$
which is not positive definite. Here below we treat the general
stationary non static case [1]. 

First we prove that $-g^{00}>0$ and $g^{kl}>0$.
The time flow vector $\frac{\partial}{\partial t}$ i.e. $t^\mu=(1,0,0,0)$ is
always assumed to be time like, i.e. in the ADM metric $N^2-N^jN_j>0$ which
implies $N^2>0$ and thus $-g^{00}= \frac{1}{N^2}>0$.

We have
$$
g^{ij} = h^{ij}-\frac{N^i N^j}{N^2}~.
$$
Then by Schwarz inequality we have
$$
v_i g^{ij}v_j = v_i h^{ij}v_j -\frac{(v_i N^i)(N^j v_j)}{N^2}\geq
v_ih^{ij}v_j -\frac{(v_i h^{ij}v_j)(N_ih^{ij}N_j)}{N^2}=
v_ih^{ij}v_j -\frac{(v_i h^{ij}v_j)(N^iN_i)}{N^2}
\geq~0~.
$$
Let us consider the ``action'' 
functional of two fields $\phi_1$ and $\phi_2$ (not necessarily an
hermitean action)
$$
S=-\int \sqrt{-g} d^nx
(g^{\mu\nu}\partial_\mu\phi^*_2\partial_\nu\phi_1+m^2\phi^*_2 \phi_1)~.
$$
Variation w.r.t. $\phi^*_2$ and $\phi_1$ give the two equations
$$
\frac{1}{\sqrt{-g}}\partial_\mu(\sqrt{-g}~
g^{\mu\nu}\partial_\nu\phi_1)-m^2\phi_1 =0
$$ 
$$
\frac{1}{\sqrt{-g}}\partial_\mu(\sqrt{-g}~
g^{\mu\nu}\partial_\nu\phi^*_2)-m^2\phi^*_2 =0~.
$$ 
From the variation of $S$ w.r.t. $g_{\mu\nu}$ we obtain a symmetric ${\bf
T}^{\mu\nu}$ which on the equations of motion is covariantly conserved
$$
\nabla_\mu{\bf T}^{\mu\nu}=0~.
$$
If $\xi^\mu$ is a Killing vector we have the conserved current ${\bf
T}^\mu_\nu\xi^\nu$ 
$$
\nabla_\mu({\bf T}^\mu_\nu\xi^\nu)=0~.
$$
If $\xi^\mu$ is time-like we choose it as time flow vector $t^\mu=\xi^\mu$.
Then we have the conserved quantity 
$$
{\bf E} = \int \sqrt{-g}~{\bf T}^0_0 ~d^{n-1}x = 
\int \sqrt{-g} ~g^{0\beta}~{\bf T}_{\beta 0} ~d^{n-1}x = {\rm const}
$$
where explicitly
$$
{\bf T}^0_0 =
-g^{00}\partial_0\phi^*_2\partial_0\phi_1+
g^{kl}\partial_k\phi^*_2\partial_l\phi_1+m^2\phi^*_2\phi_1 
$$
i.e. the conserved quantity is
\begin{equation}\label{ashtekarmagnon}
\int d^{n-1}x(-g^{00})\sqrt{-g}[\partial_0\phi^*_2\partial_0\phi_1+
\phi^*_2K\phi_1] = (\partial_0\phi_2,\partial_0\phi_1)+
(\phi_2,K\phi_1) \equiv \langle \phi_2,\phi_1\rangle 
\end{equation}
where $K$ is given again by
$$
K =-\frac{1}{(-g^{00})\sqrt{-g}}\partial_m\sqrt{-g}g^{ml}\partial_l+
\frac{m^2}{(-g^{00})}~.
$$
Eq.(\ref{ashtekarmagnon}) 
defines a positive definite scalar product on the solutions of the KG
equation. 

\bigskip

For a direct derivation of the above equation from the equation
of motion, without any appeal to the variational procedure, notice that
the equation
$$
g^{00}\partial_0^2\phi =
-\frac{1}{\sqrt{-g}}\partial_m\sqrt{-g}g^{ml}\partial_l\phi-
 \frac{1}{\sqrt{-g}}\partial_0\sqrt{-g}g^{0l}\partial_l\phi-
\frac{1}{\sqrt{-g}}\partial_m\sqrt{-g}g^{m0}\partial_0\phi+m^2\phi
$$
taking into account the time independence of $g_{\mu\nu}$ can be written as
\begin{equation}\label{KAequation}
-\partial_0^2\phi = K\phi + A \partial_0\phi~.
\end{equation}
$K$ is hermitean and positive definite in the metric
$(-g^{00})\sqrt{-g}d^{n-1}x = \frac{\sqrt{h}}{N}~d^{n-1}x$
and
$$
A=- \frac{1}{(-g^{00})\sqrt{-g}}\big(\sqrt{-g}g^{0m}\partial_m+
\partial_m\sqrt{-g}g^{m0}\big)
$$
is anti-hermitean, again in the metric $(-g^{00})\sqrt{-g}~ d^{n-1}x$.
Multiplying the equation (\ref{KAequation}) 
for $\phi_1$ by $\partial_0 \phi_2^*$, performing the
same by exchanging $1$ with $2$ and adding we obtain
$$
-\partial_0(\partial_0\phi_2^*\partial_0\phi_1)=
\partial_0\phi_2^* K\phi_1+\partial_0\phi_1 K\phi^*_2+
\partial_0\phi_2^* A\partial_0\phi_1+\partial_0\phi_1 A\partial_0\phi^*_2~.
$$
Integrating in $(-g^{00}\sqrt{-g})~d^{n-1}x$ we obtain
$$ 
-\partial_0(\partial_0\phi_2,\partial_0\phi_1) =\partial_0(\phi_2,K\phi_1)
$$
i.e.
$$
\langle\phi_2,\phi_1\rangle=(\partial_0\phi_2,\partial_0\phi_1) +
(\phi_2,K\phi_1)
$$
is a conserved positive definite scalar product on the solutions of the KG
equation. This coincides with (\ref{ashtekarmagnon}) i.e. the 
Ashtekar and Magnon scalar product [1].

For periodic solutions 
$\phi = \varphi e^{i\omega t}$ the above becomes
$$
\langle\phi_2,\phi_1\rangle=e^{i(\omega_1-\omega_2)t}
(\omega_2\omega_1(\varphi_2,\varphi_1)+(\varphi_2,K\varphi_1))
$$
implying for $\omega_1\neq \omega_2$
$$
\omega_2\omega_1(\varphi_2,\varphi_1)+(\varphi_2,K\varphi_1)=0
$$
i.e. for $\omega_1\neq \omega_2$ we have orthogonality in the positive
definite scalar product 
$$
\omega_2\omega_1 (\varphi_2,\varphi_1)+(\varphi_2,K\varphi_1)~.
$$
Notice that for positive frequency solutions $\phi = \varphi
e^{i\omega t}$ we have
\begin{equation}\label{10geneigenvalue}
\omega^2 \varphi = K\varphi+\omega iA \varphi~.
\end{equation}
For the negative frequency solution $\phi = \varphi_1 e^{-i\omega t}$
we have
\begin{equation}
\omega^2 \varphi_1 = K\varphi_1-\omega iA \varphi_1
\end{equation}
satisfied by $\varphi_1 = \bar\varphi$.
One should prove completeness for the solutions of 
Eq.(\ref{10geneigenvalue}) [2].

For $\phi_2=\phi_1=\phi$ we have the energy 
$$
E= \int d^{n-1}x(-g^{00}\sqrt{-g})(\partial_0\phi^*\partial_0\phi +\phi^* K\phi)
$$ 
so that with normalization $1$ for $\langle,\rangle$ for the modes $\phi_n$
we obtain with the decomposition
$$
\phi = \sum_{\omega_n>0} {\bf c}_n \varphi_n e^{-i\omega_n t}+
{\bf c}^+_n \bar\varphi_n e^{i\omega_n t}
$$
$$
E = \sum_{\omega_n>0} ({\bf c}^+_n{\bf c}_n
+{\bf c}_n{\bf c}^+_n)
$$
i.e. 
$$
{\bf c}_n = {\bf a}_n \sqrt\frac{\omega_n}{2}~~~~,
~~~~{\bf c}^+_n = {\bf a}^+_n \sqrt\frac{\omega_n}{2}
$$
with ${\bf a}^+_n{\bf a}_n$ the occupation number operator.

The scalar product $\langle,\rangle$ obviously applies also to the
static case giving rise to the normalization for $\phi=
e^{i\omega t}\varphi$
$$
\langle \phi,\phi\rangle = \omega^2 (\varphi,\varphi)+(\varphi,K\varphi) =
2\omega^2(\varphi,\varphi)~.
$$

\bigskip

References

\smallskip

[1] A. Ashtekar and A. Magnon,``Quantum fields in curved space-time'',
 Proc. R. Soc. Lond. A. 346 (1975) 375

\smallskip

[2] B.S. Kay,``Linear spin-zero quantum fields in external
gravitational and scalar fields'', Commun. math. Phys. 62 (1978) 62


\section{Schwarzschild solution and its maximal extension}\label{10schwarzschildSec}

Imposing a surface orthogonal time-like Killing vector field and
$SO(3)$ symmetry of the space sections one reaches the form
$$
ds^2 = -e^\nu  c^2 dt^2+ e^\lambda dr^2+r^2(d\theta^2+\sin^2\theta
d\phi^2)
$$
from which
$$
G_{rr} = \frac{1-e^\lambda +r \nu'}{r^2};~~~~G_{tt} =
\frac{e^{-\lambda+\nu}}{r^2}(-1+e^\lambda + r \lambda'). 
$$
$G_{rr}=G_{tt}=0$ give
$$
\nu'+\lambda'=0;~~~~r \lambda'  = 1-e^\lambda.
$$
The second solved gives
$$
e^\lambda = \frac{1}{1+\frac{1}{kr}}
$$
while the first gives
$$
e^\nu = \alpha e^{-\lambda} = \alpha (1+\frac{1}{kr}).
$$
Normalizing time so as to have asymptotically Minkowski metric  and
comparing with the already derived 
$$
g_{00} = -1-\frac{2\phi}{c^2}
$$
for weak fields, one has
$$
\frac{1}{k} = -\frac{2GM}{c^2}.
$$
Then
$$
ds^2 =-(1-\frac{2M G}{c^2r})c^2 dt^2+\frac{1}{1-\frac{2MG}{c^2r}}dr^2
+r^2 d\Omega^2. 
$$
The length $r_s=\frac{2GM}{c^2}$ is the Schwarzschild radius. For the sun 
$r_s= 3~km$. We recall that precession of Mercury is $43''$ per
century and the deflection of grazing light $1.75''$. 
All general relativistic effect are of order (Schwarzschild~radius)/radius.

The following remarkable Birkoff theorem exists: All solution with an
$SO(3)$ isometry in absence of matter are locally equivalent to the
Schwarzschild metric [1]; no stationarity assumption is made.

Despite the metric is singular at $r=\frac{2GM}{c^2}$ all invariant
polynomials remain finite at the Schwarzschild singularity. Instead
computation of the square of the Riemann tensor gives
$$
R_{\mu\nu\lambda\rho} R^{\mu\nu\lambda\rho}=\frac{6 r_s^2}{r^6}
$$ 
and as such the sub-manifold $r=0$ is a singular sub-manifold.

To eliminate the Schwarzschild singularity one goes over to the
Kruskal coordinates. In the following by $M$ we shall understand
$GM/c^2$ and by $t$ we shall understand $ct$. One starts by going over
to light-like coordinates. The equation for radial light ray 
$$
\dot t^2 =\left(\frac{\dot r}{1-\frac{2M}{r}}\right)^2 
$$
is solved by
$$
\pm t = r+2M\ln(\frac{r}{2M}-1) + {\rm const} ~\equiv  r_*+{\rm const}~.
$$
We see that infinite time $t$ is needed for a light-pulse to reach the
horizon. With 
$$
u=t-r_*;~~~~v=t+r_*
$$
the metric becomes
$$
ds^2 = -(1-\frac{2M}{r})~du ~dv+r^2 d\Omega^2
$$
$r$ being now defined by
$$
e^{\frac{r}{2M}}(\frac{r}{2M}-1) = e^{\frac{v-u}{4M}}.
$$
Go over to
$$
V= e^{\frac{v}{4M}}>0;~~~~U= -e^{-\frac{u}{4M}}<0
$$
to have the metric
$$
ds^2 = -\frac{32 M^3}{r}e^{-\frac{r}{2M}} dUdV+r^2 d\Omega^2
$$
where
$$
e^{\frac{r}{2M}}\big(\frac{r}{2M}-1\big) = -UV
$$
and up to now $-UV>0$. The extension occurs by realizing that 
$r\geq 0$ is uniquely defined in a larger range for $-UV$ i.e. for
$-VU>-1$ being the l.h.s. monotonically increasing in $r$ from $-1$ to
$\infty$.
This remark completes the Kruskal maximal extension of the
Schwarzschild metric.

Without adding anything new, it is useful to change again coordinates
$$
X=\frac{M}{2}(V-U);~~~~T=\frac{M}{2}(V+U)
$$
the metric becomes
$$
ds^2 = 32 M\frac{e^{-\frac{r}{2M}}}{r}(-dT^2+ dX^2)+r^2 d\Omega^2
$$
where
$$
-UV = \frac{1}{M^2}(X^2-T^2) 
$$
is restricted to be $\geq -1$. The remaining is a non physical region
as $r=0$ is a real singularity as there time geodesics are incomplete
and the square of the Riemann tensor diverges. For a discussion of
regions I, II, III, IV see [1], [2]. 
The relation to the original coordinates is 
$$
\frac{t}{2M} = 2~ {\rm arctanh}\frac{T}{X}
$$
$$
e^{\frac{r}{2M}}(\frac{r}{2M}-1) =\frac{X^2-T^2}{M^2}.
$$

\bigskip

References

\smallskip

[1] [HawkingEllis] Chap. 5

\smallskip

[2] [Wald] Chap. 6

\section{Proper acceleration of stationary observers}
\hskip 2cm {\bf \Large in Schwarzschild
metric and surface gravity}

\bigskip

The Minkowskian formula for the proper acceleration can be written as
$$
a^\mu = \frac{d^2 x^\mu}{d\tau^2} = \frac{d u^\mu}{d\tau}
=u^\lambda\frac{\partial u^\mu}{\partial x^\lambda}  
$$
is covariantly generalized for a field of four velocities $u^\mu$ to 
$$
a^\mu = u^\lambda \nabla_\lambda{u^\mu}
$$
with norm
$$
g^2 = a^\mu a_\mu \geq 0
$$
being $a^\mu$ space-like.
If the metric possesses a time like Killing vector we can ask for the
proper acceleration of observers moving along the integral lines of
the Killing vector field. If we set $\xi^\mu\xi_\mu = -V^2$ we have
$$
u^\mu = \frac{\xi^\mu}{V}
$$
and the covariant acceleration is given by
$$
a_\nu = u^\lambda\nabla_\lambda
u_\nu=\frac{\xi^\lambda}{V^2}\nabla_\lambda \xi_\nu
+\frac{\xi^\lambda}{V}\xi_\nu \nabla_\lambda\frac{1}{V} 
=\frac{\xi^\lambda}{V^2}\nabla_\lambda \xi_\nu=-
\frac{\xi^\lambda}{V^2}\nabla_\nu \xi_\lambda=      
\nabla_\nu
\ln V 
$$
where we used the fact that the norm of the Killing vector is constant
along the integral lines of the Killing vector field.

For Schwarzschild we have
$$
\xi^\mu\xi_\mu = - (1-\frac{2M}{r})
$$
from which
$$
g = \frac{1}{(1-\frac{2M}{r})^{1/2}} \frac{M}{r^2}
$$
which diverges on the horizon.

If we multiply the proper acceleration $g$ by the norm of the Killing
vector normalized to $1$ at space infinity we obtain
\begin{equation}\label{10surfacegravity}
\sqrt{-\xi^\mu \xi_\mu} ~g = \frac{M}{r^2}
\end{equation}
Eq.(\ref{10surfacegravity}) computed at the horizon is a finite quantity
and is called the surface gravity
\begin{equation}\label{10surfacegravity2}
\kappa = \frac{M}{r_s^2} = \frac{1}{4 M}~.
\end{equation}
An interpretation of $\kappa$ is the following [1]
consider an 
inextensible massless rope of length $L$
$$
L=\int_{r_1}^{r_2} \sqrt{g_{rr}} ~dr = \int_{r_1}^{r_2}
\frac{1}{\sqrt{1-\frac{2M}{r}}} ~dr 
$$
which holds the unit mass placed at $r_1$ against falling to decreasing
values of $r$. If the observer at $r_2$ pulls the rope by $dl_2$, the
end of the rope at $r_1$ moves by $dl_1$ given by
$$
\frac{dl_1}{\sqrt{1-\frac{2M}{r_1}}}=\frac{dl_2}{\sqrt{1-\frac{2M}{r_2}}}~.
$$
The force the observer $O_1$ has to apply in order to keep the
mass at fixed $r=r_1$ is $g(r_1)$. If we equate the work done by $O_2$ by the
energy gained by $O_1$ we have
$$
f_2 = g \frac{dl_1}{dl_2} = g
\frac{\sqrt{1-\frac{2M}{r_1}}}{\sqrt{1-\frac{2M}{r_2}}}~. 
$$
Taking $r_2$ to infinity and $r_1=r_s$ we have the surface gravity.

\bigskip

References

\smallskip

[1] [Wald] Chap. 6 problem 6.4.b

\section{The accelerated detector}\label{10acceldetect}

The detector is given by a point particle with discrete energy levels
$E_0, E_1, \dots$ coupled to the scalar field $\phi(x)$ by the
interaction $c(\tau)~m(\tau)~\phi(x(\tau))$. $\tau$ is the proper time
while $c(\tau)$ is an exposure function which is equal to 1 for 
$-\Omega/2<\tau<\Omega/2$ and goes to zero smoothly outside that interval.

Using perturbation theory and working in
the interaction picture we have for the transition amplitude
from $E_0$ to $E$ and for the field from the Minkowski vacuum
$|0_M\rangle$ to the state $|\psi\rangle$

$$ 
A = i \int_{-\infty}^{\infty} c(\tau) d\tau \langle
E|m(\tau)|E_0\rangle \langle \psi|\phi(x(\tau))|O_M\rangle 
$$
$$
=i \int_{-\infty}^{\infty} c(\tau) d\tau \langle
E|m(0)|E_0\rangle e^{i(E-E_0)\tau}\langle \psi|\phi(x(\tau))|O_M\rangle~. 
$$

The total transition probability from $E_0$ to $E$, found by summing
$\bar A A$ over all $|\psi\rangle$, is
$$
P =|\langle E|m(0)|E_0\rangle|^2
\int_{-\infty}^{\infty} d\tau~c(\tau) \int_{-\infty}^{\infty} d\tau'
c(\tau')
e^{-i(E-E_0)(\tau-\tau')}W(x(\tau),x(\tau'))
$$
being $W(x,x')$ the Wightman function
$$
W(x,x')=\langle 0_M|\phi(x)\phi(x')|0_M\rangle~.
$$
Due to the spectral condition $E_n\geq 0$ we have
$$
W(x,x')=\sum_n\langle 0_M|\phi(x)|n\rangle\langle
n|\phi(x')|0_M\rangle=\sum_n e^{-i E_n(t-t')}
\langle 0_M|\phi({\bf x},0)|n\rangle\langle
n|\phi({\bf x}',0)|0_M\rangle= 
$$
$$
= \lim_{\varepsilon \rightarrow +0}\sum_n e^{-i E_n(t-t'-i\varepsilon)}
\langle 0_M|\phi({\bf x},0)|n\rangle\langle
n|\phi({\bf x}',0)|0_M\rangle.
$$
In the massless case one has
\begin{equation}\label{10twopointmassless}
W(x,x')= \frac{1}{4\pi^2}\frac{1}{({\bf x}-{\bf x}')^2 -
(t-t'-i\varepsilon)^2}~. 
\end{equation}
For the inertial trajectory one has 
$$
P = - |\langle E|m(0)|E_0\rangle|^2 \int_{-\infty}^{\infty} c(\tau') d\tau'
\int_{-\infty}^{\infty} c(\tau) ~d\tau e^{-i(E-E_0)(\tau-\tau')}\frac{1}{4\pi^2
(\tau-\tau'-i\varepsilon)^2}~. 
$$
The integral in $d\tau$ for $\Omega\rightarrow \infty$ converges to a
finite limit and, as $E-E_0>0$, such a limit can be computed by
closing the integration contour in $\tau$ in the lower half plane. 
Due to the absence of poles in the lower half plane, the result is
zero. This is an expected result.

We consider now a uniformly accelerated motion (hyperbolic motion)
written as
$$
t=\frac{1}{g}\sinh g\tau,~~~~~~~~
x=\frac{1}{g}\cosh g\tau
$$
where $\tau$ is the proper time and $a_\mu a^\mu = g^2$. We find
$$
({\bf x}-{\bf x}')^2 -(t-t'-i\varepsilon)^2= -\frac{4}{g^2}\sinh^2(
g\frac{\tau-\tau'-i\varepsilon}{2})~.
$$
Now use the formula
$$
\frac{1}{\sinh^2 \pi
y}=\frac{1}{\pi^2}\sum_{k=-\infty}^{\infty}\frac{1}{(y+ik)^2} 
$$
which gives
$$
-\frac{g^2}{4\sinh^2( g\Delta\tau/2 -i\varepsilon)} =
-\sum_{k=-\infty}^{\infty}\frac{1}{(\Delta\tau -i\varepsilon +2i\pi k/g)^2}~.
$$
Again for $\Omega\rightarrow \infty$ the integral in $d\tau$
converges to a finite limit and such a limit can be computed
by closing the integration contour in the lower half plane. 
Only $k=1,2\dots$ contribute giving the result
\begin{equation}\label{10acceleratedprobability}
P = \int c(\tau') d\tau' \frac{1}{2\pi} |\langle E|m(0)| E_0\rangle
|^2 \frac{(E-E_0)}{e^{2\pi(E-E_0)/g}-1}~.
\end{equation}
The factor $\int c(\tau')d\tau'$ behaves like $\Omega$ for $\Omega
\rightarrow \infty$ which is the effective time of
exposure of the detector. 

Eq.(\ref{10acceleratedprobability}) shows that the
detector behaves as immersed in a thermal radiation of temperature
$T=g/(2\pi)$ where we ascribe the prefactor to the 
particular type of coupling of our detector to the field $\phi$.
Taking into account $\hbar$ and $c$ (which were put equal to $1$) we
have

$$
\frac{\hbar \omega_0}{k_B T}=\frac{2\pi c ~\omega_0}{g}
$$
i.e.
\begin{equation}\label{10accelerationtemperature}
k_B T = \frac{g\hbar}{2\pi c}
\end{equation}
giving for $g=10 m/sec^2$ the temperature  $T = 3.84 ~~10^{-20} K$.


\section{Elements of causal structure of space time}

We give below the definitions of some special sets 
which are related to the causal structure of space-time.
Some of them will be used in the following.

\bigskip

Future Cauchy development $D^+(S)$ of a closed $\it set$ $S$: the
points such that any backward directed inextensible non space like
trajectory starting from them hits $S$.

Similarly one defines the past Cauchy development $D^-(S)$ of a closed
set $S$.

\bigskip
Partial  Cauchy  surface:  a   space-like  $\it  hypersurface$  with  no
inextensible   non-spacelike   curve   intersecting   it   more   than
once. 

\bigskip
Global Cauchy surface: a partial Cauchy surface with the property 
$D^+(S)\cup D^-(S)=M$ [1] 

\bigskip
A space which admits a global Cauchy surface is called globally
hyperbolic.

\bigskip
Chronological future of an event $I^+(p)$: the events which can be
reached starting from $p$ with a time like curve.

\bigskip
Causal future of an event $J^+(p)$: the events which can be
reached starting from $p$ with a non space like curve.

\bigskip
Similarly one defines the chronological and causal past.

\bigskip Given a world line $x(\lambda)$, i.e. a time like curve we
define ``Future event horizon'' $H^+(x(\lambda))$ the boundary of the
past of the world line, being the past of the word line the union of
the past light cones of the events belonging to the world line. 
I.e. $H^+(x(\lambda))= {\rm boundary} (\cup_\lambda J^-(x(\lambda)))$
It is the boundary of the events by which the observer can be
influenced.

\bigskip
Past event horizon $H^-(x(\lambda)$ for a world line (observer) is the
boundary of the future of the world line $H^-(x(\lambda))= {\rm
boundary} (\cup_\lambda J^+(x(\lambda)))$ . It is the boundary of the
events which the observer can influence.

\bigskip
Future null infinity ${\cal F}^+$: (qualitatively) the endpoints of the
light-like geodesics which reach the asymptotic region [2]

\bigskip
Event horizon $H$: the boundary of the boundary of the causal past of
the future null infinity $H= {\rm boundary} (J^-({\cal F}^+))$

\bigskip Space which is future asymptotically predictable from a
partial Cauchy surface $S$: if ${\cal F}^+ \subset
\overline{D^+(S)}$. Similar definition for a space which is past
asymptotically predictable.

\bigskip

\bigskip

Examples

1. Spacelike hyperboloid in Minkowski is a partial but not global
Cauchy surface.

2. Anti de Sitter space admits a foliation in spacelike surfaces but
has no global Cauchy surface as given any space like surface there are
geodesics which never intersect it [3]

3. Horizons for a stationary Schwarzschild observer.

4. Horizons for a stationary Rindler observer.
 
\bigskip

The favorable situation is the following: 

\smallskip

1) Asymptotically, for large negative and positive times there exist,
time like Killing vector fields. This will allow a particle
interpretation of the quantum fields.

2) There exists a partial Cauchy surface $S$ for which the space is
future and past asymptotically predictable.

This will allow to perform predictions both at the classical
and quantum level.

At large negative times we can expand the field
as
$$
\phi = \sum_i {\bf a}_if_i+{\bf a}_i^+\bar f_i 
$$  
and at large positive times we can expand the field
as
$$
\phi = \sum_i {\bf b}_i p_i +{\bf b_i}^+~ \bar p_i~.
$$  

We propagate back in time the solutions $p_i$ to the partial Cauchy
surface $S$ and we propagate forward the solutions $f_j$ and $\bar
f_j$ to the same partial Cauchy surface $S$. Due to the completeness
of the $f_j,~\bar f_j$ we can express $p_i$ as a superposition of $f_j$ and
$\bar f_j$ 
$$ 
p_i = \sum_j \alpha_{ij}f_j+\beta_{ij}\bar f_j $$ and
due to the linearity of the problem such relation holds for all times
with constant coefficients $\alpha$ and $\beta$. By equating now the
two expressions of the same field $\phi$ we can extract the relation between
the operators ${\bf a}, {\bf a}^+$ and ${\bf b}, {\bf b}^+$.

\bigskip

References

\smallskip

[1] [HawkingEllis] Chap. 6 p.201 and following

\smallskip

[2] [HawkingEllis] Chap. 9 p.310

\smallskip

[3] [HawkingEllis] Chap. 6 p.133

\section{Rindler metric}

Perform on Minkowski the transformation
$$
x= z\cosh (a \eta);~~~t=z \sinh (a \eta)
$$
where $a$ has the dimension of ${\rm length}^{-1}$ and $\eta$ the
dimension of a length.
$$
0<z=(x^2-t^2)^{1/2};~~~~-\infty<a \eta={\rm arctanh}(\frac{t}{x})<\infty
$$
$$
ds^2 = - z^2 a^2 d\eta^2+dz^2 
$$
This space is not geodesically complete.
With $z=a^{-1}e^{a \xi}$ can be taken to a form conformal to Minkowski.
$$
ds^2 = e^{2a\xi}( d\xi^2-d\eta^2)
$$

The motion with constant (proper) acceleration $a^\mu a_\mu =g^2$ in
Minkowski space is given by
$$
\frac{du^\mu}{d\tau} u_\mu =a^\mu u_\mu=0.
$$
where for the four velocity we have $u^\mu
u_\mu=-1$.

In the rest frame $u^i=0$ and thus $a^0=0$ and then $a^\mu a_\mu = a^i
a_i = g^2={\rm const}$.
Then we have
$$
-u^0a^0+u^1a^1=0;~~~~-u^0u^0+u^1u^1=-1;~~~~
-a^0a^0+a^1a^1=g^2.
$$
From the first equation $a^0=f u^1$, $a^1 = fu^0$. Substitute in  
the third equation to obtain $-f^2 (u^1)^2+f^2 (u^0)^2 = f^2=g^2$ 
and thus, choosing $f=g$
$$
a^0 = g u^1 = g\frac{dx^1}{d\tau},~~~~a^1 = g u^0 = g\frac{dx^0}{d\tau}~.
$$
The solution with $u^0>0$ is
$$
u^0 = \cosh g(\tau-\tau_0),~~~~ u^1 = \sinh g(\tau-\tau_0)~. 
$$
Then properly normalizing the origin of space and time
$$
x^0 = \frac{1}{g}\sinh g(\tau-\tau_0),~~~~x^1 = \frac{1}{g}\cosh
g(\tau-\tau_0)~.
$$
This shows that the motion of Rindler observers (particles) with $z={\rm
const.}$ is a uniformly accelerated motion with acceleration $g=1/z$. 

\bigskip

``Rigidity'' of Rindler space: The Rindler space is locally rigid. By
this we mean that the distance of two nearby
points $z_2={\rm const}$, $z_1={\rm const}$, $\Delta z = z_2-z_1$ as
measured by a fixed observer $x={\rm const}$,  contracts according 
the Lorentz contraction law. In fact the speed of the point is given by
$$
V=\tanh (a\eta)
$$ 
while measuring the distance in the rest frame means solving the
equations
$$
\Delta x = \Delta z \cosh (a\eta) + z \sinh (a\eta) \Delta (a\eta)
$$
$$
0= \Delta z \sinh (a\eta) + z \cosh (a\eta) \Delta (a\eta)
$$
and thus
$$
\Delta x = \Delta z (\cosh (a\eta) - \frac{\sinh^2(a\eta)}{\cosh
(a\eta)}) = \Delta z 
\frac{1}{\cosh (a\eta)} = \Delta z (1-(\tanh (a\eta))^2)^{1/2} = \Delta z
(1-V^2)^{1/2}.
$$

We have something like an accelerating skyscraper; at $\eta=0$ we have
$\Delta x = z_2-z_1 =\Delta z$  and the skyscraper is at rest. At
different times one measures the correct Lorentz contraction. The
skyscraper cannot extend below indefinitely, i.e. for negative $z$, 
as the acceleration is
diverging for $z=0$. The acceleration of the $n-$th floor is $1/n$. 
The time like Killing vector field for the Rindler metric is
$\frac{\partial}{\partial \eta}$ whose square norm is $-V^2 =
-a^2 z^2$. Applying the formula for an observer which follows the Killing
integral lines, given by $z={\rm const.}$ we have
$$
a_\eta =0,~~~~a_z = \nabla_z \log (a z) = \frac{1}{z}
$$ 
and we re-obtain the expression for the acceleration.

\section{Eigenfunctions of K in the Rindler metric}

From

$$
ds^2 = dz^2-z^2 a^2 d\eta^2
$$
we have in the non zero mass case
$$
K = - a^2 z^2\frac{\partial^2}{\partial z^2}-a^2 z\frac{\partial}{\partial
z} +m^2 a^2 z^2 =
-a^2\left(z\frac{\partial}{\partial z}\right)^2 +m^2a^2z^2= -a^2
\left(\frac{\partial}{\partial \ln (a z)}\right)^2 +m^2 a^2 z^2. 
$$
while in the $\xi,\eta$ coordinates takes the form
$$
K = -(\frac{\partial}{\partial \xi})^2 + m^2 e^{2 a \xi}.
$$
For zero mass the right-moving field and left-moving field in (1+1)
dimensions are local independent quantum fields. Moreover for zero
mass the K.G. equation is conformal invariant and in the $\xi,~\eta$
coordinates takes the form
$$
\left((\frac{\partial}{\partial \xi})^2-(\frac{\partial}{\partial
\eta})^2\right) \phi = 0. 
$$ 

The right moving solutions are
$$
e^{\pm i\omega(\xi-\eta)}=e^{\pm i\omega(a^{-1}\log(az)-\eta)}=
e^{\pm i\omega ~a^{-1}\log(a(x-t))}
$$ 
and thus the right-moving field is expanded as
$$
\phi(\xi,\eta) = \int
d\omega\left(\frac{e^{i\omega(\xi-\eta)}}{\sqrt{2 \omega}}{\bf b}_\omega +
\frac{e^{-i\omega(\xi-\eta)}}{\sqrt{2 \omega}}{\bf b}^+_\omega \right)
$$
or using the $z,\eta$ coordinates
$$
\phi(z,\eta) = \int
d\omega\left(\frac{e^{i\omega(a^{-1}\log (az)-\eta)}}{\sqrt{2
\omega}}{\bf b}_\omega + 
\frac{e^{-i\omega(a^{-1}\log(az)-\eta)}}{\sqrt{2 \omega}}{\bf b}^+_\omega \right).
$$
In the $z,\eta$ coordinates the metric is $(-g^{00})\sqrt{-g}dz =
a^{-1}z^{-2}~ z ~dz = a^{-1}d\ln(a z)$ and 
$$
\frac{a^{-1}}{2\pi}\int e^{\pm i\omega(a^{-1}\log(az)-\eta)}e^{\mp
i\omega'(a^{-1}\log(az)-\eta)}d\ln (az) = \delta(\omega-\omega').  
$$
The form 
$$
e^{\pm i \omega ~a^{-1}\ln(a(x-t))}
$$ 
clarifies the nature of the solution; we have infinite oscillations
when we approach the horizon.

\section{The Bogoliubov transformation}\label{10bogoliubov}

We have already pointed out that in presence of a partial 
Cauchy surface, as  at large negative times the
solutions $f_\omega$, $\bar f_\omega$ are a complete set, it must be
possible to express the outgoing solutions $p_\omega$ in terms of
$f_\omega$ and $\bar 
f_\omega$. I.e. using the discrete notation
$$
p_i = \sum_j\alpha_{ij}f_j +\beta_{ij}\bar f_j 
$$
and thus
$$
{\bf a}_j = \sum_i {\bf b}_i\alpha_{ij}+
{\bf b}^+_i \bar\beta_{ij} \equiv ( {\bf b} A + {\bf b}^+ \bar B )_j
$$
$$
{\bf a}^+_j = \sum_i {\bf b}^+_i\bar\alpha_{ij}+
{\bf b}_i \beta_{ij} \equiv ( {\bf b}  B+{\bf b}^+\bar A)_j 
$$
in matrix notation with
$$
A_{ij} = \alpha_{ij}; ~~~~B_{ij}= \beta_{ij}.
$$
The ${\bf b},{\bf b}^+$ will obey the commutation relations
$$
[{\bf b}_i,{\bf b}^+_j] =\delta_{ij}.
$$
Then we have
$$
\delta_{jk}=[{\bf a}_j,{\bf a}^+_k] = (A^T \bar A - B^+ B)_{jk}
$$
i.e.
\begin{equation}\label{canonical}
A^T \bar A - B^+ B=I
\end{equation}
and from $[{\bf a}_j,{\bf a}_k]=0$ we have
$$
A^T \bar B -B^+ A = 0~.
$$
Thus using as illustration a finite dimensional space
$${\cal A}=
\begin{pmatrix}
A^T&B^+\\
B^T& A^+
\end{pmatrix}
\in U(N,N)
$$
i.e.
$$
{\cal A} S {\cal A}^+ = S~~~~
{\rm with}~~~~S=
\begin{pmatrix}
I&0\\
0&-I
\end{pmatrix}.
$$
It follows that the inverse is
\begin{equation}\label{10inverse}
{\cal A}^{-1}=
\begin{pmatrix}
\bar A &-\bar B\\
-B & A
\end{pmatrix}
\in U(N,N)
\end{equation}
from which in addition to Eq.(\ref{canonical}) we have 
$$
\bar A A^T-\bar B B^T = I
$$
giving rise respectively to the sum rules
$$
\sum_i |\alpha_{ij}|^2-|\beta_{ij}|^2 =1
$$
and
\begin{equation}\label{10sumrule}
\sum_i |\alpha_{ji}|^2-|\beta_{ji}|^2=1~.
\end{equation}
If the state $\Phi$ is such that 
$$
{\bf a}_i\Phi =0
$$
i.e. no incoming particle, we have for the mean value of the number 
of outgoing particles
$$
(\Phi, {\bf b}^+_i {\bf b}_i\Phi) = \sum_j |\beta_{ij}|^2~.
$$


\section{Quantum field theory in the Rindler wedge}

In this section we shall work out the Bogoliubov transformation for
the passage from Minkowski to Rindler space for a massless field.
The advantage of treating the massless field is that all computations
can be performed exactly.

In the massless case the scalar field splits in a Lorentz invariant
way into right-moving and left-moving fields.
$\phi(x) = \phi_R(x)+\phi_L(x)$. In fact with 
$$
\phi_R(x) = \int_0^\infty (e^{ik(x-t)} \frac{{\bf a}_R(k)}{\sqrt{2\omega}}+
e^{-ik(x-t)} \frac{{\bf a}^+_R(k)}{\sqrt{2\omega}})dk
$$
$$
\phi_L(x) = \int_{-\infty}^0 (e^{ik(x+t)} \frac{{\bf a}_L(k)}{\sqrt{2\omega}}+
e^{-ik(x+t)} \frac{{\bf a}^+_L(k)}{\sqrt{2\omega}})dk
$$
with $\omega=|k|$ we see that under the Lorentz transformation
$$
x=x' \cosh\alpha-t' \sinh\alpha
$$
$$
t = - x' \sinh\alpha+t' \cosh\alpha
$$
the two fields transform independently. E.g.
\begin{eqnarray}
\phi_R(x) = \int_0^\infty (e^{ik e^\alpha (x'-t')} \frac{{\bf a}_R(k)}{\sqrt{2\omega}}+
e^{-ik e^\alpha (x'-t')} \frac{{\bf a}^+_R(k)}{\sqrt{2\omega}})dk\\
= \int_0^\infty (e^{ik'(x'-t')} \frac{{\bf a}'_R(k')}{\sqrt{2\omega}}+
e^{-ik'(x'-t')} \frac{{\bf a}'^+_R(k')}{\sqrt{2\omega'}})dk'
\end{eqnarray}
with ${\bf a}'_R(k') = {\bf a}_R(e^{-\alpha}k')$ and similarly for 
the left-moving field. 
Thus in the massless case one can treat the two fields
independently. 
It corresponds to the fact that in two dimensions for a massless
particle it has an absolute meaning to say that the particle is moving in
the right of left direction.

Here to stay in touch with the general formalism we shall not perform
this splitting and work with the full field $\phi$. We shall exploit
only the Rindler wedge $z\geq 0$.

We can extract ${\bf b}^+(k)$ and ${\bf b}(k)$ from
$\phi(z,\eta)$ by computing the space Fourier transform of
$\phi(z,0)$ and $\dot\phi(z,0)$ where the dot stays for the
derivative w.r.t. the Rindler time $\eta$. 
We use the notation $\xi = a^{-1}\log(az)$ and $\omega =|k|$. We have
$$
\phi(z,0) = \int d k \left( 
e^{ik\xi} \frac{{\bf b}(k)}{\sqrt{2\omega}} +
e^{-ik\xi}\frac{{\bf b}^+(k)}{\sqrt{2\omega}} \right)
$$
$$
\dot \phi(z,0) = \int d k \left( 
-i\sqrt{\frac{\omega}{2}} e^{ik\xi} {\bf b}(k) +
i\sqrt{\frac{\omega}{2}}
e^{-ik\xi}{\bf b}^+(k) \right) ~.
$$
Thus
\begin{equation}\label{10bdagger}
{\bf b}^+(k) = \frac{1}{2\pi}\left(
\sqrt{\frac{\omega}{2}} \int\phi(\xi,0)e^{ik\xi}d\xi-i
\sqrt{\frac{1}{2\omega}} \int\dot\phi(\xi,0)e^{ik\xi}d\xi\right)  
\end{equation}
which result can be obtained also from the scalar product 
(\ref{fphiscalarproduct}). 
${\bf b}(k)$ is obtained by hermitean conjugation.

On the Rindler wedge the field $\phi$ and the standard Minkowski field
$$
\phi_M(x,t) = \int
dk' \left( \frac{{\bf a}(k')}{\sqrt{2\omega'}}e^{i (k'x -\omega't)}  +
\frac{{\bf a}^+(k')}{\sqrt{2\omega'}} e^{-i (k'x -\omega't)}\right)
$$
coincide and the Minkowski vacuum is characterized by 
${\bf a}(k)|0_M\rangle =0$. 
$\dot \phi_M(x,0)$ is computed by using
$$
\left .\frac{\partial}{\partial \eta}\right|_{\eta=0} = 
\left. az\frac{\partial}{\partial t}\right|_{t=0}~~~~{\rm and}~~~~
x=z \cosh a\eta 
$$
where $\partial/ {\partial} \eta$ means the derivative at
fixed $\xi$ and $\partial/ {\partial} t$ means the derivative at
fixed $x$.

Keeping in mind that at $\eta=0$ we have $x=z$ and inserting into 
Eq.(\ref{10bdagger}) we have
\begin{equation}\label{10bdagger2}
{\bf b}^+(k) = \frac{\sqrt{\omega}}{4\pi a}\int_{-\infty}^\infty d\log(az)
e^{i\frac{k}{a} \log
az}\int(\frac{{\bf a}(k')}{\sqrt{\omega'}}e^{ik'z}+\frac{{\bf a}^+(k')}
{\sqrt{\omega'}} e^{-ik'z})dk'
\end{equation}
$$
-\frac{i}{4\pi a \sqrt{\omega}}\int_{-\infty}^\infty a z ~d\log(az)
e^{i\frac{k}{a} \log
az}\int(-i {\bf a}(k')\sqrt{\omega'}e^{ik'z}+i {\bf a}^+(k')
\sqrt{\omega'} e^{-ik'z})dk'.  
$$
Thus we need the integrals
$$
I)~~~~ \int_0^\infty e^{i\kappa \log\zeta} ~e^{i\kappa' \zeta} ~\zeta^{-p}
~d\zeta; ~~~~\kappa'>0 
$$

$$
II)~~~~ \int_0^\infty e^{i\kappa \log\zeta} ~e^{-i\kappa' \zeta} ~\zeta^{-p}
~d\zeta; ~~~~\kappa'>0   
$$
where we have defined $\kappa = k/a$ and $\zeta = az$ and $p$ can take
the values $0$ or $1$. 
Integrals $I$ and $II$ are computed by rotation of the integration in the
complex plane; for $I$ one uses $\zeta = e^{i\frac{\pi}{2}}\rho$  and
for $II$ one uses $\zeta = e^{-i\frac{\pi}{2}}\rho$ to obtain
$$
I)~~~~ \int_0^\infty e^{i\kappa \log\zeta} ~e^{i\kappa' \zeta} ~\zeta^{-p}
~d\zeta = e^{i\frac{\pi}{2}(1-p)} e^{-\frac{\pi\kappa}{2}}
(\kappa')^{-1-i\kappa+p}~\Gamma(i\kappa+1-p);~~~~\kappa'>0   
$$
$$
II)~~~~ \int_0^\infty e^{i\kappa \log\zeta} ~e^{-i\kappa' \zeta} ~\zeta^{-p}
~d\zeta=e^{i\frac{\pi}{2}(p-1)} e^{\frac{\pi\kappa}{2}}
(\kappa')^{-1-i\kappa+p}~\Gamma(i\kappa+1-p);~~~~\kappa'>0~.
$$
Substituting now in Eq.(\ref{10bdagger2}) we obtain
\begin{equation}\label{10bogol+}
\frac{{\bf b}^+(k>0)}{\sqrt{2\omega}} =\frac{\Gamma(\frac{i\omega}{a})}{2\pi a}
\left(e^{\frac{\pi\omega}{2a}}\int_0^\infty\frac{{\bf a}^+(k')}  
{\sqrt{2\omega'}} \bigg(\frac{k'}{a}\bigg)^{-i\frac{\omega}{a}}dk'
+e^{-\frac{\pi\omega}{2a}}\int_0^\infty\frac{{\bf a}(k')} 
{\sqrt{2\omega'}} \bigg(\frac{k'}{a}\bigg)^{-i\frac{\omega}{a}}dk'\right)  
\end{equation}

\begin{equation}\label{10bogol-}
\frac{{\bf b}^+(k<0)}{\sqrt{2\omega}}
=\frac{\Gamma(-\frac{i\omega}{a})}
{2\pi a}
\left(e^{\frac{\pi\omega}{2a}}\int_{-\infty}^0\frac{{\bf a}^+(k')}  
{\sqrt{2\omega'}} \bigg(-\frac{k'}{a}\bigg)^{i\frac{\omega}{a}}dk'
+e^{-\frac{\pi\omega}{2a}}\int_{-\infty}^0\frac{{\bf a}(k')} 
{\sqrt{2\omega'}} \bigg(-\frac{k'}{a}\bigg)^{i\frac{\omega}{a}} dk'\right)  
\end{equation}
and ${\bf b}(k)$ is obtained by hermitean conjugation.

The two equations (\ref{10bogol+})(\ref{10bogol-}) show explicitly
the splitting of the field in right and left moving parts.

The mean value on the Minkowski vacuum, 
defined by ${\bf a}(k)|O_M\rangle=0$, of the number operator in the 
Rindler space ${\bf b}^+(k){\bf b}(k)$, $k\geq 0$, is given by
$$
\frac{1}{2 \omega}\langle 0_M|{\bf b}^+(k){\bf b}(k)|0_M\rangle = 
\frac{1}{4\pi^2
a^2}\Gamma(\frac{i\omega}{a})
\Gamma(-\frac{i\omega}{a})e^{-\frac{\pi\omega}{a}}
\int_0^\infty\frac{dk'}{2\omega'}~.  
$$ 
Recalling now the relation
\begin{equation}\label{10sin}
\Gamma(z)\Gamma(-z) = -\frac{\pi}{z \sin(\pi z)}
\end{equation}
we obtain
\begin{equation}\label{10unruh}
\langle 0_M|{\bf b}^+(k){\bf b}(k)|0_M\rangle = \frac{1}{\pi
a}\frac{1}{e^{2\pi \omega/a}-1}\int_0^\infty\frac{dk'}{2\omega'}~.  
\end{equation}
The UV divergence is due to the continuum normalization of the vectors
and can be avoided by working with wave packets as will be done in
section (\ref{10packets}) in full generality.

The frequency $\omega$ which appears in Eq.(\ref{10unruh}) is the
frequency measured w.r.t. the coordinate time $\eta$ as it appears in
the decomposition of the Rindler field. To relate it to the frequency
measured by a Rindler stationary observer i.e. one which moves with
$z={\rm const.}$ one has to relate the time $\eta$ with the proper
time $\tau$ of the observer
$$
\omega ~d\eta = \omega_o ~d\tau
$$  
with $d\tau = z a ~d\eta$. Thus $\omega = az~\omega_o$ and we have
\begin{equation}\label{10spectrum}
\langle 0_M|{\bf b}^+(k){\bf b}(k)|0_M\rangle = \frac{1}{\pi
a}\frac{1}{e^{2\pi \omega_o z}-1}\int_0^\infty\frac{dk'}{2\omega'}  
\end{equation}
which recalling that the acceleration of the observer is $g=1/z$ agrees
with the result of the accelerated detector
Eq.(\ref{10accelerationtemperature}).


\section{Rindler space with a reflecting wall}\label{10reflectingwall}

We shall consider in this section a simple two dimensional example [1]
which reproduces many of the features of Hawking's theory [2].  We
place a reflecting wall at a fixed distance $x=c$ in the Minkowski
coordinates.  This problem is interesting in several respects: It
gives a concrete example of a geometry which has only asymptotic
time-like Killing vector fields at the Rindler time $\eta=-\infty$ and
$\eta=+\infty$ and the reflecting wall also simulate the reflection of
the partial wave modes by the center of the collapsing star as
examined by Hawking [2]. It reproduces well the physical situation for
which the Hawking radiation is a late time effect and it shows how
such radiation is emitted at the boundary of the horizon.

In fact the relevant mathematics turns out to
be identical to that of the Schwarzschild-Kruskal case. 
Not to
overburden the notation we shall set the parameter $a$ of the previous
section with the dimension of a ${\rm length}^{-1}$ equal to $1$. Thus
e.g. $\log z$ has to be read as $\log (az)$.

Due to the presence of the reflecting wall, Lorentz invariance is
broken and thus the scalar field non longer can be
invariantly decomposed in left and right moving fields.

We will start with a field which at large negative Rindler times
describes incoming particles i.e. at large negative Rindler times
can be written, using for clarity the discrete notation, as
$$
\phi = \sum_{j} {\bf a}_j f_j+ {\bf a}^+_j \bar f_j
$$
with $f_j$ reducing at large negative Rindler times to superposition
of left-moving negative frequency mode 
$$
\frac{1}{\sqrt{2\omega}} e^{-i\omega(\eta+\ln z)}
=\frac{1}{\sqrt{2\omega}} e^{i\omega \ln(x+t)}~.
$$
Thus ${\bf a}_j$ and ${\bf a}^+_j$ are the annihilation 
and creation operator for the incoming particles. 
The same field will also be described by
$$
\phi= \sum_m {\bf b}_m p_m+{\bf b}^+_m \bar p_m+
\sum_n {\bf c}_n q_n+{\bf c}^+_n \bar q_n
$$
where at large positive Rindler times $p_m$ becomes a superposition of
right-moving negative frequency modes
$$
\frac{1}{\sqrt{2\omega_m}} e^{-i\omega_m(\eta-\ln z)}
=\frac{1}{\sqrt{2\omega_m}} e^{i\omega_m \ln(x-t)}
~~~~{\rm for}~~x>t~~~~{\rm and}~~~ 0~~~~{\rm for}~~~ x<t
$$
and thus ${\bf b}_m$ and ${\bf b}^+_m$ are the annihilation and 
creation operators
of the particles which flow to $+\infty$. $q_i$ and $\bar q_i$ 
are the modes representing the particles which cross the horizon
$x=t$. We shall not need the
explicit form of the $q_i$. We can repeat the procedure of section 
\ref{10bogoliubov}, where now the pair $p_m$, $q_n$ replaces the $p_i$
of section (\ref{10bogoliubov}),
and the translation of Eq.(\ref{10inverse}) to the present notation is
\begin{equation}\label{10bogob}
{\bf b}_m = \sum_j \bar\alpha_{mj} {\bf a}_j-\bar\beta_{mj} {\bf a}^+_j
\end{equation}
\begin{equation}
{\bf c}_n = \sum_j \bar\gamma_{nj} {\bf a}_j-\bar\eta_{nj} {\bf a}^+_j
\end{equation}
from which we can compute ${\bf b}^+_m$ and ${\bf c}^+_n$ by hermitean
conjugation. 
From the commutation relations we obtain the sum rules
\begin{equation}\label{10completeness}
\sum_j|\alpha_{mj}|^2-|\beta_{mj}|^2 = \delta_{mm}=1
\end{equation}
and
\begin{equation}
\sum_j|\gamma_{nj}|^2-|\eta_{nj}|^2 = \delta_{nn}=1~.
\end{equation}

Hawking's method [2] to compute the Bogoliubov coefficients, is to
consider the wave packet $p_m$ (which at large positive time is a
superposition of right-moving negative frequency modes) and to
propagate it backward in time. It is reflected by the wall and it
emerges as a superposition of left-moving modes with negative {\it and}
positive frequency.

As suggested by the notation it is much better to work with discrete
modes than with modes with continuum normalization. The explicit
decomposition of the field in discrete modes will be given in the next
section. Here we still use the continuum normalization which is
formally simpler even if it is more difficult to be interpreted
physically.  The resolution in wave packets in given in the next
section.
 
In order to satisfy the boundary condition on the reflecting wall $x=c$,
the solution
$$
\frac{1}{\sqrt{2\omega}}e^{i\omega\ln(x-t)}~~~~{\rm
  for}~~~~x>t~~~~{\rm and}~~~~0~~~~{\rm for}~~~~x<t
$$
has to be matched with
$$
\int_0^\infty \left(\alpha_{\omega\omega'}\frac{1}{\sqrt{2\omega'}}
e^{-i\omega'\ln(x+t)}+\beta_{\omega\omega'}\frac{1}{\sqrt{2\omega'}}
e^{i\omega'\ln(x+t)}\right)d\omega'
$$
with the condition 
\begin{equation}
\frac{1}{\sqrt{2\omega}}e^{i\omega\ln(c-t)}\theta(c-t)= \int_0^\infty
\left(\alpha_{\omega\omega'}\frac{1}{\sqrt{2\omega'}} 
e^{-i\omega'\ln(c+t)}+\beta_{\omega\omega'}\frac{1}{\sqrt{2\omega'}}
e^{i\omega'\ln(c+t)}\right)d\omega'.
\end{equation}
To find $\alpha_{\omega\omega'}$ and $\beta_{\omega\omega'}$ we need
to invert such an integral representation.
We can do it in a workable fashion
only for small $c-t$. This is the region of interest to analyze the
late time radiation as the radiation emitted from $x=c$ and time $t$
reaches the Rindler observes placed in $z$ at Rindler time  
$\displaystyle{\eta =-\frac{1}{a}\ln\frac{c-t}{z}}$.
This aspect will be further clarified in the next section
where we give the explicit resolution of the field in discrete modes
(wave packets).  

For small $c-t$ we have
$$
\frac{c-t}{2c}=\frac{2c - (t+c)}{2c}\approx \ln(2c)-\ln(t+c).
$$
Taking the logarithm and setting $\tau = \ln(c+t)$ and $\tau_0 =
\ln(2c)$ for small $c-t$ we have  
$$
\ln(c-t) \approx \tau_0 +\ln(\tau_0 - \tau). 
$$
Thus
$$
\frac{1}{\sqrt{2\omega}}e^{i\omega(\ln(\tau_0-\tau)+\tau_0)}\theta(\tau_0-\tau)
= 
\int_0^\infty \left(\alpha_{\omega\omega'}\frac{1}{\sqrt{2\omega'}}
e^{-i\omega'\tau}+\beta_{\omega\omega'}\frac{1}{\sqrt{2\omega'}}
e^{i\omega'\tau}\right)d\omega'
$$
from which inverting
$$
2\pi\frac{\beta_{\omega\omega'}}{\sqrt{2\omega'}} = 
\frac{1}{\sqrt{2\omega}}\int_{-\infty}^{\tau_0}e^{i\omega\ln(\tau_0-\tau)}
e^{i\omega\tau_0}  e^{-i\omega'\tau} d\tau ~.
$$
We recall that $\omega\geq 0$, $\omega'\geq 0$ always.

The integral is one of the integrals already computed in the previous
section. 
Again the integral is performed by rotation in the complex plane
putting this 
time $\tau =-i\rho$ and we have
$$
2\pi\frac{\beta_{\omega\omega'}}{\sqrt{2\omega'}}= i
\frac{e^{i(\omega-\omega')\tau_0}}{\sqrt{2\omega}}\int_0^\infty
(i\rho)^{i\omega} e^{-\omega'\rho}  d\rho=
i \frac{e^{i(\omega-\omega')\tau_0}}{\sqrt{2\omega}}
e^{-\frac{\pi\omega}{2}}(\omega')^{-1-i\omega}\int_0^\infty x^{i\omega}
e^{-x}  dx= 
$$
\begin{equation}\label{10beta}
=i
\frac{e^{i(\omega-\omega')\tau_0}}{\sqrt{2\omega}}
e^{-\frac{\pi\omega}{2}}(\omega')^{-1-i\omega}\Gamma(1+i\omega)~.
\end{equation}
The performed analysis of the solution $p_n$ at large negative Rindler 
times in terms of the $f_j$ gives to $\alpha(\omega,\omega')$ and 
$\beta(\omega,\omega')$ the meaning of the Bogoliubov coefficients.
Using the formula (\ref{10sin}) we have
\begin{equation}
\langle O_M|{\bf b}^+(\omega){\bf b}(\omega)|O_M\rangle =\frac{1}{\pi}
\frac{1}{e^{2\pi\omega}-1}\int_0^\infty\frac{d\omega'}{2\omega'}
\end{equation}
which agrees with the result (\ref{10unruh}). The divergence of the
integral on the r.h.s. at $0$ is an infrared divergence due to the use
of fields of zero mass. The divergence at infinity is due to the
continuum normalization of the modes and has a physical interpretation
. In the next section this problem will be solved by going over to
discrete $p_m$ modes.

\bigskip

References

\smallskip

[1] P.C.W. Davies,``Scalar particle production in Schwarzschild and
Rindler metrics'' J. Phys. A: Math. Gen. 8 (1975) 609

\smallskip

[2] S.W. Hawking,````Particle creation by black holes'',
Comm. Math. Phys. 43 (1975) 199

\smallskip

[3] N.D. Birrel and P.C.W. Davies, ``Quantum fields in curved space''
paragraphs 4.5, 8.1.

\smallskip

[4] [Wald] 14.2, 14.3.

\smallskip

[5] R.M. Wald, ``Quantum field theory in curved space-time and black hole 
thermodynamics''


\bigskip
\section{Resolution in discrete wave packets}\label{10packets}

From Eq.(\ref{10beta}) we see that the U.V. behavior of
$|\beta_{\omega\omega'}|^2$ as a function of $\omega'$ is ${\rm
const.}/\omega'$ which makes the integral (\ref{10spectrum})
logarithmically 
divergent for large $\omega'$. This has nothing to do
with the divergences of quantum field theory but it simply a
kinematical problem. The meaning of such a divergence is that the
amount of radiating energy measured by the observer with $z={\rm
const.}$ in an infinite amount of time is infinite. This is a
reflection of the fact that the Unruh-Hawking radiation is not a
transient phenomenon but a permanent one i.e. it lasts for infinite
time.

In order to understand this feature in detail, following Hawking [1]
we shall give a resolution of the quantum field $\phi$ for large
positive times (the out field) in discrete wave packets instead of a
continuous spectral resolution.

In general given a continuous complete set of vectors $|\omega\rangle$,
$0<\omega<\infty$ one can construct the discrete set 
$$
|j,n\rangle =
\frac{1}{\sqrt{\varepsilon}}\int_{j\varepsilon}^{(j+1)\varepsilon}
e^{2\pi i n\omega/\varepsilon}~|\omega\rangle~ d\omega~.
$$
It is immediately checked that
$$
\langle j',n'|j, n\rangle=\delta_{n'n}\delta_{j'j}~.
$$
In addition the discrete set $|j,n\rangle$ is complete; in
fact
\begin{eqnarray}\label{10completenesspackets}
\int\phi(\omega)|\omega\rangle d\omega&=&
\sum_j\int_{j\varepsilon}^{(j+1)\varepsilon}\phi(\omega)
|\omega\rangle d\omega= \sum_j \sum_n
\int_{j\varepsilon}^{(j+1)\varepsilon}\frac{e^{2\pi i\omega 
n/\varepsilon}}{\sqrt\varepsilon}c_{jn}|\omega\rangle d\omega\nonumber\\
&=& \sum_j \sum_n c_{jn}|j,n\rangle
\end{eqnarray}
with
$$
c_{jn} =
\frac{1}{\sqrt\varepsilon}\int_{j\varepsilon}^{(j+1)\varepsilon}e^{-2\pi 
i \omega n/\varepsilon}\phi(\omega)d\omega~.
$$
Similarly in our case we have 
$$
p_{jn}(\eta,z) =\frac{1}{\sqrt{\varepsilon}}
\int_{j\varepsilon}^{(j+1)\varepsilon}~ 
e^{2\pi in\omega/\varepsilon} p_\omega(\eta,z)~  d\omega~.
$$
To understand the meaning of such wave packet we notice that
for large positive $n$, it is peaked at the retarded time
$$
\ln(x-t)=\eta-\ln z = 2\pi n/\varepsilon
$$ 
as it is seen by the stationary phase method or from the explicit 
computation of Eq.(\ref{10betajnomega1}) below.

Thus for an observer at $z=$const, large $n$ means large coordinate
time $\eta$ and large proper time for the Rindler observer.

Then we have
$$
\int_0^\infty ({\bf b}_\omega p_\omega+{\bf b}^+_\omega \bar p_\omega)=
\sum_{jn} ({\bf b}_{jn}p_{jn}+ {\bf b}^+_{jn}\bar p_{jn})
$$
and using (\ref{10completenesspackets})
$$
{\bf b}_{jn} = \frac{1}{\sqrt{\varepsilon}}\int_{j\varepsilon}^{(j+1)\varepsilon}
e^{-2\pi i n\omega/\varepsilon}{\bf b}_\omega d\omega~,~~~~~~~~
{\bf b}^+_{jn} =  \frac{1}{\sqrt{\varepsilon}}\int_{j\varepsilon}^{(j+1)\varepsilon}
e^{2\pi i n\omega/\varepsilon}{\bf b}^+_\omega d\omega~.
$$
Starting now from the continuous version of (\ref{10bogob}) 
we used in (\ref{10bdagger2})
$$
{\bf b}_\omega = \int_0^\infty (\bar\alpha_{\omega\omega'}{\bf a}_{\omega'}-
\bar\beta_{\omega\omega'}{\bf a}^+_{\omega'})~d\omega'
$$
we have for the ${\bf b}_{jn}$
$$
{\bf b}_{jn} =
\int_0^\infty
(\bar\alpha_{jn\omega'}{\bf a}_{\omega'}-
\bar\beta_{jn\omega'}{\bf a}^+_{\omega'})d\omega'
$$
with
$$
\alpha_{jn\omega'}=\frac{1}{\sqrt{\varepsilon}}\int_{j\varepsilon}^{(j+1)\varepsilon} 
d\omega
e^{2\pi i n\omega/\varepsilon}\alpha_{\omega,\omega'}
$$
$$
\beta_{jn\omega'}=\frac{1}{\sqrt{\varepsilon}}\int_{j\varepsilon}^{(j+1)\varepsilon} 
d\omega
e^{2\pi i n\omega/\varepsilon}\beta_{\omega,\omega'}~.
$$
From Eq.(\ref{10beta})
$$
{\beta_{\omega\omega'}}=i
\frac{e^{i(\omega-\omega')\tau_0}}{2\pi\sqrt{\omega\omega'}}
e^{-\frac{\pi\omega}{2}}(\omega')^{-i\omega}\Gamma(1+i\omega)
$$
we have
$$
\beta_{jn\omega'}
=i\frac{e^{-i\omega'\tau_0}}{2\pi\sqrt{\varepsilon}\sqrt{\omega'}}
\int_{j\varepsilon}^{(j+1)\varepsilon}
d\omega e^{2\pi i n\omega/\varepsilon}
\frac{e^{i\omega\tau_0} e^{-\frac{\pi\omega}{2}}
(\omega')^{-i\omega}\Gamma(1+i\omega)}{\sqrt{\omega}}~.
$$
For a narrow frequency interval $\varepsilon$ and large $n$ (large
arrival times) we have
\begin{equation}\label{10betajnomega1}
\beta_{nj\omega'}\approx i\frac{e^{-i\omega'\tau_0}e^{-\frac{\pi\tilde\omega}{2}}}
{2\pi\sqrt{\varepsilon\tilde\omega\omega'}} \Gamma(1+i\tilde\omega)
~ e^{i\Delta(j+1/2)\varepsilon}~2~\frac{\sin\frac{\Delta\varepsilon}{2}}{\Delta}
\end{equation}

with $\tilde\omega = (j+1/2)\varepsilon$ and 
$\Delta= 2\pi n/\varepsilon -\ln \omega'+\tau_0$ . 

Thus $|\beta_{jn\omega'}|^2$ for large $n$ (large arrival
times) will be peaked at exponentially large values of $\omega'$. This
justifies the intuitive reasoning of section (\ref{10reflectingwall}) 
about the dominance in the Fourier
expansion of the small values of $c-t$. Moreover taking the square of
Eq.(\ref{10betajnomega1}) we see that the large $\omega'$ behavior is
$$
|\beta_{nj\omega'}|^2\sim\frac{1}{\omega'\ln^2\omega'} 
$$
which integrated in $d\omega'$ converges at infinity. Thus, after
curing the infrared divergence, the number
of quanta which reaches the observer in a finite time is finite.

The result is interesting in several respects. By going over to
discrete modes, we found the UV divergence in Eq.(\ref{10unruh})
was due to the unphysical continuum normalization. We found that if
the Rindler observer tunes his measuring apparatus to a certain frequency
range, he measures a steady flux of particles and that the flux of
such particles depends on their energy (frequency) according to
Planck's law. The flux is permanent. The divergence obtained in
Eq.(\ref{10unruh}) in the continuum formulation was an attempt of the
formalism to include the fact that in an infinite time the observer
detects an infinite number of particles in a given frequency interval.
 
In the next section we shall consider a more general
decomposition in wave packets.


\bigskip

References

\smallskip

[1] S.W. Hawking,``Particle creation by black holes'', Comm. Math. Phys. 43
(1975) 199, paragraphs 1, 2.

\section{Hawking radiation in 4-dimensional Schwarzschild space}

We give here the complete treatment of Hawking radiation for a field
obeying a linear equation of motion (free field) in 4-dimensional
Schwarzschild space following [1]. One decomposes the field $\phi({\bf
  x},t)$ in partial waves exploiting the rotational symmetry of the
problem. We shall be concerned only with the $l=0$ wave, the extension
to higher waves being completely similar. 
For spherical symmetry one has
$$
\frac{1}{\sqrt{-g}}\partial_\mu\sqrt{-g}g^{\mu\nu}\partial_\nu =
\frac{1}{r^2}~ \partial_r~ r^2(1-\frac{r_s}{r})\partial_r
-(1-\frac{r_s}{r})^{-1}\partial_t^2 ~.
$$
We shall build the most general solution of the K.G. equation by
superposing solutions periodic in time. It is preferable to write
$\phi = \frac{\psi}{r} e^{\pm i\omega t}$ and go over to the new coordinates
$r_* = r+r_s\log(\frac{r}{r_s}-1)$. With a simple computation one
reaches for $m=0$ the equation
\begin{equation}\label{10reducedeq}
-\frac{d^2}{d r_*^2} \psi(r_*) + V(r) \psi(r_*) =
\omega^2 \psi(r_*) 
\end{equation}
where 
$$
V(r) = \frac{r_s}{r^3}(1-\frac{r_s}{r})
$$
and $r$ has to be thought as function of $r_*$. As
$$
\frac{d r_*}{d r}=\frac{1}{1-\frac{r_s}{r}}
$$
never vanishes for $r>r_s$, $r$ is well defined. In case of $m^2\neq
0$ and $l\neq 0$, $V(r)$ goes over to
$$
V(r) = (1-\frac{r_s}{r})\left(\frac{r_s}{r^3}+\frac{l(l+1)}{r^2}\right)+m^2
(1-\frac{r_s}{r})~. 
$$
For simplicity we shall use $m^2=0$ but there is no real difficulty in
working with $m^2\neq 0$. The shape of $V$, for $l=0$ is depicted in 
fig.10 . $V$
vanishes on the horizon $r=r_s, ~r_*=-\infty$ while at $r_*= +\infty$,
i.e. $r=\infty$ it goes over to $m^2$. For $m^2=0$, $V$ reaches its
maximum at 
$r=4r_s/3$ and its value at that point is $V(r_M)= 27/(256 r_s^2)$. The
eigenvalue equations (\ref{10reducedeq}) due to the nature of $V$ has continuum
spectrum bounded from below by $0$. 

\begin{figure}[htb]
\begin{center}
\includegraphics{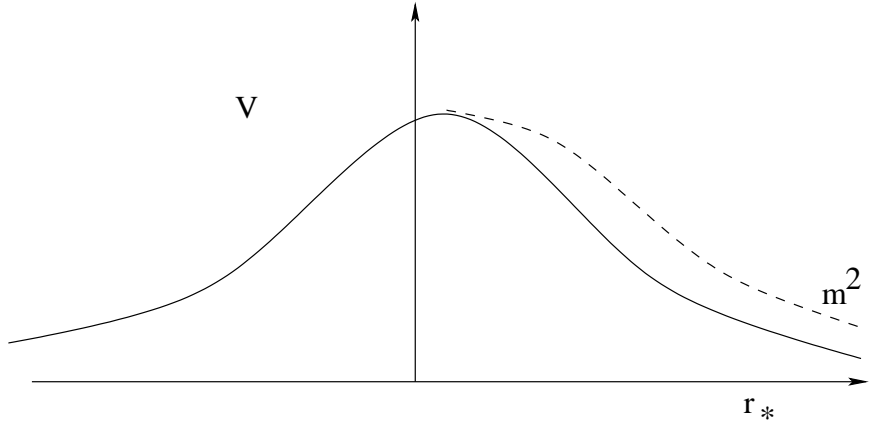}
\end{center}
\caption{Effective Schwarzschild potential}
\end{figure}

The detector placed in a
region ${\cal O}$ of 4-dimensional space will be described by the positive 
operator $Q^+Q$ 
where 
$$
Q= \int \phi(x) h(x) \sqrt{-g}d^4x
$$
and ${\cal O}$ is the support of the smooth
function $h$. Thus the detector occupies a finite region of
3-dimensional space and is switched on for a finite time. Given the
state $\Phi$ describing the system the result of the measurement is
$$
(\Phi,Q^+Q\Phi)
$$ 
which we have to compute. There is no need in this treatment to
introduce creation and destruction operators, at any rate for
clearness sake we point out that on the Fock vacuum 
${\bf a}({\bf k})\Phi_M=0$ we have
$$
(\Phi_M,Q^+Q\Phi_M) = \int \frac{|\tilde h({\bf k},
\omega)|^2}{2\omega}d{\bf k};~~~~~~~~\omega = +\sqrt{{\bf k}^2+m^2}
$$
where
$$
h({\bf x},t) = \frac{1}{(2\pi)^4} \int \tilde h({\bf
k},k_0)e^{i({\bf k}{\bf x}- k_0 t)} d^4 k~.
$$
In the limit in which $h({\bf x},t)=h({\bf x})$ we have
$(\Phi_M,Q^+Q\Phi_M)=0$ because in this case the support of
$\tilde h({\bf k},k_0)$ is in $k_0=0$ while $\omega \geq m>0$ for
$m>0$. 

Otherwise in general the result will be non zero (and positive) also
on the vacuum. This is due to the fact that the abrupt switching on
and off of the detector will produce particle also out of the
Minkowski vacuum. At the end we will specify $h({\bf x},t)$ so as to
avoid such a production.

\bigskip

In Schwarzschild coordinates the stellar collapse takes an infinite
time. The reason is that the Schwarzschild time $t_c$ of the collapse
is related to the Kruskal coordinates by  
(see Section \ref{10schwarzschildSec})
$$
\frac{t}{r_s}= 2~{\rm arctanh}\frac{T_c}{X_c}~.
$$
But as at the collapse $T_c=X_c$ we have $t=+\infty$. We recall also
the definition of the null coordinates
$$
v=t+r_*;~~~~ u=t-r_*
$$
$v={\rm const.}$ and $v={\rm const.}$ represent incoming and outgoing
null geodesics. It will be useful to go over to the new coordinates 
$$
r;~~~~\tau = v-r=t+r_*-r.
$$
For $\tau$ fixed and $r\rightarrow r_s$ we have $t\rightarrow
+\infty$, $v\rightarrow {\rm finite}$ and $u\rightarrow
+\infty$. Recalling that
$$
U=-e^{-\frac{u}{4M}}
$$
and that the horizon is given by $U=0$ we see that the horizon is
crossed at a finite value of $\tau$.

Even though not strictly necessary, the space support  
${\cal O}$ of $h$  will be chosen at a large value
of $r$ (where the metric can be taken as Minkowski) 
and we shall consider a sequence of $Q^T$ which are the $Q$ 
translated by $T$ along the Schwarzschild Killing time like vector
$\frac{\partial}{\partial t}$. The reason is that we are interested in
the measurements at large times after the gravitational collapse. Due
to the linear field equations obeyed by $\phi$, $Q^T$ can be expressed
in terms of $\phi$ and its time derivative on an arbitrary partial
Cauchy surface for which the exterior Schwarzschild region is future
asymptotically predictable as we shall see explicitly below.

As already saw, given two partial Cauchy surfaces $\Sigma_1$ and
$\Sigma_0$ and an $f$ solution of
\begin{equation}\label{10KG}
\frac{1}{\sqrt{-g}}\partial_\mu(g^{\mu\nu}\partial_\nu f(x))-m^2 f(x)
=(g^{\mu\nu}\nabla_\mu\nabla_\nu -m^2)f(x) =0
\end{equation}
we have
\begin{equation}\label{10backforward}
\int_{\Sigma_1} (\phi(x)\partiallr_\mu f(x)) g^{\mu\nu}\Sigma_\nu=
\int_{\Sigma_0} (\phi(x)\partiallr_\mu f(x)) g^{\mu\nu}\Sigma_\nu
\end{equation}
and we shall choose for $\Sigma_0$ the partial Cauchy surface
$\tau=0$. The idea is to bring back the computation of $Q^T$, which is
extended in time to the single partial Cauchy surface $\tau=0$.

We start from the computation of $Q$ writing it as sum (integral) of
contributions at constant Schwarzschild time $t$.
\begin{equation}
Q=\int \phi(x) h(x)\sqrt{-g}d^4x =
\int dt_0\int\phi({\bf x},t_0) h({\bf x},t_0)\sqrt{-g}~d{\bf x}~.
\end{equation}
We write 
\begin{equation}
\int\phi({\bf x},t_0) h({\bf x},t_0)\sqrt{-g}~d{\bf x}=
\int\phi({\bf x},t_0) \partial_0 f_{t_0}({\bf x},t_0)\sqrt{-g}~d{\bf x}
\end{equation}
where $\partial_0$ means derivative w.r.t. the fourth argument and
$f_{t_0}({\bf x},t)$ is a solution of the Eq.(\ref{10KG}) with the
following initial conditions
$$
\partial_0 f_{t_0}({\bf x},t_0) = h({\bf x},t_0),~~~~~~~~
f_{t_0}({\bf x},t_0) = 0~.
$$
We have
$$
\int \phi({\bf x},t_0) h({\bf x},t_0) \sqrt{-g} ~d{\bf x} =
\int (\phi({\bf x},t_0) \partiallr_0 f_{t_0}({\bf x},t_0)) \sqrt{-g} ~d{\bf x}
$$
which using (\ref{10backforward}) can be rewritten as
$$
\int_{\Sigma_0} (\phi(x)\partiallr_\mu f_{t_0}(x))
g^{\mu\nu} ~\Sigma_\mu
$$
i.e. as an integral on $\Sigma_0$ (the surface
$\tau = 0$). Then we have
$$
Q= \int_{\Sigma_0} (\phi(x)\partiallr_\mu f(x))
g^{\mu\nu}~\Sigma_\mu
$$
where
$$
f(x) = \int f_{t_0}(x) dt_0 ~.
$$
We shall  be interested in $Q^T$ given by
$$
\int \phi(x) h({\bf x}, t-T)\sqrt{-g}~ d^4x~.
$$
This is obtained from the solution $f^T_{t_0}({\bf x},t)$ with
$$
\partial_0f^T_{t_0}({\bf x},t_0) = h({\bf x}, t_0-T),~~~~~~~~~
f^T_{t_0}({\bf x},t_0)=0
$$
and thus
$$
Q^T = \int_{\Sigma_0} (\phi(x)\partiallr_\mu f^T(x))
g^{\mu\nu}\Sigma_\nu
$$
where
$$
f^T(x) =\int f^T_{t_0}(x) dt_0 
$$
and $\Sigma_0$ again is the surface given by $\tau=0$. Notice that
this is possible because according Huygens principle $f^T(x)$
propagates back in time completely to the partial Cauchy surface
$\Sigma_0$ and nowhere outside it.

In the new coordinates $r,\tau$ we have
$$
ds^2 = -(1-\frac{r_s}{r})d\tau^2+2\frac{r_s}{r}dr d\tau
+(1+\frac{r_s}{r})dr^2+r^2 d\Omega^2 ~.
$$
The inverse of this metric is easily computed and we have in particular
\begin{equation}\label{rtaumetric}
g^{\tau\tau} = -(1+\frac{r_s}{r});~~~~g^{\tau r}=
-\frac{r_s}{r};~~~~g^{rr}= 1-\frac{r_s}{r}
\end{equation}
so that
$$
g^{\tau\tau}\partial_\tau+g^{\tau r}\partial_r =
-(1+\frac{r_s}{r})\partial_\tau+ \frac{r_s}{r}\partial_r \equiv D
$$ 
while
$$
\Sigma_\tau = \epsilon_{\tau r\theta\phi}dr\wedge d\theta \wedge d\phi
=r^2 \sin\theta ~dr\wedge d\theta \wedge d\phi.
$$

\begin{figure}[htb]\label{10collapse}
\begin{center}
\includegraphics{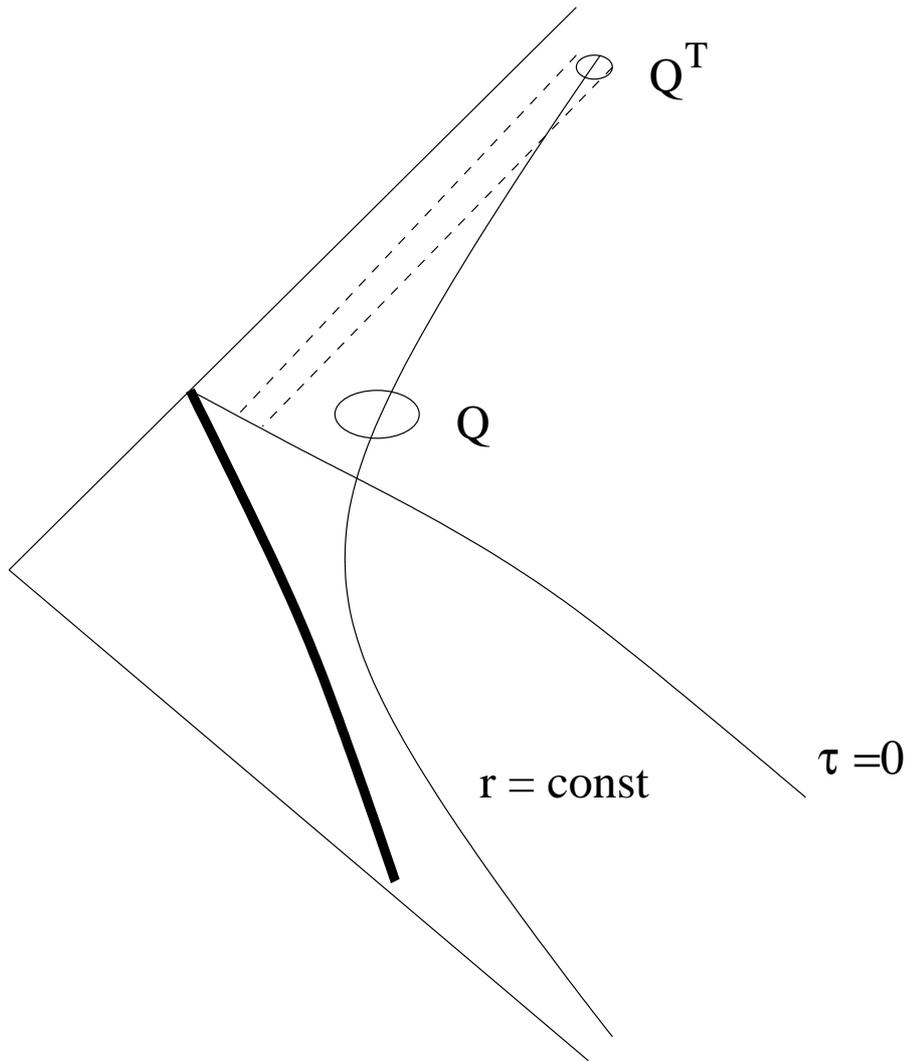}
\end{center}
\caption{Gravitational collapse}
\end{figure}

Taking into account also the contribution of $Q^{T+}$ we have the
exact result
\begin{equation}\label{QQexact}
(\Phi,Q^{T+} Q^T\Phi)=
\end{equation}
$$
\int_{\Sigma_0}\int_{\Sigma_0}(\Phi,\phi(x_1)\phi(x_2)\Phi) \DD_1
\DD_2 \bar{f}^T(x_1) f^T(x_2) ~r_1^2  dr_1 d\Omega_1 ~r_2^2
~dr_2 d\Omega_2
$$
This result is immediately extended to $m^2\neq 0$ and to $l\neq0$.

A complete set of solutions of Eq.(\ref{10reducedeq}) is given by the
$\psi_+(\pm\omega,r_*)$ with the following asymptotic behavior at
large $r_*$  (where our detector is placed)
$$
\psi_+(r^*,\pm\omega) = e^{\pm i \omega r_*};~~~~r_*\simeq +\infty
$$
while an alternative complete set is classified by the behavior of the
solution at $r_*\simeq -\infty$
$$
\psi_-(r^*,\pm\omega) = e^{\pm i \omega r_*};~~~~r_*\simeq -\infty~.
$$
The $\psi_+(r^*,\pm\omega)$ form a complete set. Thus we can write
$$
h(r^*,t) = \int_0^\infty d\omega[\psi_+(r^*,\omega)\tilde h(\omega,t)+
\psi_+(r^*,-\omega)\tilde h(-\omega,t)]
$$
where the above equation defines $\tilde h(\pm\omega,t)$

The explicit form of $f_{t_0}(r_*,t)$ is easily obtained from
the above Fourier transform
\begin{eqnarray}\label{10ft0}
f_{t_0}(r_*,t)= i\int_0^\infty \frac{d\omega}{2\omega}
&&[~\psi_+(r^*,\omega) \tilde h(\omega,t_0)(e^{-i\omega(t-t_0)}-
e^{i\omega(t-t_0)})\\
&-&\psi_+(r^*,-\omega) \tilde h(-\omega,t_0)(e^{i\omega(t-t_0)}-
e^{-i\omega(t-t_0)})]~.\nonumber
\end{eqnarray}
In fact as we have 
$$
f_{t_0}(r^*,t_0)=0~~~~~~~~{\rm and}~~~~~~~~
\partial_0 f_{t_0}(r^*,t_0)=h(r^*,t_0)~.
$$
$f^T_{t_0}(r^*,t)$ is simply obtained by replacing in Eq.(\ref{10ft0}) 
$t-t_0$ with $t-t_0-T$ and $f^T(r^*,t)$ is obtained by integrating in
$t_0$.
$$
f^T(r^*,t) = \int dt_0 ~f^T_{t_0}(r^*,t)~.
$$
The result is with obvious notation for $\tilde h(\pm\omega,\pm\omega)$
\begin{eqnarray}\label{10fT}
f^T(r^*,t)= i\int_0^\infty \frac{d\omega}{2\omega}
& &[~\psi_+(r^*,\omega) (\tilde h(\omega,\omega)e^{-i\omega(t-T)}-
\tilde h(\omega,-\omega)e^{i\omega(t-T)})\\
&-&\psi_+(r^*,-\omega) (\tilde h(-\omega,-\omega) e^{i\omega(t-T)}-
\tilde h(-\omega,\omega)e^{-i\omega(t-T)})]~.\nonumber
\end{eqnarray}
Of the four solutions appearing in (\ref{10fT}) the first and the third move
initially to increasing values of $r^*$ (right-moving) while the
second and the
fourth move initially to decreasing values $r^*$ (left-moving).
We are now interested in $f^T$ on the partial Cauchy surface $\tau=0$
for large positive $T$. We see from the form of the exponential that
increasing $T$ corresponds to propagating the solutions backward in
time. In so doing the ``left-moving'' solution move right
without encountering the potential and for large $T$ 
will meet the partial Cauchy surface $\tau=0$  at large $r$. Thus such
wave packets correspond to particle coming from the space infinity and
not from the black hole and we are not interested in them. Thus we are
left with
\begin{equation}\label{10fTright}
f^T(r^*,t)= i\int_0^\infty \frac{d\omega}{2\omega}
 [~\psi_+(r^*,\omega) \tilde h(\omega,\omega)e^{-i\omega(t-T)}-
\psi_+(r^*,-\omega)\tilde h(-\omega,-\omega)e^{i\omega(t-T)}]~.
\end{equation}
Let us consider first the term
\begin{equation}
i\int_0^\infty \frac{d\omega}{2\omega}
~\psi_+(r^*,\omega) \tilde h(\omega,\omega)e^{-i\omega(t-T)}~.
\end{equation}
We recall that for increasing $t$ the wave packet described by it
moves to $+\infty$ as it does not encounter any potential. On the
other hand taking $T$ positive and large corresponds to evolving the
same wave packet backward in time. In the process part of it will be
reflected by the potential barrier and part will be transmitted with
a transmission coefficient given by the identity
$$
D(\omega)\psi_-(r^*,\omega)= \psi_+(r^*,\omega)- B(\omega)\psi_+(r^*,-\omega)~.
$$
The computation of $D(\omega)$ is an elementary problem in quantum
mechanics even if one does not succeed in performing the computation
analytically. As for large negative $r^*$ the $\psi_-(r^*,\omega)$
becomes $e^{i\omega r^*}$ (with coefficient $1$) we have that at large
positive $T$ the contribution becomes
\begin{equation}
i\int_0^\infty \frac{d\omega}{2\omega} D(\omega)
\tilde h(\omega,\omega)e^{i\omega (r^*-t+T)}~.
\end{equation}
In exactly the same way one deals with second term in Eq.(\ref{10fTright})
and calling $f_-^T(r^*,t)$ the limit of $f(r^*,t)$ for large positive
$T$ we have 
\begin{equation}\label{10fTminus}
f_-^T(r^*,t) = i\int_0^\infty \frac{d\omega}{2\omega} [~D(\omega)
\tilde h(\omega,\omega)e^{i\omega (r^*-t+T)}-
D(-\omega)
\tilde h(-\omega,-\omega)e^{-i\omega (r^*-t+T)}~]
\end{equation}
and we can thus write for large $T$
\begin{equation}\label{QQasym}
(\Phi,Q^{T+} Q^T\Phi)=
\end{equation}
$$
\int_{\Sigma_0}\int_{\Sigma_0}(\Phi,\phi(x_1)\phi(x_2)\Phi) \DD_1
\DD_2 \bar{f}_-^T(x_1) f^T_-(x_2) ~r_1^2  dr_1 d\Omega_1 ~r_2^2
~dr_2 d\Omega_2~.
$$
The difference between Eq.(\ref{QQexact}) and Eq.(\ref{QQasym}) is that while
Eq.(\ref{QQexact}) is exact Eq.(\ref{QQasym}) holds for large positive $T$;
however the advantage of Eq.(\ref{QQasym}) is that $f^T_-$ is simply
known through (\ref{10fTminus})
in terms of the transmission coefficient $D(\omega)$. To understand the
physics of the problem it is useful to notice that
$$
f^T_-(r^*,t) =F(-u+T) =F(2 r^*-\tau-r+T) 
$$ 
which near the horizon becomes
$$
f^T_-(r_*,t) \simeq F(2 r_s\ln\frac{r-r_s}{r_s} -\tau+r_s+T) 
$$
It shows that for large $T$ the $f^T_-$ is concentrated in an
exponentially small region around the horizon and thus the result is
determined by the short distance behavior of the Wightman function
$(\Phi, \phi(x_1)\phi(x_2)\Phi)$ (see fig. 10.2).

The following simplifying features intervene in the computation of
Eq.(\ref{QQasym}) 
$$
D = -(1+\frac{r_s}{r})\partial_\tau +\frac{r_s}{r}\partial_r
$$
$$
Du = \frac{\partial u}{\partial r},~~~~~~~~
Dv = -\frac{\partial v}{\partial r}= -1
$$
$$
f_-^T(r_*,t) = F(-u+T),~~~~~~~~
D f^T_- = \frac{\partial f^T_-}{\partial r}
$$
which allow for large $T$ two integrations by parts and we reach
$$
\langle Q^{T+}Q^T\rangle =\int_{\Sigma_0}\int_{\Sigma_0} [~\hat D_1 \hat D_2 
(\Phi, \phi(x_1)
\phi(x_2)\Phi)]~ \bar f_-^T(x_1) f_-^T(x_2)
r_1^2 dr_1 d\Omega_1 r_2^2 dr_2 d\Omega_2
$$
with
$$
\hat D = 2(\frac{\partial}{\partial\tau}-\frac{\partial}{\partial r})~.
$$
We shall assume that the short distance behavior of the Wightman
function on $\tau=0$ which is a regular region of space time
not casually connected with the crossing of of the horizon by the
surface of the star, is given by the usual free field value. Arguments
[2] can be given also to support that fact that
the short distance behavior of such a function in independent of the
state.
The covariant translation of Eq.(\ref{10twopointmassless}) using the
metric (\ref{rtaumetric}) is
$$
(\Phi, \phi(x_1) \phi(x_2)\Phi)\approx\frac{1}{4\pi^2
\sigma_\varepsilon} 
$$
with $\sigma$ given at short distances 
by $(x_1-x_2)^\mu g_{\mu\nu}(x_1-x_2)^\nu$.  
Near the horizon it assumes the form
$$
\sigma_\varepsilon = 2(r_1-r_2)(r_1-r_2+\tau_1-\tau_2 -i\varepsilon)
+r^2_s \theta^2_{12}
$$
$\theta_{12}$ being the angle between the directions $1$ and $2$.
One obtains on the surface $\tau=0$
$$
\frac{1}{4\pi^2}\hat D_1\hat D_2 \frac{1}{\sigma_\varepsilon}=
-\frac{8}{\pi^2}\frac{(r_1-r_2-i\varepsilon)}{[2(r_1-r_2-i\varepsilon)^2+
r_s^2\theta^2_{12}]^3}~.
$$
Integrating over $d\Omega_1$, $d\Omega_2$ one obtains
$$
-\frac{4}{r_s^2}\frac{1}{(r_1-r_2-i\varepsilon)^2}
$$
which is the most singular part, plus a term which is irrelevant for
$T\rightarrow \infty$.
Thus we reached the result
$$
\langle Q^{T+}Q^T\rangle = -4 r_s^2
\int_{r_s}^\infty \frac{1}{(r_1-r_2-i\varepsilon)^2} 
\bar F(2 r_s\ln\frac{r_1-r_s}{r_s}+r_s+T)
F(2 r_s\ln\frac{r_2-r_s}{r_s}+r_s+T) dr_1 dr_2 ~.
$$
Using the new variables $\rho_1=\frac{r_1-r_s}{r_s}~e^{\frac{T}{r_s}+1}$ 
$\rho_2=\frac{r_2-r_s}{r_s}~e^{\frac{T}{r_s}+1}$ it can be rewritten as
$$
\langle Q^{T+}Q^T\rangle = -4 r_s^2
\int_0^\infty \frac{1}{(\rho_1-\rho_2)^2} \bar F(2 r_s\ln\rho_1)
F(2 r_s\ln\rho_2) d\rho_1 d\rho_2 ~.
$$
Now we go over to the variable $y_1= \ln\rho_1$, $y_2= \ln\rho_2$
to obtain
$$
\langle Q^{T+}Q^T\rangle = -4 \pi r_s\int_{-\infty}^{\infty}
\frac{1}{\sinh^2\frac{y-i\varepsilon}{2}}\frac{|D(\omega)|^2 
|\tilde h(\omega,\omega)|^2}{\omega^2}e^{-2r_s\omega y} ~dy~.
$$
Such an integral can be computed by using the identity
$$
\frac{1}{(\sinh \pi z-i\varepsilon
)^2}=\frac{1}{\pi^2}\sum_{k=-\infty}^{\infty} \frac{1}{(z+i k
-i\varepsilon)^2} 
$$
as done in section (\ref{10acceldetect})
obtaining from the first part
of Eq.(\ref{10fTminus})
\begin{equation}\label{10finalhawking}
\langle Q^{T+}Q^T\rangle = 64\pi^2r_s^2\int_0^\infty \frac{|D(\omega)|^2}
{e^{4\pi r_s\omega}-1} 
|h(\omega,\omega)|^2 \frac{d\omega}{\omega}~.
\end{equation}
To this we should add the contribution in Eq.(\ref{10fTminus}) of 
$h(-\omega,-\omega)$. This gives
$$ 
-64\pi^2r_s^2\int_0^\infty \frac{|D(-\omega)|^2}
{e^{-4\pi r_s\omega}-1} |h(-\omega,-\omega)|^2 \frac{d\omega}{\omega}~.
$$
which can be added to (\ref{10finalhawking}) to obtain
\begin{equation}\label{10finalhawking2}
\langle Q^{T+}Q^T\rangle = 64\pi^2r_s^2\int_{-\infty}^\infty \frac{|D(\omega)|^2}
{e^{4\pi r_s\omega}-1} 
|h(\omega,\omega)|^2 \frac{d\omega}{\omega}~.
\end{equation}

The factor $|D(\omega)|^2$ is the gray body factor, which in the free
field theory poses no problem to be computed; one has to keep in mind
that the height of the potential $V$ is of the order of magnitude
$1/r_s^2$ which is the order of magnitude of the $\omega^2$ of the typical
quantum emitted from the black hole. Thus $|D(\omega)|^2$ plays an
important role. In an interacting theory the computation of such a
factor poses a real problem.

The variable $T$ has disappeared in the integration process; this mean
that the Hawking radiation is a ``permanent'' phenomenon; the
violation of energy conservation has to be ascribed to the fact that
we are working in an external field. At the end the energy has to be
supplied by the black hole itself.  The factor $|h(\omega,\omega)|^2$
is just the description of the counter. However the $h(\omega,\omega)$
of a physical counter should have support of on positive values of
$\omega$ [3] and we go back to Eq.(\ref{10finalhawking}).

In this case $h(x)$ cannot be of compact support in space time but one
has to widen the class of test function beyond those of compact support. 

Finally the key point in producing the Planck factor is the short
distance behavior of the two point Wightman function; any modification
to it would imply a change in the emission spectrum.

We compare now the obtained formula with the Rindler spectrum. There
the temperature was given by the argument of the exponential
$$
e^{2\pi\omega/g} =e^{2\pi \omega c/g}=
e^{2\pi \omega \hbar c/g \hbar}=e^{\omega \hbar/k_B T}
$$
i.e. $k_B T = g\hbar/2\pi c$. We can now compute the surface gravity for a
black hole i.e. the force exerted at the horizon on a unit mass
as measured by an observer at infinity. We found (see
Eq.(\ref{10surfacegravity})) that it is given by
$$
g = \frac{MG_N}{(r_s)^2} =\frac{G_N M c^4}{4 M^2 G^2_N} =\frac{c^4}{4 G_N
M}
$$
and thus we expect a temperature $k_B T = \hbar c^3/8\pi G_N M$ as
found in Eq.(\ref{10finalhawking}).

\bigskip

Thus, according to Hawking's treatment, black holes are not eternal but
they evaporate. The lifetime of a black hole of mass $M$ can be
estimated to be
\begin{equation}
t =\frac{5120 \pi G^2 M^3}{\hbar c^4}
\end{equation}
with a typical initial wave length of the radiation of
$\lambda/(2\pi)\approx 4\pi r_s$ being $r_s$ the radius of the black
hole.

However for macroscopic black holes such lifetime is much larger than
the age of our universe. For a black hole of the mass of the sun $M=
1.989~10^{30}~kg$ the lifetime is $5~10^{74}~s$. For a black hole of
$M=200~000~kg$ we have a lifetime of $0.67~s$ with an initial
temperature of $50~TeV$. For a black hole of the Planck mass ($M= 1.21
~10^{19} GeV /c^2 = 2.17~10^{-8}~kg$) the lifetime is of $8.6~10^{-40} s$
of the order of magnitude of the Planck time.

\bigskip

References

\smallskip

[1] K. Fredenhagen, R. Haag, ``On the derivation of Hawking radiation
associated with the formation of a black hole'',
Comm. Math. Phys. 127 (1990) 273

\smallskip

[2] R.Haag, ``Local quantum physics'', II edition, Chap.VIII,  Springer 1992

\smallskip

[3] R.Haag, ``Local quantum physics'', II edition, Chap.VI,  Springer 1992

\vfill



\chapter{N=1 Supergravity}\label{supergravityChap}

\section{The Wess-Zumino model}

We first discuss the operation of charge conjugation for the Dirac
field. Given the Dirac equation
$$
(\gamma^\mu\partial_\mu + m)\psi =0
$$
(we always use $\gamma^j$ hermitean and $\gamma^0$ antihermitean, thus
$\gamma^4 \equiv i \gamma^0$ hermitean) 
we look for a unitary operator $C$ such that
\begin{equation}\label{conjugation}
\psi_c \equiv C\psi C^+ = A \psi^+_T\equiv A \psi^* 
\end{equation}
which again satisfies the Dirac equation.
We have
$$
(\gamma^{*\mu}\partial_\mu + m)\psi^+_T =0
$$
and thus $A$ must be such that 
\begin{equation}\label{Arelation}
A\gamma^{*\mu} A^{-1}=\gamma^\mu~.
\end{equation}
There
exists an $A$ with does the job i.e. $A=\gamma^2$ in the
representation adopted in Chapter 1. 
$A$ is unique up to a factor because if $B = X A$ also
satisfies (\ref{Arelation})
$X$ commutes with all the gammas and by Schur
lemma $X$ is a multiple of the identity $c$, being our representation of
the gammas irreducible. If we require that $C^2$ brings $\psi$ back
to itself apart a phase factor, we must have
$$
C(c\gamma^2\psi^+_T)C^+ = c c^*\psi = u\psi
$$
with $u$ phase factor, from which $c^*c=1$. Redefining the $\psi$ by a
phase factor we can write simply
\begin{equation}\label{chargeconjugation}
C\psi C^+ = \gamma^2 \psi^+_T
\end{equation}
The current $i\bar\psi\gamma^\mu\psi$, which is hermitean, under $C$ 
changes sign
$$
C i\bar\psi\gamma^\mu\psi C^+= i\bar\psi_c\gamma^\mu\psi_c =
-i\bar\psi\gamma^\mu\psi
$$ 
as it is easily verified, taking into account the anticommutative nature of
$\psi$.

We define a Majorana spinor as a spinor which coincides with its
charge conjugate $\psi_c=C\psi C^+ =\gamma^2 \psi^+_T = \psi$.
The relation with the form of the equation for a Majorana particle
given in Chapter 1 is the following. Given the Dirac equation in the form
\begin{equation}\label{11dirac}
\begin{pmatrix}
\sigma^\mu p_\mu&0\\
0&\tilde\sigma^\mu p_\mu
\end{pmatrix} 
\begin{pmatrix}
\psi\\
\phi
\end{pmatrix}=
mc
\begin{pmatrix}
0&1\\
1&0
\end{pmatrix} 
\begin{pmatrix}
\psi\\
\phi
\end{pmatrix} 
\end{equation}
for a Majorana spinor from (\ref{chargeconjugation}) we have 
$$
-i\sigma_2\phi^+_T= \psi
$$
equivalent to
$$
i\sigma_2\psi^+_T= \phi
$$
and thus the lower pairs of equation in (\ref{11dirac}) can be written as
$$
\tilde\sigma^\mu p_\mu \phi = mc \psi = -i mc \sigma_2\phi^+_T\equiv
-i mc \sigma_2\phi^*
$$
which is Eq.(\ref{1majorana}) while the upper pair of equations in 
(\ref{11dirac}) is equivalent to the above.

In dealing with Majorana spinors it is useful to adopt the so called 
Majorana representation of the gamma matrices: $\gamma^i$ will be
again hermitean and $\gamma^0$ antihermitean but now all will be
real.
$$\gamma^1 =
\begin{pmatrix}
0&-1\\
-1&0
\end{pmatrix},~ 
\gamma^2 =
\begin{pmatrix}
0&-i\sigma_2\\
i\sigma_2&0
\end{pmatrix},~ 
\gamma^3 =
\begin{pmatrix}
1&0\\
0&-1
\end{pmatrix}, 
\gamma^0 =
\begin{pmatrix}
0&-\sigma_3\\
\sigma_3&0
\end{pmatrix}\equiv-i\gamma^4. 
$$
\begin{equation}
\gamma_5  = \gamma^1\gamma^1\gamma^1\gamma^4= i
\begin{pmatrix}
0&\sigma_1\\
-\sigma_1&0
\end{pmatrix}~. 
\end{equation}
The following relations valid in the Majorana representation will be
useful 
$$
\gamma^j_T = \gamma^j;~~~~\gamma^0_T =
-\gamma^0;~~~~\gamma^4_T=-\gamma^4;~~~~\gamma_{5T} =
-\gamma_5~. 
$$
In the Majorana representation the matrix $A$ of (\ref{conjugation}) 
is simply the identity and thus a Majorana spinor in the Majorana
representation is described by an hermitean spinor.

\bigskip

To get acquainted with supersymmetry we examine now the Wess-Zumino
model. In such a model we have two boson fields, a scalar $A$ and a
pseudoscalar $B$ and one Majorana fermion $\lambda$, all massless.

The Lagrangian of the Wess-Zumino model is
$$
L_{WZ}=-\frac{1}{2}\partial_\mu A\partial^\mu A -\frac{1}{2} \partial_\mu
B\partial^\mu B -\frac{1}{2} \bar\lambda  \dsl \lambda.
$$
As supersymmetry transformation is given by
$$
\delta A = \frac{1}{2}\bar\epsilon\lambda,~~~~\delta
B=-\frac{i}{2}\bar\epsilon\gamma_5\lambda,~~~~\delta \lambda =
\frac{1}{2}(\dsl A+i\gamma_5\dsl B)\epsilon
$$
where $\epsilon$ is a constant real classical anticommuting spinor.
Notice that $\bar\epsilon\lambda$ and $-i\bar\epsilon\gamma_5\lambda$
are hermitean fields. The variation of the Lagrangian is
$$
\delta L = \delta_A L + \delta_B L
$$
with
$$
\delta_A L = -\frac{1}{2}\partial^\mu A ~\bar\epsilon \partial_\mu
\lambda -\frac{1}{4}\bar\lambda\dsl \dsl A~\epsilon + 
\frac{1}{4}\bar\epsilon \dsl A \dsl \lambda.
$$
$$
\delta_B L = \frac{i}{2}\partial^\mu B ~\bar\epsilon \gamma_5\partial_\mu
\lambda +\frac{i}{4}\bar\lambda\gamma_5 \dsl \dsl B~\epsilon - 
\frac{i}{4}\bar\epsilon \gamma_5\dsl B \dsl \lambda.
$$

Thus the variation is not zero but is equal to a divergence; i.e. it
is immediately shown that
$$
\delta_A L = \partial_\mu K^\mu
$$
with
$$
K^\mu = -\frac{1}{4}\bar\epsilon\gamma^\mu\dsl A \lambda
$$
and similarly one finds that 
$$
\delta_B L = \partial_\mu K_5^\mu
$$
with
$$
K_5^\mu = -\frac{i}{4}\bar\epsilon\gamma^\mu\gamma_5 \dsl B\lambda.
$$
Thus the action is invariant.

The chiral doublets $A+iB, ~(1+\gamma_5)\lambda$ and $A-iB,
~(1-\gamma_5)\lambda$ under a SUSY transformation, transform into each
other  
$$
\delta(A+iB)=\frac{1}{2}\bar\epsilon(1+\gamma_5)\lambda
$$
$$
\delta(A-iB)=\frac{1}{2}\bar\epsilon(1-\gamma_5)\lambda
$$
$$
\delta(1+\gamma_5)\lambda=\frac{1}{2}(1+\gamma_5) \dsl(A+iB)\epsilon
$$
$$
\delta(1-\gamma_5)\lambda=\frac{1}{2}(1-\gamma_5) \dsl(A-iB)\epsilon.
$$
One notices that 

1) the SUSY transformations relate bosons with fermions. 

2) Irreducible representation must involve at least a boson and a fermion. 

3) $\epsilon$ is a (constant) spinor under Lorentz transformation if
we have to maintain invariance under Lorentz transformations.

4) In $n=4$ the boson field has dimension $A\sim B\sim l^{-1} \sim  m$
while the fermion $\lambda \sim m^{3/2}\sim l^{-3/2}$ and thus from
$\delta A = \frac{1}{2}\bar\epsilon\lambda$ we have $\epsilon\sim
m^{-1/2} \sim l^{1/2}$ 

5) In  $\delta \lambda = {\rm boson} \times \epsilon$ the gap in
dimensions is filled by the derivative $\partial \sim m \sim l^{-1}$. 

Most important is the following result: The commutator of two SUSY
transformations is a (global) space time translation. In a sense a
SUSY transformation is the square root of a space time translation.

We have in fact
$$
\delta_2(\delta_1 A) = \delta_2(\frac{1}{2}\bar\epsilon_1\lambda) =
\frac{1}{4}\bar\epsilon_1(\dsl A +i\gamma_5\dsl B)\epsilon_2 
$$
$$
\delta_1(\delta_2 A) = \delta_1(\frac{1}{2}\bar\epsilon_2\lambda) =
\frac{1}{4}\bar\epsilon_2(\dsl A +i\gamma_5\dsl B)\epsilon_1 =
 \frac{1}{4}\bar\epsilon_1(-\dsl A +i\gamma_5\dsl B)\epsilon_2
$$
and thus
$$
(\delta_2\delta_1-\delta_1\delta_2)A =
\frac{1}{2}\bar\epsilon_1\gamma^\mu\epsilon_2\partial_\mu A.
$$
Similarly
$$
(\delta_2\delta_1-\delta_1\delta_2)B =
\frac{1}{2}\bar\epsilon_1\gamma^\mu\epsilon_2\partial_\mu B
$$
and thus the commutator is a space time translation by
$\frac{1}{2}\bar\epsilon_1\gamma^\mu\epsilon_2$. 
Now the question is
\begin{equation}\label{commutatorsym}
(\delta_2\delta_1-\delta_1\delta_2)\lambda =
\frac{1}{4}\partial_\mu (\bar\epsilon_2\lambda) \gamma^\mu\epsilon_1
-(1\leftrightarrow 2)+
\frac{1}{4}\partial_\mu (\bar\epsilon_2\gamma_5 \lambda)
\gamma_5\gamma^\mu\epsilon_1 
-(1 \leftrightarrow 2)
=\frac{1}{2}(\bar\epsilon_1\gamma^\mu\epsilon_2)~\partial_\mu\lambda ~? 
\end{equation}
Not quite, as we shall see.

To proceed we shall need the Fierz rearrangement identity: 
Given spinors $\lambda,\chi,\psi,\phi$
we have
\begin{equation}\label{fierz}
(\bar\lambda\chi)(\bar\psi\phi) = -\frac{1}{4}\sum_j (\bar\lambda
O^j\phi )(\bar\psi O^j \chi) 
\end{equation}
with 
$$
O^j=\{I,\gamma^a,2i\sigma^{ab},i\gamma_5\gamma^a,\gamma_5\}
~~~~a,b =1,2,3,4,~~~~a< b~.
$$
The $O^j$ are the 16 independent Dirac $4\times 4$ matrices, with
$$
\sigma^{ab} = \frac{1}{4}[\gamma^a,\gamma^b] \equiv \frac{1}{2}\gamma^{ab}.
$$
$O^j$ are hermitean and orthonormal in the
trace norm 
$$
\frac{1}{4}{\rm Tr}(O^j O^l) = \delta^{jl}.
$$
$O^j$ being a complete orthonormal set we have for any matrix $M$
$$
M = \frac{1}{4}\sum_j O^j {\rm Tr}(O^jM).
$$
Given the matrix $M^\gamma_{~\beta} =
\delta^\gamma_\delta\delta^\alpha_\beta$
we have
$$
\delta^\gamma_\delta\delta^\alpha_{\beta}=\frac{1}{4}\sum_j
(O_j)^\gamma_{~\beta} (O^j)^\alpha_{~\delta} 
$$
because ${\rm Tr}(O^j M) = (O^j)^\alpha_\delta$. 
Multiplying the above by the spinors
$\bar\lambda_\alpha,\chi^\beta,\bar\psi_\gamma,\phi^\delta$, and
keeping in mind anticommutativity we obtain
Eq.(\ref{fierz}). Eq.(\ref{fierz}) is immediately generalized to
\begin{equation}\label{generalizedFierz}
(\bar\lambda M\chi)(\bar\psi N\phi) =-\frac{1}{4}\sum_j(\bar\lambda
MO^jN\phi)(\bar\psi O^j\chi) 
\end{equation}
by letting $\bar\lambda \rightarrow \bar\lambda M$  and $\phi
\rightarrow N\phi$.

We shall now prove by using the Fierz transformation that
Eq.(\ref{commutatorsym}) holds on the equation of motion for $\lambda$
( i.e. ``on shell'').

For Majorana spinors (classical fields or operators) we have
$$
\bar\epsilon_2\epsilon_1=\bar\epsilon_1\epsilon_2
$$
$$
\bar\epsilon_2\gamma^a\epsilon_1=-\bar\epsilon_1\gamma^a\epsilon_2
$$
$$
\bar\epsilon_2\gamma_5
\gamma^a\epsilon_1=\bar\epsilon_1\gamma_5\gamma^a\epsilon_2  
$$
$$
\bar\epsilon_2\gamma_5\epsilon_1=\bar\epsilon_1\gamma_5\epsilon_2
$$
\begin{equation}\label{11exchanges}
\bar\epsilon_2\sigma^{ab}\epsilon_1=-\bar\epsilon_1\sigma^{ab}\epsilon_2.
\end{equation}

We go back now to Eq.(\ref{commutatorsym}). With $\bar \chi$ a 
dummy spinor
$$
\frac{1}{4}(\bar\chi \gamma^\mu\epsilon_1)(\bar\epsilon_2
\partial_\mu\lambda)
-\frac{1}{4}(\bar\chi \gamma^\mu\gamma_5\epsilon_1)(\bar\epsilon_2
\gamma_5\partial_\mu\lambda)=
$$
$$
=-\frac{1}{16}(\bar\epsilon_2 O^j\epsilon_1) (\bar\chi\gamma^\mu
O^j\partial_\mu\lambda) +\frac{1}{16}(\bar\epsilon_2 O^j\epsilon_1)
(\bar\chi\gamma^\mu \gamma_5 O^j\gamma_5 \partial_\mu\lambda).  
$$
where in the first we used (\ref{generalizedFierz}) with 
$M=\gamma^\mu$, $N=I$ and in the second again (\ref{generalizedFierz}) with
$M=\gamma^\mu\gamma_5$ and $N=\gamma_5$.
 
Due to the relations (\ref{11exchanges}) and antisymmetrization 
only $O^j = \gamma^\mu$
and $O^j= 2 i \sigma^{\mu\nu}$ contribute and we are left with
$$
-\frac{1}{16}(\bar\epsilon_2
\gamma^\nu\epsilon_1)(\bar\chi\gamma^\mu\gamma_\nu\partial_\mu\lambda) 
+\frac{1}{4}(\bar\epsilon_2
\sigma^{\alpha\beta}\epsilon_1)
(\bar\chi\gamma^\mu\sigma^{\alpha\beta}\partial_\mu\lambda)   
$$
$$
+\frac{1}{16}(\bar\epsilon_2
\gamma^\nu\epsilon_1)(\bar\chi\gamma^\mu\gamma_5\gamma_\nu
\gamma_5\partial_\mu\lambda)   
-\frac{1}{4}(\bar\epsilon_2
\sigma^{\alpha\beta}\epsilon_1)
(\bar\chi\gamma^\mu\gamma_5\sigma^{\alpha\beta}\gamma_5\partial_\mu\lambda).   
$$
The second and the fourth term cancel, while the first and the third
sum together and we are left with
$$
-\frac{1}{4} (\bar\epsilon_2\gamma^\nu\epsilon_1) (\bar\chi\partial_\nu
\lambda) + \frac{1}{8} (\bar\epsilon_2\gamma^\nu\epsilon_1)
(\bar\chi\gamma_\nu\dsl \lambda) -(1\leftrightarrow 2) =  
\frac{1}{2} (\bar\epsilon_1\gamma^\nu\epsilon_2) (\bar\chi\partial_\nu
\lambda) - \frac{1}{4} (\bar\epsilon_1\gamma^\nu\epsilon_2)
(\bar\chi\gamma_\nu\dsl \lambda). 
$$
Removing the dummy spinor $\chi$ Eq.(\ref{commutatorsym}) becomes
$$
\frac{1}{2} (\bar\epsilon_1\gamma^\nu\epsilon_2) \partial_\nu
\lambda - \frac{1}{4} (\bar\epsilon_1\gamma^\nu\epsilon_2)
\gamma_\nu\dsl \lambda 
$$
and thus it is the same translation on $\lambda$ as on $A$ and $B$
provided we work on shell i.e. for $\dsl\lambda=0$.

\section{Noether currents}

Suppose that under a global i.e. $\varepsilon(x) = {\rm const.}$
transformation the Lagrangian varies by a divergence i.e.
$$
\delta L = \varepsilon ~\partial_\mu K^\mu.
$$
Thus the action is invariant and the transformation is a symmetry
transformation. We want to compute how the action varies when
$\varepsilon$ is made space-time dependent.
We have
$$
\delta L = \frac{\partial L}{\partial \phi} \varepsilon
\phi_\varepsilon+
\frac{\partial L}{\partial \partial_\mu\phi} \partial_\mu (\varepsilon
\phi_\varepsilon)= ({\rm via~eq.~motion})~ 
$$
\begin{equation}\label{nvariation}
=
\frac{\partial}{\partial x^\mu}(\frac{\partial L}{\partial
\partial_\mu \phi} ~\varepsilon ~\phi_\varepsilon) =
\varepsilon~\frac{\partial}{\partial x^\mu}(\frac{\partial L}{\partial 
\partial_\mu \phi}\phi_\varepsilon) +\frac{\partial
L}{\partial\partial_\mu \phi}\phi_\varepsilon \partial_\mu\varepsilon.    
\end{equation}
We know that for $\partial_\mu \varepsilon =0$ we have 
$$
\delta L = \varepsilon\partial_\mu K^\mu
$$
i.e. the conserved current 
$$
J^\mu_N = \frac{\partial L}{\partial \partial_\mu \phi}\phi_\varepsilon-K^\mu
$$
and
$$
\partial_\mu\left(\frac{\partial L}{\partial \partial_\mu
\phi}\phi_\varepsilon\right)-\partial_\mu K^\mu=0~.
$$
Coming back to Eq.(\ref{nvariation}) for non constant $\varepsilon$ we have 
$$
\delta L = \varepsilon \partial_\mu K^\mu +\frac{\partial
L}{\partial\partial_\mu\phi}\phi_\varepsilon \partial_\mu\varepsilon 
$$
and thus
$$
\delta I =\int d^4x \delta L  = \int d^4 x(\varepsilon \partial_\mu K^\mu
+\frac{\partial L}{\partial\partial_\mu\phi}\phi_\varepsilon
\partial_\mu\varepsilon)  =
\int d^4 x(\varepsilon \partial_\mu K^\mu+K^\mu\partial_\mu\varepsilon
+J_N^\mu
\partial_\mu\varepsilon)  
=\int d^4 x ~\partial_\mu \varepsilon ~J^\mu_N
$$
i.e.
$$
\delta I = \int d^4x~ \partial_\mu\varepsilon~  J^\mu_N
$$
where $J^\mu_N$ is the Noether current.  

Thus given a theory invariant under a global transformation the
obtained result gives a hint on how to render the theory invariant 
under local transformations. In
fact $\partial_\mu\varepsilon$ is the first term in the variation of a
gauge field 
(1-form) under a gauge transformation. Thus it is suggested to couple
the previous theory to a gauge field $A_\mu$ described by a gauge
invariant Lagrangian $L_A$ and coupled to the field $\phi$ by the term
$$
-\int d^nx J_N^\mu A_\mu.
$$ 

\section{$N=1$ supergravity}

If we want to render a theory which is invariant under global
supersymmetry transformations, invariant under local supersymmetry
transformation the suggestion is to couple it to a gauge field with
the nature of a spinor 1-form, i.e. $\psi_\mu dx^\mu$. We have already
discussed in Section 1.15 such a structure which describes, going over
to irreducible representations, a spin $3/2$ particle, which will be
the gravitino.

However it is immediately checked that the Rarita-Schwinger Lagrangian
in presence of mass is not invariant under the gauge transformation
$$
\psi_\mu \rightarrow \psi_\mu +\partial_\mu\epsilon.
$$
Thus we must deal with the massless Rarita-Schwinger field. 
In order to put to zero the two unwanted spin $1/2$ components of the
field we have to impose a gauge condition. We shall choose
$\gamma^\mu\psi_\mu=0$ which we proved to be always attainable. Such
condition implies through the equations of motion
$\partial_\mu\psi^\mu=0$.  In fact from
$$
\varepsilon^{\mu\nu\rho\sigma}\gamma_5\gamma_\nu\partial_\rho\psi_\sigma=0 
$$
we have with $\varepsilon^{0123}=-1$
$$
0=\frac{1}{2}
\varepsilon^{\mu\nu\rho\sigma}\gamma_5(\gamma_\mu
\gamma_\nu-\gamma_\nu \gamma_\mu) 
\partial_\rho\psi_\sigma
=-i
(\gamma^\rho
\gamma^\sigma-\gamma^\sigma \gamma^\rho) 
\partial_\rho\psi_\sigma= -2i(\gamma^\rho
\gamma^\sigma- \eta^{\rho\sigma}) 
\partial_\rho\psi_\sigma= 2i\partial_\mu\psi^\mu.
$$

We saw in Chapter 5 how one can formulate gravity both in the second
and first order formalism. The formalism which provides the simplest
proof of the invariance of $N=1$ gravity under supersymmetric
transformations is the so called $1.5$ formalism which we are going to
explain.

We start from the first order action
\begin{equation}\label{supergravity}
S_{SG}= S_{EC} +S_{RS}
\end{equation}
with
$$
S_{EC} = -\frac{1}{8}\int R^{ab}\wedge e^c\wedge e^d \varepsilon_{abcd}=
-\frac{1}{4}\int e ~d^4x~ R^{ab}_{\mu\nu} e^\mu_{~a}e^\nu_{~b}
$$
($\varepsilon_{0123}= 1$)
and $S_{RS}$ is the covariant transcription of the massless
Rarita-Schwinger action 
$$
S_{RS} = \frac{i}{2}\int d^4x~ \varepsilon^{\mu\nu\rho\sigma}\bar\psi_\mu
\gamma_\nu\gamma_5 D_\rho \psi_\sigma
$$
where $\gamma_\nu = \gamma_a e^a_\mu$  and for 
the Levi-Civita antisymmetric tensor we have according to section (4.25) 
$\varepsilon^{0123}=-1$. For computational reasons the constants in
front of the Einstein-Cartan action is different from the one adopted
in Section (5.4). The only thing that matters is the ratio between the
Einstein-Cartan and Rarita-Schwinger terms.
$$
D_\rho = \partial_\rho+\omega_\rho
$$
with as in Section 5.8
$$
\omega_\rho = \Gamma^{ab}_\rho\Sigma_{ab}\equiv \omega^{ba}_\rho\Sigma_{ab}
$$
and
$$
\Sigma_{ab}= \frac{1}{8}(\gamma_a\gamma_b-\gamma_b\gamma_a)\equiv
\frac{1}{4}\gamma_{ab}
$$
and thus
$$
\omega_\rho = \frac{1}{4}\Gamma^{ab}_\rho\gamma_{ab}~.
$$
We already know that $S_{EC}$ is invariant under local Lorentz
rotations and diffeomorphisms. By construction $S_{RS}$ is invariant
under local Lorentz rotations. With regard to diffeomorphisms, due to
the presence of the antisymmetric symbol we have
$$
\varepsilon^{\alpha\beta\gamma\delta}\bar\psi'_\alpha\gamma'_\beta\gamma_5
D'_\gamma\psi'_\delta = \frac{\partial x^\mu}{\partial x'^{\alpha}}
\frac{\partial x^\nu}{\partial x'^{\beta}} \frac{\partial
x^\rho}{\partial x'^{\gamma}} \frac{\partial x^\sigma}{\partial
x'^{\delta}}\varepsilon^{\alpha\beta\gamma\delta}
\bar\psi_\mu\gamma_\nu\gamma_5 D_\rho\psi_\sigma= \det\left(\frac{\partial
x}{\partial x'}\right)  \varepsilon^{\mu\nu\rho\sigma}
\bar\psi_\mu\gamma_\nu\gamma_5 D_\rho\psi_\sigma
$$
which makes the Rarita-Schwinger action invariant.

The torsion equation is
$$
\delta_\Gamma S_{EC}+\delta_\Gamma S_{RS}=0
$$
$$
\delta_\Gamma S_{EC}=-\frac{1}{8}\int D \delta\Gamma^{ab}\wedge
e^c\wedge e^d \varepsilon_{abcd }= -\frac{1}{4}\int\delta \Gamma^{ab}\wedge
S^c\wedge e^d \varepsilon_{abcd}=
$$
$$  
=  
-\frac{1}{8}\int e~ d^4x ~\delta\Gamma^{ab}_r 
S^c_{mn} ~\varepsilon_{abcd} \varepsilon^{rmnd}  
=-\frac{1}{8}\int e~d^4x~\delta\Gamma^{ab}_r 
S^c_{ms} \delta_{abc}^{rms}  
$$
while
$$
\delta_\Gamma S_{RS} =
\frac{i}{8}\int ~e~ d^4 x~ \delta
\Gamma^{ab}_r\bar\psi_m\gamma_n\gamma_5\gamma_{ab}\psi_s
\varepsilon^{mnrs} 
$$
and thus
$$
S^c_{mn}\delta^{rmn}_{abc}=-i\bar\psi_m\gamma_5\gamma_n\gamma_{ab}\psi_s
\varepsilon^{mnrs}
=-\frac{i}{2}\bar\psi_m\gamma_5\{\gamma_n,\gamma_{ab}\}\psi_s 
\varepsilon^{mnrs}
$$
where the last expression is obtained by summing the second to the
transposed and dividing by $2$.

Using the identity
\begin{equation}\label{gammaidentity}
\gamma_5 \{\gamma_n,\gamma_{ab}\} = -2i~ \varepsilon_{nabc}~\gamma^c
\end{equation}

the above equation becomes
$$
S^c_{mn}\delta^{rmn}_{abc}= \bar\psi_m\gamma^c\psi_n 
\delta^{rmn}_{abc}
$$
trivially solved by
\begin{equation}\label{torsionconstraint}
S^c_{mn}\equiv D_m e^a_n-D_n e^a_m= \bar\psi_m\gamma^c\psi_n 
\end{equation}
or
$$
S^c = \bar\psi_\mu\gamma^c\psi_\nu \frac{dx^\mu\wedge dx^\nu}{2}.
$$

But we know the solution of the torsion in terms of the torsion source to
be unique and thus the found solution is the unique solution.

Recalling an old result of Section 5.9 we have
\begin{equation}\label{soltorsion}
\Gamma_{abc}=\Gamma[e]_{abc}+ \frac{1}{2}(S_{cab}+S_{acb}-S_{bca}).
\end{equation}

The so called 1.5 formalism consists in the following: Substitute the
expression (\ref{soltorsion}), where now $\Gamma$ is function of
the $e^a_\mu$ and of the field $\psi_\mu$, in the $R^{ab}$ of the
$L_{EC}$ i.e. 
$$
R^{ab}= d\Gamma^{ab}+ (\Gamma\wedge\Gamma)^{ab}
$$
and in the $RS$ action.
Thus the 1.5 formalism is really a {\it second order formalism with a
properly chosen Lagrangian}. We can forget now how we arrived to the
written Lagrangian. The real point is to show that such a Lagrangian
is invariant under 1) diffeomorphisms; 2) local Lorentz
transformations; 3) properly defined local supersymmetry transformations.

The invariance under 1) and 2) has already been proved. We come now to
the local supersymmetry transformations.

The transformations are given by
\begin{equation}\label{supertr1}
\delta_Q e^a_\mu =\bar\epsilon\gamma^a\psi_\mu
\end{equation}
and
\begin{equation}\label{supertr2}
\delta_Q \psi_\mu = D_\mu \epsilon\equiv (\partial_\mu
+\frac{1}{4}\gamma_{ab} \Gamma^{ab}_\mu)\epsilon
\end{equation}
where the last transformation is expected form the reasoning on the Noether
currents.

The great advantage of the 1.5 formalism is the following: In varying
the action (\ref{supergravity}) under (\ref{supertr1},\ref{supertr2})
we can ignore the 
variations of the 
connection induced by such variations because the connection solves the
stationarity equation.

The variation of $S_{EC}$ under a variation of $e^a_\mu$ is given by
\begin{eqnarray}
\delta S_{EC}&=&-\frac{1}{4}\int R^{ab}\wedge e^c\wedge \delta e^d 
\varepsilon_{abcd}=-\frac{1}{4}\int R^{ab}\wedge e^c\wedge e^f
\varepsilon_{abcd} ~\delta e^d_\mu~ e^\mu_f\\
&=&
\frac{1}{2} \int e G^a_b e^\mu_a\delta e^b_\mu d^4x=
\frac{1}{2} \int e G^a_b ~\bar\epsilon \gamma^b\psi_a d^4x
\end{eqnarray}
and as already discussed we have ignored the variation induced on
$\Gamma$.

With regard to the variation of the $RS$ action we have
$$
\delta_Q L_{RS} =\delta_Q
(\frac{i}{2}\varepsilon^{\mu\nu\rho\sigma}\bar\psi_\mu \gamma_a e^a_\nu
\gamma_5 D_\rho\psi_\sigma)  =
$$
$$
=\frac{i}{2}\varepsilon^{\mu\nu\rho\sigma}\delta_Q(\bar\psi_\mu) 
\gamma_a e^a_\nu
\gamma_5 D_\rho\psi_\sigma +
\frac{i}{2}\varepsilon^{\mu\nu\rho\sigma}\bar\psi_\mu \gamma_a
\delta_Q(e^a_\nu) 
\gamma_5 D_\rho\psi_\sigma+
\frac{i}{2}\varepsilon^{\mu\nu\rho\sigma}\bar\psi_\mu \gamma_a e^a_\nu
\gamma_5 D_\rho\delta_Q \psi_\sigma.
$$
Performing an integration by parts in the first term we obtain for the
previous expression
\begin{eqnarray}
&-&\frac{i}{2}\varepsilon^{\mu\nu\rho\sigma}\bar\epsilon\gamma_a (D_\mu
e^a_\nu)\gamma_5 D_\rho \psi_\sigma -
\frac{i}{2}\varepsilon^{\mu\nu\rho\sigma}\bar\epsilon\gamma_a  
e^a_\nu\gamma_5 D_\mu D_\rho \psi_\sigma \nonumber\\
&+&
\frac{i}{2}\varepsilon^{\mu\nu\rho\sigma}\bar\psi_\mu \gamma_a
\delta_Q(e^a_\nu) 
\gamma_5 D_\rho\psi_\sigma
+\frac{i}{2}\varepsilon^{\mu\nu\rho\sigma}\bar\psi_\mu \gamma_a e^a_\nu
\gamma_5 D_\rho\delta_Q \psi_\sigma~.
\end{eqnarray}
The first and the third term due to the torsion constraint
(\ref{torsionconstraint}) cancel i.e. we have
\begin{equation}\label{useoffierz}
-\varepsilon^{\mu\nu\rho\sigma}~(\bar\psi_\mu\gamma^a
\psi_\nu) ~(\bar\epsilon\gamma_a\gamma_5D_\rho\psi_\sigma)+
2 \varepsilon^{\mu\nu\rho\sigma}~(\bar\epsilon\gamma^a
\psi_\nu) ~(\bar\psi_\mu\gamma_a\gamma_5D_\rho\psi_\sigma)=0
\end{equation}
as can be proved by using the Fierz identity (\ref{fierz}) on the
second term of Eq.(\ref{useoffierz}) with $\bar\lambda\rightarrow \bar
\epsilon$, $\chi\rightarrow\gamma^a\psi_\nu$,
$\bar\psi\rightarrow\bar\psi_\mu$ and
$\phi\rightarrow\gamma_a\gamma_5D_\rho\psi_\sigma$ keeping into
account of the presence of the antisymmetric symbol 
$\varepsilon^{\mu\nu\rho\sigma}$ and of the properties (\ref{11exchanges}). 
This is the only point in the proof where the Fierz rearrangement
occurs.

Thus the variation of the $RS$ action,
exploiting the antisymmetry of the $\varepsilon$ symbol, and taking
into account the spinor nature of $\epsilon$ and on the $\psi_\sigma$,
is given by
\begin{eqnarray}
& &\delta_Q L_{RS}=\frac{i}{4}\varepsilon^{\mu\nu\rho\sigma}
\bar\psi_\mu\gamma_\nu\gamma_5
[D_\rho,D_\sigma]\epsilon- 
\frac{i}{4}\varepsilon^{\mu\nu\rho\sigma}\bar\epsilon \gamma_\nu\gamma_5
[D_\mu,D_\rho]\psi_\sigma \nonumber\\
&=&
\frac{i}{4}\varepsilon^{\mu\nu\rho\sigma}\bar\psi_\mu\gamma_\nu\gamma_5
\frac{1}{4}\gamma_{ab}\epsilon ~R^{ab}_{\rho\sigma}- 
\frac{i}{4}\varepsilon^{\mu\nu\rho\sigma}\bar\epsilon \gamma_\nu\gamma_5
\frac{1}{4}\gamma_{ab}\psi_\sigma ~R^{ab}_{\mu\rho}\nonumber\\
&=& 
\frac{i}{16}\varepsilon^{\mu\nu\rho\sigma}\bar\epsilon \gamma_5\gamma_\nu
\gamma_{ab}\psi_\sigma ~R^{ab}_{\mu\rho} +
\frac{i}{16}\varepsilon^{\mu\nu\rho\sigma}\bar\epsilon \gamma_5
\gamma_{ab}\gamma_\nu\psi_\sigma ~R^{ab}_{\mu\rho}=
\frac{i}{16}\varepsilon^{\mu\nu\rho\sigma}\bar\epsilon \gamma_5\{\gamma_\nu,
\gamma_{ab}\}\psi_\sigma ~R^{ab}_{\mu\rho}\nonumber\\
&=&\frac{ie}{16}\varepsilon^{mnrs}\bar\epsilon\gamma_5
\{\gamma_n,\gamma_{ab}\}\psi_s R^{ab}_{mr} ~.
\end{eqnarray}
Using again the identity (\ref{gammaidentity}) we have
$$
\delta L_{RS}= \frac{e}{8}\varepsilon^{mars}\bar\epsilon \gamma^b
 \psi_s~R^{cd}_{mr} \varepsilon_{acdb} = \frac{e}{8}
R^{cd}_{mr}\delta^{mrs}_{cdb} ~\bar\epsilon \gamma^b \psi_s =
-\frac{e}{2}G^s_b ~\bar\epsilon \gamma^b \psi_s
$$
which cancels the variation of $S_{EC}$.

In conclusion we have proven the invariance of $S_{SG}$ also under
local supersymmetry transformations.

\bigskip

References

\smallskip

[1] P. van Nieuwenhuizen, ``Supergravity'', Phys.Rep.68 (1981) 189

\smallskip
                                             
[2] Y. Tanii, ``Introduction to supergravity in diverse dimension''
hep-th/9802138

\vfill


\chapter{Appendix}

{\bf \Large Derivation of the hamiltonian equations of motion and of the Poisson
algebra of the constraints}

\bigskip

We give here the details of the derivation of the equations of motion
and of the Poisson algebra of the constraints in the hamiltonian
formulation of general relativity. We saw in Section
\ref{7canonicalactionSec} that the action in
hamiltonian form is
\begin{eqnarray}\label{A1action}
S_H &=& \int dt \int_{\Sigma_t} d^D x (\pi^{ij}\dot h_{ij} - NH - N^i
H_i) + 2\int 
dt\int_{B_t}\sqrt{\sigma}~d^{D-1}x~(N k - r_i
\frac{\pi^{ij}}{\sqrt{h}} N_j)\nonumber\\ 
&=&\int dt~L
\end{eqnarray}
with
\begin{equation}
H=\sqrt{h}~\left(-{\cal R}+\frac{{\rm
Tr}(\pi\pi)}{h}-\frac{\pi^2}{(D-1)h}\right),    
~~~~H_i= -2\sqrt{h}~D_j\left(\frac{\pi^j_{~i}}{\sqrt{h}}\right)\nonumber\\
\end{equation}
$r_i$ being the outward pointing unit normal to $B_t$ as a 
sub-manifold of $\Sigma_t$
and $k$ the extrinsic curvature of intersection of $\Sigma_t$
with $B$ considered as a sub-manifold of $\Sigma_t$. The equations of
motion are obtained by setting to zero the variations w.r.t. the
independent variables which are $N,~N^i,~\pi^{ij},~h_{ij}$. We recall
that $\pi^i_j$ has to be understood as $\pi^{il}h_{lj}$ and thus it
depends on $h_{ij}$ and similarly for $\pi_{ij}$ and that $\pi^{ij}$
is not a tensor in $D$ dimensions but a tensorial density.

We recall that in the hamiltonian dynamics the variation with respect
to the ``coordinates'' $h_{ij}, N, N_i$ has to be performed with vanishing
variation at the boundary, while non such restriction can be imposed
on the variation of the conjugate momenta $\pi^{ij}$.

The variations w.r.t. $N$ and $N^i$ give rise to the constraints
$$
H=0,~~~~H_i=0
$$
To compute the variation w.r.t. $\pi^{ij}$ and $h_{ij}$ it is better
to revert to the form
\begin{equation}\label{Aoriginalaction}
S_H = \int dt \int_{\Sigma_t} d^D x (\pi^{ij}\dot h_{ij} - NH
-2\pi^{ij}D_i N_j) + 2\int 
dt\int_{B_t}\sqrt{\sigma}~d^{D-1}x~N k
\end{equation}
The variation w.r.t. $\pi^{ij}$ gives
\begin{equation}\label{deltaLoverdeltapi}
0=\frac{\delta L}{\delta \pi^{ij}}=
\dot h_{ij}-\frac{2N}{\sqrt{h}}(\pi_{ij}-\frac{\pi h_{ij}}{D-1})-
D_i N_j-D_j N_i~.
\end{equation}
Equation (\ref{deltaLoverdeltapi}) is just the inversion of 
Eq.(\ref{7conjugatemom}).

We come now to the variation w.r.t. $h_{ij}$.
Repeating in our
$D$-dimensional case the procedure applied in Section 5.2 for the
$n$-dimensional case, we have
\begin{equation}\label{firstcontribution}
\delta(N \sqrt{h} {\cal R}) =-\delta h_{ij}N\sqrt{h}({\cal
  R}^{ij}-\frac{h^{ij}}{2}{\cal R})+N\sqrt{h} D_i v^i
\end{equation}
with
\begin{equation}
v^i = \delta \Gamma^{i}_{kl} h^{kl}-\delta \Gamma^{l}_{kl} h^{ki}\nonumber
\end{equation}
where the connections appearing here are the connections on $\Sigma$
i.e. computed in terms of the metric $h_{ij}$.

The change in the connection induced by the change of the metric can
be computed by the following general procedure.

Let $D_l$ be the covariant derivative with the metric $h_{ij}$ and 
$D_l+\Delta\Gamma_l$ the covariant derivative with the metric $\bar h_{ij}=h_{ij}+
\Delta h_{ij}$. Metric compatibility gives
$$
0 = (D_l+\Delta\Gamma_l) \bar h_{ij} = D_l \Delta h_{ij}- 
\Delta\Gamma^k_{il} \bar h_{kj}  -\Delta\Gamma^k_{jl} \bar
h_{ik}\equiv
D_l \Delta h_{ij}- \Delta\Gamma_{jil} -\Delta\Gamma_{ijl}~.  
$$
Then exploiting zero torsion i.e. the symmetry
$\Delta\Gamma_{ijl}=\Delta\Gamma_{ilj}$ we have the exact formula
$$
\Delta\Gamma^{l}_{ij} = 
\frac{1}{2}\bar h^{lk}(D_j \Delta h_{ik}+D_i \Delta h_{kj} -D_k \Delta
h_{ij})
$$
and for infinitesimal $\Delta h_{ij}$ 
\begin{equation}\label{A1deltaGamma}
\delta\Gamma^{l}_{ij} = 
\frac{1}{2} h^{lk}(D_j \delta h_{ik}+D_i \delta h_{kj} -D_k \delta
h_{ij})
\end{equation}
giving
$$
v^i = D_k \delta h^{ik}-h^{ik} h^{mn}D_k \delta h_{mn}~.
$$

Substituting we have
\begin{equation}
\int N\sqrt{h}D_i v^i d^Dx= \int\partial_i(N\sqrt{h}~ v^i) dx^D-
\int \sqrt{h}(D_k \delta h^{ik}-h^{ik} h^{mn}D_k \delta h_{mn})D_i N d^Dx~.
\nonumber
\end{equation}
The first contribution is canceled by the variation of the boundary
term in (\ref{Aoriginalaction}) as it happens in the $n$-dimensional
case of Section \ref{6TrKaction}, while the second, integrated by
parts, gives
\begin{equation}\label{firstsubstitution}
\int \sqrt{h} ~\delta h_{ij} (D^i D^j N- h^{ij} D^lD_l N) d^Dx~.
\end{equation}

We come now to the term 
\begin{equation}
-\frac{N}{\sqrt{h}}\left(\pi^{ij}h_{il}h_{jm}\pi^{lm}-\frac{(\pi^{ij}h_{ij})^2}{D-1}
\right)~.\nonumber
\end{equation}
The dependence on $h_{ij}$ here is algebraic (no derivatives are
present) and the variation is easily computed to be
\begin{equation}\label{secondcontribution}
\delta h_{ij}\left(\frac{N h^{ij}}{2\sqrt{h}}(\pi^{kl}\pi_{kl}-\frac{\pi^2}{D-1})
-\frac{2N}{\sqrt{h}}(\pi^{ik}\pi^j_k-\frac{\pi \pi^{ij}}{D-1})\right)~.
\end{equation}
With regard to the term
\begin{equation}
-2\pi^{ij}D_i N_j \nonumber
\end{equation}
using Eq.(\ref{A1deltaGamma}) for the variation of $D_i$ and using
the diffeomorphism constraint $H_i=0$ we have
\begin{equation}\label{thirdcontribution}
\delta h_{ij}\left(\sqrt{h}D_l\left(\frac{N^l\pi^{ij}}{\sqrt{h}}\right)-
\pi^{lj}D_lN^i-\pi^{li}D_lN^j\right)~.
\end{equation}
Summing (\ref{firstcontribution}) after using
(\ref{firstsubstitution}), to (\ref{secondcontribution}) and 
(\ref{thirdcontribution}) we obtain Eq.(\ref{7pidotequation}) of
Section 7.4.

\bigskip\bigskip

We come now to the algebra of constraints.
\begin{equation}
\{A,B\}= \int d^Dy\left(\frac{\delta A}{\delta h_{ij}(y)}
\frac{\delta B}{\delta \pi^{ij}(y)}-
\frac{\delta A}{\delta \pi^{ij}(y)}
\frac{\delta B}{\delta h_{ij}(y)}\right)~.
\nonumber
\end{equation}
It is simpler to work with the constraints $H_i$ and $H$ weighted by
test functions ( $C_0^\infty$ functions of $x^1\dots x^D$) which
we shall denote by $N^i(x),N(x)$ and ${N'}^i(x),N'(x)$ like e.g.
\begin{equation}
\{\int N^i(x) H_i(x)d^Dx,\int N^l(x')H_l(x') d^Dx'\}~.
\nonumber
\end{equation}

We begin computing
\begin{eqnarray}
& &\{h_{ij}(x),\int N^l(y)H_l(y) d^Dy\}= 
-2\frac{\delta}{\delta \pi^{ij}(x)}\int d^Dy ~\sqrt{h}~ N^l~
D_k\left(\frac{\pi^k_l}{\sqrt{h}}\right)\nonumber\\
&=&
2\frac{\delta}{\delta \pi^{ij}(x)}\int d^Dy ~\pi^{ml}(y) h_{lc}(y) D_m
N^c= h_{il}(x)D_j N^l+h_{jl}(x)D_i N^l = {\cal L}_{{\bf N}} h_{ij}(x)~.
\nonumber
\end{eqnarray}
In addition
\begin{eqnarray}
& &\{\pi^{ij}(x),\int N^l(y)H_l(y) d^Dy\}= 
2\frac{\delta}{\delta h_{ij}(x)}\int d^Dy N^l\sqrt{h}
D_k\bigg(\frac{\pi^k_l}{\sqrt{h}}\bigg)\nonumber\\
&=&
-2\frac{\delta}{\delta h_{ij}(x)}\int d^Dy ~h_{nl}\pi^{mn}D_m N^l~.
\nonumber
\end{eqnarray}
Under a variation $\delta h_{ij}$ of $h_{ij}$ we have using the
identity (\ref{A1deltaGamma})
\begin{equation}
\delta (-2\int d^Dy h_{nl}\pi^{mn}D_m N^l)=
-2\int d^Dy (\delta h_{nl}\pi^{mn}D_m N^l+
\frac{1}{2}\pi^{ms}N^l D_l\delta h_{sm})
\nonumber
\end{equation}
which after an integration by parts gives
\begin{eqnarray}
& &\int d^Dy \delta h_{mn}
\sqrt{h}(N^lD_l\frac{\pi^{mn}}{\sqrt{h}}-\frac{\pi^{ln}}{\sqrt{h}}
D_l N^m-\frac{\pi^{lm}}{\sqrt{h}}D_l N^n
+\frac{\pi^{mn}}{\sqrt{h}} D_l N^l)\nonumber\\
&=& \int d^Dy \delta h_{mn}{\cal L}_{{\bf N}}\pi^{mn}~.\nonumber
\end{eqnarray}
The reason for the last equality is that $\pi^{ij}$ is not a tensor
in $D$ dimensions but a tensorial density. Thus using
\begin{equation}
{\cal L}_{\bf N}\sqrt{h}=\sqrt{h}D_lN^l
\end{equation}
we have
\begin{equation}
{\cal L}_{{\bf N}}\pi^{ij} = \sqrt{h}
{\cal L}_{{\bf N}}\frac{\pi^{ij}}{\sqrt{h}} +
\frac{\pi^{ij}}{\sqrt{h}}{\cal L}_{{\bf N}} \sqrt{h} =
\sqrt{h}
{\cal L}_{{\bf N}}\frac{\pi^{ij}}{\sqrt{h}} +\pi^{ij}D_lN^l~.
\nonumber
\end{equation}
Thus we found that $\int d^Dy N^l(y) H_l(y)$ generates the
space diffeomorphisms on $h_{ij}$ and on $\pi^{ij}$ 
and thus on any functional of $h_{ij}$ and
$\pi^{ij}$, in particular on the constraints $H_i$ and $H$.
For the $H_i$ we have
\begin{equation}
{\cal L}_{{\bf N}} H_i = \sqrt{h}{\cal L}_{{\bf N}}\frac{H_i}{\sqrt{h}}
+ H_i D_lN^l=
N^l \partial_l H_i+H_l \partial_i N^l+H_i \partial_l N^l ~.
\nonumber
\end{equation}
Summarizing we found
\begin{equation}
\{H_i(x),\int d^Dy ~N^l H_l\}=
N^l \partial_l H_i+H_l \partial_i N^l+H_i \partial_l N^l
\nonumber
\end{equation}
which can be rewritten as
\begin{equation}
\{H_i(x),H_j(x')\}=
H_i(x')\partial_j\delta(x,x')-H_j(x)\partial'_i\delta(x,x')~.
\nonumber
\end{equation}
With regard to 
\begin{equation}
\{H(x),\int d^Dy ~N^l H_l\}= {\cal L}_{{\bf N}}H(x)
\nonumber
\end{equation}
we notice that $H$ is a scalar density and thus
\begin{equation}
{\cal L}_{{\bf N}}H(x) =
\sqrt{h}N^l\partial_l\frac{H}{\sqrt{h}}+\frac{H}{\sqrt{h}}
{\cal L}_{{\bf N}}\sqrt{h}= N^l\partial_l H+H\partial_lN^l 
\nonumber
\end{equation}
from which we have
\begin{equation}
\{ H(x), H_i(x')\} = -H(x')\partial'_i\delta(x',x)~.
\nonumber
\end{equation}
The result is that the Poisson brackets with $\int d^Dy ~N^l H_l$ is
a canonical representation of the algebra of the diffeomorphisms in
$D$ dimensions.

We come now to the P.B.
\begin{equation}
\{ \int d^Dx N(x)H(x), \int d^Dx' N'(x') H(x')\}~.
\nonumber
\end{equation}
We note that $h_{ij}$ intervenes in $H$ both in algebraic way and also
in non algebraic way i.e. through derivatives. On the other hand
$\pi^{ij}$ intervenes in $H$ only in algebraic way. Moreover in the
computation of the P.B. the algebraic-algebraic variation contribute
always zero because for
\begin{equation}
\frac{\delta H(x)}{\delta h_{ij}(y)} = f_{h_{ij}}(x) \delta(x,y),~~~~~~~~
\frac{\delta H(x)}{\delta \pi^{ij}(y)} = f_{\pi^{ij}}(x) \delta(x,y)
\nonumber
\end{equation}
we have
\begin{eqnarray}
& &\int d^Dy( f_{h_{ij}}(x) \delta(x,y)f_{\pi^{ij}}(x') \delta(x',y) N(x)
N'(x')\nonumber\\
&-&f_{\pi^{ij}}(x)\delta(x,y)f_{h_{ij}}(x')\delta(x',y)
N(x) N'(x'))d^Dx~d^Dx' =0~.
\nonumber
\end{eqnarray}
Thus only the non algebraic contribution of the variation of $h_{ij}$
combined with the (algebraic) contributions of the variation of
$\pi^{ij}$ contribute. The non algebraic contribution of the
variation of $h_{ij}$ is given by Eq.(\ref{firstsubstitution})
\begin{equation}
-\int d^Dx \sqrt{h}~\delta h_{ij}(D^iD^j N-
 h^{ij}D^lD_l N)
\end{equation}
and thus we have
\begin{eqnarray}
& &\int d^Dy \frac{\delta\int d^Dx N H}{\delta h_{ij}(y)}~~
\frac{\delta\int d^Dx' N' H}{\delta \pi^{ij}(y)}\nonumber\\
&=&-\int d^Dy~\sqrt{h}~(D^iD^j N-h^{ij}D^lD_lN)~
2\left(\frac{\pi_{ij}}{\sqrt{h}}-\frac{\pi
  ~h_{ij}}{(D-1)\sqrt{h})}\right) N'
\nonumber\\
&=&2\int d^Dy\sqrt{h}D_iN~D_j\left(\frac{\pi^{ij}}{\sqrt{h}}N'\right)~.
\nonumber
\end{eqnarray}
Summing the second part of the P.B. (i.e. subtract exchanging $N$ with
$N'$) and integrating by parts we have
\begin{eqnarray}
& &\{ \int d^Dx N(x)H(x) \int d^Dx' N'(x') H(x')\}\nonumber\\
&=&2\int d^Dy~\sqrt{h}~(N' D_i N- N D_iN')
  D_j\left(\frac{\pi^{ij}}{\sqrt{h}}\right)=
\int d^Dy~(-N' D_i N+ N D_iN') H^i\nonumber
\end{eqnarray}
equivalent to
\begin{equation}
\{H(x),H(x')\}= H^i(x)\partial_i\delta(x,x')-
H^i(x')\partial'_i\delta(x,x')~.
\nonumber
\end{equation}


\end{document}